\documentclass[11pt,openright,twoside]{report}

\pdfoutput=1
\usepackage{epsfig}
 \usepackage{epstopdf} 
\usepackage{graphicx}
\usepackage{dcolumn}
\usepackage{bm}
\usepackage{longtable}

\usepackage{amssymb}
\usepackage{amsfonts}
\usepackage[margin=1.0in]{geometry}
\usepackage{amsmath}
\usepackage{multirow}

\usepackage[linktocpage=true]{hyperref}

\usepackage{xspace}

\usepackage[titles]{tocloft}

\usepackage{slashed}

\usepackage{hyperref}

\usepackage{tgtermes}

\allowdisplaybreaks[1]

\usepackage{fancyhdr}
\fancypagestyle{plain}{%
\fancyhead{} 
}

\def\dd{\textrm{d}}  				     
\def\wa{\wedge\ast}				     

\newcommand{\be}{\begin{equation}}
\newcommand{\ee}{\end{equation}}
\newcommand{\one}{{\rm 1\kern -.9mm l}} 

\newcommand{\ft}[2]{{\textstyle\frac{#1}{#2}}}

\def\a{\alpha}
\def\b{\beta}
\def\g{\gamma}

\def\d{\delta}

\def\e{\epsilon}

\def\p{\partial}

\def\k{\kappa}
\def\l{\lambda}
\def\L{\Lambda}
\def\m{\mu}
\def\n{\nu}
\def\r{\rho}
\def\s{\sigma}
\def\t{\theta}
\def\om{\omega}
\def\w{\omega}

\def\CL{{\cal L}}

\def\zetab{\boldsymbol{\zeta}}

\def\bN{\,\mathbf{N}}
\def\bM{\,\mathbf{M}}
\def\bK{\,\mathbf{K}}
\def\bV{\,\mathbf{V}}
\def\bB{\mathbf{B}}
\def\bC{\mathbf{C}}
\def\bF{\mathbf{F}}
\def\bH{\mathbf{H}}

\def\cD{ {\cal D} }

\def\dd{\textrm{d}}
\def\wa{\wedge\ast}

\def\half{{1\over 2}}
\def\ra{\rightarrow}

\def\ac{+\nonumber\\&}
\def\acI{+\right.\nonumber\\& \left.}
\def\acII{+\right.\right.\nonumber\\& \left.\left.}
\def\acIII{+\right.\right.\right.\nonumber\\& \left.\left.\left.}

\def\accI{\right.\nonumber\\& \left.}

\usepackage{tikz}
\usetikzlibrary{arrows,shapes}

\tikzstyle{format} = [draw, thin, shading=radial, outer color=blue!20, inner color=white]
\tikzstyle{medium} = [ellipse, draw, thin, shading=radial, outer color=green!20, inner color=white, minimum height=2.5em]
\tikzstyle{palla}=[ball color=red,circle,text=white]

\begin{document}



\newcommand{\myTitle}{Fermionic Variations on AdS/CFT Themes\xspace}

\newcommand{\mySubtitle}{Subtitle or whatever \xspace}
\newcommand{\myName}{{\bfseries Andrea Mezzalira}\xspace}
\newcommand{\myProf}{Prof.~Pietro Antonio Grassi\xspace}
\newcommand{\rel}{{\bf Relatore} \xspace}
\newcommand{\corel}{{\bf Controrelatore} \xspace}
\newcommand{\myaldo}{{Prof.~Giuseppe Policastro} \xspace}
\newcommand{\com}{{\bf Commissione}\xspace}
\newcommand{\comdue}{Prof.~Giuseppe Policastro\xspace}
\newcommand{\comtre}{Prof.~Marco Bill\`o\xspace}
\newcommand{\comuno}{Prof.~Carlos Nu\~nez\xspace}

\newcommand{\mySupervisora}{Prof.~Giuseppe Policastro\xspace}

\newcommand{\contrel}{{\bf Controrelatore}\xspace}

\newcommand{\myFaculty}{Facolt\`a di Scienze Matematiche, Fisiche e Naturali \xspace}
\newcommand{\myDepartment}{Dipartimento di Fisica Teorica \xspace}
\newcommand{\myUni}{\protect{\bf Universit\`a degli Studi di Torino}\xspace}

\newcommand{\myTime}{March $26^{\textrm{th}}$, $2013$}

\begin{titlepage}
\pagestyle{empty}

\begin{center}
\begin{large}
{\bf Scuola di Dottorato in Scienza ed Alta Tecnologia} \\
{Indirizzo in Fisica ed Astrofisica} \\
\end{large}

\hrulefill
        \large  

        \hfill

        \vfill

        \begingroup
	\huge{ {\scshape Fermionic Variations on A}d{\scshape S/CFT Themes}}\\ \bigskip
        \endgroup
\bigskip
        {\scshape \myName}

        \vfill

        \includegraphics[width=4cm]{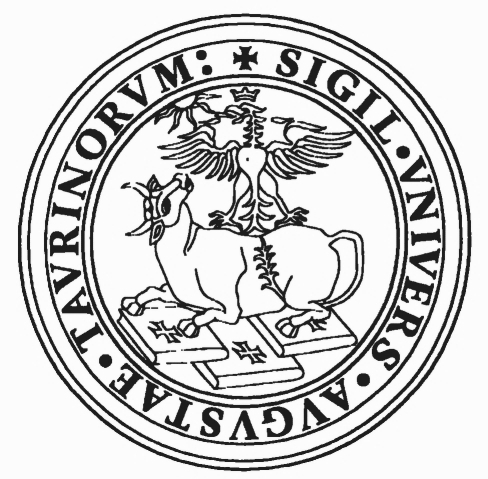} 
        \bigskip \\
	{\bfseries Universit\`a degli Studi di Torino} 
\bigskip

        \vfill

\bigskip
\bigskip

\noindent{\rel \hfill \com}\\
\noindent{\myProf \hfill \comtre}\\
\noindent{\corel \hfill \comuno}\\
\noindent{\myaldo \hfill \comdue}\\
\bigskip
	\vfill
\bigskip
	\myDepartment \\                            
        \myFaculty \\
        \myUni 
\vfill

        \myTime

    \end{center}  


\newpage
\hfill
\thispagestyle{empty}


\newpage
\hfill
\thispagestyle{empty}
\vspace{25mm}

\begin{flushright}
\textit{to memories}
\end{flushright}


\newpage
\hfill
\thispagestyle{empty}


\end{titlepage}

\pagenumbering{roman}

\tableofcontents

\chapter{Introduction}


\pagenumbering{arabic}
\setcounter{page}{1}
\setcounter{footnote}{0}

\section{Prologue}

The present Thesis aims to explore the effects caused by the presence of Grassmann numbers in two different frameworks, purely fermionic $\sigma$--models and fluid/gravity correspondence; they are both derived from  one of the most important achievements in theoretical physics of the last few decades, the $AdS/CFT$ correspondence.

The first analysis we present deals with fermionic $\sigma$--models constructed by means of purely fermionic coset spaces,
namely we consider models with anticommuting target--space fields only. 
We notice that the Grassmanian behaviour of the fields implies several simplifications when we derive the action and the radiative corrections, however it provides obstructions to some supercoordinates transformations.

In the second part of this Thesis we deal with a particular limit of $AdS/CFT$ correspondence.
It is named fluid/gravity correspondence for it permits to describe the hydrodynamics of a strongly--coupled fluids from an analysis of black hole solutions in pure gravity.
We extend the correspondence by considering supergravity theories in the bulk and computing the black hole superpartners.
More in detail, we dress a pure bosonic solution with contributions obtained from acting with a particular supersymmetry transformation on the black hole.
The resulting metric is a black hole solution of the supergravity equations of motion and its associated charges and entropy depend upon Grassmann numbers.
Moreover by fluid/gravity techniques we find that also the fluid dynamics parameters, such as the temperature and the velocity, are in general affected by Grassmann numbers.

Since the modified quantities are bosonic, Grassmann numbers appear in the corrections in terms of bilinears.
The interpretation of the fermionic bilinears contribution is delicate. 
To evaluate them, in this work we adopt the approach used in computing --for instance--  fermionic condensates, namely we consider their vacuum expectation value.


\section{T--duality}

T--duality is a transformation of non--linear sigma models which connects different geometries of the target space leaving the partition function invariant \cite{Giveon:1994fu}.
As a first example, we consider a bosonic string which propagates in a spacetime with
one dimension compactified on a circle of radius $R$.
The spectrum of the model is then invariant under the target space transformation
which maps the radius of compactification $R$ to its inverse $\frac{1}{R}$.
In addition to this simple case, T--duality plays a key r\^ole in string theory by proving the equivalence between different superstring models.

It was shown \cite{Berkovits:2008ic} that the $AdS_5 \times S^5$ background of superstring theory is invariant under the 
combined action of bosonic T--duality and a similar transformation applied to a part of the fermionic 
superstring coordinates, which is then called fermionic T--duality. 
This invariance turns out to be the reason causing the emergence of the dual superconformal symmetry of scattering amplitudes in $\mathcal{N}=4$ super Yang--Mills.

In \cite{Berkovits:2007zk}, the author pointed out that there exists an interesting limit where the $AdS_5 \times S^5$ pure spinor string model shows a decoupling between the fermionic and bosonic coordinates. 
In
particular, in that limit, the model appears to be purely fermionic and it can be viewed as a coset
sigma model (obtained by a gauged linear sigma model) based on a fermionic coset. The latter
is naturally obtained by dividing with respect to the complete bosonic subgroup.

Motivated by these results, we study the interplay between the fermionic T--duality and the radiative corrections to the sigma models. 
Our work is a preliminary account on these problems and we clarify some issues both at the 
theoretical level (determination of the T-dual models) and at the computational level (radiative corrections at higher loops). 

We first review the T--duality for generic $\sigma$-models \cite{Giveon:1994fu,Buscher:1987qj} and we 
extend it to supersymmetric models with target--space spinors. This is the way to embrace 
the Green-Schwarz formalism for string theory, p-branes and the pure spinor string theory. 
The presence of target space spinors allows us to consider generic 
super-isometries which encompass the fermionic
T-duality \cite{Berkovits:2008ic,Beisert:2008iq}.\footnote{
See also the new developments in \cite{Adam:2009kt,Hao:2009hw,Bakhmatov:2009be,ChangYoung:2011rs,Dekel:2011qw,Sorokin:2010wn,Adam:2010hh,Bakhmatov:2010fp,Sfetsos:2010xa}.}

The first issues we encounter are the obstructions for constructing the T-dual models in the case 
of superisometries. Indeed, the Grassmannian nature of the spinorial variables implies that 
some coordinate changes appear to be either trivial (redefinition by an overall constants) or 
impossible. That prevents us from finding the holonomy bases where the super-isometry appears as 
a shift in the fermionic coordinates. In addition, the gauging procedure which is a well-paved 
way to perform the T-duality for $\sigma$-models gets obstructed by non-invertible fermionic matrices. 
We review those problems and in particular we adopt  a recently discussed 
simple model \cite{Berkovits:2007zk} as a playground. 

We consider generalizations and simplifications of the example in \cite{Berkovits:2007zk} and in particular we 
take into account models based on the $OSp(n|m)$ supergroup \cite{Fre:2009ki}. 
In those coset models the super-isometries are non-abelian (they close on the bosonic 
subgroup) and they are realized non-linearly \cite{CastellaniDAuriaFre}. For these two reasons, 
the usual gauging procedure cannot be performed.  These issues are discussed for 
generic $OSp(n|m)$ models and, in particular, we study the simplest one, namely $OSp(1|2)$, 
which already exhibits such characteristics. 

We first convert the non-linear symmetries into linear ones by introducing additional bosonic coordinates to the $\sigma$-model. 
In the simplest case, namely $OSp(1|2)$, this can be easily done by adding a single bosonic 
coordinate constrained by a quadratic algebraic equation. That equation is invariant under the action 
of the isometries which are linearly represented. In this way, the $\sigma$-model can be 
easily written and the gauging procedure can be employed.  Since the superisometries are non-abelian, 
we use the construction provided by \cite{de la Ossa:1992vc,Alvarez:1994zr} and we introduce the gauge fields for all isometries.  Then, a two-step process leads us to a dual model which contains the 
dual fermionic coordinates, a dual bosonic field (which appears to be dynamical in the dual model) 
and a ghost field associated to the gauge fixing of the local isometries. Therefore, we have bypassed 
the obstructions encountered in the theoretical analysis of T-dualities and we provided a dual 
lagrangian. 

For a generic model, this procedure can be also applied with technical difficulties. The first one is 
to discover the correct set of bosonic coordinates to implement the superisometries as linear 
representations. The second step is finding suitable algebraic constraints (a similar procedure as the construction of the Pl\"ucker relations in projective geometry).  Finally, a conventional 
gauging procedure can be applied and the gauge fields integrated. However, 
we notice that the gauge fixing procedure suggested in \cite{de la Ossa:1992vc}  leads 
to a cumbersome action which turns out not to be very useful for loop computations. On the 
other side, a clever gauge choice yields remarkable simplifications and a good starting point 
for loop computations. 

At the quantum level we compute the one--loop and two--loop corrections 
to the action and we check the conformal invariance up to that order. This is a preliminary 
account on the problem of conformal invariance of $\sigma$-models based on orthosymplectic 
groups $OSp(n|m)$. Indeed, even though there is a fairly amount of literature on the 
$PSU$-type of supergroups and the conformal invariance of their $\sigma$-models, 
there is no proof based on the orthosymplectic ones.  A related problem is checking the T-duality at the quantum 
level as pointed out in a series of papers \cite{Balog:1996im,balog2,balog3,balog4}, but at the moment 
no check for fermionic T-duality has been done. 

In \cite{Sethi:1994ch}, for the first time, the analysis of  $\sigma$-models on supermanifolds has been performed. It has been observed that 
for some supermanifolds viewed as supergroup manifolds, the vanishing of the dual Coxeter number (or the quadratic Casimir in a given representation) might lead to a conformal invariant theory. In  \cite{Bershadsky:1999hk}, the renormalization of Principal Chiral Model on $PSL(n|n)$ is studied. 
They assert the conformal invariance of the model by looking at one-loop and by using symmetry arguments 
(based on Background Field Method BFM) for higher loops. They also discuss the presence of WZ terms and 
how does the conformal invariance depend upon it. In paper \cite{Berkovits:1999zq}, the $\sigma$-model based on $AdS_2 \times S^2$ is discussed using the hybrid formalism for superstrings. 
It has been shown by explicit computation that the one--loop beta function vanishes because of the vanishing of the dual Coxeter number. 
A discussion about the vanishing of the beta function depending on the structure of the coset is given. In the same paper, for the first 
time the WZ term is written as a quadratic term in the worldsheet currents. In paper \cite{Berkovits:1999im}, the proof of the conformal invariance to all orders in the case of $AdS_3 \times S^3 \times CY_2$ is provided. 
It is discussed how the proof can be implemented to all orders. They refer to the situation of the supergroups $U(n|n)$. 
In paper \cite{Read:2001pz}, the models based on supersphere $OSp(2n + m|2n )/OSp(2n + m-1|2n) \sim S^{2n+m-1|2n +m}$ and the superprojective spaces $U(n+m|n)/U(1)\times U(n+m-1|n)\sim {\mathbb CP}^{n+m-1|n}$ are considered.  In particular, they claim: ``In most cases (in a suitable range for $m$, and for $n$ sufficiently large), the beta function for the
coupling in the nonlinear $\sigma$-model is non--zero, and
there is a single non--trivial renormalization-group (RG) fixed-point theory for each model.'' Other 
discussions can be found in \cite{Kagan:2005wt,Wolf:2009ep}. 

Recently, the regained interested into $AdS_4 \times X$ models 
\cite{Nilsson:1984bj,Aharony:2008ug,Arutyunov:2008if,Stefanski:2008ik,Fre:2008qc} brought 
the attention on the conformal invariance of those models. Here, we extend the computation 
in a specific limit of fermionic coset models and we found that up to two loops the condition on the 
target manifold for being a super-Calabi-Yau seems to be sufficient for the conformal invariance. In addition, we explored the dual models and we discover that they have the 
same type of unique interaction terms leading to the same loop computations. 

From the technical point of view, we adopt two methods for computation: 1) we expand the action around a trivial vacuum  and we perform the computation at one--loop, 2) we use the Background Field Method to expand the action around a non-trivial background and we compute the corrections at two-loops. A complete all-loop proof is still missing.

%
%
%
%

\section{Fluid Dynamics from Gravity}

Gauge/gravity correspondence is a conjectured equivalence between theories of string/gravity on a curved spacetime and quantum field theories on the boundary of such space.
Although it relates two models defined in different dimensions  
remarkably the information encoded in a $d+1$ dimensional theory can be described entirely by analysing a system in a lower number of dimensions, an idea which takes the name of holographic principle.

Gauge/gravity correspondence is often called $AdS/CFT$ correspondence because at first it was formulated as a duality between superstring theory in a spacetime with negative curvature (Anti de-Sitter space) and a supersymmetric conformal field theory.
Further studies showed evidence of the duality also for many different models, even among conformal field theories without supersymmetry.
This raised the interest of a part of the physics community, inspiring a number of different applications in different physical systems.
Among them, important results were obtained in particle physics and condensed matter physics.


A key feature of the conjecture is that it maps the strongly--coupled regime of one theory with the weakly--coupled regime of the dual model.
For this reason, besides the theoretical aspects, gauge/gravity correspondence revealed to be an important tool to understand the dynamics of strongly--coupled quantum field theories.

In this context, one of the most important application of the gauge/gravity conjecture is in the heavy-ion physics.
The experiments at the Large Hadron Collider (LHC) and at the Relativistic Heavy Ion Collider (RHIC) produce a plasma of quarks and gluon (quark--gluon plasma or QGP) by heavy--ion collisions.
After the collisions, the components of the QGP rapidly come into local thermal equilibrium and then their behaviour can be studied by a hydrodynamic model.
In particular, the QGP evolution is characterized by a set of transport coefficients, the most relevant of them being the so--called shear viscosity.
For weakly--coupled theories, the transport coefficients can be computed by standard perturbative calculation.
However, the temperature of the QGP is estimated to be approximately 170 MeV, which is near the confinement scale of QCD.
Hence, QGP is deep inside the non--perturbative regime of quantum chromodynamics (QCD) and then perturbative techniques can not be applied.
Moreover, the standard numerical approach to strong interacting QCD (computations on Lattice), which gives a precise analysis of thermodynamical quantities, is not well suited for computing transport coefficients.

On the other hand, gauge/gravity correspondence offers a theoretical framework to investigate strongly interacting systems by considering the simpler dual weakly--coupled (super)gravity models.
To compute the transport coefficients in the near--equilibrium regime of quantum field theory, in the dual side of the conjecture one considers deformations of black holes in anti de Sitter space.

It should be mentioned that the quantum field theories studied by the conjecture are quite different from QCD in their vacuum.
However, it is suspected that at finite temperature the different models would show several analogies. 
In addition, the study of such theories may increase our understanding of the universal properties of strongly coupled dynamics.


One of the first computations in this field is performed by G. Policastro, D.T. Son and A.O. Starinets in \cite{Policastro:2001yc}, where they derive the shear viscosity for a finite temperature $\mathcal{N}=4$ supersymmetric Yang--Mills theory through the computation of the absorption cross section of gravitons by black--three branes. 
Further works deeply investigate the subject, leading to the computations of other transport coefficents and to other different analysis such as the phase transition of Fermi liquids or sound waves. 
One fundamental result is that the ratio between shear viscosity and density entropy current $\frac{\eta}{s}$  turns out to be the same for all theories which can be constructed from gauge/gravity correspondence. 
Remarkably, the value of that ratio for QGP obtained at RHIC approximates this theoretical value. 

These results led to the idea that fluid dynamics is the correct long wavelength effective description of the strongly--coupled field theory dynamics. 
Therefore a direct map between black hole solutions in asymptotically $AdS$ spacetimes and fluids in strongly interacting field theories on the boundary has been conjectured. This is named fluid/gravity correspondence \cite{Bhattacharyya:2008jc,Rangamani:2009xk,Hubeny:2011hd}. 

In one of the simplest examples, the authors of \cite{Bhattacharyya:2008jc,Rangamani:2009xk,Hubeny:2011hd} perturb a 
solution of a gravity theory on a $AdS$ background (such as black holes or 
black branes) by means of isometry transformations whose infinitesimal parameters depend on $AdS$--boundary coordinates.
The transformed expression
is no longer a solution to the equations of motion, unless those local parameters 
satisfy some differential equations on the boundary which turn out to be
the linearized Navier--Stokes equations for the boundary field theory \cite{Bhattacharyya:2008jc,Rangamani:2009xk,Hubeny:2011hd}. 
Starting from this, they provide an iterative procedure to construct the gravity counterpart of the fluid dynamic system in terms of a boundary derivative expansion. They compute hydrodynamic stress--energy momentum tensor and thus the transport coefficients up to second order. 
In particular, the result for the viscosity to entropy density ratio $\frac{\eta}{s}$ coincides with the one derived in \cite{Policastro:2001yc}.

We study in detail the procedure described in  \cite{Bhattacharyya:2008jc,Rangamani:2009xk,Hubeny:2011hd}. 
We notice that the complete $AdS/CFT$ correspondence can only be fully established between 
the supergravity extension of the general relativity and its holographic dual. 
This would correspond to a supersymmetric extension of Navier-Stokes equations. 
Two possible ways are available: a top--down point of view by generalizing the procedure of \cite{Bhattacharyya:2008jc,Rangamani:2009xk,Hubeny:2011hd}
or from a bottom--up approach by constructing a supersymmetric effective action using superfield formalism. 
Note that the effective action formalism does not allow us to study the dissipative terms.
Nevertheless it gives the supersymmetric generalization of the perfect fluid theory, which is needed in order to 
study supersymmetric hydrodynamic.
In both approaches we obtain interesting results. 

By extending the construction of \cite{Bhattacharyya:2008jc,Rangamani:2009xk,Hubeny:2011hd} it is possible to 
derive the fermionic corrections to Navier-Stokes equations in terms of fermion bilinears generated by bulk fermions \cite{Gentile:2011jt}. 
In particular, we consider as bulk fermions the superpartners of the zero modes of an uncharged black hole in $AdS$.

Our analysis in \cite{Gentile:2011jt} is original although
 limited to the linear approximation since we do not possess the complete expression (obtained from a finite superisometry). 
For that reason,
in \cite{Gentile:2012jm} we construct the complete supersymmetric extension of classical
solutions of the $AdS$-Schwarzschild type. We denote the complete solution as ``fermionic wig'', 
to underline the anticommuting nature of these ``hairs''. 
The wig is constructed by acting with supersymmetry transformations upon
the supergravity fields (vielbein, gravitino and gauge field) and that expansion naturally truncates 
at some order in the fermionic zero modes.

We apply this procedure for non--extremal Schwarzschild--like solution of
$\mathcal{N}=2$, $D=5$ and $D=4$ supergravity in $AdS$ background.
Having the full metric solution we compute the boundary stress--energy tensor using Brown--York procedure. 
We perform the same analysis in the simpler set-up of BTZ black holes \cite{Banados:1992wn,Banados:1992gq,Gentile:2012tu,Gentile:2013nha} 
for $\mathcal{N}=2$, $D=3$ supergravity. 
In this case we compute analytically all charges associated to the BTZ black hole with all fermionic
contributions. 
We notice that also the entropy of the black hole is modified in terms of
the fermionic bilinears. Moreover, using fluid/gravity techniques we derive linearized Navier--Stokes equations and a set of new differential equations from Rarita--Schwinger equation.

In relation to the construction of an action for supersymmetric fluids, there are 
several candidates and our proposal is based on linear/chiral superfield for $4$-- and $3$--dimensional model \cite{Grassi:2011wt}.
We deduce the equations of motion in two ways: by imposing the null-divergence condition of the energy-momentum tensor and by using a suitable parametrization of the auxiliary fields.
We develop algorithms written in \verb FORM  language to derive the complete component expansion and we give a preliminary analysis of the physics of this supersymmetric fluid.


\section{Structure of the Thesis}

This Thesis is organized as follows.

Part I is divided in two main chapters: the first one explores
the classical structure of the theory and its T-duals and the second one studies the quantum
corrections.

In sec.~2.1, we review the T-duality. In sec.~2.2, we construct the $\sigma$--models used in the rest
of the analysis  by three different methods. Sec.~2.3 deals with the possible obstructions in constructing
the T-dual models. Finally, in sec.~2.4 we provide a T-dualization of our fermionic cosets. 

At the level of quantum analysis, in sec.~3.1  we deal with one--loop computations and in sec.~3.2 the two-loop
analysis with BFM is completed.

Part II deals with supersymmetric fluid dynamics, first by extending the fluid/gravity correspondence and then by considering a suitable supersymmetric action.

In Chapter~4 the basic concepts to develop supersymmetric fluid dynamics are presented.
Sec.~4.1 reviews relativistic fluid dynamics, sec~4.2 introduces anti--de Sitter spaces and in sec.~4.3 we describe the procedure to obtain the boundary energy--momentum tensor from a given metric.
The fluid/gravity procedure is reviewed in sec.~4.4 and it is extended to include supergravity in sec.~4.5.

In Chapter~5 we derive the boundary fluid dynamics for the BTZ black hole by constructing the complete black hole superpartner (the wig).
The basic setup is discussed in sec.~5.1, the fermionic wig is constructed in sec.~5.2 and we discuss the linearized boundary equations in sec.~5.3. 
Sec.~5.4 is devoted to analyze the physical implications of the fermionic wig.

The wig procedure is applied to $AdS_{5}$ model in chapter~6.
The action for $\mathcal{N}=2$, $D=5$ gauged supergravity is presented in sec.~6.1 and in sec.~6.2 and 6.3 we compute the main ingredients to perform the computation.
The results are presented in sec.~6.4.

Chapter~7 deals with the construction of a supersymmetric action to describe supersymmetric fluid dynamics.
The action principle for bosonic ideal fluids is discussed in sec.~7.1 and it is generalized in sec.~7.2.
In this section we present the supersymmetric action and we discuss  the equations of motion and the energy--momentum tensor in some limits.

In the appendix some auxiliary material is collected.


\section{Disclaimer}

The contents and in particular the bibliographic references of this thesis are updated to March 26th 2013, date of the defense.
All the results obtained afterwards do not appear here.


\part{Aspects of Fermionic T--duality}


\chapter{Classical Analysis}
\label{chT1}

\section{Fermionc Extension of T-duality}\label{fermExt}

\subsection{Review of Bosonic T-duality}

This section provides a short review of the T-duality construction method for $\s$-models with a single abelian isometry \cite{Giveon:1994fu,Alvarez:2000bh,Alvarez:2000bi}. Let us introduce a $D$-dimensional $\s$-model
\begin{equation}\label{SMTD1}
 S=\int G_{A B}(X)\dd X^{A}\wa \dd X^{B}
=
\int\dd^{2}x\sqrt{-\gamma}G_{AB}\gamma^{\mu\nu}\partial_{\mu}X^{A}\partial_{\nu}X^{B}
\ ,
\end{equation}
where $A,B= 1,\dots,D$ and the set of $\{X^{A}\}$ are bosonic coordinates.
If the $\s$-model has a translational isometry, then the metric $G$ is independent of one coordinate ({\it i.e.} $X^{d}$). The (\ref{SMTD1}) becomes then
\begin{equation}\label{sm1}
 S=\int \left[  G_{a b}(X)\dd X^{a} \wa\dd X^{b}+G_{a d}\dd X^{a} \wa\dd X^{d} +G_{d d}\dd X^{d}\wa \dd X^{d} \right] 
\ ,
\end{equation}
where $a,b=1,\cdots,D-1$. To construct the T-dual $\s$-model we introduce the gauge field $A$ via the covariant derivative $\dd X^{d}\ra \nabla X^{d}=\dd X^{d}+A$. The new action is now invariant under the local gauge transformation and therefore we can choose a suitable gauge where $X^{d}=0$.\footnote{We use the BRST formalism
\begin{eqnarray}
sX^{d}=c,
\quad\quad\quad
sA=-\dd c
\ .
\label{RBTD_BRST}
\end{eqnarray}} The new action is then
\begin{equation}\label{sm2}
 S=\int \left[  G_{a b} \dd X^{a}\wa\dd X^{b}+  G_{d d}\left( A \wa A \right)+G_{a d}\dd X^{a} \wa A \right] 
\ .
\end{equation}
Now we can add in (\ref{sm2}) the $2$-form $F=\dd A$, weighted by a Lagrange multiplier $\tilde X^{d}$
\begin{equation}\label{SMTD8}
 S=\int \left[  G_{a b} \dd X^{a}\wa\dd X^{b}+  G_{d d}\left( A \wa A \right)+G_{a d}\dd X^{a} \wa A+2\tilde{X}^{d}\dd A  \right] 
 \ .
 \end{equation}
The equation of motion for the new parameter $\tilde X^{d}$ shows that (\ref{SMTD8}) is equivalent to (\ref{sm2}). 
Otherwise, from the equation of motion of $A$ we compute\footnote{Recall that $\ast$ is the Hodge dual operator defined in the $\sigma$-model $2$-dimensional worldsheet  equipped by the metric $\gamma_{\mu\nu}$. Then
\begin{eqnarray}
\ast\ast A=-\det \gamma A
\ .
\label{HodgeDualastast}
\end{eqnarray}
}
\begin{equation}
 A
=
\frac{1}{G_{dd}}
\left( 
-G_{ad}\dd X^{a}
+
\frac{1}{\det \gamma}\ast\dd\tilde X^{d}
 \right)\ .
\end{equation}
The T-dual model is then obtained substituting this result into (\ref{SMTD8})
\begin{eqnarray}
 S_{Dual}
&=&
\int \left[   
\left(
G_{a b}
-\frac{G_{ad}G_{bd}}{G_{dd}}
  \right)
 \dd X^{a}\wa\dd X^{b}+
\right.\nonumber\\&&\left.
-\frac{G_{ad}}{G_{dd}}\dd X^{a}\wedge \dd\tilde X^{d}
-
\frac{1}{G_{dd}\det \gamma}\dd\tilde X^{d}\wa\dd\tilde X^{d}
  \right]
  \ .
\end{eqnarray}
Notice that this simple formulation is guaranteed by the trivial action of the isometry. For a generic bosonic $\sigma$-model one can choose a set of coordinates such that the isometry appears as a translation along a single coordinate (holonomic coordinate). Nevertheless, one can in principle perform a T-duality along any transformation of the isometry group. In general the isometry group could be non-abelian and the corresponding Killing vectors are non-trivial expression of the coordinates of the manifold, therefore the above derivation can not be used any longer. For that, we refer to the work of de la Ossa and Quevedo \cite{de la Ossa:1992vc} where they study such a situation in detail.

At the classical level, the above derivation is correct, but at the quantum level in the case of string models, we have to recall that the  integration measure of the Feynman integral gets an additional piece which can be reabsorbed by a dilaton shift
\begin{equation}
\phi'=\phi-\ln \det f
\ ,
\label{RBTDdilaton1}
\end{equation}
where $f$ is the Jacobian of the field redefinition \cite{de la Ossa:1992vc,DeJaegher:1998pp}. We also recall that the most general quantum corrections for abelian T-duality are the dilaton shift and the zero-mode determinant  \cite{Schwarz:1992te}.


\subsection{Review of Fermionic T-duality}\label{revFer}

Here we review the fermionic T-duality for an abelian isometry \cite{Berkovits:2008ic,Beisert:2008iq}. The above procedure can be followed through verbatim changing the dictionary and the statistical nature of the ingredients. The bosonic fields $X^{A}=\left( X^{a},X^{d} \right)$  becomes fermionic  $\theta^{A}=\left( \theta^{a},\theta^{d-1},\theta^{d} \right)$, the symmetric metric $G_{AB}=\left( G_{ab},G_{ad},G_{dd} \right)$ is replaced by a super-metric where $G_{AB}=-G_{BA}$. 
Notice that, due to the antisymmetric nature of $G_{AB}$, we need two translational isometries. Therefore we consider a super-metric which is independent of the two fields $\theta^{d-1},\theta^{d}$. The action is written as
\begin{eqnarray}
S
&=& 
\int \left[ G_{ab}\left( \theta \right)\dd\theta^{a}\wedge\ast\dd\theta^{b}
+2G_{ad-1}\dd\theta^{a}\wedge\ast\dd\theta^{d-1} 
+
\right.\nonumber\\&&\left.
+2G_{ad}\dd\theta^{a}\wedge\ast\dd\theta^{d}
+2G_{d-1d}\dd\theta^{d-1}\wedge\ast\dd\theta^{d}
 \right]\ .
\label{RFTDaction}
\end{eqnarray}
As in the previous section, we promote the derivatives of holonomic coordinates $\theta^{d-1}$ and $\theta^{d}$ to covariant ones, introducing two (fermionic) gauge fields $A^{d-1}$ and $A^{d}$. Therefore we  add to (\ref{RFTDaction}) the field strengths $F=\dd A$  weighted by the dual coordinates: $\tilde\theta^{d-1}$ and $\tilde\theta^{d}$. Using again BRST technique, we fix the gauge to set $\theta^{d-1}$ and $\theta^{d}$ to zero. The resulting action is the following
\begin{eqnarray}
S
&=& 
\int 
\left[ 
G_{ab}\left( \theta \right)\dd\theta^{a}\wedge\ast\dd\theta^{b}
+2G_{ad-1}\dd\theta^{a}\wedge\ast A^{d-1}
+2G_{ad}\dd\theta^{a}\wedge\ast A^{d}+
\right.\nonumber\\&&\left.
+
2G_{d-1d} A^{d-1}\wedge\ast A^{d}
+
\tilde\theta^{d-1}\dd A^{d-1}+\tilde\theta^{d} \dd A^{d}
 \right]
=
\nonumber\\
&=& 
\int \left[
G_{ab}\left( \theta \right)\dd\theta^{a}\wedge\ast\dd\theta^{b}
+
A^{d}\wedge
\left( 
-2G_{ad}\ast\dd\theta^{a}-\dd\tilde\theta^{d}
 \right)
+
\right.\nonumber\\&&\left.
+
A^{d-1}\wedge
\left( 
-2G_{ad-1}\ast\dd\theta^{a}
+
2G_{d-1d}\ast A^{d}
+
\dd\tilde\theta^{d-1}
 \right)
 \right] 
 \ .
\label{RFTDaction2}
\end{eqnarray}
The computation of the EoM for $A^{d-1}$ gives
\begin{eqnarray}
A^{d}
=
\frac{1}{G_{d-1d}}
\left( 
	G_{ad-1}\dd\theta^{a}-\frac{1}{2\det \gamma}\ast\dd\tilde\theta^{d-1}
\right)\ .
\label{RFTDEoM1}
\end{eqnarray}
The dual model is finally obtained inserting this solution back in (\ref{RFTDaction2})
\begin{eqnarray}
S_{Dual}
&=& 
\int 
\left[ 
\left( 
G_{ab}\left( \theta \right)-2\frac{G_{ad-1}G_{bd}}{G_{d-1d}}
 \right)\dd\theta^{a}\wa\dd\theta^{b}
+
\right.\nonumber\\&&\left.
-
\frac{G_{ad-1}}{G_{d-1d}}\dd\theta^{a}\wedge\tilde\theta^{d}
-
\frac{G_{ad}}{G_{d-1d}}\dd\theta^{a}\wedge\tilde\theta^{d-1}
+
\right.\nonumber\\&&\left.
-\frac{1}{2G_{d-1d}\det \gamma}\dd\tilde\theta^{d-1}\wa\dd\tilde\theta^{d}
 \right]\ .
\label{RFTDactionDual}
\end{eqnarray}
Notice that the fermionic nature of the fields might lead to some problems (see the following examples). In particular, we were able to find two obstructions in the construction of T-dual model: the first is connected to the non existence of holonomic coordinates, and the second deals with the non invertibility of the equation (\ref{RFTDEoM1}).

Notice that fermionic T-duality is valid only at tree level, since an interpretation of a compact Grassmannian direction is missing.


\subsection{Geometry of T-duality}

In order to illustrate the possible obstructions in performing the T-duality in the case of fermionic isometries, we derive some general conditions for T-duality for coset models \cite{Stern:1999dh}. In particular we show that there is an algebraic  and a differential condition. In the following we present two explicit examples to which this analysis applies.

We want to generalize the procedure reviewed in the first section to $\sigma$-models with an arbitrary number of isometries for which we can not use the holonomic coordinates. We consider a (super) group $G$ and one of its subgroup $H$. The generators of the associated Lie algebra $\mathfrak{g}$ are divided as follows
\begin{equation}
\mathfrak{g}=\mathfrak{k}+\mathfrak{h}\ ,
\label{TD00}
\end{equation}
where $\mathfrak{h}$ is the super-algebra associated to $H$ and $\mathfrak{k}$ is the coset vector space. We consider the case of symmetric and reductive coset
\begin{eqnarray}
&&
\left[ H_{I},H_{J} \right]=C_{I J}^{\phantom{I J}K}H_{K}\ ,
\nonumber\\&&
\left[ H_{I},K_{A} \right]=C_{I A}^{\phantom{I A}B} K_{B}\ ,
\nonumber\\&&
\left[ K_{A},K_{B} \right]=C_{A B}^{\phantom{A B}I}H_{I}\ ,
\label{TDcommrules}
\end{eqnarray}
where $H\in\mathfrak{h}$ and $K\in\mathfrak{k}$. The vielbeins $V^{A}$ of the coset manifold $G/H$ are obtained expanding the left invariant $1$-form $g^{-1}\dd g$ on the $\mathfrak{g}$ generators
\begin{equation}
g^{-1}\dd g=V^{A}K_{A}+\Omega^{I}H_{I}\ ,
\label{TD0}
\end{equation}
where $\O^{I}$ are the connections associated to the $H$-subgroup. Differentiating (\ref{TD0}) and using (\ref{TDcommrules}) we obtain the Maurer-Cartan equations
\begin{eqnarray}
&&
\dd V^{A}=-C_{B I}^{\phantom{A I} A}V^{B}\wedge \Omega^{I}\ ,
\nonumber\\&&
\dd \Omega^{I}=-\frac{1}{2}C_{A B}^{\phantom{A B} I} V^{A}\wedge V^{B}-\frac{1}{2}C_{J K}^{\phantom{A B} I}\Omega^{J}\wedge\Omega^{K}\ .
\label{TDMCeq}
\end{eqnarray}
These equations are rewritten defining the torsion $2$-form $T^{A}$ and the curvature $2$-form $R$
\begin{eqnarray}
&&
T^{A}=\dd V^{A}+C_{B I}^{\phantom{A I} A}V^{B}\wedge \Omega^{I}=0\ ,
\nonumber\\&&
R^{I}=\dd \Omega^{I}+\frac{1}{2}C_{J K}^{\phantom{A B} I}\Omega^{J}\wedge\Omega^{K}=-\frac{1}{2}C_{A B}^{\phantom{A B} I} V^{A}\wedge V^{B}\ ,
\label{TDMCeq2}
\end{eqnarray}
{\it i.e.} the coset manifold is a Einstein symmetric space. The metric is defined as
\begin{equation}
 G=V^{A}\otimes V^{B}\k_{AB}\ ,
\end{equation}
where $\k_{AB}=Str\left( K_{A}K_{B} \right)$ is the  Killing metric restricted to coset generators. 

To define a $\sigma$-model we need the pull-back
\begin{eqnarray}
V^{A}=V^{A}_{\mu}\partial_{i}Z^{\mu}\dd z^{i}\ ,
\label{geomViel0}
\end{eqnarray}
where $z^{i}$ are the coordinates on the  $2$-dimensional manifold $\Sigma$ and $Z^{\mu}=Z^{\mu}\left( z \right)$ are the embeddings of $\Sigma$ in the target space $G/H$, the action of the $\sigma$-model is then
\begin{equation}\label{sm5}
 S=\int_{\Sigma}\, V^{A}\wa V^{B}\k_{AB}\ .
\end{equation}

We can now focus on Killing vectors $K_{\L}=K^{\mu}_{\L}\frac{\p}{\p Z^{\mu}}$. They  generate the isometries $\Lambda$ that 
act on the coordinates as follows
\begin{equation}
 Z^{\mu}\rightarrow Z^{\mu}+\l^{\Lambda} K_{\Lambda}^{\mu}\ ,
\label{TDKilX}
\end{equation}
where $\l_{\Lambda}$ denote a set of infinitesimal parameters and $K^{\mu}_{\Lambda}$ is a function of $Z$.
By definition the Killing vectors satisfy $\CL_{K}G=0$ which reads
\begin{equation}\label{TDI1}
 \CL_{K_{\L}}\left(V^{A}\wa V^{B}\k_{AB} \right)= 2\left( \CL_{K_{\L}}V^{A}\right)\wa V^{B}\k_{AB}=0\ .
\end{equation}
The general solution to (\ref{TDI1}) is
\begin{equation}\label{TDI2}
 \CL_{K_{\L}}V^{A}=(\Theta_{\L})^{A}_{\phantom{A}B}V^{B}\ .
\end{equation}
where, because of the symmetries of the reduced Killing metric,  $(\Theta_{\L})_{AB}$ is {antisymmetric} if $V^{A}$ are {bosonic}. Otherwise, if the vielbeins are {fermionic} (anticommutant) and $\k_{AB}$ is antisymmetric and then $(\Theta_{\L})_{AB}$ is {symmetric}. Using $
 \CL_{X}\omega=i_{X}\dd \omega+ \dd (i_{X}\omega)$ we get
\begin{eqnarray}\label{TDI3}
 \CL_{K_{\L}}V^{A}&=& \dd\, i_{K_{\L}}V^{A}+ i_{K_{\L}}\dd V^{A} =
{}\nonumber\\
	&=& {}
\dd\, i_{K_{\L}}V^{A}-i_{K_{\L}}\left( \Omega^{A}_{\phantom{A}B}\wedge V^{B} \right)=
{}\nonumber\\
	&=& {}
\dd\, i_{K_{\L}}V^{A}-\left( i_{K_{\L}}\Omega^{A}_{\phantom{A}B} \right) V^{B}+\Omega^{A}_{\phantom{A}B} \left( i_{K_{\L}}V^{B} \right)\ ,
\end{eqnarray}
where $\Omega^{A}_{\phantom{A}B}=\Omega^{I}C^{A}_{\phantom{A}IB}$. 
Then the condition (\ref{TDI2}) can be rewritten as
\begin{equation}\label{TDI4}
 \nabla \left( i_{K_{\L}}V^{A}\right)=\left( \Theta_{\L}+i_{K_{\L}}\Omega\right)^{A}_{\phantom{A}B}V^{B} \ ,
\end{equation}
where the covariant derivative is defined by
\begin{equation}
 \nabla \left( i_{K_{\L}}V^{A}\right) =\dd \left(  i_{K_{\L}}V^{A}\right) +\Omega^{A}_{\phantom{A}B}\wedge \left( i_{K_{\L}}V^{B} \right)\ .
\end{equation}
This relation will be useful to search for the holonomy basis.

It is important to understand how the vielbein $V^{A}=V^{A}_{\mu}\dd Z^{\mu}$ transforms under (\ref{TDKilX}). 
First of all, we notice that (\ref{TDKilX}) shall be rewritten using the contraction operator $ i_{K_{\Lambda}}$ as follows
\begin{equation}
 Z^{\mu}\ra Z^{\mu}+\l^{\Lambda} i_{K_{\Lambda}}\dd Z^{\mu}\ .
\label{TDKilX2}
\end{equation}
Using the fact that the components $V^{A}$ are functions of $Z$ we can obtain, expanding $V^{A}$ in the first order of $\l$ 
\begin{eqnarray}\label{CTDtrasfK2}
V^{A}&\ra& V^{A}_{\a}\left(\{Z+ \lambda^{\Lambda}\,i_{K_{\Lambda}}\dd Z\} \right)   \dd\left(Z^{\a}+ \lambda^{\Lambda}\,i_{K_{\Lambda}}\dd Z^{\a}\right) =\phantom{\bigg |}
{}\nonumber\\
	&=& {}\phantom{\bigg |}
\left[ V_{\a}^{A}(\{Z\})+\lambda^{\Lambda} i_{K_{\Lambda}}\dd Z^{\b}\,\p_{\b}V^{A}_{\a} \right]\left[ \dd Z^{\a}+\dd \lambda^{\Lambda}\, i_{K_{\Lambda}}\dd Z^{\a}+\lambda^{\Lambda}\,\dd(i_{K_{\Lambda}}\dd Z^{\a}) \right] =
{}\nonumber\\
	&=& {}\phantom{\bigg |}
V^{A}+\dd \lambda^{\Lambda}\,V^{A}_{\a}\,i_{K_{\Lambda}} \dd Z^{\a}+\lambda^{\Lambda}\left[ V_{\a}^{A}\dd(i_{K_{\Lambda}}\dd Z^{\a})+i_{K_{\Lambda}}\dd Z^{\b}\,\p_{\b}V^{A}_{\a}\dd Z^{\a} \right] =
{}\nonumber\\
	&=& {}\phantom{\bigg |}
V^{A}+\dd \lambda^{\Lambda}\,i_{K_{\Lambda}} V^{A}+\lambda^{\Lambda}\left[ V_{\a}^{A}\dd(i_{K_{\Lambda}}\dd Z^{\a})+i_{K_{\Lambda}}\dd Z^{\b}\,\p_{\b}V^{A}_{\a}\dd Z^{\a} \right] \ .
\end{eqnarray}
Consider now
\begin{eqnarray}
 i_{K_{\Lambda}}\dd Z^{\b}\,\p_{\b}V^{A}_{\a}\dd Z^{\a}&=&i_{K_{\Lambda}}\left(\dd Z^{\b}\, \p_{\b}V^{A}_{\a} \right)\dd Z^{\a}=\phantom{\bigg |}
{}\nonumber\\
	&=& {}
\phantom{\bigg |}
i_{K_{\Lambda}}\left(\dd V^{A}_{\a} \right)\dd Z^{\a} \ ,
\end{eqnarray}
and
\begin{eqnarray}
i_{k}(\dd V^{A})&=&
i_{K_{\Lambda}}\left( \dd V^{A}_{\a}\wedge\dd Z^{\a} \right)= \phantom{\bigg |}
{}\nonumber\\
	&=& {}
\phantom{\bigg |}
i_{K_{\Lambda}}\left(\dd V^{A}_{\a} \right)\dd Z^{\a} - \dd V_{\a}^{A}i_{K_{\Lambda}}\dd Z^{\a}=
{}\nonumber\\
	&=& {}
\phantom{\bigg |}
i_{K_{\Lambda}}\left(\dd V^{A}_{\a} \right)\dd Z^{\a} +V^{A}_{\a}\dd \left(i_{K_{\Lambda}}\dd Z^{\a} \right) - \dd \left( i_{K_{\Lambda}}V^{A}\right)\ ,
\end{eqnarray}
we then obtain the final relation
\begin{equation}\label{sm4}
 V^{A}\ra V^{A}+ \dd \lambda^{\Lambda}\,i_{K_{\Lambda}} V^{A}+\lambda^{\Lambda}\,\CL_{K_{\Lambda}}V^{A}\ .
\end{equation}
Using (\ref{TDI2}) this becomes
\begin{equation}
 V^{A}\ra V^{A}+ \dd \lambda^{\Lambda}\,i_{K_{\Lambda}} V^{A}+\lambda^{\Lambda}\,\left(\Theta_{\Lambda}\right)^{A}_{\phantom{A}B}V^{B}\ .
\end{equation}
This relation expresses the  transformation of the vielbeins induced by (\ref{TDKilX2}).

We are now ready to generalize the construction method of the T-duality . First of all we gauge the action (\ref{sm5}) via the following shift of the vielbeins
\begin{equation}\label{sm6}
 V^{A}\ra V^{A}+A^{A}
\end{equation}
for a not-yet-specified number of vielbeins and gauge fields $A$. After this, the action is invariant under the gauge transformations
\begin{equation}\label{sm7}
 \Bigg\{
 \begin{array}{c}
    V^{A}\ra V^{A}+ \dd \l^{\L}\,i_{K_{\L}} V^{A}+\l^{\L}\,\left(\Theta_{\L}\right)^{A}_{\phantom{A}B}V^{B} \ ,\\
\\
\hspace{-3cm}A^{A}\ra A^{A}-\dd \l^{\L}\,i_{K_{\L}} V^{A}\ ,
\end{array}
\end{equation}
and so we can gauge some vielbeins to zero. For that we have to solve the following equations
\begin{equation}\label{sm8}
\dd \l^{\L}=-V^{\L}\left( i_{K_{\L}}V^{A}\right)^{-1}\ .
\end{equation}
Notice that, thanks to the symmetries of the reduced Killing metric, the term $\l^{\L}\,\left(\Theta_{\L}\right)^{A}_{\phantom{A}B}V^{B}$ can be omitted. 
The condition (\ref{sm8}) implies two constraints on the matrix\footnote{Notice we have restricted the set of indices in order to find a minor satisfying the two conditions.} $M^{\phantom{L}\hat{S}}_{\hat{L}}\equiv \left( i_{K_{\hat{L}}}V^{\hat{S}}\right)$: which must be invertible
\begin{equation}
 \textrm{det}M\neq 0\ ,
\label{TDconstraint1}
\end{equation}
and 
\begin{equation}
\dd \l^{\hat G}=-V^{\hat S}\left(M^{-1} \right)^{\phantom{L}\hat G}_{\hat S}\ ,
\end{equation}
which locally is equivalent to
\begin{equation}
 \dd \left[  V^{\hat S} (M^{-1})_{\hat S}^{\phantom{\S}\hat G} \right] =0\ ,
\label{TDconstraint2}
\end{equation}
because of the Poincar\'e Lemma.
Being constraints (\ref{TDconstraint1}) and (\ref{TDconstraint2})  satisfied, we are able to construct the T-dual model: first, we add to the action the field strength weighted with the Lagrange multipliers $\tilde Z_{L}\dd A^{L}$ (Chern-Simons term in $2$d), then we substitute the expression of $A^{L}$ as functions of the $\tilde X$ obtained by solving the equations of motion of $A^{L}$.


\subsection{Gauge Fixing and Cyclic Coordinates}

Dealing with the generic isometry-T-duality construction, we discuss the connection between the possibility of fixing the gauge (and performing the T-duality) and the existence of a system of coordinates in which the generic isometry is reduced to a translational one. To do this, we first focus on a simple bosonic-coordinates system.

Consider the following isometry of  action $S=\int\mathcal{L}\left( x \right)$
\begin{equation}
Z^{0}\rightarrow Z^{0}+\lambda K^{0}\ .
\label{KilGauisom1}
\end{equation}
We note that to fix the gauge we must have that
\begin{eqnarray}
Z^{0}+\lambda K^{0}=0
\quad\quad\Rightarrow\quad\quad
\lambda=-Z^{0} \left[ K^{0} \right]^{-1}\ .
\label{KilGaugf1}
\end{eqnarray}
Then we try to find a system of coordinates (the holonomy base) in which (\ref{KilGauisom1}) is reduced to
\begin{equation}
\tilde Z^{0}\rightarrow \tilde Z^{0}+\lambda\ .
\label{KilGauisom2}
\end{equation}
Introducing a new variable $\tilde Z^{0}\left( Z \right)$ and imposing condition (\ref{KilGauisom2}) we get
\begin{eqnarray}
&&
\tilde Z^{0}\left( Z+\lambda K \right)=\tilde Z^{0}+\lambda\ ,
\nonumber\\&&
\tilde Z^{0}\left( Z \right)+\lambda K \frac{\partial \tilde Z^{0}}{\partial Z}\Big{|}_{\tilde Z=Z}
=\tilde Z^{0}+\lambda\ ,
\nonumber\\&&
 \frac{\partial \tilde Z^{0}}{\partial Z}\Big{|}_{\tilde Z=Z}= \left[ K \right]^{-1}\ ,
\label{KilGaubla}
\end{eqnarray}
then
\begin{equation}
\tilde Z^{0} =\int \left[ K \right]^{-1} \dd Z\ .
\label{KilGauinttildex}
\end{equation}
From (\ref{KilGaugf1}) and (\ref{KilGauinttildex}) we see that the non-existence of the inverse of the Killing vector invalidates both the gauge fixing and the redefinition of cyclic coordinate.

This conclusion changes dramatically if we include also fermionic coordinates $\theta$. For sake of simplicity let us consider a purely fermionic lagrangian and following isometry
\begin{equation}
\theta^{\alpha}\rightarrow \theta^{\alpha}+\varepsilon^{\beta}K^{\alpha}_{\beta}\left( \theta \right)\ .
\label{KilGauisomfermZx1}
\end{equation}
Condition (\ref{KilGaubla}) reads
\begin{equation}
\frac{\partial\tilde\theta^{\rho}}{\partial\theta^{\alpha}}
=
\left[ K^{-1}\left( \theta \right) \right]^{\rho}_{\alpha}\ .
\label{KilGauKilGau1}
\end{equation}
This differential equation can not be integrated in Berezin sense.
We can find the solution defining the more general combination of $\theta$
\begin{equation}
\tilde\theta^{\rho}
=
\sum_{i=0}^{n}\left( \frac{1}{2i+1}c^{\rho}_{\phantom{\rho}\sigma_{1}\cdots\sigma_{2i+1}}\theta^{\sigma_{1}}\cdots\theta^{\sigma_{2i+1}} \right)\ ,
\label{KilGautildetheta}
\end{equation}
where in the most general case, $c^{\rho}_{\phantom{\rho}\sigma_{1}\cdots\sigma_{i}}$ are function of the bosonic coordinates. Equation (\ref{KilGauKilGau1}) becomes
\begin{equation}
\sum_{i=0}^{n}\left( 
c^{\rho}_{\phantom{\rho}\alpha\sigma_{2}\cdots\sigma_{2i+1}}
\theta^{\sigma_{2}}\cdots\theta^{\sigma_{2i+1}} \right)
=
\left[ K^{-1}\left( \theta \right) \right]^{\rho}_{\alpha}\ ,
\label{KilGau2}
\end{equation}
where, for $i=0$ we have $c^{\rho}_{\alpha}$. In conclusion, to construct the holonomic base , the Killing vector component $K$ has to be invertible and (\ref{KilGau2}) must be solvable.


\section{Fermionic Coset Models}

Before applying the above considerations, we present a set of models which become of interest recently \cite{Berkovits:2007rj,Berkovits:2008qc,Bonelli:2008us}. We mainly deal with fermionic coset models based on the orthosymplectic supergroup $OSp(n|m)$ where we quotient  by its maximal bosonic subgroup $SO(n)\times Sp(m)$. These models are obtained as a certain limit of $AdS_{5}\times S^{5}$ in \cite{Berkovits:2007rj} and as a limit of $AdS_{4}\times\mathbb{P}^{3}$ in \cite{Bonelli:2008us}. 

We take into account only the principal part without any WZ term and we study its conformal invariance. To construct the model, we do not proceed from a string theory and taking its limit, but we use three independent methods to construct such simple models. Since we are interested in studying the (super) isometries, we focus on the symmetry constraints.

The first method is based on a specific choice of the coset representative, on the nilpotency  of the supercharges and their anticommutative properties. As examples, we construct the $OSp(1|2)/Sp(2)$ and the $OSp(2|2)/SO(2)\times Sp(2)$ models.
This method is very powerful and advantageous in the case of small supergroups.
The second method is based on the geometric construction of the vielbeins and H-connection. We follow the book \cite{CastellaniDAuriaFre} for the derivation and we adapt their formulas for our purposes. Finally, the third method is based on the symmetric requirements. The latter can be implemented perturbatively and it allows more general models for which only the conformal invariance seems to discriminate among them. 

\subsection{Nilpotent Supercharges Method}\label{appOsp22}

Given the supercharges $Q_{\alpha}$ we impose an ordering $Q_{1},Q_{2},\cdots$ and we construct the coset representative $L$ as the product of exponentials
\begin{equation}
L\left( \theta\right)=
e^{\theta_{1}Q_{1}}
e^{\theta_{2}Q_{2}}
\cdots\ .
\label{MIOmethod1}
\end{equation}
By the fermionic statistic of the $\theta$'s and the anticommutation relations of the  super-algebra we can compute the complete expansion of $L\left( \theta \right)$ and we easily derive the action for the models.

\subsubsection{$OSp(1|2)/Sp(2)$}

This simple model has $2$ anticommuting coordinates $\theta_{1}$ and $\theta_{2}$. Notice that they form a vector of $\mathfrak{sp}\left( 2 \right)$. We can write the coset representative $L\left( \theta \right)$ as (\ref{MIOmethod1}) and we can expand in power of $\theta$
\begin{eqnarray}
L\left( \theta \right)
&=& 
e^{\theta_{1}Q_{1}}e^{\theta_{2}Q_{2}}=
\nonumber\\&=& 
\left( 1+\theta_{1}Q_{1} \right)
\left( 1+\theta_{2}Q_{2} \right)\ ,
\label{MIOosp12eq1}
\end{eqnarray} 
then, the inverse $L^{-1}$ and the $1$-form $\dd L$ are defined as follows
\begin{eqnarray}
L^{-1}&=& \left( 1-\theta_{2}Q_{2} \right)\left( 1-\theta_{1}Q_{1} \right)\ ,
\nonumber\\
\dd L
&=& 
\dd\theta_{1}Q_{1}\left( 1+\theta_{2}Q_{2} \right)
+
\left( 1+\theta_{1}Q_{1} \right)\dd\theta_{2}Q_{2}\ .
\label{MIOosp12L-1dL}
\end{eqnarray}
Therefore, the left invariant $1$-form is
\begin{eqnarray}
L^{-1}\dd L
&=& 
\left( 1-\theta_{2}Q_{2} \right)\left( 1-\theta_{1}Q_{1} \right)\dd\theta_{1}Q_{1}\left( 1+\theta_{2}Q_{2} \right)+
\nonumber\\&&
+\left( 1-\theta_{2}Q_{2} \right)\dd\theta_{2}Q_{2}
\nonumber\\
&=& 
\dd\theta_{1}Q_{1}+\dd\theta_{2}Q_{2}
+\nonumber\\&&
-\frac{1}{2}\theta_{1}\dd\theta_{1}\{Q_{1},Q_{1}\}
-\frac{1}{2}\theta_{2}\dd\theta_{2}\{Q_{2},Q_{2}\}
-\theta_{2}\dd\theta_{1}\{Q_{1},Q_{2}\}
+\nonumber\\&&
-\frac{1}{2}\theta_{2}\theta_{1}\dd\theta_{1}\left[ Q_{2},\{Q_{1},Q_{1}\} \right]\ .
\label{MIOosp12leftinv1form}
\end{eqnarray}
Using the (anti)commutation relations given in app.~\ref{appendixA}, the left invariant $1$-form can be expanded into the $\mathfrak{osp}\left( 1|2 \right)$ generators, obtaining the vielbein $V_\alpha$ (the $1$-form associated the coset generators $Q_{\alpha}$) and the H-connection (the $1$-form associated to the generators of the isotropy subalgebra $\mathfrak{sp}(2)$)
\begin{eqnarray}
L^{-1}\dd L
&=& 
\left( 1+\theta_{1}\theta_{2} \right)\dd\theta_{1}Q_{1}
+
\dd\theta_{2}Q_{2}
+
\textrm{H-connection}\ .
\label{MIOosp12leftinv1form2}
\end{eqnarray} 
The vielbeins are then
\begin{eqnarray}
V_{1}
&=& 
\left( 1+\theta_{1}\theta_{2} \right)\dd\theta_{1}\ ,
\nonumber\\
V_{2}
&=& 
\dd\theta_{2}\ .
\label{MIOosp12vielbein}
\end{eqnarray}
The action reads
\begin{eqnarray}
S
&=&
\int_{\Sigma}
k^{\alpha\beta} 
V_{\alpha}\wa V_{\beta}\ ,
\end{eqnarray}
where $k^{\alpha\beta}$ is the Killing metric reduced to the coset. Here, $k^{\alpha\beta}=\varepsilon^{\alpha\beta}$. Then we obtain
\begin{eqnarray}
S
&\propto&
\int_{\Sigma}
\left( 1+\theta_{1}\theta_{2} \right)\dd\theta_{1}\wa\dd\theta_{2}\ .
\label{MIOosp12action}
\end{eqnarray}
We can also derive the same action (up to a field redefinition) from the Maurer Cartan equations (\ref{TDMCeq})
\begin{eqnarray}
&&
\dd V^{\alpha}-\varepsilon_{\gamma\beta}V^{\alpha\beta}\wedge V^{\gamma}=0\ ,
\nonumber\\&&
\dd V^{\mu\nu}-\frac{1}{2}V^{\mu}\wedge V^{\nu}-2\varepsilon_{\alpha\beta}V^{\alpha\mu}\wedge V^{\beta\nu}=0\ .
\label{osp12MC0}
\end{eqnarray}
From these equations we obtain the vielbeins
\begin{eqnarray}
&&
V^{\alpha}=\left( 1+\frac{1}{4}\theta^{\rho}\varepsilon_{\rho\sigma}\theta^{\sigma} \right)\dd \theta^{\alpha}\ ,
\nonumber\\&&
V^{\alpha\beta}=-\frac{1}{4}\left( \theta^{\alpha}\dd\theta^{\beta}+\theta^{\beta}\dd\theta^\alpha \right)\ ,
\label{osp12vilebeins}
\end{eqnarray}
then the action is
\begin{eqnarray}\label{OSp12sigmaT}
 S&\propto&\int_{\Sigma}\,\left(1+\frac{1}{2}\,\t^{\r}\varepsilon_{\r\s}\t^{\s} \right)\varepsilon_{\a\b}\dd \t^{\a}\wa\dd \t^{\b}=
{}\nonumber\\
	&=& {}
\int_{\Sigma}\,\dd^{2}z\left(1+\frac{1}{2}\,\t^{\r}\varepsilon_{\r\s}\t^{\s} \right)\varepsilon_{\a\b}\partial\theta^{\alpha}\bar{\partial}\theta^{\beta}\ .
\end{eqnarray}
The action is invariant under the isometries discussed in sec. \ref{obstr}.

\subsubsection{$OSp(2|2)/SO(2)\times Sp(2)$}\label{OSP22etc}

The procedure described in the previous section can be used  also for the present model, but we show that an alternative choice of the generators of super-algebra $\mathfrak{osp}\left( n|m \right)$, with $n$  even, (see for example \cite{LieDict})  leads to a further simplification.

We redefine the generators to make the fermionic ones nilpotent ({\it i.e.} $\left\{ Q_{i},Q_{i} \right\}=0$). To do this, we first define the following matrices
\begin{equation}
G_{IJ}=
\left( \begin{array}{ccc|ccc}
 0 & &\mathbb{I}_{m}  & &   \\
&& &&0&\\
 \mathbb{I}_{m} && 0   && &\\
\hline
& &&0&&\mathbb{I}_{n}\\
&0 &&&&\\
& &&-\mathbb{I}_{n}&&0
\end{array}
\right)
\quad\quad\quad\textrm{if } M=2m\ ,
\end{equation}
\begin{equation}
G_{IJ}=
\left( \begin{array}{ccc|ccc}
 0 & \mathbb{I}_{m}&0  & &   \\
 \mathbb{I}_{m} &0& 0   && 0&\\
0&0 &1&&&\\
\hline
& &&0&&\mathbb{I}_{n}\\
&0 &&&&\\
& &&-\mathbb{I}_{n}&&0
\end{array}
\right)
\quad\quad\quad\textrm{if } M=2m+1\ .
\end{equation}
We introduce a new set of matrices $e_{IJ}$ by components
\begin{eqnarray}
	\left( e_{IJ} \right)_{KL}&=& \delta_{IL}\delta_{JK}\ .
\end{eqnarray}
By these ingredients we can introduce the generators of $\mathfrak{osp}\left( n|m \right)$
\begin{eqnarray}
	E_{ij}&=& G_{ik}e_{kj}-G_{jk}e_{ki}\ ,
	\nonumber\\
	E_{i'j'}&=& G_{i'k'}e_{k'j'}+G_{j'k'}e_{k'i'}\ ,
	\nonumber\\
	E_{ij'}&=& E_{j'i}=G_{ik}e_{kj'}\ ,
	\label{osp22Gen2}
\end{eqnarray}
where we have splitted the capital indices $\{I,J\}$ in $\{i,j\}=1\cdots M$ and $\{i',j'\}=M+1\cdots N$. 
They satisfy the (anti)commutation relations
\begin{equation}\label{OSPMNr}
  \begin{array}{c}
\phantom{\bigg |}
 \left[E_{ij} \,,\, E_{kl}\right]= G_{jk}E_{il}+G_{il}E_{jk}-G_{ik}E_{jl}-G_{jl}E_{ik}\ ,
\\\phantom{\bigg |}
\left[E_{i'j'} \,,\, E_{k'l'}\right] =
-G_{j'k'}E_{i'l'}-G_{i'l'}E_{j'k'}-G_{i'k'}E_{j'l'}-G_{j'l'}E_{i'k'}\ ,
\\\phantom{\bigg |}
\left[E_{ij} \,,\, E_{k'l'}\right] =0\ ,
\\\phantom{\bigg |}
\left[E_{ij} \,,\, E_{kl'}\right] = G_{jk}E_{il'}-G_{ik}E_{jl'}\ ,
\\\phantom{\bigg |}
\left[E_{i'l'} \,,\, E_{kl'}\right]= -G_{j'l'}E_{kj'}-G_{j'l'}E_{ki'}\ ,
\\\phantom{\bigg |}
\left\lbrace E_{ij'}\,,\,E_{kl'}\right\rbrace =G_{ik}E_{j'l'}-G_{j'l'}E_{ik}\ ,
\end{array}
\end{equation}
where $E_{ij}$ are generators of $\mathfrak{so}(n)$, the $E_{i'j'}$ of $\in \mathfrak{sp}(m)$ and $E_{ij'}$ are the supercharges
Notice that, with this choice, the supercharges are nilpotent
\begin{equation}
 \left(E_{ij'} \right)^{2}=0\quad\quad\quad\quad \forall\,\, i,j'\ ,
\end{equation}
and this simplifies the computation.
We set
\begin{equation}
 Q_{1}=Q_{1'}^{1}\quad\quad Q_{2}=Q_{2'}^{1}\quad\quad Q_{3}=Q_{1'}^{2}\quad\quad Q_{4}=Q_{2'}^{2}\ ,
\end{equation}
\begin{equation}
 E_{1'}=T_{1'1'}\quad\quad E_{2'}=T_{1'2'}=T_{2'1'}\quad\quad E_{3'}=T_{2'2'}\quad\quad\ ,
\end{equation}
and
\begin{equation}
 E_{0}=T_{12}=-T_{21}\ ,
\end{equation}
where the prime indices corresponds to the $\mathfrak{sp}$  indices. The reduced Killing metric is
($A=\{i,i'\}$)
\begin{equation}\label{OSP22KMR}
\k_{AB}= 4\left(\begin{array}{cccc}
  0 & 0 & 0 & 1 \\
0 &  0 & -1 & 0 \\
 0 &  1 & 0 & 0 \\
-1 & 0 & 0 & 0  
\end{array}\right)\ .
\end{equation}
The complete computation is derived in app.~\ref{appendixAA}. We choose the coset representative as in (\ref{MIOosp12eq1})
\begin{equation}\label{OSP22L}
 L(\t)=e^{\t_{1}Q_{1}}\,e^{\t_{2}Q_{2}}\,e^{\t_{3}Q_{3}}\,e^{\t_{4}Q_{4}}\ ,
\end{equation}
which expanded in series becomes 
\begin{equation}\label{OSP22L1}
 L(\t)=\left(1+\t_{1}Q_{1} \right)\left(1+\t_{2}Q_{2} \right) \left(1+\t_{3}Q_{3} \right) \left(1+\t_{4}Q_{4} \right) \ .
\end{equation}
Then, the left-invariant $1$-form reads:
\begin{eqnarray}\label{OSP22conti1XX}
 L^{-1}\dd L&=&
\phantom{\Big|}
\left(1-\t_{4}Q_{4} \right)\left(1-\t_{3}Q_{3} \right) \left(1-\t_{2}Q_{2} \right) \left(1-\t_{1}Q_{1} \right)\times
{}\nonumber\\
	&& {}\phantom{\Big|}
\times \dd\t_{1}\,Q_{1}\left(1+\t_{2}Q_{2} \right) \left(1+\t_{3}Q_{3} \right) \left(1+\t_{4}Q_{4} \right)+
{}\nonumber\\\phantom{\Big|}
	&& {}
+
\left(1-\t_{4}Q_{4} \right)\left(1-\t_{3}Q_{3} \right) \left(1-\t_{2}Q_{2} \right)\dd\t_{2}\,Q_{2}\left(1+\t_{3}Q_{3} \right) \left(1+\t_{4}Q_{4} \right)+
{}\nonumber\\\phantom{\Big|}
	&& {}
+
\left(1-\t_{4}Q_{4} \right)\left(1-\t_{3}Q_{3} \right) \dd\t_{3}\,Q_{3}  \left(1+\t_{4}Q_{4} \right)+\left(1-\t_{4}Q_{4} \right)\dd\t_{4}\,Q_{4} \ .
\end{eqnarray}
Notice that only an even number of commutators of $Q$ gives again $Q$. Therefore, to obtain the vielbeins we compute only this kind of terms. We get (see app.~\ref{appendixAA})
\begin{equation}
 L^{-1}\dd L = Q_{1}\dd\t_{1}+Q_{2}\dd\t_{2}+Q_{3}\left(-2\t_{3}\t_{4}\dd\t_{1}+\dd\t_{3} \right) +Q_{4}\left(-2\t_{3}\t_{4}\dd\t_{2}+\dd\t_{4} \right) + \Omega^{I}H_{I}\ ,
\end{equation}
hence, the vielbeins are
\begin{equation}\label{OSP22viel}
  \begin{array}{cc}
\phantom{\Big|}
V^{1}=\dd\t_{1} &  V^{3}=  -2\t_{3}\t_{4}\dd\t_{1}+\dd\t_{3}\ ,
\\\phantom{\Big|}
V^{2}=\dd\t_{2} &    V^{4}=  -2\t_{3}\t_{4}\dd\t_{2}+\dd\t_{4}\ ,
\end{array}
\end{equation}
So, the $\sigma$-model action is
\begin{eqnarray}\label{OSP22sigmamod}
 S&=&\int_{\Sigma}\,\textrm{tr}\left(V\wa V \right)= \int_{\Sigma} k_{AB}V^{A}\wa V^{B}=
{}\nonumber\\
	&=& {}
4\int_{\Sigma}\Big\{
\dd\t_{1}\wa\left( -2\t_{3}\t_{4}\dd\t_{2}+\dd\t_{4}\right) -
\dd\t_{2}\wa\left(-2\t_{3}\t_{4}\dd\t_{1}+\dd\t_{3} \right)+
{}\nonumber\\
	&& {}
+
\left(-2\t_{3}\t_{4}\dd\t_{1}+\dd\t_{3} \right)\wa\dd\t_{2}
-\left( -2\t_{3}\t_{4}\dd\t_{2}+\dd\t_{4}\right)\wa\dd\t_{1}\Big\}\ .
\end{eqnarray}
But $\dd\theta^{\alpha}\wa\dd\theta^{\beta}$ is antisymmetric, then:
the action of $OSp(2|2)/SO(2)\times Sp(2)$ is
\begin{equation}\label{SMOSP22ferm1}
 S=8\int_{\Sigma}\Big\{
\dd\t_{1}\wa\dd\t_{4}-\dd\t_{2}\wa\dd\t_{3}-4\t_{3}\t_{4}\dd\t_{1}\wa\dd\t_{2}
\Big\}\ ,
\end{equation}
or, explicitly:
 \begin{equation}\label{SMOSP22ferm2}
	 S=8\int_{\Sigma}\dd^{2}x\,\sqrt{-\g}\gamma^{\mu\nu }\Big\{
 \p_{\mu}\t_{1}\p_{\nu}\t_{4}-\p_{\mu}\t_{2}\p_{\nu}\t_{3}-4\t_{3}\t_{4}\p_{\mu}\t_{1}\p_{\nu}\t_{2}
\Big\}\ .
\end{equation}


\subsection{Vielbein Construction Method}\label{vielbeinconstrmeth}


In this section we construct the $OSp(n|m)/SO(n)\times Sp(m)$ action through the coset  vielbeins. The method used is similar to the one described in \cite{CastellaniDAuriaFre}.

Let $L$ be the coset element
\begin{equation}
L=\exp{\hat\theta_{a}^{\alpha}Q_{\alpha}^{a}}\ ,
\end{equation}
where  $Q^{a}_{\alpha}\in\mathfrak{osp}(n|m)/\mathfrak{so}(n)\times\mathfrak{sp}(m)$ (see app.~\ref{appendixA}). The vielbeins $V^{\alpha}_{a}$ are obtained by expanding the left-invariant $1$-form $L^{-1}\dd L$
\begin{eqnarray}
L^{-1}\dd L=V^{\alpha}_{a}Q_{a}^{\alpha}+\textrm{H-connection}\ .
\label{VCM1}
\end{eqnarray}
Consider now the matrix realization in fundamental representation of the generators $Q^{a}_{\alpha}$
\begin{equation}
\left[ Q^{a}_{\alpha} \right]^{I}_{\phantom{I}J}=\delta^{a I}\varepsilon_{\alpha J}+\delta^{a}_{J}\varepsilon_{\alpha}^{\phantom{\alpha}I}\ .
\end{equation}
Notice that $\varepsilon_{\alpha}^{\phantom{\alpha}I}=\hat\delta^{\phantom{\alpha}I}_{\alpha}$ where $\hat\delta$ is the Kronecker delta in $m$ dimensions. We write the generators as block matrices
\begin{eqnarray}
\hat\theta_{a}^{\alpha}Q_{\alpha}^{a}=
\left( 
\begin{array}{cc}
0 & b \\
\tilde b & 0
\end{array}
 \right)\ ,
\label{VCM2}
\end{eqnarray}
where
\begin{eqnarray}
\left\{ \begin{array}{l}
\left[ b^{\alpha}_{a} \right]^{I}_{\phantom{I}J}=\hat\theta^{\alpha}_{a}\delta^{aI}\varepsilon_{\alpha J}
\\
\\
\left[ \tilde b^{\alpha}_{a} \right]^{I}_{\phantom{I}J}=\hat\theta^{\alpha}_{a}\delta^{a}_{J}\varepsilon_{\alpha }^{\phantom{\alpha}I}
\end{array} \right.\ .
\label{VCM3}
\end{eqnarray}
The group element is then
\begin{eqnarray}
L(\hat\theta)&=&
\left( 
\begin{array}{c|c}
\delta^{I}_{\phantom{I}J}+\frac{1}{2}b^{I}_{\phantom{I} K} \tilde b^{K}_{\phantom{I} J}+ \cdots 
& 
b^{I}_{\phantom{I}J} +\frac{1}{3!}b^{I}_{\phantom{I} K} \tilde b^{K}_{\phantom{I} L}b^{L}_{\phantom{I}J}+ \cdots \\
& \\\hline\\
\tilde b^{I}_{\phantom{I}J} +\frac{1}{3!}\tilde b^{I}_{\phantom{I} K} b^{K}_{\phantom{I} L}\tilde b^{L}_{\phantom{I}J}+ \cdots 
& 
\varepsilon^{I}_{\phantom{I}J}+\frac{1}{2}\tilde b^{I}_{\phantom{I} K}  b^{K}_{\phantom{I} J}+ \cdots
\end{array}
 \right)=
\nonumber\\&=&
\left( 
\begin{array}{cc}
\cosh \sqrt{b\tilde b}
  &
b\frac{\sinh\sqrt{\tilde b b}}{\sqrt{\tilde b b}}
\\
\\
\frac{\sinh\sqrt{b\tilde b }}{\sqrt{b \tilde b }}\tilde b
 &
\cosh \sqrt{\tilde b b}
\end{array}
 \right)\ .
\label{VCM4}
\end{eqnarray}
We shall now perform the following change of variable
\begin{eqnarray}
\theta^{\alpha}_{a}\equiv b\frac{\sinh\sqrt{\tilde b b}}{\sqrt{\tilde b b}}\ ,
\label{VCM5}
\end{eqnarray}
then the group element becomes
\begin{eqnarray}
L\left( \theta \right)=
\left( 
\begin{array}{cc}
\left( 
\delta_{ab}+\theta^{\alpha}_{a}\varepsilon_{\alpha\beta}\theta^{\beta}_{b}
 \right)^{\frac{1}{2}}\delta^{aI}\delta^{b}_{J}
  &
\theta^{\alpha}_{a}\delta^{aI}\varepsilon_{\alpha J}
 \\
& \\
\theta^{\alpha}_{a}\delta^{a}_{J}\varepsilon_{\alpha}^{\phantom{\alpha} I}
&
\left( \varepsilon^{\alpha\beta}+\theta^{\alpha}_{a}\delta^{ab}\theta_{b}^{\beta} \right)^{\frac{1}{2}}\varepsilon_{\alpha}^{\phantom{\alpha} I}\varepsilon_{\beta J}\end{array}
 \right)\ .
\label{VCML}
\end{eqnarray}
The inverse is then
\begin{eqnarray}
L^{-1}\left( \theta \right)=
\left( 
\begin{array}{cc}
\left( 
\delta_{ab}+\theta^{\alpha}_{a}\varepsilon_{\alpha\beta}\theta^{\beta}_{b}
 \right)^{\frac{1}{2}}\delta^{aI}\delta^{b}_{J}
  &
-\theta^{\alpha}_{a}\delta^{aI}\varepsilon_{\alpha J}
 \\
& \\
-\theta^{\alpha}_{a}\delta^{a}_{J}\varepsilon_{\alpha}^{\phantom{\alpha} I}
&
\left( \varepsilon^{\alpha\beta}+\theta^{\alpha}_{a}\delta^{ab}\theta_{b}^{\beta} \right)^{\frac{1}{2}}\varepsilon_{\alpha}^{\phantom{\alpha} I}\varepsilon_{\beta J}\end{array}
 \right)\ ,
\label{VCMLI}
\end{eqnarray}
and the $1$-form $\dd L$
\begin{eqnarray}
\dd L\left( \theta \right)=
\left( 
\begin{array}{c|c}
E & F \\
\hline
G & H
\end{array}
 \right)\ ,
\end{eqnarray}
where
\begin{equation}
\begin{array}{l}
E^{I}_{\phantom{I}J}=
\frac{1}{2}\left( \delta_{ru}+\theta^{\rho}_{r}\varepsilon_{\rho\sigma}\theta^{\sigma}_{u} \right)^{-\frac{1}{2}}\delta^{uv}\left[ \dd\theta^{\tau}_{v}\varepsilon_{\tau\lambda}\theta^{\lambda}_{s}+\theta^{\tau}_{v}\varepsilon_{\tau\lambda}\dd\theta^{\lambda}_{s} \right]\delta^{rI}\delta^{s}_{J}\ ,
\\
\\
F^{I}_{\phantom{I}J}=\dd \theta^{\alpha}_{a}\delta^{aI}\varepsilon_{\alpha J}\ ,
\\
\\
G^{I}_{\phantom{I}J}=\dd \theta^{\alpha}_{a}\delta^{a}_{J}\varepsilon_{\alpha}^{\phantom{\alpha}I}\ ,
\\
\\
H^{I}_{\phantom{I}J}=\frac{1}{2}\left( \varepsilon^{\rho\tau}+\theta^{\rho}_{r}\delta^{rs}\theta^{\tau}_{s} \right)^{-\frac{1}{2}}\varepsilon_{\tau\lambda}\left[ \dd\theta^{\lambda}_{u}\delta^{uv}\theta^{\sigma}_{v}+\theta^{\lambda}_{u}\delta^{uv}\dd\theta^{\sigma}_{v} \right]\varepsilon_{\rho}^{\phantom{\rho}I}\varepsilon_{\sigma J}\ .
\end{array}
\label{VCMdL}
\end{equation}


We shall write the left-invariant $1$-form as
\begin{eqnarray}
L^{-1}\left( \theta \right)\dd L\left( \theta \right)=
\left( 
\begin{array}{c|c}
A & B \\
\hline
C & D
\end{array}
 \right)\left( 
\begin{array}{c|c}
E & F \\
\hline
G & H
\end{array}
 \right)=
\left( 
\begin{array}{c|c}
 & AF+BH \\
\hline
CE+DG & 
\end{array}
 \right)\ .
\label{VCM6}
\end{eqnarray}
In order to obtain the vielbeins, we compute only the off-diagonal blocks. We gets
\begin{eqnarray}
V^{\sigma}_{a} \delta^{aI}\varepsilon_{\sigma J} &=& AF+BH=
\nonumber\\&=& 
\left( \delta_{ar}+\theta^{\alpha}_{a}\varepsilon_{\alpha\beta}\theta^{\beta}_{r} \right)^{-\frac{1}{2}}\delta^{rs}\left[ \dd\theta^{\sigma}_{s}+\theta^{\mu}_{s}\varepsilon_{\mu\nu}\theta^{\nu}_{b}\delta^{bz}\dd\theta^{\sigma}_{z}+
\right.\nonumber\\&&\left. 
\hspace{3.6cm}
-\frac{1}{2}\theta^{\alpha}_{s}\varepsilon_{\alpha\lambda}\dd\theta^{\lambda}_{u}\delta^{uv}\theta^{\sigma}_{v}
\right]\delta^{aI}\varepsilon_{\sigma J}\ ,
\label{VCME1}
\end{eqnarray}
and
\begin{eqnarray}
\hat V^{s}_{\alpha} \varepsilon_{\alpha}^{\phantom{\alpha}I}\delta^{s}_{J}&=& CE+DG=
\nonumber\\&=& 
\left( \varepsilon^{\alpha\rho}+\theta^{\alpha}_{a}\delta^{ab}\theta^{\rho}_{r} \right)^{-\frac{1}{2}}\varepsilon_{\rho\sigma}\left[ \dd\theta^{\sigma}_{s}-\frac{1}{2}\theta^{\sigma}_{u}\delta^{uv}\dd\theta^{\tau}_{v}\varepsilon_{\tau\lambda}\theta^{\lambda}_{s}+
\right.\nonumber\\&&\left. 
\hspace{3.8cm}
+\theta^{\sigma}_{u}\delta^{uv}\theta^{\tau}_{v}\varepsilon_{\tau\lambda}\dd\theta^{\lambda}_{s}
\right]\varepsilon_{\alpha}^{\phantom{\alpha}I}\delta^{s}_{J}\ .
\label{VCME2}
\end{eqnarray}
The $\sigma$-model is then
\begin{eqnarray}
S&=& \int_{\Sigma}\textrm{Str}\left( V\wa \hat V \right)=
\nonumber\\&=& 
\int_{\Sigma}
\left(
V_{a}^{\sigma} \delta^{aI}\varepsilon_{\sigma J} \wa
\hat V^{s}_{\alpha} \varepsilon_{\alpha}^{\phantom{\alpha}I}\delta^{s}_{J}\right)=
\nonumber\\&=& 
\int_{\Sigma}\left(\varepsilon_{\sigma\alpha}\delta^{as} V_{a}^{\sigma}\wa\hat V^{\alpha}_{s} \right)\ .
\label{VCMaction}
\end{eqnarray}
Dealing with fermionic fields, the expansion of (\ref{VCMaction}) leads to a polynomial action in $\theta$. We obtain
\begin{eqnarray}
S
&\sim&
\int
\dd^2 z \sqrt{\det \gamma}\left[ 
\partial_{\mu}\theta^{\alpha}_{a}\partial^{\mu}\theta^{\beta}_{b}\varepsilon_{\alpha\beta}\delta^{ab}+
\right.\nonumber\\&&\left.
+
\theta_{a}^{\alpha}\theta^{\beta}_{b}\partial_{\mu}\theta_{c}^{\gamma}\partial^{\mu}\theta^{\delta}_{d}\left( 
-2\delta^{ac}\delta^{bd}\varepsilon_{\alpha\delta}\varepsilon_{\beta\gamma}
+
\delta^{ab}\delta^{cd}\varepsilon_{\alpha\delta}\varepsilon_{\beta\gamma}
+
\delta^{ad}\delta^{bc}\varepsilon_{\alpha\beta}\varepsilon_{\gamma\delta}
 \right)
+\cdots\right] \ .
\nonumber\\&&
\label{VCMaction2}
\end{eqnarray}


\subsection{Supersymmetry Construction Method}\label{SUSYconstrMethod}

Here we derive the $4$-field terms ({\it i.e.} $\t\t\partial\t\partial\t$) for $OSp(n|m)/SO(n)\times Sp(m)$ action using  supersymmetry invariance. To perform this computation we have to build the supersymmetry transformation up to the second-order. Now, the variation of the zero-order term of the action must be canceled by the zero-order variation of the $\t\t\partial\t\partial\t$ term. With this observation we are able to reconstruct  the second-order contribution to the action.


The first-order generators of $Sp(m)$ and $SO(n)$ are
\begin{eqnarray}
  M^{\a\b} &=& \t_{a}^{\a}\varepsilon^{\b\r}\frac{\partial}{\partial\t^{\r}_{a}}+\t^{\b}_{a}\varepsilon^{\a\r}\frac{\partial}{\partial\t^{\r}_{a}}\ , \nonumber \\
  M_{ab} &=& \t^{\r}_{a}\d_{b c}\frac{\partial}{\partial\t^{\r}_{a}}-\t^{\r}_{b}\d_{a c}\frac{\partial}{\partial\t^{\r}_{a}}\ .
\end{eqnarray}
To find the second-order supersymmetry generators $Q^{\a}_{a}$ we use the closure relation
\begin{equation}\label{closure}
  \{ Q^{\a}_{a}\,,\,Q^{\a'}_{a'}\} =
  \varepsilon^{\a\a'}M_{aa'}+\d_{aa'}M^{\a\a'}\ .
\end{equation}
The generators  $Q^{\a}_{a}$ can be written as
\begin{equation}
  Q^{\a}_{a}= G^{\a\b}_{ab}\p_{\t^{\b}_{b}}
,\quad\quad\quad\quad
\partial_{\t^{\b}_{b}}\equiv\frac{\partial}{\partial\theta^{\beta}_{b}}\ ,
\end{equation}
and the relation (\ref{closure}) becomes
\begin{eqnarray}\label{closure2}
  &&
  \left[ G^{\a\b}_{ab}\left(\p_{\t^{\b}_{b}}G^{\a'\r}_{a'r}\right)+
      	 G^{\a'\b'}_{a'b'}\left(\p_{\t^{\b'}_{b'}}G^{\a\r}_{ar}\right)
	 \right]\,\p_{\t^{\r}_{r}}
	 = \nonumber \\ &=&
	\left[
	\d_{aa'}\ \t^{\a}_{r}\varepsilon^{\a'\r}+\d_{aa'}\t^{\a'}_{r}\varepsilon^{\a\r}
	+\varepsilon^{\a\a'}\t^{\r}_{a}\d_{a'r}-\varepsilon^{\a\a'}\t^{\r}_{a'}\d_{ar}
	\right]\p_{\t_{r}^{r}}\ .
\end{eqnarray}
Consider now $G^{\a\b}_{ab}$: at zero-order it is  $\varepsilon^{\a\b}\d_{ab}$. To find the exact second-order structure, we construct the most general term
\begin{equation}
  G^{\a\b}_{ab}=x\,\t^{\a}_{a}\t^{\b}_{b}+y\,\t_{b}^{\a}\t_{a}^{\b}+c\,\t^{\a}_{c}\d^{cd}\t_{d}^{\b}
  +d\,\t^{\g}_{a}\varepsilon_{\g\d}\t^{\d}_{b}+e\,\t^{\g}_{c}\varepsilon_{\g\d}\d^{cd}\t_{d}^{\d}\varepsilon^{\a\b}\d_{ab}\ .
  \label{}
\end{equation}
Using the zero- and second-order in (\ref{closure2}) we set the coefficient ${a,b,c,d,e}$. The computation yields the following results
\begin{equation}
  x=2e\quad,\quad
  c=d\quad,\quad
  d-y=1\ ,
\end{equation}
where we used the following relations
\begin{eqnarray}
  \varepsilon^{12}=1\quad\quad \varepsilon^{\a}_{\phantom{\a}\b}=\varepsilon^{\phantom{\b}\a}_{\b}\quad\quad \t^{\a}=\varepsilon^{\a\b}\t_{\b}\ .
\end{eqnarray}
Then, the supersymmetric generators are
\begin{eqnarray}
  Q^{\a}_{a}&=&\left[ \varepsilon^{\a\b}\d_{ab}+x\,\t^{\a}_{a}\t^{\b}_{b}+y\,\t_{b}^{\a}\t_{a}^{\b}+(1+y)\,\t^{\a}_{c}\d^{cd}\t_{d}^{\b}+\right.
  \nonumber\\&&\left.
  +(1+y)\,\t^{\g}_{a}\varepsilon_{\g\d}\t^{\d}_{b}+\frac{x}{2}\,\t^{\g}_{c}\varepsilon_{\g\d}\d^{cd}\t_{d}^{\d}\varepsilon^{\a\b}\d_{ab}
  +O(4)\right]\frac{\p}{\p\t^{\b}_{b}}\ ,
  \label{supergenerator2}
\end{eqnarray}
up to second-order.
Now we have to perform the second-order variation of the zero-order  lagrangian density
\begin{equation}
  \mathcal{L}_{0}=\gamma^{\mu\nu}\p_{\mu}\t^{\a}_{a}\p_{\nu}\t^{\b}_{b}\varepsilon_{\a\b}\d^{ab}\ .
  \label{zeroaction}
\end{equation}
The supersymmetric transformation generated by (\ref{supergenerator2}) is
\begin{equation}
  \d_{\e}=\e^{\a}_{a}Q^{a}_{\a}\ ,
   \label{supertransform1}
\end{equation}
explicitly
\begin{eqnarray}
 \d_{\e}\t^{\rho}_{r}&=&\e^{\rho}_{n}+\e^{a}_{\a}G^{\a\b}_{ab}\p_{\t^{\b}_{b}}\t^{\rho}_{r}=
 \nonumber\\
 &=&
 \e^{\rho}_{r}+x\,\e_{\a}^{a}\t^{\a}_{a}\t^{\rho}_{r}+y\,\e^{a}_{\a}\t^{\a}_{r}\t^{\rho}_{a}+
 (1+y)\,\e^{a}_{\a}\t^{\a}_{c}\d^{cd}\t^{\rho}_{d}\d_{ar}+
 \nonumber\\&&
 +(1+y)\,\e^{a}_{\a}\t^{\g}_{a}\varepsilon_{\g\d}\t^{\d}_{r}\varepsilon^{\a\rho}+
 \frac{x}{2}\,\e^{a}_{\a}\t^{\g}_{c}\varepsilon_{\g\d}\d^{cd}\t^{\d}_{d}\varepsilon^{\a\rho}\d_{ar}\ .
  \label{supertransform2}
\end{eqnarray}
The second-order transformation of (\ref{zeroaction}) is then
\begin{eqnarray}
&&
  \d(\mathcal{L}_{0})){\big|}_{II}=
  \nonumber\\&=&
  x\,\e^{\b}_{b}\p^{\mu}\t^{\a}_{a}\t^{\rho}_{r}\p_{\mu}\t^{\sigma}_{s}\varepsilon_{\b\a}\d^{ba}\varepsilon_{\rho\sigma}\d^{rs}+
  x\,\e^{\b}_{b}\t^{\a}_{a}\p^{\mu}\t^{\rho}_{r}\p_{\mu}\t^{\sigma}_{s}\varepsilon_{\b\a}\d^{ba}\varepsilon_{\rho\sigma}\d^{rs}+
  \nonumber\\&&
  +y\,\e^{\b}_{b}\p^{\mu}\t^{\a}_{r}\t^{\rho}_{a}\p_{\mu}\t^{\sigma}_{s}\varepsilon_{\b\a}\d^{ba}\varepsilon_{\rho\sigma}\d^{rs}
  +y\,\e^{\b}_{b}\t^{\a}_{r}\p^{\mu}\t^{\rho}_{a}\p_{\mu}\t^{\sigma}_{s}\varepsilon_{\b\a}\d^{ba}\varepsilon_{\rho\sigma}\d^{rs}+
  \nonumber\\&&
  +(1+y)\,\e^{\b}_{b}\p^{\mu}\t^{\a}_{c}\t^{\rho}_{d}\p_{\mu}\t^{\sigma}_{s}\varepsilon_{\b\a}\d^{cd}\varepsilon_{\rho\sigma}\d^{bs}
  +(1+y)\,\e^{\b}_{b}\t^{\a}_{c}\p^{\mu}\t^{\rho}_{d}\p_{\mu}\t^{\sigma}_{s}\varepsilon_{\b\a}\d^{cd}\varepsilon_{\rho\sigma}\d^{bs}+
  \nonumber\\&&
  +(1+y)\,\e^{\b}_{b}\p^{\mu}\t^{\g}_{a}\t^{\d}_{r}\p_{\mu}\t^{\sigma}_{s}\varepsilon_{\b\sigma}\d^{ba}\varepsilon_{\g\d}\d^{rs}
  +(1+y)\,\e^{\b}_{b}\t^{\g}_{a}\p^{\mu}\t^{\d}_{r}\p_{\mu}\t^{\sigma}_{s}\varepsilon_{\b\sigma}\d^{ba}\varepsilon_{\g\d}\d^{rs}+
  \nonumber\\&&
  +\frac{x}{2}\,\e^{\b}_{b}\p^{\mu}\t^{\g}_{c}\t^{\d}_{d}\p_{\mu}\t^{\sigma}_{s}\varepsilon_{\b\sigma}\varepsilon_{\g\d}\d^{bs}\d^{cd}
  +\frac{x}{2}\,\e^{\b}_{b}\t^{\g}_{c}\p^{\mu}\t^{\d}_{d}\p_{\mu}\t^{\sigma}_{s}\varepsilon_{\b\sigma}\varepsilon_{\g\d}\d^{bs}\d^{cd}\ .
  \nonumber\\&&
  \label{Szerotrasf}
\end{eqnarray}
Finally
\begin{eqnarray}
  \d(\mathcal{L}_{0}){\big|}_{II}&=&
  +2x\, \e^{\a}_{a}\p^{\mu}\t^{\rho}_{r}\t^{\b}_{b}\p_{\mu}\t^{\sigma}_{s}\varepsilon_{\a\rho}\d^{ar}\varepsilon_{\b\sigma}\d^{bs}+
\nonumber\\&&
-x\, \e^{\a}_{a}\p^{\mu}\t^{\rho}_{r}\t^{\b}_{b}\p_{\mu}\t^{\sigma}_{s} \varepsilon_{\a\b}\d^{ab}\varepsilon_{\rho\sigma}\d^{rs}+
\nonumber\\&&
+(1+2y)\,\e^{\a}_{a}\p^{\mu}\t^{\rho}_{r}\t^{\b}_{b}\p_{\mu}\t^{\sigma}_{s} \varepsilon_{\a\sigma}\d^{ab}\varepsilon_{\rho\b}\d^{rs}+
\nonumber\\&&
-(1+2y)\,\e^{\a}_{a}\p^{\mu}\t^{\rho}_{r}\t^{\b}_{b}\p_{\mu}\t^{\sigma}_{s} \varepsilon_{\a\b}\d^{as}\varepsilon_{\rho\sigma}\d^{rb}+
\nonumber\\&&
+2(1+y)\, \e^{\a}_{a}\p^{\mu}\t^{\rho}_{r}\t^{\b}_{b}\p_{\mu}\t^{\sigma}_{s} \varepsilon_{\a\sigma}\d^{ar}\varepsilon_{\rho\b}\d^{bs}\ .
\label{Szerotrasf1}
\end{eqnarray}
As we have already said, this variation must be compensated by the zero-order variation of the second-order lagrangian density $\mathcal{L}{\big|}_{II}$. Imposing this we obtain
\begin{eqnarray}
  \mathcal{L}{\big|}_{II}&=&
  +2x \t^{\a}_{a}\p^{\mu}\t^{\rho}_{r}\t^{\b}_{b}\p_{\mu}\t^{\sigma}_{s}\varepsilon_{\a\rho}\d^{ar}\varepsilon_{\b\sigma}\d^{bs}+
\nonumber\\&&
-x \t^{\a}_{a}\p^{\mu}\t^{\rho}_{r}\t^{\b}_{b}\p_{\mu}\t^{\sigma}_{s} \varepsilon_{\a\b}\d^{ab}\varepsilon_{\rho\sigma}\d^{rs}+
\nonumber\\&&
+(1+2y)\t^{\a}_{a}\p^{\mu}\t^{\rho}_{r}\t^{\b}_{b}\p_{\mu}\t^{\sigma}_{s} \varepsilon_{\a\sigma}\d^{ab}\varepsilon_{\rho\b}\d^{rs}+
\nonumber\\&&
-(1+2y)\t^{\a}_{a}\p^{\mu}\t^{\rho}_{r}\t^{\b}_{b}\p_{\mu}\t^{\sigma}_{s} \varepsilon_{\a\b}\d^{as}\varepsilon_{\rho\sigma}\d^{rb}+
\nonumber\\&&
+2(1+y) \t^{\a}_{a}\p^{\mu}\t^{\rho}_{r}\t^{\b}_{b}\p_{\mu}\t^{\sigma}_{s} \varepsilon_{\a\sigma}\d^{ar}\varepsilon_{\rho\b}\d^{bs}\ .
\label{S2}
\end{eqnarray}
Notice that for $x=0$ and $y=-\frac{2}{3}$ we obtain the $4$-field term derived in sec.~\ref{vielbeinconstrmeth}.

Since this method is merely perturbative, to each orders some freedom is left by the constants leading different models. If we were able to pursue it till to the end (namely in the case of a limited number of $\theta$'s coordinates) then we would have seen a unique solution. Hence this falls in the same class of problems known as {\it gauge completion} in supergravity and supersymmetry where starting from the bosonic components of a given superfield, the constraints would permit the construction of the full superfield. However, this is in general not achievable (see \cite{Peeters:2000qj} and reference therein.).


\section{Obstructions to Conventional T-duality}\label{obstr}

Here, as we announced in sec. \ref{fermExt}, we present two typical obstructions in the T-duality construction. To do this we apply the procedures outlined there to the simple models discussed above. 

\subsection{$OSp(1|2)/Sp(2)$}

This first case refers to the coset space $OSp(1|2)/Sp(2)$, characterized by two fermionic coordinates $\t_{1}$ and $\t_{2}$. We recall the action (\ref{OSp12sigmaT})
\begin{eqnarray}\label{OSp12sigmaT2}
 S&\propto&
\int_{\Sigma}\,\dd^{2}z\left(1+\frac{1}{2}\,\t^{\r}\varepsilon_{\r\s}\t^{\s} \right)\varepsilon_{\a\b}\partial\theta^{\alpha}\bar{\partial}\theta^{\beta}\ .
\end{eqnarray}
This model is not invariant under $\theta\rightarrow\theta+c$, but it possesses -- besides the $Sp\left( 2 \right)$ invariance under $\theta^{\alpha}\rightarrow\Lambda^{\alpha}_{\phantom{\alpha}\beta}\theta^{\beta}$ with $\Lambda_{\alpha\beta}=\Lambda_{\beta\alpha}$ -- also the following isometry\footnote{It is easy to demonstrate that it does not exist a coordinate transformation that reduces this isometry to  $\theta\rightarrow\theta+c$.}
\begin{equation}
\theta^{\alpha}\rightarrow\theta^{\alpha}+\left( 1+\frac{1}{2}\theta^{\rho}\varepsilon_{\rho\sigma}\theta^{\sigma} \right)\varepsilon^{\alpha}\ ,
\label{osp12isom0}
\end{equation}
{\it i.e.} the action (\ref{OSp12sigmaT}) has the following Killing vectors
\begin{equation}
K_{\left( \alpha \right)}^{\beta}=\left( 1+\frac{1}{2}\theta^{\rho}\varepsilon_{\rho\sigma}\theta^{\sigma} \right)\delta^{\beta}_{\alpha}\ .
\label{osp12Kilvec}
\end{equation}
To demonstrate this  we use the fermionic Killing equation
\begin{equation}
K_{\left( \alpha \right)}^{\lambda}\partial_{\lambda}G_{\rho\sigma}-\partial_{\rho}K_{\left( \alpha \right)}^{\lambda}G_{\lambda\sigma}-\partial_{\sigma}K_{\left( \alpha \right)}^{\lambda}G_{\rho\lambda}=0\ ,
\label{FermKilEq}
\end{equation}
an alternative proof is found in app.~\ref{AppNonLinIsom}. To construct the T-dual model we have to determine the matrix $i_{K_{\L}}V^{A}$, find a invertible minor, and check eq.~(\ref{TDconstraint2}). We obtain
\begin{equation}\label{OSp12matrx}
i_{K_{\left( \alpha \right)}}V^{A} = \left(\begin{array}{cc}
\left(1+\frac{3}{4}\t_{1}\t_{2} \right) &  0 \\
 0 &  \left(1+\frac{3}{4}\t_{1}\t_{2} \right)
\end{array}\right)\ ,
\end{equation}
and this matrix is invertible, so we do not need to find a minor. Otherwise, the (\ref{TDconstraint2}) becomes
\begin{equation}
 \Bigg\{\begin{array}{c}
  \t_{2}\dd\t_{1}\wedge\dd\t_{1}+\t_{1}\dd\t_{1}\wedge\dd\t_{2}=0\phantom{\big |}\ ,\\
 \t_{1}\dd\t_{2}\wedge\dd\t_{2}+\t_{2}\dd\t_{1}\wedge\dd\t_{2}=0\phantom{\big |}\ ,
\end{array}
\end{equation}
and these two condition are not in general true. 
So, the dual model can not be constructed in the conventional way. In the following we will show a way to bypass this step by first linearizing the isometries and then by gauging them.


\subsection{$OSp(2|2)/SO(2)\times Sp(2)$}\label{obstructionosp22}

The action for this new coset space is derived in sec. \ref{appOsp22}
\begin{equation}\label{OSP22modellononduale}
 S=8\int_{\Sigma}\Big\{
\dd\t_{1}\wa\dd\t_{4}-\dd\t_{2}\wa\dd\t_{3}-4\t_{3}\t_{4}\dd\t_{1}\wa\dd\t_{2}
\Big\}\ ,
\end{equation}
notice that there are two translational isometries, referred to $\t_{1}$ and $\t_{2}$
\begin{equation}
 \t_{1}\ra \t_{1}+\l_{1}\quad\quad\textrm{and}\quad\quad \t_{2}\ra \t_{2}+\l_{2}\ ,
\end{equation}
where $\l$ is a \textit{fermionic} parameter (i.e. $\l\t=-\t\l$). To find the dual model we then use the first procedure. First of all, we introduce the gauge field $A_{i}$ via the covariant derivative, obtaining
\begin{eqnarray}\label{TDIOSP22action}
S'&=&8\int_{\S}\Big\{
\left(\dd\t_{1}+A_{1}\right)\wa\dd\t_{4}-\left(\dd\t_{2}+A_{2}\right)\wa\dd\t_{3}+
{}\nonumber\\
	&& {}
\phantom{\bigg |}
-4\t_{3}\t_{4}\left(\dd\t_{1}+A_{1}\right)\wa\left(\dd\t_{2}+A_{2}\right)
\Big\}\ .
\end{eqnarray}
The new action is then invariant under a local transformation which allows us to choose the gauge
\begin{equation}
 \t_{1}=\t_{2}=0\ .
\end{equation}
After doing this, we introduce the $2$-forms and the Lagrange multipliers $\tilde\t_{i}$
\begin{equation}\label{TDIOSP22actionT}
S''=8\int_{\S}\Big\{
A_{1}\wa\dd\t_{4}-A_{2}\wa\dd\t_{3}-4\t_{3}\t_{4}A_{1}\wa A_{2}+\tilde\t_{1}\dd A_{2}+\tilde\t_{2}\dd A_{1}\Big\}\ .
\end{equation}
Now we calculate the equation of motion for $A_{1}$, obtaining
\begin{equation}
 4\t_{3}\t_{4}\ast A_{2}=\ast\dd\t_{4}-\dd\tilde\t_{2}\ .
\end{equation}
We have to factor $A_{2}$ but this is not possible, considering that $\theta_{3}\theta_{4}$ is not invertible. This problem hinders the construction of the dual model. 

However we can modify the original action (\ref{OSP22modellononduale}) in this way
\begin{equation}\label{SMOSP22ferm0001}
 \hat{S}\propto \lim_{\e\ra0}\int_{\Sigma}\Big\{
\dd\t_{1}\wa\dd\t_{4}-\dd\t_{2}\wa\dd\t_{3}-(\e+4\t_{3}\t_{4})\dd\t_{1}\wa\dd\t_{2}
\Big\}\ .
\end{equation}
In the limit $\e\ra0$ this action is equivalent to the original, but in this form we are able to invert the equations of motion. The results are
\begin{equation}\label{OSp22eqA1}
  A_{2}=\frac{1}{\e+4\t_{3}\t_{4} }\left[\dd\t_{4} +\frac{1}{\textrm{det}\g}\ast\dd\tilde\t_{2}\right] \ ,
\end{equation}
and
\begin{equation}\label{OSp22eqA2}
  A_{1}=\frac{1}{\e+4\t_{3}\t_{4} }\left[\dd\t_{3} -\frac{1}{\textrm{det}\g}\ast\dd\tilde\t_{1}\right] \ .
\end{equation}
Through the substitution of  these equations in (\ref{SMOSP22ferm0001}), we obtain the dual model
\begin{equation}\label{OSP22modelloduale}
   \hat{S}_{T}\propto \lim_{\e\ra0}\int_{\S}\frac{1}{\e+4\t_{3}\t_{4} }\Big\{
\dd\t_{3}\wa\dd\t_{4}-\frac{1}{\textrm{det}\g}\dd\tilde\t_{1}\wa\dd\tilde\t_{2}-\dd\tilde\t_{1}\wedge\dd\t_{4}-\dd\tilde\t_{2}\wedge\dd\t_{3}
\Big\}\ .
\end{equation}
In order to analyze the connection between the actions, we compute the curvature for both models. From the torsion equation we derive the spin connection
\begin{equation}\label{torsion}
 \dd V^{A}-\k_{BC}\hat{\w}^{AB}\wedge V^{C}=0\ ,
\end{equation}
then, from the definition of the curvature $2$-form,  we obtain the curvature components
\begin{equation}\label{curvature}
 R^{AB}=\dd\hat{\w}^{AB}-\k_{CD}\hat{\w}^{AC}\wedge\hat{\w}^{DB}\ .
\end{equation}
However, the results obtained for the dual model depend on the term $\frac{1}{\varepsilon+\theta\theta}$, for instance
\begin{equation}
 R^{11}=-4\frac{1}{(\e+4\t_{3}\t_{4})^{3}}\t_{3}\t_{4}\dd\t_{1}\wedge\dd\t_{1}\ .
\end{equation}
This shows that a physical quantities such as the curvature of the dual model is not defined for $\varepsilon\rightarrow0$.


\section{New Methods for T-duality}

Now, we decided to go for another path. Since in general the holonomic coordinates can not be found, we use  de la Ossa-Quevedo method \cite{de la Ossa:1992vc}, which is suitable for non-abelian T-dualities, for constructing the T-dual. 
Therefore we add new gauge fields and we perform the integration of them as suggested in \cite{de la Ossa:1992vc}.

We would like to mention that a possible issue for non-abelian T-duality is the non-equivalence of the actions at the quantum level  \cite{Elitzur:1994ri}.

Another  important point is that the model for the coset space is written in terms of a given parametrization. 
If some isometries act non linearly, we might encounter several problems in the duality construction.
To avoid them we choose a new set of coordinates subject to some algebraic equations (Pl\"ucker relations \cite{GriffithsHarris}) in terms of which 
the original model can be written. In this way the isometries act linearly and therefore they can be easily gauged in the conventional way \cite{Mirrsimm200308}.


\subsection{BRST Transformations for $OSp(1|2)$}

Consider the following lagrangian density
\begin{eqnarray}
&&
\mathcal{L}_{0}=
-\nabla\phi\bar\nabla\phi
+\varepsilon_{\alpha\beta}\nabla\theta^{\alpha}\bar\nabla\theta^{\beta}
+\alpha\left( \phi^{2}-\theta^{2}-1 \right)\ .
\label{L1}
\end{eqnarray}
The covariant derivatives are defined as
\begin{eqnarray}
&&
\nabla\theta^{\alpha}=\partial\theta^{\alpha}-A^{\alpha}\phi-A^{\alpha}_{\phantom{\alpha}\beta}\theta^{\beta}
\label{covder}\ ,
\nonumber\\&&
\nabla\phi=\partial\phi-A^{\alpha}\theta_{\alpha}\ .
\end{eqnarray}
The equation of motion for $\alpha$ 
reduces the lagrangian to the usual form (\ref{OSp12sigmaT}).
Notice that we use $
A_{\alpha}=\varepsilon_{\alpha\rho}A^{\rho}
$ and $
A^{\alpha}=\varepsilon^{\alpha\rho}A_{\rho}
$ as raising-lowering convection. 
The nilpotence of BRST operator $s$ implies the following BRST transformations
\begin{eqnarray}
&&
s\theta^{\alpha}
=\eta^{\alpha}\phi
+c^{\alpha\gamma}\varepsilon_{\gamma\beta}\theta^{\beta}\ ,
\nonumber\\&&
s\phi
=\eta^{\alpha}\theta_{\alpha}\ ,
\nonumber\\&&
s\eta^{\alpha}
=c^{\alpha\beta}\varepsilon_{\beta\gamma}\eta^{\gamma}\ ,
\nonumber\\&&
sc^{\alpha\beta}
=-\eta^{a}\eta^{\beta}
+c^{\alpha\rho}\varepsilon_{\rho\sigma}c^{\sigma\beta}\ ,
\label{BRST1}
\end{eqnarray} 
where the ghosts denoted by a latin letter are anticommuting while those denoted by a greek letter are commuting quantities.
Last, they have the following symmetries
\begin{eqnarray}
&&
c_{\alpha\beta}=c_{\beta\alpha}\quad\quad\quad\quad A_{\alpha\beta}=A_{\beta\alpha}\ .
\label{symmCon}
\end{eqnarray}
We  fix the transformations rules for the gauge fields by requiring the covariance of covariant derivatives
\begin{eqnarray}
&&
s\left( \nabla\theta^{\alpha} \right)
=\eta^{\alpha}\nabla \phi-c^{\alpha}_{\phantom{\alpha}\beta}\nabla\theta^{\beta}
=\eta^{\alpha}\nabla \phi+c^{\alpha\beta}\varepsilon_{\beta\gamma}\nabla\theta^{\gamma}\ ,
\nonumber\\&&
s\left( \nabla\phi \right)
=\eta^{\alpha}\nabla\theta_{\alpha}
=\eta^{\alpha}\varepsilon_{\alpha\beta}\nabla\theta^{\beta}\ .
\label{scovder}
\end{eqnarray}
From the second equation of (\ref{scovder}) we obtain
\begin{eqnarray}
s A^{\alpha}&=& 
\partial\eta^{\alpha}
+A^{\gamma}\varepsilon_{\gamma\rho}c^{\rho\alpha}
-\eta^{\rho}\varepsilon_{\rho\sigma}A^{\sigma\alpha}
=\nonumber\\&=& 
\partial\eta^{\alpha}
+c^{\alpha\beta}\varepsilon_{\beta\gamma}A^{\gamma}
+A^{\alpha\rho}\varepsilon_{\rho\sigma}\eta^{\sigma}\ ,
\label{sA1}
\end{eqnarray}
and from $s^{2}A^{\alpha}=0$ we get
\begin{equation}
sA^{\alpha\beta}=
-\partial c^{\alpha\beta}
-\eta^{\alpha}A^{\beta}
-A^{\alpha}\eta^{\beta}
+c^{\alpha\lambda}\varepsilon_{\lambda\gamma}A^{\gamma\beta}
-A^{\alpha\lambda}\varepsilon_{\lambda\rho}c^{\rho\beta}\ .
\label{sA2}
\end{equation}
We define the following field strengths
\begin{eqnarray}
	&&
F^{\alpha}=
\partial\bar A^{\alpha}-\bar\partial A^{\alpha}
+A^{\alpha\beta}\varepsilon_{\beta\gamma}\bar A^{\gamma}
-\bar A^{\alpha\beta}\varepsilon_{\beta\gamma} A^{\gamma}\ ,
\nonumber\\&&
F^{\alpha\beta}=
\partial\bar A^{\alpha\beta}-\bar\partial A^{\alpha\beta}
+A^{\alpha\gamma}\varepsilon_{\gamma\delta}\bar A^{\delta\beta}
-\bar A^{\alpha\gamma}\varepsilon_{\gamma\delta} A^{\delta\beta}
+A^{\alpha}\bar A^{\beta}
-\bar A^{\alpha}A^{\beta}\ ,
	\label{FieldStrenght0}
\end{eqnarray}
which transform as follows
\begin{eqnarray}
	&&
	s F^{\alpha}=
	c^{\alpha\beta}\varepsilon_{\beta\gamma}F^{\gamma}
	+F^{\alpha\beta}\varepsilon_{\beta\gamma}\eta^{\gamma}\ ,
	\nonumber\\&&
	s F^{\alpha\beta}=
	c^{\alpha\gamma}\varepsilon_{\gamma\delta}F^{\delta\beta}
	-F^{\alpha\gamma}\varepsilon_{\gamma\delta}c^{\delta\beta}
	-\eta^{\alpha}F^{\beta}-F^{\alpha}\eta^{\beta}\ ,
	\label{FieldStrenghtTransf}
\end{eqnarray}


\subsection{Performing T-duality}

In order to construct the T-dual model we consider the gauged form of lagrangian (\ref{L1})
\begin{eqnarray}
\mathcal{L}_{gauging}&=&
-\nabla\phi\bar\nabla\phi
+\varepsilon_{\alpha\beta}\nabla\theta^{\alpha}\bar\nabla\theta^{\beta}
+\alpha\left( \phi^{2}-\theta^{2}-1 \right)
+i\tilde\theta^{\alpha}\varepsilon_{\alpha\beta}F^{\beta}
+i\tilde\phi^{\alpha\beta}\varepsilon_{\beta\gamma}\varepsilon_{\alpha\delta}F^{\gamma\delta}=
\nonumber\\&=&
\alpha\left( \phi^{2}-\theta^{2}-1 \right)
+\left( \partial\phi-A^{\alpha}\varepsilon_{\alpha\beta}\theta^{\beta} \right)\left( \bar\partial\phi-\bar A^{\gamma}\varepsilon_{\gamma\delta}\theta^{\delta} \right)+
\nonumber\\&&+
\left( \partial\theta^{\alpha}-A^{\alpha}\phi+A^{\alpha\beta}\varepsilon_{\beta\gamma}\theta^{\gamma} \right)
\varepsilon_{\alpha\beta}
\left( \bar\partial\theta^{\beta}-\bar A^{\beta}\phi+\bar A^{\beta\gamma}\varepsilon_{\gamma\delta}\theta^{\delta} \right)
+\nonumber\\&&+
i\tilde\theta^{\alpha}\varepsilon_{\alpha\beta}
\left( 
\partial\bar A^{\beta}-\bar\partial A^{\beta}
+A^{\beta\rho}\varepsilon_{\rho\gamma}\bar A^{\gamma}
-\bar A^{\beta\rho}\varepsilon_{\rho\gamma} A^{\gamma} \right)
+\nonumber\\&&+
i\tilde\phi^{\alpha\beta}\varepsilon_{\beta\gamma}\varepsilon_{\alpha\delta}
\left( \partial\bar A^{\gamma\delta}-\bar\partial A^{\gamma\delta}
+A^{\gamma\rho}\varepsilon_{\rho\sigma}\bar A^{\sigma\delta}
-\bar A^{\gamma\rho}\varepsilon_{\rho\sigma} A^{\sigma\delta}
+A^{\gamma}\bar A^{\delta}
-\bar A^{\gamma}A^{\delta} \right)\ .
\nonumber\\&&
\label{L2}
\end{eqnarray}
Notice we gauged the whole isometry group $OSp\left( 1|2 \right)$. Following the procedure described in \cite{de la Ossa:1992vc} we rewrite (\ref{L2}) as
\begin{eqnarray}
\mathcal{L}_{gauging}&=&
\mathcal{L}_{0}+\left( h^{\alpha}+f^{\alpha\beta}A_{\beta}+g^{\alpha\left( \beta\gamma \right)}A_{\beta\gamma} \right)\bar A_{\alpha}+
\nonumber\\&&
+\left( l^{\left( \alpha\beta \right)}+m^{\left( \alpha\beta \right)\rho}A_{\rho}+n^{\left( \alpha\beta \right)\left( \rho\sigma \right)}A_{\rho\sigma} \right)\bar A_{\alpha\beta}+
\nonumber\\&&
+\bar h^{\alpha} A_{\alpha}+\bar l^{\left( \alpha\beta \right)} A_{\alpha\beta}\ ,
\label{L3}
\end{eqnarray}
where
\begin{eqnarray}
&&
\begin{array}{lcl}
h^{\alpha}=-\partial\left( \phi\theta^{\alpha} \right)-i\partial\tilde\theta^{\alpha}\ ,
& 
&
l^{\left( \alpha\beta \right)}=\partial\theta^{\left( \alpha \right.}\theta^{\left. \beta \right)}-i\partial\tilde\phi^{\alpha\beta}\ ,
 \\
&&\\
f^{\alpha\beta}=\phi^{2}\varepsilon^{\alpha\beta}+\theta^{\alpha}\theta^{\beta}+2i\tilde\phi^{\alpha\beta}\ ,
& 
&
m^{\left( \alpha\beta \right)\rho}=\varepsilon^{\left( \alpha | \rho \right.}\theta^{\left. \beta \right)}\phi+i\tilde\theta^{\left( \alpha \right.}\varepsilon^{\left. \beta \right)\rho}\ ,
\\
&&\\
g^{\alpha\left( \beta\gamma \right)}=i\varepsilon^{\alpha\left( \beta \right.}\tilde\theta^{\left. \gamma \right)}-\varepsilon^{\alpha\left( \beta \right.}\theta^{\left. \gamma \right)}\phi\ ,
&
&
n^{\left( \alpha\beta \right)\left( \rho\sigma \right)}=-i\tilde\phi^{\alpha\left( \rho \right.}\varepsilon^{\left. \sigma \right)\beta}-i\tilde\phi^{\beta\left( \rho \right.}\varepsilon^{\left. \sigma \right)\alpha}\ ,
\\
&&
\\
\bar h^{\alpha}=-\bar\partial\left( \phi\theta^{\alpha} \right)+i\bar\partial\tilde\theta^{\alpha}\ ,
& 
&
\bar l^{\left( \alpha\beta \right)}=\bar\partial\theta^{\left( \alpha \right.}\theta^{\left. \beta \right)}+i\bar\partial\tilde\phi^{\alpha\beta}\ .
\end{array}
\label{hfglmn}
\end{eqnarray}
Notice that
\begin{eqnarray}
\varepsilon^{\alpha\beta}\varepsilon_{\beta\gamma}=\delta^{\alpha}_{\gamma}\ ,
\quad\quad\quad\quad
\delta^{\alpha\beta}\varepsilon_{\beta\gamma}=-\varepsilon^{\alpha\beta}\delta_{\beta\gamma}
\neq\delta^{\alpha}_{\gamma}\ .
\label{metricsnote}
\end{eqnarray}
This model has $8$ degrees of freedom: $2$ for $\theta^{\alpha}$, $1$ for $\phi$, $2$ for $\tilde\theta^{\alpha}$ and $3$ for $\tilde\phi^{\alpha\beta}$. By gauge fixing we eliminate some degrees of freedom among them. Notice that it is not possible to set a symmetric $2\times2$ tensor field to a constant by a $Sp\left( 2 \right)$-transformation: we can not set  all the three components of $\tilde\phi^{\alpha\beta}$ to a constant.  Nevertheless, we choose
\begin{equation}
\tilde\phi^{\alpha\beta}=\left( \textrm{det} \tilde\phi \right)^{\frac{1}{2}}\delta^{\alpha\beta}\ ,
\label{gf1}
\end{equation}
thus, only one degree of freedom survives and there is an $Sp(2)$-gauge isometry left. 
Now, we  set $\theta^{\alpha}$ to zero via the $OSp(1|2)/Sp(2)$ gauge transformation. Consequently, the constraint in $\mathcal{L}_{0}$ impose that $\phi=1$. This reduces the degrees of freedom from $8$ to $3$. The remained fields are
\begin{equation}
\textrm{det}\tilde\phi
,\quad\quad\quad\quad
\tilde\theta^{\alpha}\ ,
\label{remfields}
\end{equation}
and we have a one-parameter residual $Sp\left( 2 \right)$ symmetry. If we rename
\begin{equation}
\left( \textrm{det} \tilde\phi \right)^{\frac{1}{2}}=\hat\phi\ ,
\label{redefdet}
\end{equation}
definitions (\ref{hfglmn}) become
\begin{eqnarray}
&&
\begin{array}{lcl}
h^{\alpha}=-i\partial\tilde\theta^{\alpha}\ ,
& 
&
l^{\left( \alpha\beta \right)}=-i\partial\hat\phi\delta^{\alpha\beta}\ ,
 \\
&&\\
f^{\alpha\beta}=\varepsilon^{\alpha\beta}+2i\hat\phi\delta^{\alpha\beta}\ ,
& 
&
m^{\left( \alpha\beta \right)\rho}=+i\tilde\theta^{\left( \alpha \right.}\varepsilon^{\left. \beta \right)\rho}\ ,
\\
&&\\
g^{\alpha\left( \beta\gamma \right)}=i\varepsilon^{\alpha\left( \beta \right.}\tilde\theta^{\left. \gamma \right)}\ ,
&
&
n^{\left( \alpha\beta \right)\left( \rho\sigma \right)}=-i\hat\phi\delta^{\alpha\left( \rho \right.}\varepsilon^{\left. \sigma \right)\beta}-i\hat\phi\delta^{\beta\left( \rho \right.}\varepsilon^{\left. \sigma \right)\alpha}\ ,
\\
&&
\\
\bar h^{\alpha}=+i\bar\partial\tilde\theta^{\alpha}\ ,
& 
&
\bar l^{\left( \alpha\beta \right)}=+i\bar\partial\hat\phi\delta^{\alpha\beta}\ .
\end{array}
\label{hfglmn1}
\end{eqnarray}
To derive the T-dual model we compute from (\ref{L3}) the equation of motion for $\bar A_{\alpha}$, obtaining
\begin{eqnarray}
	&&
	A_{\beta}=
	-[f^{-1}]_{\beta\lambda}\left( h^{\lambda}+g^{\lambda\left( \rho\sigma \right)}A_{\rho\sigma} \right)\ .
	\label{EoMA1}
\end{eqnarray}
The lagrangian becomes 
\begin{eqnarray}
	\mathcal{L}_{gauging}&=& 
	\mathcal{L}_{0}+
	\left[ l^{\left( \alpha\beta \right)} 
	-m^{\left( \alpha\beta \right)\rho}\left[ f^{-1} \right]_{\rho\lambda}h^{\lambda}
	+
	\right.\nonumber\\&&\left.
	+\left( 
	n^{\left( \alpha\beta \right)\left( \rho\sigma \right)	}
	-m^{\left( \alpha\beta \right)\zeta}\left[ f^{-1} \right]_{\zeta\lambda}
	g^{\lambda\left( \rho\sigma \right)}
	\right)A_{\rho\sigma}	
	\right]\bar A_{\alpha\beta}+
	\nonumber\\&&
	-\bar h^{\alpha}\left[ f^{-1} \right]_{\alpha\lambda}h^{\lambda}
	+\left[ \bar l^{\left( \rho\sigma \right)}- \bar h^{\alpha}\left[ f^{-1} \right]_{\alpha\lambda}g^{\lambda\left( \rho\sigma \right)}\right]A_{\rho\sigma}\ .	
		\label{L4}
\end{eqnarray}
That is
\begin{equation}
	\mathcal{L}_{A\bar A}=\Xi+\bar\Omega^{\left( \alpha\beta \right)}A_{\alpha\beta}+\Omega^{\left( \alpha\beta \right)}\bar A_{\alpha\beta}+\Pi^{\left( \alpha\beta \right)\left( \rho\sigma \right)}A_{\alpha\beta}\bar A_{\rho\sigma}\ ,
	\label{L5}
\end{equation}
where
\begin{eqnarray}
	\Xi&=& 	-\bar h^{\alpha}\left[ f^{-1} \right]_{\alpha\lambda}h^{\lambda}
		=
		-\frac{1}{1-4\hat\phi^2}\partial\tilde\theta^{\alpha}\varepsilon_{\alpha\beta}\partial\tilde\theta^{\beta}\ ,
		\nonumber\\&&
		\nonumber\\
		\Omega^{\left( \alpha\beta \right)}&=& 
		 l^{\left( \alpha\beta \right)} 
		-m^{\left( \alpha\beta \right)\rho}\left[ f^{-1} \right]_{\rho\lambda}h^{\lambda}=
		\nonumber\\&=& 
		-i\partial\hat\phi\delta^{\alpha\beta}-\frac{1}{1-4\hat\phi^{2}}\left( 
		\tilde\theta^{\left( \alpha \right.}\partial\tilde\theta^{\left. \beta \right)}
		+2i\hat\phi
		\tilde\theta^{\left( \alpha \right.}\varepsilon^{\left. \beta \right)\rho}\delta_{\rho\lambda}\partial\tilde\theta^{\lambda}
			 \right)\ ,
		\nonumber\\&&
			 \nonumber\\
		\Pi^{\left( \alpha\beta \right)\left( \rho\sigma \right)}&=& 
		 n^{\left( \alpha\beta \right)\left( \rho\sigma \right)}-m^{\left( \alpha\beta \right)\rho}\left[ f^{-1} \right]_{\rho\sigma}g^{\sigma\left( \rho\sigma \right)} =
\nonumber\\&=& 
\frac{\tilde\theta^{2}}{4\left( 1-4\hat\phi^{2} \right)}
\left[ 
\varepsilon^{\alpha\sigma}\varepsilon^{\beta\rho}
+\varepsilon^{\alpha\rho}\varepsilon^{\beta\sigma}\right]+
\nonumber\\&&+
i\hat\phi\frac{ 1-4\hat\phi^{2}-\tilde\theta^{2}}{4\left( 1-4\hat\phi^{2} \right)} 
\left[ \varepsilon^{\alpha\sigma}\delta^{\beta\rho}+\varepsilon^{\alpha\rho}\delta^{\beta\sigma}+\varepsilon^{\beta\sigma}\delta^{\alpha\rho}+\varepsilon^{\beta\rho}\delta^{\alpha\sigma} \right]\ .
	\label{CapGredef1}
\end{eqnarray}
Now, we can find the equation of motion (EoM) of the last gauge fields. Substituting it back into the lagrangian  gives the T-dual model. To do this we have to compute the inverse of $\Pi$. Consider the following $4$-indices tensor
\begin{equation}
T^{\left( \alpha\beta \right)\left( \gamma\delta \right)}=
A
\left[ 
\varepsilon^{\alpha\delta}\varepsilon^{\beta\gamma}
+\varepsilon^{\alpha\gamma}\varepsilon^{\beta\delta}\right]+B
\left[ \varepsilon^{\alpha\delta}\delta^{\beta\gamma}+\varepsilon^{\alpha\gamma}\delta^{\beta\delta}+\varepsilon^{\beta\delta}\delta^{\alpha\gamma}+\varepsilon^{\beta\gamma}\delta^{\alpha\delta} \right]
\ .\label{findInv1v2nt}
\end{equation}
To find its inverse we impose the following definition of inverse tensor
\begin{eqnarray}
	\left[ M^{-1} \right]_{\alpha\beta\,\,\mu\nu}
	M^{\alpha\beta\,\,\rho\sigma}
	&=& \varepsilon_{\left( \mu \right.}^{\phantom{\mu}\left( \rho \right.}\varepsilon_{\left.\nu \right)}^{\phantom{\nu}\left.\sigma \right)}
	\ ,\label{osp12inverseDEF}
\end{eqnarray}
and then we can fix the coefficient of the following generic tensor
\begin{eqnarray}
\left[ T^{-1} \right]_{\left( \alpha\beta \right)\left( \gamma\delta \right)}&=& L
\left[ 
\varepsilon_{\alpha\delta}\varepsilon_{\beta\gamma}
+\varepsilon_{\alpha\gamma}\varepsilon_{\beta\delta}\right]
+
P
\left[ 
\delta_{\alpha\delta}\delta_{\beta\gamma}
+\delta_{\alpha\gamma}\delta_{\beta\delta}\right]
+\nonumber\\&&+
M
\left[ \varepsilon_{\alpha\delta}\delta_{\beta\gamma}+\varepsilon_{\alpha\gamma}\delta_{\beta\delta}+\varepsilon_{\beta\delta}\delta_{\alpha\gamma}+\varepsilon_{\beta\gamma}\delta_{\alpha\delta} \right]\equiv
\nonumber\\&\equiv& 
L<\varepsilon\varepsilon>+M<\varepsilon\delta>+P<\delta\delta>
\ .\label{findInv2v2nt}
\end{eqnarray}
We find that
\begin{eqnarray}
L=\frac{A^{2}+2B^{2}}{A\left( A^{2}+4B^{2} \right)}
\ ,\quad
M=\frac{B}{ A^{2}+4B^{2}}
\ ,\quad
P=\frac{2B^{2}}{A\left( A^{2}+4B^{2} \right)}
\ .
\label{findInv3v2nt}
\end{eqnarray}
Here,  $A\propto\tilde\theta^{2}$, then it is impossible to invert.


\subsection{Residual Gauge Fixing}

We want to fix the residual gauge invariance, via BRST method: we introduce a set of lagrangian multipliers $b_{\alpha\beta}$  and the corresponding ghosts $\bar c_{\alpha\beta}$ such that
\begin{eqnarray}
s\bar c_{\alpha\beta}=b_{\alpha\beta},\quad\quad\quad\quad sb_{\alpha\beta}=0\ .
\label{BRSTgf1nt}
\end{eqnarray}
Notice that the metric in this model is $\varepsilon_{\alpha\beta}$ and  we use it to raise and lower the indices.
To fix the gauge we introduce a new term in (\ref{L5})
\begin{eqnarray}
\mathcal{L}_{A\bar A}\rightarrow \mathcal{L}_{g.f.}&=& \mathcal{L}_{A\bar A}+s\left[ \bar c_{\alpha\beta}\varepsilon^{\left( \alpha\beta \right)\left( \rho\sigma \right)}A_{\rho\sigma}+\frac{1}{2\xi}\bar c_{\alpha\beta}\varepsilon^{\left( \alpha\beta \right)\left( \rho\sigma \right)}b_{\rho\sigma}+h.c. \right]=
\nonumber\\&=& 
\mathcal{L}_{A\bar A}
+
b_{\alpha\beta}\varepsilon^{\left( \alpha\beta \right)\left( \rho\sigma \right)}A_{\rho\sigma}
+
\bar b_{\alpha\beta}\varepsilon^{\left( \alpha\beta \right)\left( \rho\sigma \right)}\bar A_{\rho\sigma}+
\nonumber\\&&
+\frac{1}{2\xi}\bar b_{\alpha\beta}\varepsilon^{\left( \alpha\beta \right)\left( \rho\sigma \right)}b_{\rho\sigma}
+f\left( \left\{ c \right\} \right) \ ,
\label{gfL2nt}
\end{eqnarray}
where
\begin{eqnarray}
\varepsilon^{\left( \alpha\beta \right)\left( \rho\sigma \right)}
=
\frac{1}{2}\left(
\varepsilon^{\alpha\sigma}\varepsilon^{\beta\rho}
+\varepsilon^{\alpha\rho}\varepsilon^{\beta\sigma}
  \right)\ .
\label{defepsSymnt}
\end{eqnarray}
We collect the ghost term into the symbol $f\left( \left\{ c \right\} \right) $.
Computing the EoM for $b$ and $\bar b$ we obtain
\begin{equation}
b_{\alpha\beta}=-\xi\bar A_{\alpha\beta}\ ,
\quad\quad\quad\quad
\bar b_{\alpha\beta}=-\xi A_{\alpha\beta}\ .
\label{bbarbnt}
\end{equation}
Then we have
\begin{eqnarray}
\mathcal{L}_{g.f.}&=& 
\Xi+\bar\Omega^{\left( \alpha\beta \right)}A_{\alpha\beta}+\Omega^{\left( \alpha\beta \right)}\bar A_{\alpha\beta}
+
\nonumber\\&&
+\left[ \Pi^{\left( \alpha\beta \right)\left( \rho\sigma \right)}-\xi\varepsilon^{\left( \alpha\beta \right)\left( \rho\sigma \right)} \right]A_{\alpha\beta}\bar A_{\rho\sigma}
+f\left( \left\{ c \right\} \right)\ .
\label{gfL3nt}
\end{eqnarray}
Defining $\left[ \Pi^{\left( \alpha\beta \right)\left( \rho\sigma \right)}-\xi\varepsilon^{\left( \alpha\beta \right)\left( \rho\sigma \right)} \right]=\hat\Pi^{\left( \alpha\beta \right)\left( \rho\sigma \right)}$, we get 
\begin{equation}
\mathcal{L}_{g.f.}= \Xi+\bar\Omega^{\left( \alpha\beta \right)}A_{\alpha\beta}+\Omega^{\left( \alpha\beta \right)}\bar A_{\alpha\beta}
+ \hat\Pi^{\left( \alpha\beta \right)\left( \rho\sigma \right)}A_{\alpha\beta}\bar A_{\rho\sigma}+f\left( \left\{ c \right\} \right)\ .
\label{gfL4nt}
\end{equation}
Now, we fix $\xi$ to make $\hat\Pi$  invertible.
Comparing with (\ref{CapGredef1}) we have
\begin{eqnarray}
A=\frac{\tilde\theta^{2}}{4\left( 1-4\hat\phi^{2} \right)}+\xi
,\quad\quad\quad\quad
B=i\hat\phi\frac{1-4\hat\phi^{2}-\tilde\theta^{2}}{4\left( 1-4\hat\phi^{2} \right)}\ ,
\label{def00}
\end{eqnarray}
or, more simply
\begin{eqnarray}
A=\frac{\tilde\theta^{2}+\xi'}{4\left( 1-4\hat\phi^{2} \right)}
,\quad\quad\quad\quad
B=i\hat\phi\frac{1-4\hat\phi^{2}-\tilde\theta^{2}}{4\left( 1-4\hat\phi^{2} \right)}\ .
\label{def01}
\end{eqnarray}
Then
\begin{equation}
L=\frac{4\left( 1-4\hat\phi^2 \right)\left( 2\hat\phi^{2}\left( -1+4\hat\phi^{2}+\theta^{2} \right)^{2}+\frac{\left( \theta^{2}+\xi \right)^{2}}{16\left( 1-4\hat\phi^{2} \right)^{2}} \right)}{\left( \theta^{2}+\xi \right)\left( 4\hat\phi^{2}\left( -1+4\hat\phi^{2}+\theta^{2} \right)^{2} +\frac{\left( \theta^{2}+\xi \right)^{2}}{16\left( 1-4\hat\phi^{2} \right)^{2}}\right)}
\ ,\label{coefL1}
\end{equation}
\begin{equation}
M=\frac{i\hat\phi\left( 1-4\hat\phi^{2}-\theta^{2} \right)}{4\hat\phi^{2}\left( -1+4\hat\phi^{2}+\theta^{2} \right)^{2}+\frac{\left( \theta^{2}+\xi \right)^{2}}{16\left( 1-4\hat\phi^{2} \right)^{2}}}
\ ,\label{coefM1}
\end{equation}
\begin{equation}
P=\frac{8\hat\phi^{2}\left( 1-4\hat\phi^{2} \right)\left( -1+4\hat\phi^{2}+\theta^{2} \right)^{2}}{\left( \theta^{2}+\xi \right)\left( 4\hat\phi^{2}\left( -1+4\hat\phi^{2}+\theta^{2} \right)^{2} +\frac{\left( \theta^{2}+\xi \right)^{2}}{16\left( 1-4\hat\phi^{2} \right)^{2}}\right)}
\ .\label{coefP1}
\end{equation}
The EoM for $\bar A_{\rho\sigma}$ is then
\begin{eqnarray}
A_{\alpha\beta}&=& 
-\left[ \hat\Pi^{-1} \right]_{\left( \alpha\beta \right)\left( \rho\sigma \right)}\Omega^{\left( \rho\sigma \right)}
\ .\label{EoMbarA}
\end{eqnarray}
Finally the dual lagrangian is
\begin{eqnarray}
\mathcal{L}_{dual}&=& \Xi-\bar\Omega^{\left(\alpha\beta\right)}
\left[ \hat\Pi^{-1} \right]_{\left( \alpha\beta \right)\left( \rho\sigma \right)}\Omega^{\left( \rho\sigma \right)}+f\left( \left\{ c \right\} \right)
\ .\label{DUAL1}
\end{eqnarray}
With simple algebraic manipulations (see app.~\ref{detailsosp12}), the dual lagrangian becomes
\begin{eqnarray}
\mathcal{L}_{dual}&=& \Xi- 
\bar\partial\hat\phi\partial\hat\phi
-\frac{2\left( L+P \right)}{1-4\hat\phi^{2}}
\left[ 
-i\bar\partial\hat\phi\theta^{\alpha}\partial\theta^{\beta}\delta_{\alpha\beta}
+i\theta^{\alpha}\bar\partial\theta^{\beta}\delta_{\alpha\beta}\partial\hat\phi
+\right.\nonumber\\&&
\left.+
2\hat\phi\bar\partial\hat\phi \theta^{\alpha}\partial\theta^{\beta}\varepsilon_{\alpha\beta}+
2\theta^{\alpha}\bar\partial\theta^{\beta}\varepsilon_{\alpha\beta}\hat\phi\partial\phi
 \right]+
\nonumber\\&&
-\frac{\theta^{2}}{2\left( 1-4\hat\phi^{2} \right)^{2}}
\left[ 
\bar\partial\theta^{\alpha}\partial\theta^{\beta}\delta_{\alpha\beta}\left( -4M\left( 1+4\hat\phi^{2} \right)
+4i\hat\phi\left( 3L-P \right)
\right)
+\right.\nonumber\\&&+\left.
\bar\partial\theta^{\alpha}\partial\theta^{\beta}\varepsilon_{\alpha\beta}
\left(- \left( 3L-P \right)\left( 1+4\hat\phi^{2} \right)-8iM\hat\phi \right)
\right]+f\left( \left\{ c \right\} \right)
\ .\label{grosso2}
\end{eqnarray}
Notice that exist just two combinations of $L$ and $P$. Using (\ref{findInv3v2nt}), we have
\begin{eqnarray}
L+P
&=&
\frac{1}{A}=\frac{4\left( 1-4\hat\phi^{2} \right)}{\theta^{2}+\xi}\ ,
\nonumber\\
3L-P&=& 
 \frac{1}{A}+\frac{2A}{A^{2}+4B^{2}}=
\nonumber\\&=& 
\frac{4\left( 1-4\hat\phi^{2} \right)\left( \theta^{2}+\xi \right)}{\left( \theta^{2}+\xi \right)^{2}+4\hat\phi^{2}\left( 1-4\hat\phi^{2}-\theta^{2} \right)}+\frac{4\left( 1-4\hat\phi^{2} \right)}{\theta^{2}+\xi}\ .
\label{LPred}
\end{eqnarray}
The form of the lagrangian is rather cumbersome and therefore it might be rather awful to proceed with a loop analysis from this expression. Of course, it can be expanded in power of $\hat\phi$,  suitable for $1$-loops analysis.




\subsection{Another Gauge Fixing for $OSp(m|n)/SO(n)\times Sp(m)$}

The method presented above can not be used in general: even in a slightly more extended example as $OSp(4|2)/SO(4)\times Sp(2)$ the computation becomes quite prohibitive. We  then found an alternative gauge fixing condition that leads to a simpler treatment. 

The coset model $OSp\left( m|n \right)/SO\left( n \right)\times Sp\left( m \right)$ is built from the following fields
\begin{itemize}
	\item $\Lambda_{\left( ij \right)}$ bosonic $SO\left( n \right)$ fields;
	\item $\Phi_{\left[ \alpha\beta \right]}$ antisymmetric $Sp\left( m \right)$ fields;
	\item $\Theta_{i\alpha}$ fermionic fields.
\end{itemize}
To these are associated ghost fields
\begin{itemize}
	\item $d_{\left[ ij \right]}$: fermionic $SO\left( n \right)$ ghosts;
	\item $c_{\left( \alpha\beta \right)}$: fermionic $Sp\left( m \right)$ ghosts;
	\item $\eta_{i\alpha}$: bosonic ghosts
\end{itemize}
The BRST transformations read
\begin{eqnarray}
	s\Theta_{i\alpha}
	&=& 
	c_{\alpha\beta}\varepsilon^{\beta\gamma}\Theta_{i\gamma}
	+
	d_{ij}\delta^{jk}\Theta_{k\alpha}
	+
	\eta_{i\beta}\varepsilon^{\beta\gamma}\Phi_{\gamma\alpha}
	+
	\eta_{j\alpha}\delta^{jk}\Lambda_{ki}\ ,
	\nonumber\\
	s\Lambda_{\left( ij \right)}
	&=& 
	\eta_{\left( i \right|\alpha}\varepsilon^{\alpha\beta}\Theta_{\left. j \right)\beta}
	+
	d_{\left( i \right|k}\delta^{kl}\Lambda_{l\left|j \right)}\ ,
	\nonumber\\
	s\Phi_{\left[ \alpha\beta \right]}
	&=& 
	\eta_{i\left[ \alpha \right.}\delta^{ij}\Theta_{j\left|\beta \right]}
	+
	c_{\left[ \alpha \right|\gamma}\varepsilon^{\gamma\delta}\Phi_{\delta\left|\beta \right]}\ .
	\label{ospmnBRSTfields}
\end{eqnarray}
In order to construct a gauged principal chiral model, we introduce the following gauge fields
\begin{itemize}
	\item $A_{\left[ ij \right]}$: antisymmetric $SO\left( n \right)$ gauge fields;
	\item $A_{\left( \alpha\beta \right)}$: symmetric $Sp\left( m \right)$ gauge fields;
	\item $A_{i\alpha}$: fermionic gauge fields.
\end{itemize} 
Their associated field strengths are
\begin{eqnarray}
	F_{\left[ ij \right]}
	&=& 
	\partial\bar A_{ij}
	-
	\bar\partial A_{ij}
	+
	A_{\left[ i \right| b}\delta^{bc} \bar A_{c\left|j \right]}
	+
	\nonumber\\&&
	-
	\bar A_{\left[ i \right| b}\delta^{bc} A_{c\left|j \right]}
	+
	A_{\left[ i \right| \alpha}\varepsilon^{\alpha\beta} \bar A_{\left.j \right]\beta}
	-
	\bar A_{\left[ i \right| \alpha}\varepsilon^{\alpha\beta} A_{\left.j \right]\beta}\ ,
	\nonumber\\
	F_{\left( \alpha\beta \right)}
	&=& 
	\partial\bar A_{\alpha\beta}
	-
	\bar\partial A_{\alpha\beta}
	+
	A_{\left(\alpha \right| \gamma}\varepsilon^{\gamma\delta} \bar A_{\delta\left|\beta \right)}
	+
	\nonumber\\&&	
	-
	\bar A_{\left(\alpha \right| \gamma}\varepsilon^{\gamma\delta} A_{\delta\left|\beta \right)}
	+
	A_{i\left( \alpha \right.}\delta^{ij}\bar A_{j\left|\beta \right)}
	-
	\bar A_{i\left( \alpha \right.}\delta^{ij} A_{j\left|\beta \right)}\ ,
	\nonumber\\
	F_{i\alpha}
	&=& 
	\partial\bar A_{i\alpha}
	-
	\bar\partial A_{i\alpha}
	+
	A_{ij}\delta^{jk}\bar A_{k\alpha}
		+
	\nonumber\\&&
	-
	\bar A_{ij}\delta^{jk} A_{k\alpha}
	+
	A_{\alpha\gamma}\varepsilon^{\gamma\delta}\bar A_{i\delta}
	-
	\bar A_{\alpha\gamma}\varepsilon^{\gamma\delta} A_{i\delta}\ .
	\label{ospmnF1}
\end{eqnarray}

\subsection{Construction Method}

The lagrangian for the coset model is constructed starting from the whole model $OSp\left( m|n \right)$ lagrangian. The supergroup representative $L$ is
\begin{equation}
	L=
	\left( 
	\begin{array}{cc}
		\Lambda^{i}_{\phantom{i}j}
		&
		\Theta^{i}_{\phantom{i}\alpha}
		\\
		\Theta^{\alpha}_{\phantom{\alpha}j}
		&
		\Phi^{\alpha}_{\phantom{\alpha}\beta}
	\end{array}
	\right)\ .
	\label{ospmnrepr1}
\end{equation}
The vielbein are obtained expanding $L^{-1}\partial L$ into the generators of the superalgebra $\mathfrak{osp}\left( m|n \right)$. Our final aim is the fermionic coset, so the vielbeins are the off-diagonal part of $L^{-1}\partial L$: $V^{a}_{\phantom{a}\alpha}$ and $V^{\alpha}_{\phantom{\alpha}a}$. We get
\begin{eqnarray}
	V^{a}_{\phantom{a}\alpha}
	&=& 
	A^{a}_{\phantom{a}i}\partial \Theta^{i}_{\phantom{i}\alpha}
	+
	B^{a}_{\phantom{a}\gamma}\partial \Phi^{\gamma}_{ \phantom{\gamma}\alpha}\ ,
	\nonumber\\
	V^{\alpha}_{\phantom{\alpha}a}
	&=& 
	C^{\alpha}_{\phantom{\alpha}i}\partial\Lambda^{i}_{\phantom{i}a}+
	D^{\alpha}_{\phantom{\alpha}\gamma}\partial\Theta^{\gamma}_{a}\ ,
	\label{ospmnViel1}
\end{eqnarray}
where
\begin{eqnarray}
	A^{a}_{\phantom{a}i}
	&=& 
	\left[ 
	\Lambda^{i}_{\phantom{a}j}
	-
	\Theta^{i}_{\phantom{a}\beta}
	\left[ \Phi^{\gamma}_{\phantom{\gamma}\beta} \right]^{-1}\Theta^{\gamma}_{\phantom{\gamma}j}
	\right]^{-1}\delta^{a}_{j}\ ,
	\nonumber\\
	B^{a}_{\phantom{a}\gamma}
	&=& 
	-A^{a}_{\phantom{a}i}\Theta^{i}_{\phantom{i}\beta}
	\left[ \Phi^{\alpha}_{\phantom{\alpha}\beta} \right]^{-1}\ ,
	\nonumber\\
	C^{\alpha}_{\phantom{\alpha}i}
	&=& 
	-D^{\alpha}_{\phantom{\alpha}\gamma}\Theta^{\gamma}_{\phantom{\gamma}j}
	\left[ \Lambda^{i}_{\phantom{i}j} \right]^{-1}\ ,
	\nonumber\\
	D^{\alpha}_{\phantom{\alpha}\beta}
	&=& 
		\left[ 
	\Phi^{\beta}_{\phantom{\beta}\gamma}
	-
	\Theta^{\beta}_{\phantom{\beta}j}
	\left[ \Lambda^{i}_{\phantom{i}j} \right]^{-1}\Theta^{i}_{\phantom{i}\gamma}
	\right]^{-1}\epsilon^{\alpha}_{\phantom{\alpha}\gamma}\ .
	\label{ospmnViel2}
\end{eqnarray}
The lagrangian for the coset is made of two pieces. The first one is the contraction of the vielbeins by the Killing metric and it produces the kinetic term for the fields $\Theta$, $\Lambda$ and $\Phi$. The second term deals with the so-called Pl\"ucker relations as constraints. By solving them we re-express the bosonic fields as functions of $\Theta$ and the purely-fermionic coset model is reproduced.

The first term is
\begin{eqnarray}
	\mathcal{L}_{V}
	&=& 
	V^{\alpha}_{\phantom{\alpha}i}\delta^{ij}\varepsilon_{\alpha\beta}V^{\beta}_{\phantom{\beta}j}
	+
	V^{i}_{\phantom{a}\alpha}\delta_{ij}\varepsilon^{\alpha\beta}V^{j}_{\phantom{j}\beta}\ .
	\label{ospmnL01}
\end{eqnarray}
And the second one is: 
\begin{eqnarray}
	\mathcal{L}_{P}&=&
\alpha^{\left( ij \right)}
\left( 
\Lambda_{ik}\delta^{kl}\Lambda_{lj}
-
\Theta_{i\alpha}\varepsilon^{\alpha\beta}\Theta_{j\beta} 
-
\delta_{ij}
\right)
+\nonumber\\&&
+
\beta^{\left[ \alpha\beta \right]}
\left( 
\Phi_{\alpha\gamma}\varepsilon^{\gamma\delta}\Phi_{\delta\beta}
-
\Theta_{i\alpha}\delta^{ij}\Theta_{j\beta}
-
\varepsilon_{\alpha\beta}
 \right)
+\nonumber\\&&
+
\gamma^{i\alpha}
\left( 
\Lambda_{ik}\delta^{kl}\Theta_{k\alpha}
+
\Phi_{\alpha\beta}\varepsilon^{\beta\gamma}\Theta_{i \gamma }
 \right)
\ .\label{ospmnLp1}
\end{eqnarray}
The constraints imply
\begin{eqnarray}
	\Lambda_{IJ}
	&=& 
	\sqrt{
	\delta_{IJ}
	+
	\Theta_{I\alpha}\varepsilon^{\alpha\beta}\Theta_{J\beta} 
	}\ ,
	\nonumber\\
	\Phi_{\alpha\beta}
	&=& 
	\sqrt{
	\varepsilon_{\alpha\beta}
	+
	\Theta_{I\alpha}\delta^{IJ}\Theta_{J\beta}
	}\ .
	\label{ospmnPlu}
\end{eqnarray}
It can be shown that substituting them into (\ref{ospmnL01}) will recover the original lagrangian (\ref{VCMaction}).
The $OSp(m|n)/SO(n)\times Sp(m)$ lagrangian is then
\begin{eqnarray}
	\mathcal{L}_{0}&=& \mathcal{L}_{V}+\mathcal{L}_{P}\ .
	\label{ospmnL0}
\end{eqnarray}

\subsection{T-duality}

In order to construct the T-dual model we gauge the whole isometry group. We introduce then the covariant derivatives defined as
\begin{eqnarray}
\nabla \Theta_{i\alpha}&=& 
\partial \Theta_{i\alpha}
-A_{ij}\delta^{jk}\Theta_{k\alpha}
-A_{\alpha\beta}\varepsilon^{\beta\gamma}\Theta_{i\gamma}+
\nonumber\\&&
-A_{i\beta}\varepsilon^{\beta\gamma}\Phi_{\gamma\alpha}
-A_{j\alpha}\delta^{jk}\Lambda_{ki}\ ,
\nonumber\\
\nabla \Lambda_{\left( ij \right)}&=& 
\partial\Lambda_{\left( ij \right)}
-A_{\left( i \right|\alpha}\varepsilon^{\alpha\beta}\Theta_{\left| j \right)\beta}
-A_{\left( i \right|k}\delta^{kl}\Lambda_{l\left|j \right)}\ ,
\nonumber\\
\nabla\Phi_{\left[ \alpha\beta \right]}&=& 
\partial\Phi_{\left[ \alpha\beta \right]}
-A_{i\left[ \alpha \right|}\delta^{ij}\Theta_{j\left|\beta \right]}
-A_{\left[ \alpha \right|\rho}\varepsilon^{\rho\sigma}\Phi_{\sigma\left|\beta \right]}\ ,
\label{ospmndercov}
\end{eqnarray}
and we add the field strengths (\ref{ospmnF1}) as Chern-Simons terms
\begin{eqnarray}
	\mathcal{L}_{D}
	&=& 
	i\theta^{i\alpha}F_{i\alpha}
	+
	i\lambda^{\left[ ij \right]}F_{\left[ ij \right]}
	+
	i\phi^{\left( \alpha\beta\right)}F_{\left( \alpha\beta \right)}\ .
	\label{ospmnLD1}
\end{eqnarray}
Now, we set $\Theta^{i\alpha}=0$ adding to the lagrangian the BRST gauge fixing condition
\begin{eqnarray}
	\mathcal{L}_{BRST1}=s\left[ \bar c^{i\alpha} \Theta_{i\alpha} \right]\ ,
	\label{ospmnLBRST1}
\end{eqnarray}
where $s \bar c^{i\alpha}=b^{i\alpha}$ and $s b^{i\alpha}= 0$. Solving the Pl\"ucker constraint, we get $\Lambda_{ij}=\delta_{ij}$ and $\Phi_{\alpha\beta}=\varepsilon_{_{\alpha\beta}}$. This simplifies the functions (\ref{ospmnViel2})
\begin{eqnarray}
	A^{a}_{\phantom{a}i}
	=
	\delta^{a}_{\phantom{a}i}
	,\quad\quad\quad
	B^{a}_{\phantom{a}\gamma}
	=0
	,\quad\quad\quad
	C^{\alpha}_{\phantom{\alpha}i}
	=0
	,\quad\quad\quad
	D^{\alpha}_{\phantom{\alpha}\beta}
	=\varepsilon^{\alpha}_{\phantom{\alpha}\beta}\ .
	\label{ospmnViel3}
\end{eqnarray}
The lagrangian is then
\begin{eqnarray}
	\mathcal{L}_{gf1}
	=A_{i\alpha}\delta^{ij}\varepsilon^{\alpha\beta}\bar A_{j\beta}+\mathcal{L}_{D}\ .
	\label{ospmnLgf1}
\end{eqnarray}
We can now perform another gauge fixing. We can set, analogously to (\ref{ospmnLBRST1})
\begin{eqnarray}
	\bar A_{ij} =0,\quad\quad\quad\quad \bar A_{\alpha\beta}=0\ ,
	\label{ospmngf2}
\end{eqnarray}
notice that this gauge fixing does not imply $A_{ij}=0$ and $A_{\alpha\beta}=0$. Then, (\ref{ospmnLgf1}) becomes
\begin{eqnarray}
	\mathcal{L}_{gf2}
	&=& 
	\left[ 
	\left( 
	\delta^{tl}\varepsilon^{\tau\lambda}
	+
	2i\lambda^{\left[ tl \right]}\varepsilon^{\tau\lambda}
	+
	2i\phi^{\left( \tau\lambda \right)}\delta^{tl}
	\right)A_{t\tau}
	+ 
	\right.\nonumber\\&&
	\left.
	-i\partial\theta^{l\lambda}
	+
	i\theta^{i\lambda}A_{\left[ ij \right]}\delta^{jl}
	+
	i\theta^{l\alpha}A_{ \left( \alpha\gamma \right)}\varepsilon^{\gamma\lambda}
	\right]\bar A_{l\lambda}+
	\nonumber\\&&
	+i\bar\partial \theta^{i\alpha}A_{i\alpha}
	+
	i\bar\partial\lambda^{\left[ ij \right]}A_{ij}
	+
	i\bar\partial\phi^{\left( \alpha\beta \right)}A_{\alpha\beta}\ .
	\label{ospmnLgf2}
\end{eqnarray}
We now compute the EoM for $\bar A_{i\alpha}$
\begin{eqnarray}
	A_{i\alpha}
	&=& 
	i
	\Xi_{il\,\alpha\lambda}
	\left( 	
	+
	\partial\theta^{l\lambda}
	-
	\theta^{c\lambda}A_{cj}\delta^{jl}
	-
	\theta^{l\beta}A_{ \beta\gamma}\varepsilon^{\gamma\lambda} 
	\right)\ ,
	\label{ospmnEoMbarA}
\end{eqnarray}
where $\Xi^{rm\,\rho\mu}$ is defined as follows
\begin{eqnarray}
	\Xi_{il\,\alpha\lambda}
	&=& \delta_{ir}\varepsilon_{\alpha\rho}
	\Xi^{rm\,\rho\mu}
	\delta_{ml}\varepsilon_{\mu\lambda}\ ,
	\label{ospmnXidef1}
\end{eqnarray}
and
\begin{eqnarray}
	\Xi^{rm\,\rho\mu}
	\delta_{ml}\varepsilon_{\mu\lambda}
	\left( 
		\delta^{tl}\varepsilon^{\tau\lambda}
	+
	2i\lambda^{\left[ tl \right]}\varepsilon^{\tau\lambda}
	+
	2i\phi^{\left( \tau\lambda \right)}\delta^{tl}
	\right)
	&=& 
	\delta^{rt}\varepsilon^{\rho\tau}\ .
	\label{ospmnXi}
\end{eqnarray}
Substituting (\ref{ospmnEoMbarA}) in (\ref{ospmnLgf2}) we obtain a first version of the dual lagrangian
\begin{eqnarray}
	\mathcal{L}_{Dual_{0}}
	&=&
	-\bar\partial\theta^{i\alpha}
		\Xi_{il\,\alpha\lambda}
	\partial\theta^{l\lambda}
	+
	\nonumber\\&&
	+
	i\bar\partial\theta^{i\alpha}
		\Xi_{il\,\alpha\lambda}
	\left( 
	\theta^{k\lambda}A_{\left[ kj \right]}\delta^{jl}
	+
	\theta^{l\nu}A_{ \left( \nu\gamma \right)}\varepsilon^{\gamma\lambda}
	\right)
	+
	\nonumber\\&&
	+i\bar\partial\lambda^{\left[ ij \right]}A_{ij}
	+
	i\bar\partial\phi^{\left( \alpha\beta \right)}A_{\alpha\beta}\ .
	\label{ospmnLDual0}
\end{eqnarray}
We notice that in $2$-dimensions the gauge fields $A$ are not dynamics. Therefore we can integrate them and take their EoM's  as constraints. The dual model then is composed by a lagrangian 
\begin{eqnarray}
	\mathcal{L}_{Dual}
	&=& 
	-\bar\partial\theta^{i\alpha}
		\Xi_{il\,\alpha\lambda}
	\partial\theta^{l\lambda}\ ,
	\label{ospmnLDual1}
\end{eqnarray}
and two constraints
\begin{equation}
	\left\{
	\begin{array}{l}
	\bar\partial\lambda^{\left[ ij \right]}
	+
	\bar\partial\theta^{k\alpha}
		\Xi_{kl\,\alpha\lambda}
	\theta^{\left[ i \right|\lambda}\delta^{\left. j  \right]l}=0\ ,
\\
\phantom{a}
\\
	\bar\partial\phi^{\left( \alpha\beta \right)}
	+
	\bar\partial\theta^{k\gamma}
		\Xi_{kl\,\gamma\lambda}
	\theta^{l\left( \alpha \right|}\varepsilon^{\left. \beta \right)\lambda}=0\ .
\end{array}\right.
	\label{ospmnConstraints}
\end{equation}
The fields $\lambda^{\left[ ij \right]}$ and $\phi^{\left( \alpha\beta \right)}$ are expressed in term of $\theta^{i\alpha}$. 
The EoM's for $\lambda^{\left[ ij \right]}$ and $\phi^{\left( \alpha\beta \right)}$ can be constructed by recursive application of $\partial$ and $\bar\partial$ to the constraints (\ref{ospmnConstraints}).

\subsection{Analysis}

To study the lagrangian (\ref{ospmnLDual1}) and the constraints (\ref{ospmnConstraints}) we can expand over small $\lambda$ and $\phi$. Using definition (\ref{ospmnXi}) we compute the first order of $\Xi^{rm\,\rho\mu}$ 
\begin{eqnarray}
	\Xi^{rm\,\rho\mu}
	&\sim& 
	-\delta^{rm}\varepsilon^{\rho\mu}
	-2i
	\lambda^{\left[ rm \right]}\varepsilon^{\rho\mu}
	-2i
	\phi^{\left( \rho\mu \right)}\delta^{rm}\ .
\label{ospmnXiExp}
\end{eqnarray}
We obtain then
\begin{eqnarray}
	\mathcal{L}_{Dual}
	&\sim&
	\bar\partial\theta^{i\alpha}\delta_{ij}\varepsilon_{\alpha\beta}\partial\theta^{j\beta}
	+
	2i
	\bar\partial\theta^{i\alpha}
	\delta_{ir}
	\lambda^{\left[ rm \right]}
	\delta_{ml}
	\varepsilon_{\alpha\lambda}
	\partial\theta^{l\lambda}
	+
	\nonumber\\&&
	\phantom{ 
	\bar\partial\theta^{i\alpha}\delta_{ij}\varepsilon_{\alpha\beta}\partial\theta^{j\beta}
	}+
	2i
	\bar\partial\theta^{i\alpha}
	\delta_{il}\varepsilon_{\alpha\rho}
	\phi^{\left( \rho\mu \right)}
	\varepsilon_{\mu\lambda}
	\partial\theta^{l\lambda}\ ,
	\label{ospmnLdualexp0}
\end{eqnarray}
and
\begin{equation}
	\left\{
	\begin{array}{l}
	\bar\partial\lambda^{\left[ ij \right]}
	=
	-\bar\partial\theta^{\left[ i \right|\gamma}
	\varepsilon_{\gamma\delta}
	\theta^{\left. j \right]\delta}\ ,
\\
\phantom{a}
\\
	\bar\partial\phi^{\left( \alpha\beta \right)}
	=	
	-\bar\partial\theta^{c\left( \alpha \right.}
	\delta_{cd}
	\theta^{d\left| \beta\right)}\ .
\end{array}\right.
	\label{ospmnConstraintsExp}
\end{equation}
The two interacting terms of (\ref{ospmnLdualexp0}) can be rewritten as
\begin{eqnarray}
	\bar\partial\theta^{i\alpha}
	\delta_{ir}
	\lambda^{\left[ rm \right]}
	\delta_{ml}
	\varepsilon_{\alpha\lambda}
	\partial\theta^{l\lambda}
	&=& 
	-
	\theta^{i\alpha}
	\delta_{ir}
	\bar\partial\lambda^{\left[ rm \right]}
	\delta_{ml}
	\varepsilon_{\alpha\lambda}
	\partial\theta^{l\lambda}
	+
	\nonumber\\&&
	-
	\theta^{i\alpha}
	\delta_{ir}
	\lambda^{\left[ rm \right]}
	\delta_{ml}
	\varepsilon_{\alpha\lambda}
	\bar\partial\partial\theta^{l\lambda}
	+
	\textrm{total derivative}\ .
	\label{ospmnxx}
\end{eqnarray}
The last term vanishes on-shell for the EoM of $\theta$ ({\it i.e. $\bar\partial\partial\theta^{l\lambda}=0$}). Therefore, it can be absorbed by a field redefinition and we can neglect this kind of term. The lagrangian becomes
\begin{eqnarray}
		\mathcal{L}_{Dual}
	&\sim&
	\bar\partial\theta^{i\alpha}\delta_{ij}\varepsilon_{\alpha\beta}\partial\theta^{j\beta}
	-
	2i
	\theta^{i\alpha}
	\delta_{ir}
	\bar\partial\lambda^{\left[ rm \right]}
	\delta_{ml}
	\varepsilon_{\alpha\lambda}
	\partial\theta^{l\lambda}
	+
	\nonumber\\&&
	\phantom{ 
	\bar\partial\theta^{i\alpha}\delta_{ij}\varepsilon_{\alpha\beta}\partial\theta^{j\beta}
	}
	-
	2i
	\theta^{i\alpha}
	\delta_{il}\varepsilon_{\alpha\rho}
	\bar\partial\phi^{\left( \rho\mu \right)}
	\varepsilon_{\mu\lambda}
	\partial\theta^{l\lambda}\ .
	\label{ospmnLdualexp0v1}
\end{eqnarray}
Substituting the two constraints (\ref{ospmnConstraintsExp})
\begin{eqnarray}
	\mathcal{L}_{Dual}
	&\sim&
	\bar\partial\theta^{i\alpha}\delta_{ij}\varepsilon_{\alpha\beta}\partial\theta^{j\beta}
	+
	2i
	\theta^{i\alpha}\delta_{ir}
	\bar\partial\theta^{\left[ r \right|\gamma}\varepsilon_{\gamma\delta}\theta^{\left. m \right]\delta}
	\delta_{ml}\varepsilon_{\alpha\lambda}\partial\theta^{l\lambda}
	+
	\nonumber\\&&
	\phantom{
	\bar\partial\theta^{i\alpha}\delta_{ij}\varepsilon_{\alpha\beta}\partial\theta^{j\beta}
	} 
	+2i
	\theta^{i\alpha}\delta_{il}\varepsilon_{\alpha\rho}
	\bar\theta^{c\left( \rho \right.}
	\delta_{cd}
	\theta^{d\left| \mu \right)}
	\varepsilon_{\mu\lambda}
	\partial\theta^{l\lambda}\ .
	\label{ospmnLdualexp1}
\end{eqnarray}
We obtain the following $4$-$\theta$ terms
\begin{eqnarray}
	\mathcal{L}_{Dual}{\big |}_{4\theta}
	&=& 
	2i
	\theta^{a\alpha}\theta^{b\beta}\bar\partial\theta^{c\gamma}\partial\theta^{d\delta}
	\left( 
	2
	\delta_{ac}\delta_{bd}\varepsilon_{\alpha\delta}\varepsilon_{\beta\gamma}
	-
	\delta_{ab}\delta_{cd}\varepsilon_{\alpha\delta}\varepsilon_{\beta\gamma}
	-
	\delta_{ad}\delta_{bc}\varepsilon_{\alpha\beta}\varepsilon_{\gamma\delta}
	 \right)
	 \nonumber\ ,
\label{ospmnDual4t2}
\end{eqnarray}
and this is exactly the same expression for the $4$-$\theta$ term of the original model (\ref{VCMaction2}). 

Notice that we neglected some terms proportional to the equations of motion. That is allowed at the classical level, but at the quantum one some suitable field redefinitions must be performed to achieve the equivalence. Indeed we check at one loop such field redefinitions.


\subsection{Fibration and T-duality}

Finally,  we treat a further example where the T-duality can be done as outlined in sec. \ref{revFer} for a fermionic model.
This model is obtained adding to every point of a base space a vectorial space (a fiber). This can be done adding at the metric of the base space a term like
\begin{equation}
  \nabla\psi_{1}\wa\nabla\psi_{2}\ ,
\end{equation}
where
\begin{equation}
 \dd\psi\ra\nabla\psi=\dd\psi + B \ ,
\end{equation}
$B$ is the connection from the various fibers and it depends only by the coordinates of the basic space. We use this method in the case  of $OSp(1|2)/Sp(2)$ and we get (we consider only the lagrangian density for simplicity)
\begin{equation}
 \CL_{2}\propto
\left(1+\t_{1}\t_{2}\right)\dd\t_{1}\wa\dd\t_{2}\longrightarrow  \CL_{4}\propto\left(\dd \psi_{3}+B_{3}\right)\wa\left(\dd\psi_{4}+B_{4}\right)+ \CL_{2}\ .
\label{originalFIBR}
\end{equation}
The most general form of the connection is the following
\begin{equation}\label{fibrazioneB}
 B_{i}=\left( a+b\t_{1}\t_{2}\right)\dd\t_{i} \ .
\end{equation}
The new model has four fermionic coordinates and has two translational isometries, as in $OSp(2|2)/SO(2)\times Sp(2) $, so the procedure is the same: we introduce the gauge fields, we set the coordinates to zero, we sum the $2$-forms and finally we calculate the equation of motion, from which we have
\begin{equation}
 \Bigg\{\ \begin{array}{ccc}
\phantom{\bigg |} &  A_{4} = -B_{4} -\frac{1}{\textrm{det}\g}\ast\dd\tilde\psi_{4} \ ,\\
\phantom{\bigg |}&  A_{3} = -B_{3} +\frac{1}{\textrm{det}\g}\ast\dd\tilde\psi_{3}\ .
\end{array}
\end{equation}
Notice that in contrast to the example given in sec \ref{obstructionosp22} we do not need to modify the action to solve the equations. The dual model is then
\begin{equation}\label{fibrazioneduale}
 \CL_{4\,Dual}\propto
\frac{1}{\textrm{det}\g}\dd\tilde\psi_{3}\wa\dd\tilde\psi_{4}+\left(1+\t_{1}\t_{2}\right)\dd\t_{1}\wa\dd\t_{2}+\dd\tilde\psi_{3}\wedge B_{4}+B_{3}\wedge \dd\tilde\psi_{4}\ .
\end{equation}
We shall calculate the curvature components for both the models obtained (the original (\ref{originalFIBR}) and the T-dual (\ref{fibrazioneduale})), without considering topological terms. However, it seems that does not exist a trivial connection between the two curvatures.


\chapter{Quantum Analysis for $\frac{OSp\left( n|m \right)}{SO\left( n \right)\times Sp\left( m \right)}$ models}
\label{chT2}

\section{One Loop Computation}

In this section we compute the $1$-loop correction to $\theta\theta$ propagator from the $2$-parameter dependent model derived in sec.~\ref{SUSYconstrMethod}.  
As told in  \cite{Bershadsky:1999hk}, $OSp\left( n|m \right)$ has vanishing $\beta$-function ({\it i.e.} UV-finiteness) at least at  $1$-loop if $m+2-n=0$.  We expect the same behavior also for the associated purely fermionic coset model. 

\subsection{Propagator and Vertex}

The $\theta\theta$  propagator is obtained from $\mathcal{L}_{0}$ defined in (\ref{zeroaction}) using the usual Green-functions method. We have that
\begin{equation}\label{symmpropagator}
  P_{\alpha\beta}^{ab}(p)=\frac{\varepsilon_{\beta\alpha}\delta^{ab}
	}{p^{2}}\ ,
\end{equation}
where $p$ is the $2$d-entering momentum. The $4$-vertex is obtained from (\ref{S2}) symmetrizing the fermionic $\theta$ legs (which are labelled by $A,B,C,D$)
\begin{eqnarray}
V_{4\theta}&=&
\left(4xp_{A}\cdot p_{B}+4xp_{A}\cdot p_{C}+4xp_{A}\cdot p_{D}
\right.+\nonumber\\&&\left.\quad
+4xp_{B}\cdot p_{C}+4xp_{B}\cdot p_{D}+4xp_{C}\cdot p_{D}\right)\delta^{ad}\delta^{bc}\varepsilon_{\alpha\delta}\varepsilon_{\beta\gamma}
+\nonumber\\&&
+\left(4p_{A}\cdot p_{B}+4yp_{A}\cdot p_{B}+2p_{A}\cdot p_{C}+4yp_{A}\cdot p_{C}
\right.+\nonumber\\&&\left.\quad
+2p_{A}\cdot p_{D}+4yp_{A}\cdot p_{D}
+2p_{B}\cdot p_{C}+4yp_{B}\cdot p_{C}
\right.+\nonumber\\&&\left.\quad
+2p_{B}\cdot p_{D}
+4yp_{B}\cdot p_{D}+4p_{C}\cdot p_{D}+4yp_{C}\cdot p_{D}\right)\delta^{ac}\delta^{bd}\varepsilon_{\alpha\delta}\varepsilon_{\beta\gamma}
+\nonumber\\&&
+\left(2p_{A}\cdot p_{B}+4yp_{A}\cdot p_{B}+4p_{A}\cdot p_{C}
+4yp_{A}\cdot p_{C}+2p_{A}\cdot p_{D}
\right.+\nonumber\\&&\left.\quad
+4yp_{A}\cdot p_{D}+2p_{B}\cdot p_{C}
+4yp_{B}\cdot p_{C}+4p_{B}\cdot p_{D}
\right.+\nonumber\\&&\left.\quad
+4yp_{B}\cdot p_{D}+2p_{C}\cdot p_{D}+4yp_{C}\cdot p_{D}\right)\delta^{ab}\delta^{cd}\varepsilon_{\alpha\delta}\varepsilon_{\beta\gamma}
+\nonumber\\&&
+\left(-4p_{A}\cdot p_{B}-4yp_{A}\cdot p_{B}-2p_{A}\cdot p_{C}-4yp_{A}\cdot p_{C}
\right.+\nonumber\\&&\left.\quad
-2p_{A}\cdot p_{D}-4yp_{A}\cdot p_{D}-2p_{B}\cdot p_{C}-4yp_{B}\cdot p_{C}-2p_{B}\cdot p_{D}
\right.+\nonumber\\&&\left.\quad
-4yp_{B}\cdot p_{D}-4p_{C}\cdot p_{D}-4yp_{C}\cdot p_{D}\right)\delta^{ad}\delta^{bc}\varepsilon_{\alpha\gamma}\varepsilon_{\beta\delta}
+\nonumber\\&&
+\left(-4xp_{A}\cdot p_{B}-4xp_{A}\cdot p_{C}-4xp_{A}\cdot p_{D}
\right.+\nonumber\\&&\left.\quad
-4xp_{B}\cdot p_{C}-4xp_{B}\cdot p_{D}-4xp_{C}\cdot p_{D}\right)\delta^{ac}\delta^{bd}\varepsilon_{\alpha\gamma}\varepsilon_{\beta\delta}
+\nonumber\\&&
+\left(-2p_{A}\cdot p_{B}-4yp_{A}\cdot p_{B}-2p_{A}\cdot p_{C}-4yp_{A}\cdot p_{C}
\right.+\nonumber\\&&\left.\quad
-4p_{A}\cdot p_{D}-4yp_{A}\cdot p_{D}-4p_{B}\cdot p_{C}-4yp_{B}\cdot p_{C}-2p_{B}\cdot p_{D}
\right.+\nonumber\\&&\left.\quad
-4yp_{B}\cdot p_{D}-2p_{C}\cdot p_{D}-4yp_{C}\cdot p_{D}\right)\delta^{ab}\delta^{cd}\varepsilon_{\alpha\gamma}\varepsilon_{\beta\delta}
+\nonumber\\&&
+\left(2p_{A}\cdot p_{B}+4yp_{A}\cdot p_{B}+4p_{A}\cdot p_{C}+4yp_{A}\cdot p_{C}+2p_{A}\cdot p_{D}
\right.+\nonumber\\&&\left.\quad
+4yp_{A}\cdot p_{D}+2p_{B}\cdot p_{C}+4yp_{B}\cdot p_{C}+4p_{B}\cdot p_{D}+4yp_{B}\cdot p_{D}
\right.+\nonumber\\&&\left.\quad
+2p_{C}\cdot p_{D}+4yp_{C}\cdot p_{D}\right)\delta^{ad}\delta^{bc}\varepsilon_{\alpha\beta}\varepsilon_{\gamma\delta}
+\nonumber\\&&
+\left(2p_{A}\cdot p_{B}+4yp_{A}\cdot p_{B}+2p_{A}\cdot p_{C}+4yp_{A}\cdot p_{C}+4p_{A}\cdot p_{D}
\right.+\nonumber\\&&\left.\quad
+4yp_{A}\cdot p_{D}+4p_{B}\cdot p_{C}+4yp_{B}\cdot p_{C}+2p_{B}\cdot p_{D}+4yp_{B}\cdot p_{D}
\right.+\nonumber\\&&\left.\quad
+2p_{C}\cdot p_{D}+4yp_{C}\cdot p_{D}\right)\delta^{ac}\delta^{bd}\varepsilon_{\alpha\beta}\varepsilon_{\gamma\delta}
+\nonumber\\&&
+\left(4xp_{A}\cdot p_{B}+4xp_{A}\cdot p_{C}+4xp_{A}\cdot p_{D}+4xp_{B}\cdot p_{C}
\right.+\nonumber\\&&\left.\quad
+4xp_{B}\cdot p_{D}+4xp_{C}\cdot p_{D}\right)\delta^{ab}\delta^{cd}\varepsilon_{\alpha\beta}\varepsilon_{\gamma\delta}
\ .\nonumber\\&&
\label{SymVertex4t1.0}
\end{eqnarray}
Notice that the dot product  refers to the world-sheet metric $\gamma_{ij}$ contraction 
$$p_{A}\cdot p_{B}=\left[ p_{A} \right]_{i}\gamma^{ij}\left[ p_{B} \right]_{j}\ .$$

\subsection{$1$-Loop Self Energy}\label{1loop1}

The $1$-loop correction to propagator is obtained contracting the $4\theta$ vertex (\ref{SymVertex4t1.0}) with the propagator (\ref{symmpropagator})
{
\begin{center}
\includegraphics[scale=.3]{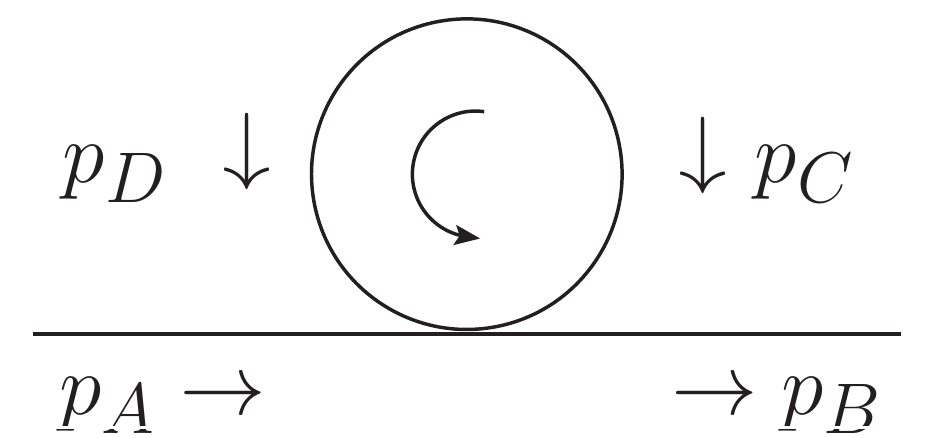}
\end{center}
}\noindent
Moreover, we impose the following momentum redefinitions
\begin{equation}
\begin{array}{ccc}
p_{A}=p\ , & & p_{B}=-p\ , \\
p_{C}=-q\ , & & p_{D}=q\ .
\end{array}
\end{equation}
We obtain then
\begin{eqnarray}
\Gamma&=& -4 (-2 + m - n - 2 x + m n x +2y(- 1  + 
    m  -  n )) \delta^{ab} \varepsilon_{\alpha\beta}
\int\dd^{2}q\frac{\left( p^2 + q^2 \right)}{q^{2}}\ .
\label{Sym12tloop1.0}
\end{eqnarray}
We add to the lagrangian a mass term  in order to avoid IR-divergences
\begin{equation}
\mathcal{L}
\longrightarrow
\mathcal{L}
+
M^{2}
\theta^{\alpha}_{a}
\varepsilon_{\alpha\beta}\delta^{ab}
\theta^{\beta}_{b}\ ,
\label{SymcorrL}
\end{equation}
then, the propagator (\ref{symmpropagator}) becomes
\begin{equation}\label{symmpropagator1}
  P_{\s\tau}^{st}(q)=\frac{\varepsilon_{\s\tau}\,\d^{st}
	}{q^{2}+M^{2}}\ .
\end{equation}
Notice that setting $x=0$ and $y=-\frac{2}{3}$, which leads to the vertex obtained in sec.~\ref{vielbeinconstrmeth}, we have
\begin{eqnarray}
\Gamma= \frac{4}{3}  (2+m-n) \delta^{ab} \varepsilon_{\alpha\beta}
\int\dd^{2}q\frac{\left( p^2 + q^2 \right)}{q^{2}+M^{2}}\ .
\label{Sym2tloop1.0}
\end{eqnarray}
It is easy to show that this is the only choice of $x$ and $y$ that leads to a $1$-loop correction depending by $2+m-n$. This vertex is obtained from the following lagrangian term
\begin{eqnarray}
\mathcal{L}\big|_{4\theta}
&=&
\theta_{a}^{\alpha}\theta^{\beta}_{b}\partial_{\mu}\theta_{c}^{\gamma}\partial^{\mu}\theta^{\delta}_{d}\left( 
-2\delta^{ac}\delta^{bd}\varepsilon_{\alpha\delta}\varepsilon_{\beta\gamma}
+
\delta^{ab}\delta^{cd}\varepsilon_{\alpha\delta}\varepsilon_{\beta\gamma}
+
\delta^{ad}\delta^{bc}\varepsilon_{\alpha\beta}\varepsilon_{\gamma\delta}
 \right)\ ,
\label{L4thetaFond}
\end{eqnarray}
as in sec.~\ref{vielbeinconstrmeth}.

It is useful to introduce the following pictorial convection:
\setlength{\unitlength}{1mm}
\begin{center}
\begin{picture}(60,14)
\put(19,11){\vector(1, 0){12}}
\put(19,11){\line(0,-1){3}}
\put(31,11){\line(0,-1){3}}
\put(21,10){\vector(1, 0){5}}
\put(21,10){\line(0,-1){2}}
\put(26,10){\line(0,-1){2}}
\put(19,2){\line(1,0){7}}
\put(19,2){\line(0,1){3}}
\put(26,2){\line(0,1){3}}
\put(21,3){\line(1,0){10}}
\put(21,3){\line(0,1){2}}
\put(31,3){\line(0,1){2}}
\put(38,11){\vector(1, 0){12}}
\put(38,11){\line(0,-1){3}}
\put(50,11){\line(0,-1){3}}
\put(40,10){\vector(1, 0){5}}
\put(40,10){\line(0,-1){2}}
\put(45,10){\line(0,-1){2}}
\put(38,3){\line(1,0){2}}
\put(38,3){\line(0,1){2}}
\put(40,3){\line(0,1){2}}
\put(45,3){\line(1,0){5}}
\put(45,3){\line(0,1){2}}
\put(50,3){\line(0,1){2}}
\put(58,10){\vector(1, 0){2}}
\put(58,10){\line(0,-1){2}}
\put(60,10){\line(0,-1){2}}
\put(65,10){\vector(1, 0){5}}
\put(65,10){\line(0,-1){2}}
\put(70,10){\line(0,-1){2}}
\put(58,2){\line(1,0){12}}
\put(58,2){\line(0,1){3}}
\put(70,2){\line(0,1){3}}
\put(60,3){\line(1,0){5}}
\put(60,3){\line(0,1){2}}
\put(65,3){\line(0,1){2}}
\put(0,5){$\mathcal{L}\big|_{4\theta}=-2\,\theta\,\theta\,\partial\theta\,\partial\theta
+\,\theta\,\theta\,\partial\theta\,\partial\theta
+\,\theta\,\theta\,\partial\theta\,\partial\theta\ ,$}
\end{picture}
\end{center}
where the upper arrow line contracts the $Sp$ indices while the lower simple line contracts the $SO$ ones. Notice that this vertex is exactly the same found via the vielbein construction method (\ref{VCMaction2}).


\section{Two Loop Computation with BFM}

\subsection{Outline of the Method}\label{bfmoutline}

The background field method (BFM) is a powerful tool that allows various simplifications to compute 1PI Green's functions \cite{Abbott:1981ke}. Here we briefly review the foundations of the method. 

Consider the generating functional for connected graphs
\begin{eqnarray}
	W\left[ J \right]=-i\ln \int\mathcal{D}\Phi \,exp\left\{ i S\left[ \Phi \right]+iJ\cdot \Phi \right\}\ ,
	\label{bfmgenfunZ}
\end{eqnarray}
where $J$ is the classic source of the field $\Phi$. We now split the $\Phi$ in a {\it background} field $B$ and in a {\it quantum} one $\varphi$, for example through a linear splitting $\Phi=B+\varphi$. The background field $B$ is seen as another classical source. We have then
\begin{eqnarray}
	\tilde{W}\left[ J,B \right]
	=
	-i\ln
	\int\mathcal{D}\varphi \,exp\left\{ i S\left[ B+\varphi \right]+iJ\cdot \varphi \right\}\ ,
	\label{bfmgenfunZ2}
\end{eqnarray}
where $J$ is now the source of the quantum field $\varphi$. Notice that $\frac{\delta^{n}}{\delta B^{n}} \tilde W{\big|}_{B=J=0}$ gives the $n$-point connected Green functions with only external $B$ fields while with $\frac{\delta^{n}}{\delta J^{n}} \tilde W{\big|}_{B=J=0}$ we obtain the $n$-points connected Green functions with external $\varphi$ fields. 

The {\it $1$-particle irreducible} ($1$PI) functional generator  is defined as 
\begin{equation}
	\Gamma\left[ Q \right] = W[J]-QJ\ ,
	\label{BFMGammaX}
\end{equation}
where $Q=\frac{\delta W}{\delta J}$. In presence of the background field splitting, it becomes
\begin{eqnarray}
	\tilde\Gamma\left[ \tilde Q,B \right] = \tilde W[J,B]-\tilde QJ\ ,
	\label{BFMGammaXB}
\end{eqnarray}
with $\tilde Q=\frac{\delta \tilde W}{\delta J}$. 

Notice that there is a class of transformations of the quantum and background fields that preserve the lagrangian. If the splitting is linear  $\Phi=B+\varphi$ the $1$PI generating functional $\tilde \Gamma$ is invariant under the following transformations
\begin{eqnarray}
   B \rightarrow B+\eta
  \quad\quad\quad\quad
  \varphi\rightarrow\varphi-\eta\ .
\label{BFMtransffieldlin}
\end{eqnarray}
Notice that from the definition of $\Gamma$, $\tilde Q$ transforms as $\varphi$. Then, we shall write
\begin{eqnarray}
0&=& \delta =\delta  \tilde Q \frac{\delta \tilde \Gamma}{\delta \tilde Q}+\delta B \frac{\delta \tilde \Gamma}{\delta B}\ .
\label{BFMlinspl}
\end{eqnarray}
Further differentiations give the {\it Ward identities} between $n$-point $1PI$ Green's function. 
These observations yield
\begin{eqnarray}
	\tilde \Gamma\left[ \tilde Q,B \right] = \Gamma\left[ \tilde Q+B \right]\ ,
	\label{BFMkey1}
\end{eqnarray}
and setting $\tilde Q=0$ we have
\begin{eqnarray}
	\tilde \Gamma\left[ 0,B \right] = \Gamma\left[ B \right]\ ,
	\label{BFMkey2}
\end{eqnarray}
thus, the $1$PI Green functions of the original field theory obtained differentiating the r.h.s. functional generator are computed by the  $1$PI Green functions with only external background legs derived from l.h.s. generator.

\subsection{BFM Lagrangian}

We define the group elements as
\begin{equation}
  g=g_{0}e^{\l X}\ ,
  \label{bfmdefgroup}
\end{equation}
where $g_{0}$ is the background field and $X$ is an element of the coset Lie algebra ( $X \in \mathfrak{g}/\mathfrak{h}$). Notice that $\l$ is a coupling constant. We can write the left-invariant $1$-form current as the following
\begin{eqnarray}
	\tilde{J}_{\mu}&=&g^{-1}\partial_{\mu} g = e^{-\l X}B_{\mu}e^{\l X} + e^{-\lambda X}\partial_{\mu}e^{\l X}=
	\nonumber\\
  &=& 
  B_{\mu}+ \l \left[B_{\mu}\,,\,X \right]+\frac{\l^{2}}{2}\left[\left[B_{\mu}\,,\,X\right]\,,\,X\right]
  +\l\partial_{\mu}X+\frac{\l^{2}}{2}\left[\partial_{\mu}X\,,\,X\right]+
  	\nonumber\\
	&&
	+\frac{\l^{3}}{3!}\left[ \left[ \left[ \partial_{\mu}X,X \right],X \right]X \right]+
	 \frac{\l^{4}}{4!}\left[ \left[ \left[ \left[ \partial_{\mu}X,X \right],X \right],X \right],X \right]+\dots\ ,
  \label{bfmdefJ}
\end{eqnarray}
where $B_{\mu}=g_{0}^{-1}\partial_{\mu} g_{0}$. 

The action is then obtained via the principal chiral sigma model construction $\int Str \left( \tilde{J}_{\mu} \tilde{J}_{\nu}\eta^{\mu\nu}\right)$
\begin{equation}
  S_{G/H}=\frac{1}{2\pi\l^{2}}\int Str \left(
  e^{-\l X}B_{\mu} e^{\l X}{\Big|}_{\mathfrak{g}/\mathfrak{h}}+e^{-\l X}\partial_{\mu}e^{\l X}{\Big|}_{\mathfrak{g}/\mathfrak{h}} 
  \right)^{2}\ .
  \label{bfmaction1}
\end{equation}
The total current $\tilde{J}_{\mu}$ can be expanded in term of algebra generators. Considering the $\mathbb{Z}_{2}$-grading of the fermionic coset algebra and the (anti-)commutation relations we can divide $\tilde{J}_{\mu}=\tilde{J}^{0}_{\mu}+\tilde{J}^{1}_{\mu}$ where
\begin{eqnarray}
\mathfrak{h}  \ni  \tilde{J}_{\mu}^{(0)}&=&  
  B^{(0)}_{\mu}+ \l \left[B^{(1)}_{\mu}\,,\,X \right]+\frac{\l^{2}}{2}\left[\left[B^{(0)}_{\mu}\,,\,X\right]\,,\,X\right]
  +\frac{\l^{2}}{2}\left[\partial_{\mu}X\,,\,X\right]+\dots
\nonumber\\
\frac{\mathfrak{g}}{\mathfrak{h}}\ni   \tilde{J}^{(1)}_{\mu} &=&  
B^{(1)}_{\mu}+ \l \left[B^{(0)}_{\mu}\,,\,X \right]+\frac{\l^{2}}{2}\left[\left[B^{(1)}_{\mu}\,,\,X\right]\,,\,X\right]
  +\l\partial_{\mu}X+\dots
  \label{z2current}
\end{eqnarray}
Notice that $B^{(1)}_{\mu}$ is the fermionic background field and $B^{(0)}_{\mu}$ is the bosonic one.
The coset formalism allows us to neglect the bosonic current $\tilde{J}^{\left( 0 \right)}$ and all the bosonic contributions (obtained from commutators). The only term which survives is then
\begin{equation}
  \frac{\mathfrak{g}}{\mathfrak{h}}
	\ni  
  \tilde{J}^{(1)}_{\mu}
  =  
B^{(1)}_{\mu}+\frac{\l^{2}}{2}\left[\left[B^{(1)}_{\mu},X\right],X\right]
+\l\partial_{\mu}X+\frac{\lambda^{3}}{3!}\left[  \left[\partial_{\mu}X,X \right],X \right]+\dots
  \label{z2current1}
\end{equation}
The action is then computed from the following
\begin{equation}
  \frac{1}{2\pi\lambda^{2}}\int Str\left( \tilde{J}\cdot\tilde{J} \right) = Str\left(  \tilde{J}^{\left( 1 \right)}\cdot\tilde{J}^{\left( 1 \right)} \right)\ .
\end{equation}
We use (\ref{bfmdefJ}) and the cyclic property of the supertrace to compute (\ref{bfmaction1})
\begin{eqnarray}
   S_{G/H}&=&\frac{1}{2\pi\l^{2}}\int Str\left( 
   B^{\left( 1 \right)}\cdot B^{\left( 1 \right)}+2\l B^{\left( 1 \right)}\cdot\partial X+ \lambda^{2}\partial X\cdot\partial X+
   \right.
\nonumber\\&&+
   \frac{2\lambda^{3}}{3!}B^{\left( 1 \right)}\cdot \left[ \left[ \partial X,X \right],X \right]+\lambda^{3}\left[ \left[ B^{\left( 1 \right)},X \right],X \right]\partial X+
   \nonumber\\&&
   \left.+\frac{2\lambda^{4}}{3!}\left[ \left[ \partial X,X \right],X \right]\partial X
   +  \lambda^{2} B^{\left( 1 \right)} \left[ \left[ B^{\left( 1 \right)},X \right],X \right]+
	\right. 
	\nonumber\\&&
   \left.
	\frac{\lambda^{4}}{4\cdot3}B\cdot \left[ \left[ \left[ \left[ B,X \right],X \right],X \right],X \right]+
	\frac{\lambda^{4}}{4}\left[ \left[ B,X \right],X \right]\left[ \left[ B, X \right],X \right]
   \right)=
\nonumber\\&=&
\frac{1}{2\pi\l^{2}}\int Str\left( 
   B^{\left( 1 \right)}\cdot B^{\left( 1 \right)}+2\l B^{\left( 1 \right)}\cdot\partial X+ \lambda^{2}\partial X\cdot\partial X+
   \right.
\nonumber\\&&+
   \frac{4}{3}\lambda^{3}B^{\left( 1 \right)} \left[ \left[ \partial X,X \right],X \right]+
   \nonumber\\&&
   \left.+\frac{2\lambda^{4}}{3!}\left[ \left[ \partial X,X \right],X \right]\partial X
   +  \lambda^{2} B^{\left( 1 \right)} \left[ \left[ B^{\left( 1 \right)},X \right],X \right]+
	\right. 
	\nonumber\\&&
   \left.
	\frac{\lambda^{4}}{4\cdot3}B\cdot \left[ \left[ \left[ \left[ B,X \right],X \right],X \right],X \right]+
	\frac{\lambda^{4}}{4}\left[ \left[ B,X \right],X \right]\left[ \left[ B, X\right],X \right]
   \right)\ .
   \nonumber\\&&
  \label{bfmaction2}
\end{eqnarray}

\subsection{Feynman Rules}

We now obtain the Feynman rules for the propagators and for the basic vertex in (\ref{bfmaction2}). Further details are in app.~\ref{appendixC}.  We expand  $X$ and the background current on the fermionic generators $X= \t^{\alpha}_{a}Q^{a}_{\alpha}\in \mathfrak{g}/ \mathfrak{h} $ , $B^{\left( 1 \right)}_{\mu}=B^{\left( 1 \right)\,\alpha}_{\mu\,\,\, a}Q^{a}_{\alpha}$.

To compute the $XX$ propagator we extract the quadratic operator from the lagrangian as follows
\begin{eqnarray}
	\mathcal{L}=\frac{1}{2}\varepsilon_{\beta\gamma}\delta^{bc}\theta^{\beta}_{b} \square \theta^{\gamma}_{c}\quad\quad\Rightarrow\quad\quad O=4\varepsilon_{\beta\gamma}\delta^{bc}\square \ .
\label{XX0.1}
\end{eqnarray}
Notice that a factor $2$ comes from the supertraces (\ref{bfmsupertraces}) and the other is due to (\ref{FRC2}). Then we define the propagator $\Delta$ as
\begin{equation}
O\left( p \right) \Delta\left( p \right) =1\ .
\label{OProp}
\end{equation}
We obtain (we omit the metrics)
\begin{equation}
	4\gamma^{\mu\nu} p_{\mu}p_{\nu} \Delta=1\ .
\label{XX0.2}
\end{equation}
The full propagator is finally
\begin{eqnarray}
\Delta^{\beta\gamma}_{cb}(\theta)=+\frac{1}{4}\frac{\varepsilon^{\gamma\beta}\delta_{cb}}{p^{2}}\ .
\label{XX}
\end{eqnarray}
With this set of conventions (no $i$ for the propagator and no $-i$ for the vertex and the (\ref{OProp})), $2$-point functions are simply defined as $\frac{1}{\Delta}$. Then the $\theta\theta$ $2$-point function is
\begin{eqnarray}
\frac{\delta^{2}\Gamma}{\delta \theta^{\beta}_{b}\left( p \right)\delta \theta_{c}^{\gamma}\left( -p \right)}=+4 p^{2} \varepsilon_{\beta\gamma}\delta^{bc}\ .
\label{XX2point}
\end{eqnarray}
The $BB$ $2$-point  function is 
\begin{eqnarray}
\frac{\delta^{2}\Gamma}{\delta B_{\mu b}^{\phantom{\mu} \beta}\left( p \right)\delta B_{\nu c}^{\phantom{\nu} \gamma}\left(- p \right)}
=+4\lambda^{-2}\gamma_{\mu\nu}\varepsilon_{\beta\gamma}\delta^{bc}\ .
\label{BB}
\end{eqnarray}
The simplest vertex we found in (\ref{bfmaction2}) is  $2\lambda Str(B\cdot \partial Q)$. It corresponds to the following Feynman rule
\begin{eqnarray}
\frac{\delta^{2}\Gamma}{\delta B_{\mu b}^{\phantom{\mu} \beta}\left( p \right)\delta \theta_{c}^{\gamma}\left( -p \right)}&=&
4\lambda^{-1}\varepsilon_{\beta\gamma}\delta^{bc}\left( -i \right)q_{\mu}=
\nonumber\\&=&
-4i\lambda^{-1}\varepsilon_{\beta\gamma}\delta^{bc}\left( -p_{\mu} \right)=
\nonumber\\&=&
4i\lambda^{-1}\varepsilon_{\beta\gamma}\delta^{bc}p_{\mu} \ .
\label{BX}
\end{eqnarray}
We compute now the $4$-legs vertex  $Str\left( B^{\left( 1 \right)}\left[ \left[ B^{\left( 1 \right)},X\right],X \right] \right)$.
Recalling the (anti)commutator rules (\ref{appendixalgebra1}), we can write the vertex as follows
\begin{eqnarray}
  Str\left( B^{\left( 1 \right)},\left[ \left[ B^{\left( 1 \right)},X \right],X \right] \right)&=&
  B^{(1)\,\alpha}_{\mu\, a}B^{(1)\,\beta}_{\nu\, b}\theta^{\gamma}_{c}\theta^{\delta}_{d}\times
  \nonumber\\&&
  \times \left( 
  -\delta^{bc}\varepsilon_{\delta\beta}Str\left( Q^{a}_{\alpha}Q^{d}_{\gamma}\right)
  -\delta^{bc}\varepsilon_{\delta\gamma}Str\left( Q^{a}_{\alpha}Q^{d}_{\beta} \right)+
  \right.\nonumber\\&&\left.
  +\delta^{cd}\varepsilon_{\beta\gamma}Str\left( Q^{a}_{\alpha}Q^{b}_{\delta} \right)
  -\delta^{bd}\varepsilon_{\beta\gamma}Str\left( Q^{a}_{\alpha}Q^{c}_{\delta} \right)\right)\ .
   \label{bfmvertex1}
\end{eqnarray}
Using the relations (\ref{bfmsupertraces}) we obtain that
\begin{eqnarray}
&&
 B^{(1)\,\alpha}_{\mu\, a}B^{(1)\,\beta}_{\nu\, b}\gamma^{\mu\nu}\theta^{\gamma}_{c}\theta^{\delta}_{d}\times
  \nonumber\\&& \times 2\left( 
	-2
	\varepsilon_{\alpha\delta}\varepsilon_{\beta\gamma}\delta^{ac}\delta^{bd} 
	+
	\varepsilon_{\alpha\delta}\varepsilon_{\beta\gamma}\delta^{ab}\delta^{cd} 
	+
	\varepsilon_{\alpha\beta}\varepsilon_{\gamma\delta}\delta^{bc}\delta^{ad} 
  \right)\ .
  \label{bfmvertex2}
\end{eqnarray}
Notice that we treat $B_{\mu}$ as a vectorial field. So  we do not associate any momentum. 
To obtain the Feynman rules we go in the momentum frame ($\partial_{\mu}\rightarrow -ip_{\mu}$) and we perform all the possible permutations of indistinguishable  quantum legs. We obtain the following expression (we consider also the constant in the action (\ref{bfmaction2}) but we skip the $\left( 2\pi \right)^{-1}$ factor)
\begin{eqnarray}
  \left[BBXX\right]^{abcd}_{\alpha\beta\gamma\delta\,.\mu\nu}&=&
V^{\left[ 2 \right]}=
	\nonumber\\&=&
\left[
-4\delta^{ac}\delta^{bd}\varepsilon_{\alpha\delta}\varepsilon_{\beta\gamma}+2\delta^{ab}\delta^{cd}\varepsilon_{\alpha\delta}\varepsilon_{\beta\gamma}+4\delta^{ad}\delta^{bc}\varepsilon_{\alpha\gamma}\varepsilon_{\beta\delta} 
+  \right.
	\nonumber\\&&
  \left.
  -2\delta^{ab}\delta^{cd}\varepsilon_{\alpha\gamma}\varepsilon_{\beta\delta}+2\delta^{ad}\delta^{bc}\varepsilon_{\alpha\beta}\varepsilon_{\gamma\delta}+2\delta^{ac}\delta^{bd}\varepsilon_{\alpha\beta}\varepsilon_{\gamma\delta}
  \right]\gamma_{\mu\nu}\ .
  \label{bfmBBXX}
\end{eqnarray}
where $K_{i}$ are the momenta associated with the background fields. Notice that we define $V^{[i]}$ as the vertex obtained symmetrizing only the metric term, without constants. The explicit structure for all the derived terms $V^{\left[ i \right]}$ are in app.~\ref{FRappendix}.
In the same way we now compute the $BXXX$ term $Str\left(B^{\left( 1 \right)} \left[ \left[ \partial X,X \right],X \right]\right)$. The lagrangian term gives
\begin{eqnarray}
&&
B^{(1)\,\alpha}_{\mu\, a}\partial_{\nu}\theta^{\beta}_{b}\gamma^{\mu\nu}\theta^{\gamma}_{c}\theta^{\delta}_{d}\times
  \nonumber\\&& \times
2\left(
-2
	\varepsilon_{\alpha\delta}\varepsilon_{\beta\gamma}\delta^{ac}\delta^{bd} 
	+
	\varepsilon_{\alpha\delta}\varepsilon_{\beta\gamma}\delta^{ab}\delta^{cd} 
	+
	\varepsilon_{\alpha\beta}\varepsilon_{\gamma\delta}\delta^{bc}\delta^{ad} 
\right)\ .
\label{bfmBXXX0.1}
\end{eqnarray}
Performing the symmetrization we have
\begin{eqnarray}
  \left[BXXX\right]^{abcd}_{\alpha\beta\gamma\delta\,\mu}&=&
-i\frac{4\lambda}{3}\left[ V^{\left[ 3 \right]} \right]_{\mu}\ .
  \label{bfmBXXX}
\end{eqnarray}
We determine the $XXXX$ vertex. From the lagrangian we have\footnote{Notice that this vertex shall be written as \ref{L4thetaFond}.}
\begin{eqnarray}
&&
\partial_{\mu}\theta^{\alpha}_{a}\partial_{\nu}\theta^{\beta}_{b}\gamma^{\mu\nu}\theta^{\gamma}_{c}\theta^{\delta}_{d}\,2\times
  \nonumber\\&& \times
\left(
-2
	\varepsilon_{\alpha\delta}\varepsilon_{\beta\gamma}\delta^{ac}\delta^{bd} 
	+
	\varepsilon_{\alpha\delta}\varepsilon_{\beta\gamma}\delta^{ab}\delta^{cd} 
	+
	\varepsilon_{\alpha\beta}\varepsilon_{\gamma\delta}\delta^{bc}\delta^{ad} 
\right)\ .
\label{bfmXXXX0.1}
\end{eqnarray}
The final term is then
\begin{eqnarray}
\left[XXXX\right]^{abcd}_{\alpha\beta\gamma\delta}&=& 
-\frac{1}{3}\lambda^{2}V^{\left[ 4 \right]}\ .
\label{XXXX1.0}
\end{eqnarray}
Finally, we calculate the $BBXXXX$ vertex. As usual, from the lagrangian we get
\begin{eqnarray}
&&
B^{\alpha}_{\mu\,a}B^{\beta}_{\nu\,b}\gamma^{\mu\nu}\theta^{\gamma}_{c}\theta^{\delta}_{d}\theta^{\rho}_{r}\theta^{\sigma}_{s}\times
  \nonumber\\&& \times
\left(
-6\delta^{ac}\delta^{bs}\delta^{dr}\varepsilon_{\alpha\sigma}\varepsilon_{\beta\rho}\varepsilon_{\gamma\delta}+6\delta^{ac}\delta^{bd}\delta^{rs}\varepsilon_{\alpha\sigma}\varepsilon_{\beta\rho}\varepsilon_{\gamma\delta}+6\delta^{ac}\delta^{br}\delta^{ds}\varepsilon_{\alpha\rho}\varepsilon_{\beta\sigma}\varepsilon_{\gamma\delta}
+  \right.
	\nonumber\\&&
  \left.
+6\delta^{ac}\delta^{br}\delta^{ds}\varepsilon_{\alpha\delta}\varepsilon_{\beta\sigma}\varepsilon_{\gamma\rho}-6\delta^{ac}\delta^{bs}\delta^{dr}\varepsilon_{\alpha\delta}\varepsilon_{\beta\rho}\varepsilon_{\gamma\sigma}+6\delta^{ac}\delta^{bd}\delta^{rs}\varepsilon_{\alpha\delta}\varepsilon_{\beta\rho}\varepsilon_{\gamma\sigma}
+  \right.
	\nonumber\\&&
  \left.
+2\delta^{ac}\delta^{br}\delta^{ds}\varepsilon_{\alpha\sigma}\varepsilon_{\beta\gamma}\varepsilon_{\delta\rho}-2\delta^{ab}\delta^{cr}\delta^{ds}\varepsilon_{\alpha\sigma}\varepsilon_{\beta\gamma}\varepsilon_{\delta\rho}-2\delta^{ac}\delta^{bd}\delta^{rs}\varepsilon_{\alpha\sigma}\varepsilon_{\beta\gamma}\varepsilon_{\delta\rho}
+  \right.
	\nonumber\\&&
  \left.
+2\delta^{ab}\delta^{cd}\delta^{rs}\varepsilon_{\alpha\sigma}\varepsilon_{\beta\gamma}\varepsilon_{\delta\rho}+6\delta^{as}\delta^{br}\delta^{cd}\varepsilon_{\alpha\gamma}\varepsilon_{\beta\sigma}\varepsilon_{\delta\rho}-6\delta^{ad}\delta^{br}\delta^{cs}\varepsilon_{\alpha\gamma}\varepsilon_{\beta\sigma}\varepsilon_{\delta\rho}
+  \right.
	\nonumber\\&&
  \left.
-2\delta^{ar}\delta^{bc}\delta^{ds}\varepsilon_{\alpha\gamma}\varepsilon_{\beta\sigma}\varepsilon_{\delta\rho}+2\delta^{ad}\delta^{bc}\delta^{rs}\varepsilon_{\alpha\gamma}\varepsilon_{\beta\sigma}\varepsilon_{\delta\rho}-2\delta^{ar}\delta^{bc}\delta^{ds}\varepsilon_{\alpha\beta}\varepsilon_{\gamma\sigma}\varepsilon_{\delta\rho}
+  \right.
	\nonumber\\&&
  \left.
+2\delta^{ad}\delta^{bc}\delta^{rs}\varepsilon_{\alpha\beta}\varepsilon_{\gamma\sigma}\varepsilon_{\delta\rho}-2\delta^{ac}\delta^{bs}\delta^{dr}\varepsilon_{\alpha\rho}\varepsilon_{\beta\gamma}\varepsilon_{\delta\sigma}+2\delta^{ab}\delta^{cs}\delta^{dr}\varepsilon_{\alpha\rho}\varepsilon_{\beta\gamma}\varepsilon_{\delta\sigma}
+  \right.
	\nonumber\\&&
  \left.
-6\delta^{ar}\delta^{bs}\delta^{cd}\varepsilon_{\alpha\gamma}\varepsilon_{\beta\rho}\varepsilon_{\delta\sigma}+6\delta^{ad}\delta^{bs}\delta^{cr}\varepsilon_{\alpha\gamma}\varepsilon_{\beta\rho}\varepsilon_{\delta\sigma}+2\delta^{as}\delta^{bc}\delta^{dr}\varepsilon_{\alpha\gamma}\varepsilon_{\beta\rho}\varepsilon_{\delta\sigma}
+  \right.
	\nonumber\\&&
  \left.
-6\delta^{ad}\delta^{bc}\delta^{rs}\varepsilon_{\alpha\gamma}\varepsilon_{\beta\rho}\varepsilon_{\delta\sigma}+6\delta^{ab}\delta^{cd}\delta^{rs}\varepsilon_{\alpha\gamma}\varepsilon_{\beta\rho}\varepsilon_{\delta\sigma}+2\delta^{as}\delta^{bc}\delta^{dr}\varepsilon_{\alpha\beta}\varepsilon_{\gamma\rho}\varepsilon_{\delta\sigma}
+  \right.
	\nonumber\\&&
  \left.
-2\delta^{ac}\delta^{bs}\delta^{dr}\varepsilon_{\alpha\delta}\varepsilon_{\beta\gamma}\varepsilon_{\rho\sigma}+2\delta^{ab}\delta^{cs}\delta^{dr}\varepsilon_{\alpha\delta}\varepsilon_{\beta\gamma}\varepsilon_{\rho\sigma}-6\delta^{ac}\delta^{br}\delta^{ds}\varepsilon_{\alpha\delta}\varepsilon_{\beta\gamma}\varepsilon_{\rho\sigma}
+  \right.
	\nonumber\\&&
  \left.
-6\delta^{as}\delta^{br}\delta^{cd}\varepsilon_{\alpha\gamma}\varepsilon_{\beta\delta}\varepsilon_{\rho\sigma}+6\delta^{ad}\delta^{br}\delta^{cs}\varepsilon_{\alpha\gamma}\varepsilon_{\beta\delta}\varepsilon_{\rho\sigma}+2\delta^{as}\delta^{bc}\delta^{dr}\varepsilon_{\alpha\gamma}\varepsilon_{\beta\delta}\varepsilon_{\rho\sigma}
+  \right.
	\nonumber\\&&
  \left.
+2\delta^{as}\delta^{bc}\delta^{dr}\varepsilon_{\alpha\beta}\varepsilon_{\gamma\delta}\varepsilon_{\rho\sigma}+6\delta^{ac}\delta^{br}\delta^{ds}\varepsilon_{\alpha\beta}\varepsilon_{\gamma\delta}\varepsilon_{\rho\sigma}
\right)\ .
\label{bfmBBXXXX0.1}
\end{eqnarray} 
The final result is
\begin{eqnarray}
\left[BBXXXX\right]^{abcdrs}_{\alpha\beta\gamma\delta\rho\sigma}&=& 
	\frac{\lambda^{2}}{12}V^{\left[ 6 \right]}\ .
\label{bfmBBXXXX}
\end{eqnarray}

\subsection{Wick Theorem}

Now that we have derived all the Feynman rules (summarized in app.~\ref{FRappendix}), we compute the Wick theorem for all the diagrams we are interested to.
The first computation will clarify the method.
\begin{itemize}
\item $1$-loop $BB$:\\
\begin{eqnarray}
B_{A}B_{B}
\quad\quad
B_{a}B_{b}\theta_{c}\theta_{d}V^{[2]}_{\left[ abcd \right]}
&=& 
-\theta_{c}\theta_{d}V^{[2]}_{\left[ ABcd \right]}=-V^{[2]}_{\left[ ABcc \right]}\ ,
\label{WT_BB}
\end{eqnarray}
the notation used is: $V^{i}$ indicates the vertex with $i$ quantum legs, capital latin index $\{A,B,\dots\}$ labels the external fields and small latin index $\{a,b,\dots\}$ the internal ones. Contractions between legs are performed using both $SO$ and $Sp$ metrics.  

\item $1$-loop $BX$:\\
Analogously, we obtain
\begin{eqnarray}
B_{A}\theta_{B}
\quad\quad
B_{a}\theta_{b}\theta_{c}\theta_{d}V^{[3]}_{\left[ abcd \right]}
&=& 
-V^{[3]}_{\left[ ABcc \right]}\ .
\label{WT_BX}
\end{eqnarray}

\item $1$-loop $XX$:\\
Again
\begin{eqnarray}
\theta_{A}\theta_{B}
\quad\quad
\theta_{a}\theta_{b}\theta_{c}\theta_{d}V^{[4]}_{\left[ abcd \right]}
&=& 
-V^{[4]}_{\left[ ABcc \right]}\ .
\label{WT_XX}
\end{eqnarray}

\item $1$-loop $BBXX=BBXX\times XXXX$:\\
This computation is more complicated
\begin{eqnarray}
&&
B_{A}B_{B}\theta_{C}\theta_{D}
\quad\quad
B_{a}B_{b}\theta_{c}\theta_{d}V^{[2]}_{\left[ abcd \right]}
\quad\quad
\theta_{e}\theta_{f}\theta_{g}\theta_{h}V^{[4]}_{\left[ efgh \right]}=
\nonumber\\&=&
B_{B}\theta_{C}\theta_{D}\left( -B_{b}\theta_{c}\theta_{d}V^{[2]}_{\left[ Abcd \right]}
\quad\quad
\theta_{e}\theta_{f}\theta_{g}\theta_{h}V^{[4]}_{\left[ efgh \right]} \right)
=
\nonumber\\&=&
B_{B}\theta_{D}\left( -B_{b}\theta_{c}\theta_{d}V^{[2]}_{\left[ Abcd \right]}
\quad\quad
\theta_{f}\theta_{g}\theta_{h}V^{[4]}_{\left[ Cfgh \right]} \right)
=
\nonumber\\&=&
\theta_{D}\left(+\theta_{c}\theta_{d}V^{[2]}_{\left[ ABcd \right]}
\quad\quad
\theta_{f}\theta_{g}\theta_{h}V^{[4]}_{\left[ Cfgh \right]} \right)
=
\nonumber\\&=&
\left(+\theta_{c}\theta_{d}V^{[2]}_{\left[ ABcd \right]}
\quad\quad
\theta_{g}\theta_{h}V^{[4]}_{\left[ CDgh \right]} \right)
=
\nonumber\\&=&
-V^{[2]}_{\left[ ABcd \right]}V^{[4]}_{\left[ CDcd \right]} \ .
\label{WT_BBXX1}
\end{eqnarray}

\item $1$-loop $BBXX=BXXX\times BXXX$:\\
\begin{eqnarray}
&&
B_{A}B_{B}\theta_{C}\theta_{D}
\quad\quad
B_{a}\theta_{b}\theta_{c}\theta_{d}V^{[3]}_{\left[ abcd \right]}
\quad\quad
B_{e}\theta_{f}\theta_{g}\theta_{h}V^{[3]}_{\left[ efgh \right]}=
\nonumber\\&=&
B_{B}\theta_{C}\theta_{D}
\left( 
\theta_{b}\theta_{c}\theta_{d}V^{[3]}_{\left[ Abcd \right]}
B_{e}\theta_{f}\theta_{g}\theta_{h}V^{[3]}_{\left[ efgh \right]}
-
B_{a}\theta_{b}\theta_{c}\theta_{d}V^{[3]}_{\left[ abcd \right]}
\theta_{f}\theta_{g}\theta_{h}V^{[3]}_{\left[ Afgh \right]}
 \right)
=
\nonumber\\&=&
\cdots\,\,\,\,=
\nonumber\\&=&
2V^{[3]}_{\left[ ACrs \right]}V^{[3]}_{\left[ BDrs \right]}-2V^{[3]}_{\left[ ADrs \right]}V^{[3]}_{\left[ BCrs \right]}\ .
\label{WT_BBXX2}
\end{eqnarray}

\item $1$-loop $BBXX=BBXXXX$:\\
\begin{eqnarray}
B_{A}B_{B}\theta_{C}\theta_{D}
\quad\quad
B_{a}B_{b}\theta_{c}\theta_{d}\theta_{e}\theta_{f}V^{[6]}_{\left[ abcdef \right]}
&=&  
-V^{[6]}_{\left[ABCDee  \right]}\ .
\label{WT_BBXX3}
\end{eqnarray}

\item $2$-loops $BB=BBXX\times XXXX$:\\
\begin{eqnarray}
B_{A}B_{B}
\quad\quad
B_{a}B_{b}\theta_{c}\theta_{d}V^{[2]}_{\left[ abcd \right]}
\quad\quad
\theta_{e}\theta_{f}\theta_{g}\theta_{h}V^{[4]}_{\left[ efgh\right]}
&=&
+V^{[2]}_{\left[ABrs  \right]}V^{[4]}_{\left[rs gg  \right]}\ .
\label{WT_BB2loop1}
\end{eqnarray}

\item $2$-loop $BB=BXXX\times BXXX$:\\
\begin{eqnarray}
B_{A}B_{B}
\quad\quad
B_{a}\theta_{b}\theta_{c}\theta_{d}V^{[3]}_{\left[ abcd \right]}
\quad\quad
B_{e}\theta_{f}\theta_{g}\theta_{h}V^{[3]}_{\left[ efgh\right]}
&=&
-2V^{[3]}_{\left[Abcd  \right]}V^{[3]}_{\left[Bbcd \right]}\ .
\label{WT_BB2loop2}
\end{eqnarray}

\item $2$-loop $BB=BBXXXX$:\\
\begin{eqnarray}
B_{A}B_{B}
\quad\quad
B_{a}B_{b}\theta_{c}\theta_{d}\theta_{e}\theta_{f}V^{[6]}_{\left[ abcd \right]}
&=&
-V^{[6]}_{\left[ABccee  \right]}\ .
\label{WT_BB2loop3}
\end{eqnarray}

\end{itemize}

\subsection{Non Linear Splitting and Ward Identities}

As already discussed in sec. \ref{bfmoutline}, the BFM is implementated by some Ward identities. In the present model the splitting  (\ref{bfmdefgroup}) is non linear and the fields transformations which make the $1PI$ functional generator invariant are not trivial. To find them we choose a simple transformation for one of the two fields and derive the transformation law for the other one imposing the invariance of the action. We set the linear field $X$ transforming linearly
\begin{equation}
  X\rightarrow X+\eta
  \Rightarrow
  e^{\lambda X}\rightarrow
  e^{\lambda\left( X+\eta \right)} \ .
  \label{Qtrasfineta}
\end{equation}
Obviously, with this notation we intend that the true field $\theta^{\alpha}_{a}$ transform linearly. Notice that for the action to be invariant it is enough that the group element or, simpler, the left invariant $1$-form is invariant.
Considering the $\lambda$ power expansion, $B$ becomes
\begin{equation}
  B\rightarrow B+\lambda\delta B^{\left[ 1 \right]}+\lambda^{2}\delta B^{\left[ 2 \right]}+\dots\ .
\label{Btrasfineta}
\end{equation}
To find the various $\delta B^{\left[ i \right]}$ we impose the invariance of $\tilde{J}^{\left( 1 \right)}$ (\ref{z2current1}) under the transformation (\ref{Qtrasfineta}) and (\ref{Btrasfineta}).
We obtain
\begin{equation}
  \begin{array}c
    \delta B^{\left[ 1 \right]}_{\mu}=-\partial_{\mu}\eta\ ,\\
    \delta B^{\left[ 2 \right]}_{\mu}=
-\frac{1}{2}\left( \left[ \left[ B \,,\,\eta \right],X \right]+
\left[ \left[ B \,,\,X \right],\eta \right]+
\left[ \left[ B \,,\,\eta \right],\eta\right] \right)\ ,
  \end{array}
  \label{bfmvariazioniB}
\end{equation}
that is
\begin{eqnarray}
\delta B^{\left[ 2 \right]}_{\mu}&=&
B_{\mu t}^{\phantom{\mu} \tau}\theta^{\l}_{l}\eta^{\rho}_{r}\,\Omega^{rlt\,\sigma}_{\tau\rho\lambda\,s}\,Q^{s}_{\s}+
B_{\mu t}^{\phantom{\mu} \tau}\eta^{\l}_{l}\eta^{\rho}_{r}\,\hat{\Omega}^{rlt\,\sigma}_{\tau\rho\lambda\,s}\,Q^{s}_{\s}
\ ,\label{deltaB2}
\end{eqnarray}
where
\begin{eqnarray}
\Omega^{rlt\,\sigma}_{\tau\rho\lambda\,s}&=&
+\frac{1}{2}\varepsilon_{\tau\rho}\delta^{rl}\varepsilon_{\lambda}^{\phantom{\lambda}\sigma }\delta_{s}^{t}
-\varepsilon_{\tau\rho}\delta^{tl}\varepsilon_{\lambda}^{\phantom{\lambda}\sigma }\delta_{s}^{r}
-\varepsilon_{\lambda\tau}\delta^{tr}\varepsilon_{\rho}^{\phantom{\lambda}\sigma }\delta_{s}^{l}
+\nonumber\\&&
-\frac{1}{2}\varepsilon_{\lambda\rho}\delta^{tr}\varepsilon_{\tau}^{\phantom{\lambda}\sigma }\delta_{s}^{l}
-\frac{1}{2}\varepsilon_{\tau\lambda}\delta^{lr}\varepsilon_{\rho}^{\phantom{\lambda}\sigma }\delta_{s}^{t}
+\frac{1}{2}\varepsilon_{\rho\lambda}\delta^{tl}\varepsilon_{\tau}^{\phantom{\lambda}\sigma }\delta_{s}^{r}
\ ,\label{OmegadeltaB2}
\end{eqnarray}
and
\begin{eqnarray}
\hat\Omega^{rlt\,\sigma}_{\tau\rho\lambda\,s}&=&
-\frac{1}{2}\varepsilon_{\tau\lambda}\delta^{lr}\varepsilon_{\rho}^{\phantom{\lambda}\sigma }\delta_{s}^{t}
+\frac{1}{2}\varepsilon_{\tau\lambda}\delta^{tr}\varepsilon_{\rho}^{\phantom{\lambda}\sigma }\delta_{s}^{l}
+\nonumber\\&&
+\frac{1}{2}\varepsilon_{\rho\tau}\delta^{tl}\varepsilon_{\lambda}^{\phantom{\lambda}\sigma }\delta_{s}^{r}
+\frac{1}{2}\varepsilon_{\rho\lambda}\delta^{tl}\varepsilon_{\tau}^{\phantom{\lambda}\sigma }\delta_{s}^{r}
\ .\label{hatOmegadeltaB2}
\end{eqnarray}

As we mentioned in sec. \ref{bfmoutline}, if the lagrangian is invariant under the simultaneous transformations (\ref{Qtrasfineta}) and (\ref{bfmvariazioniB}), the $1$PI functional generator satisfies the following relation
\begin{eqnarray}
\delta\tilde \Gamma=0\Rightarrow\delta B_{\mu}\left( x \right)\frac{\delta\tilde\Gamma}{\delta B_{\mu}\left( x \right)}+\eta\left( x \right)\frac{\delta\tilde \Gamma}{\delta \tilde X\left( x \right)}=0
\ ,\label{WIfond}
\end{eqnarray}
where $\tilde X$ is the analogous of $\tilde Q$ defined in sec. \ref{bfmoutline}. 
Obviously this equation must hold for every power of $\lambda$. If we derive (\ref{WIfond1}) by $B$ or $\tilde X$ we obtain relations between $1$PI Green functions: the {\it Ward Identities}.\\
We consider only $\delta B_{\mu}= \lambda\delta B_{\mu}^{\left[ 1 \right]}=-\lambda\partial_{\mu}\eta$. We get
\begin{eqnarray}
	-\lambda\partial^{\left[ x \right]}\eta\left( x \right)\frac{\delta\tilde \Gamma}{\delta B_{\mu}\left( x \right)}
+
\eta\left( x \right)\frac{\delta\tilde \Gamma}{\delta \tilde X\left( x \right)}
=0
\ .\label{WIfond1}
\end{eqnarray}
We now perform a Fourier transformation, recalling that
\begin{eqnarray}
&&
\partial_{\mu}\rightarrow - i p_{\mu}
\ ,\label{relationPartialP}
\end{eqnarray}
we obtain, simplifying  $\eta$, the following functional equation
\begin{eqnarray}
i \lambda p_{\mu}\frac{\delta\tilde \Gamma}{\delta B_{\mu}\left( p \right)}+\frac{\delta\tilde \Gamma}{\delta  \tilde X\left( p \right)}=0
\ .\label{WIfond2}
\end{eqnarray}
From this equation we shall obtain the Ward Identities differentiating by the fields $B$ or $\tilde X$. To be more precise, we expand $B$ or $\tilde X$ over the generators and we consider the  fields $B_{\mu a}^{\phantom{\mu}\alpha}$ and $\tilde \theta^{\alpha}_{a}$. We have
\begin{eqnarray}
i\lambda p_{\mu}\frac{\delta^{2}\tilde \Gamma}{\delta B_{\mu b}^{\phantom{\mu} \beta}\left( p \right)\delta B_{\nu c}^{\phantom{\nu} \gamma}\left(- p \right)}
+
\frac{\delta^{2}\tilde \Gamma}{\delta\tilde \theta^{\beta}_{b}\left( p \right)\delta B_{\nu c}^{\phantom{\nu} \gamma}\left( -p \right)}=0
\ .\label{WI1.2p}
\end{eqnarray}
In an analogous way we obtain a second Ward Identity
\begin{eqnarray}
i\lambda p_{\mu}\frac{\delta^{2}\tilde \Gamma}{\delta B_{\mu b}^{\phantom{\mu} \beta}\left( p \right)\delta \tilde \theta_{c}^{\gamma}\left( -p \right)}
+
\frac{\delta^{2}\tilde \Gamma}{\delta \tilde \theta^{\beta}_{b}\left( p \right)\delta \tilde \theta_{c}^{\gamma}\left( -p \right)}=0
\ .\label{WI2.0}
\end{eqnarray}
Using relations (\ref{BB}),(\ref{BX}) and (\ref{XX2point}) we get
\begin{eqnarray}
&&
4ip_{\mu}\lambda^{-1}\varepsilon_{\beta\gamma}\delta^{bc}-4ip_{\mu}\lambda^{-1}\varepsilon_{\beta\gamma}\delta^{bc}=0
\ ,\nonumber\\&&
4(i)^{2}p_{2}\varepsilon_{\beta\gamma}\delta^{bc}+4p_{2}\varepsilon_{\beta\gamma}\delta^{bc}=0
\ .\label{WI_def_0}
\end{eqnarray}
Then, the $1$-loop $2$-legs first order Ward Identities are satisfied.

\subsection{$1$-Loop Correction to $2$-Legs Green Functions}

We now construct the $1$-loop diagram for the self-energy of the background field $B^{\left( 1 \right)}_{\mu}$. 
The $1$-loop correction to the propagator is obtained contracting the indices $c,d$ and $\gamma,\delta$ with the propagator (\ref{XX}) and integrating over the loop momentum $q$. We obtain
\begin{eqnarray}
  \Gamma^{BB}_{1loop\,\mu\nu}&=&\left( \frac{1}{4} \right)\left( 1 \right)\left( -V^{[2]}_{\left[ ABcc \right]} \right)=
\nonumber\\
&=&\left( \frac{1}{4} \right)\left( 1 \right)\left( -4\left( n-m+2 \right)\varepsilon_{\alpha\beta}\delta^{ab}\int\dd^{d}q\frac{1}{q^{2}} \right)\gamma_{\mu\nu}=
\nonumber\\&=&
-\left( n-m+2 \right)\varepsilon_{\alpha\beta}\delta^{ab}\int\dd^{d}q\frac{1}{q^{2}}\gamma_{\mu\nu}
\ .  \label{bfmBB1loop1}
\end{eqnarray}
So, when $m+2-n=0$ the $1$ loop contribute is zero.
In the same way we compute the $1$-loop two point function with one external leg $B$ and one $X$. We contract the  indices $c,d$ and $\gamma,\delta$ of the term (\ref{bfmBXXX}) with (\ref{XX})\footnote{Remember that $B$ labels the external $\theta$ field and that we choose all the momenta as entering in the vertex.\label{footnotepB}} 
\begin{eqnarray}
   \Gamma^{BX}_{1loop\, \mu}&=&
   \left( \frac{1}{4} \right)\left( -i\frac{4\lambda}{3} \right)\left( -V^{[3]}_{\left[ ABcc \right]} \right)=
\nonumber\\&=&
   \left( \frac{1}{4} \right)\left( -i\frac{4\lambda}{3} \right)\left( 4\left( n-m+2 \right)\varepsilon_{\alpha\beta}\delta^{ab}\int\dd^{d}q\frac{1}{q^{2}} p_{\mu} \right)=
\nonumber\\&=&
-i\frac{4}{3}\lambda\left( 2+m-n \right)\varepsilon_{\alpha\beta}\delta^{ab}\int\dd^{d}q\frac{1}{q^{2}}p_{\mu}
  \ . \label{bfmBX1loop}
\end{eqnarray}
Finally, we calculate the $1$-loop self energy for the $XX$ propagator. As usual we contract the indices $\delta_{cd}\varepsilon^{\delta\gamma}$. We obtain\footnote{See note [\ref{footnotepB}]}
\begin{eqnarray}
\Gamma^{XX}_{1loop}&=&
\left( \frac{1}{4} \right)\left(- \frac{1}{3}\lambda^{2} \right)\left( -V^{[4]}_{\left[ ABcc \right]} \right)=
\nonumber\\&=&
\left( \frac{1}{4} \right)\left(- \frac{1}{3}\lambda^{2} \right)\left( 4\left( n-m+2 \right)\varepsilon_{\alpha\beta}\delta^{ab}\int\dd^{d}q\frac{p^{2}+q^{2}}{q^{2}} p_{\mu} \right)=
\nonumber\\&=&
-\frac{1}{3}\lambda^{2}\left( 2+m-n \right)\varepsilon_{\alpha\beta}\delta^{ab}\int\dd^{d}q\frac{p^{2}+q^{2}}{q^{2}}
\nonumber\ .
\label{bfmXX1loop}
\end{eqnarray}
To compute the UV-divergences we introduce a mass term ($M^{2}$) associated to the $\theta$ field, as we have done in sec. \ref{1loop1}. The lagrangian is then modified, becoming
\begin{eqnarray}
&&
\mathcal{L}=\frac{1}{2\pi\l^{2}} Str\left( 
   B^{\left( 1 \right)}\cdot B^{\left( 1 \right)}+2\l B^{\left( 1 \right)}\cdot\partial X+ \lambda^{2}\partial X\cdot\partial X+
   \right.
\nonumber\\&&+
M^{2}X\cdot X
  + \frac{4}{3}\lambda^{3}B^{\left( 1 \right)} \left[ \left[ \partial X,X \right],X \right]+
   \nonumber\\&&
   \left.+\frac{2\lambda^{4}}{3!}\left[ \left[ \partial X,X \right],X \right]\partial X
   +  \lambda^{2} B^{\left( 1 \right)} \left[ \left[ B^{\left( 1 \right)},X \right],X \right]+
	\right. 
	\nonumber\\&&
   \left.
	\frac{\lambda^{4}}{4\cdot3}B\cdot \left[ \left[ \left[ \left[ B,X \right],X \right],X \right],X \right]+
	\frac{\lambda^{4}}{4}\left[ \left[ B,X \right],X \right]\left[ \left[ B,x \right],X \right]
   \right)\ ,
\label{bfmLMass}
\end{eqnarray}
the new propagator is 
\begin{eqnarray}
\Delta^{\beta\gamma}_{cb}(\theta)=\frac{1}{4}\frac{\varepsilon^{\gamma\beta}\delta_{cb}}{q^{2}+M^{2}}
\ .\label{XXmass}
\end{eqnarray}
Using (\ref{UV_Pass_A})-(\ref{UV_Pass_A5}), we obtain
\begin{equation}
 \Gamma^{BB}_{1loop\,\mu\nu}\Big{|}_{UV}=
-\left( 2+m-n \right)\varepsilon_{\alpha\beta}\delta^{ab}
\frac{2\pi}{\varepsilon}\gamma_{\mu\nu}
\ ,\label{bfmBBUV}
\end{equation}
\begin{equation}
\Gamma^{BX}_{1loop\, \mu}\Big{|}_{UV}=
-i\frac{4}{3}\lambda\left( 2+m-n \right)\varepsilon_{\alpha\beta}\delta^{ab}p_{\mu}
\frac{2\pi}{\varepsilon}
\ ,\label{bfmBXUV}
\end{equation}
and
\begin{equation}
\Gamma^{XX}_{1loop}\Big{|}_{UV}=
-\frac{1}{3}\lambda^{2}\left( 2+m-n \right)\varepsilon_{\alpha\beta}\delta^{ab}\frac{2\pi}{\varepsilon}\left( p^{2}-M^{2} \right)
\ .\label{bfmXXUV}
\end{equation}
From now on we set $F=\left( m+2-n \right)$ and $\hat{F}=\left( m+2-n \right)\frac{2\pi}{\varepsilon}$. Then, skipping the metric terms
\begin{eqnarray}
&&
 \Gamma^{BB}_{1loop\,\mu\nu}\Big{|}_{UV}=-\hat{F}
\ ,\nonumber\\&&
\Gamma^{BX}_{1loop\, \mu}\Big{|}_{UV}=-i\frac{4}{3}\lambda\hat{F}p_{\mu}
\ ,\nonumber\\&&
\Gamma^{XX}_{1loop}\Big{|}_{UV}=-\frac{1}{3}\lambda^{2}\hat{F}\left( p^{2}-M^{2} \right)
\ .\label{UVcoeff}
\end{eqnarray}

\subsection{Renormalization}

In order to renormalize the theory we have to notice that
\begin{itemize}
\item we have to cancel the divergences from $BB$, $B\theta$ and $\theta\theta$ $1$-loop functions (\ref{UVcoeff});
\item to absorb such divergences we have to consider the following terms from the lagrangian (we miss the coefficient $(2\pi)^{-1}$
\begin{equation}
\frac{1}{\lambda^{2}}\varepsilon_{\alpha\beta}\delta^{ab}\gamma^{\mu\nu}B^{\alpha}_{a\,\mu} B^{\beta}_{b\,\nu}
\ ,\quad\quad\quad
\frac{2}{\lambda}\varepsilon_{\alpha\beta}\delta^{ab}\gamma^{\mu\nu}B^{\alpha}_{a\,\mu}\cdot \partial\theta^{\beta}_{b\,\nu}
\ ,\quad\quad\quad
\varepsilon_{\alpha\beta}\delta^{ab}\gamma^{\mu\nu}\partial \theta^{\alpha}_{a\,\mu}\cdot \partial\theta^{\beta}_{b\,\nu}
\ ;\label{renBare2func}
\end{equation} 
\item the classic field $B$ should not be renormalized via the wave function renormalization;
\end{itemize}
To perform the renormalization we  introduce
\begin{align}
&
\lambda=Z_{\lambda}\lambda_{R}
\ ,
&\theta=Z_{\theta}^{1/2}\theta_{R}
\ ,\label{renDef}
\end{align}
where
\begin{eqnarray}
Z_{x}=1+\lambda^{2}_{R}\delta Z_{x}
\ .\label{renDefZ}
\end{eqnarray}
The coefficient $\delta Z_{x}$ is the counterterm. Notice that it is possible  to perform the following expansion
\begin{eqnarray}
\frac{1}{\lambda^{2}}\rightarrow \frac{1}{\lambda_{R}^{2}}\frac{1}{1+2\lambda_{R}^{2}\delta Z_{\lambda}}=
\frac{1}{\lambda_{R}^{2}}\left(  1-2\lambda_{R}^{2}\delta Z_{\lambda}+O(\lambda_{R}^{4})\right)
\ .\label{renExp}
\end{eqnarray}
The first terms (\ref{renBare2func}) of the lagrangian read
\begin{eqnarray}
\mathcal{L}
&=&
\mathcal{L}_{R}+\delta\mathcal{L}=
\nonumber\\&=&
\frac{1}{\lambda^{2}}\left( 1-2\lambda^{2}_{R}\delta Z_{\lambda} \right)B\cdot B
+
\frac{2}{\lambda_{R}}\left( 1-\lambda^{2}_{R}\delta Z_{\lambda} \right) B\cdot\partial\theta_{R}\left( 1+\lambda_{R}^{2}\delta Z_{\theta} \right)^{1/2}
+
\nonumber\\&&
+
\left( 1+\lambda_{R}^{2}\delta Z_{\theta} \right)\partial\theta_{R}\cdot \partial\theta_{R}
=
\nonumber\\&=&
\frac{1}{\lambda^{2}_{R}}B\cdot B^{\beta}_{b}
+
\frac{2}{\lambda_{R}}B\cdot \partial\theta_{R}
+
\partial \theta_{R}\cdot \partial\theta_{R}
+
\nonumber\\&&
-
2\delta Z_{\lambda} B\cdot B
+
2\lambda_{R}\left( -\delta Z_{\lambda}+\frac{1}{2}\delta Z_{\theta} \right)B \cdot \partial\theta_{R}
+
\lambda_{R}^{2}\delta Z_{\theta} \partial \theta_{R}\cdot \partial\theta_{R}
\ ,\label{renL}
\end{eqnarray}
where the $\cdot$ sign here implies the  contraction between all the indices with the metrics $\varepsilon_{\alpha\beta}\delta^{ab}\gamma^{\mu\nu}$.
To absorb the coefficients we construct the counterterm diagrams. Using the same rules we obtain
\begin{eqnarray}
&&
\delta BB=-4\delta Z_{\lambda}
\ ,\nonumber\\&&
\delta BX=4i\lambda \left( -\delta Z_{\lambda} +\frac{1}{2}\delta Z_{\theta} \right) p_{\mu}
\ ,\nonumber\\&&
\delta XX=2\lambda^{2}\delta Z_{\theta}\left( p^{2}+M^{2} \right)
\ .\label{counterterms.0}
\end{eqnarray}
In order to cancel the divergences (\ref{UVcoeff}) we have to solve the following equations
\begin{eqnarray}
&&
\Gamma^{BB}_{1loop\,\mu\nu}\Big{|}_{UV}+\delta BB=0
\ ,\nonumber\\&&
\Gamma^{BX}_{1loop\, \mu}\Big{|}_{UV}+\delta BX=0
\ ,\nonumber\\&&
\Gamma^{XX}_{1loop}\Big{|}_{UV}+\delta XX=0
\ .\label{counterterms.1}
\end{eqnarray}
We have then
\begin{align}
&
\delta Z_{\lambda}=-\frac{1}{4}\hat{F}
\ ,
&\delta Z_{\theta} = \frac{1}{6}\hat{F}\ .
\label{counterterms.2}
\end{align}

\subsection{$2$-Loop Correction to $2$-Legs Green Function}

We want to compute a more complicated diagram. The $1$-loop $4B$ Green function is obtained from two vertices $V^{\left[ 2 \right]}$ but power counting assures that it is UV-finite. We shall then pass to $2$-loop correction to $2$-legs Green function.

There are three diagrams which contribute to the $2$-loop $2$-point function
{
\begin{center}
\includegraphics[scale=.3]{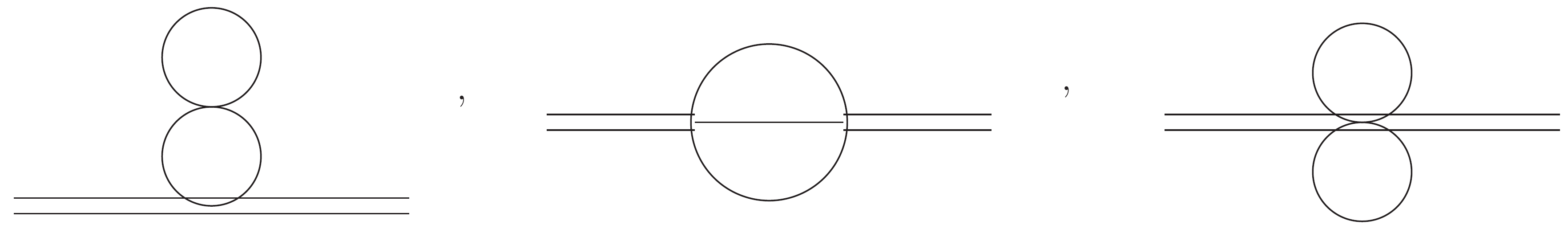}
\end{center}
}\noindent
The Wick theorem fixed the combinatorial coefficients.

\subsubsection{First Diagram}

To construct the first diagram we consider the $BBXX$ and  $XXXX$ vertices
{
\begin{center}
\includegraphics[scale=.3]{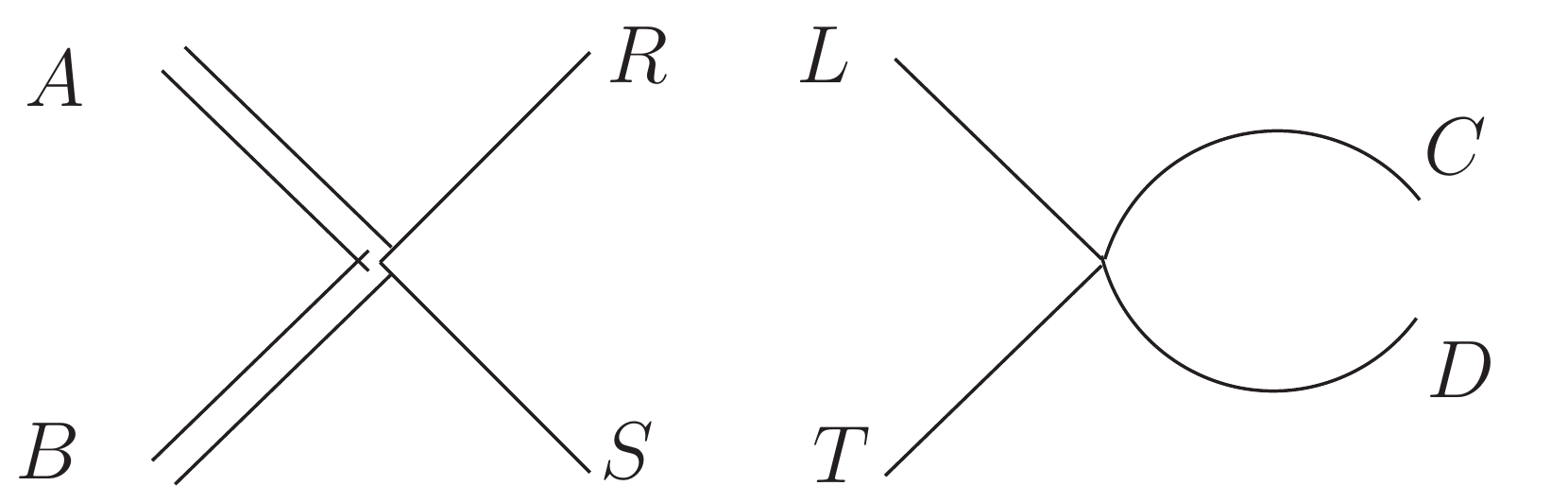}
\end{center}
}
\noindent
with the following conventions
\begin{align}
&
K_{R}=-q\ ,
&
&
K_{C}=-k\ , 
&
&
K_{L}=q\ ,
&
\nonumber\\&
K_{S}=q\ ,
&
&
K_{D}=k\ ,
&
&
K_{T}=-q\ .
&
\label{BB2loop1conv}
\end{align}
We obtain
\begin{eqnarray}
&&
\left( \frac{1}{4} \right)^{3}\left( 1 \right)\left( -\frac{1}{3}\lambda^{2} \right)\left( +V^{[2]}_{\left[ABrs  \right]}V^{[4]}_{\left[rs gg  \right]} \right)=
\nonumber\\&=&
-\frac{1}{192}\lambda^{2} \int\dd^{q}\dd^{k}\frac{1}{\left( q^{2}+M^{2} \right)^{2}\left( k^{2}+M^{2} \right)}\left( +V^{[2]}_{\left[ABrs  \right]}V^{[4]}_{\left[rs gg  \right]}\right)  \label{BB2loop1.1}=
\nonumber\\&=&
-\frac{8}{192}\lambda^{2} \left(2+m-n\right)^2 \int\dd^{d}q\dd^{d}k\frac{1}{\left( q^{2}+M^{2} \right)^{2}\left( k^{2}+M^{2} \right)}\left( +V^{[2]}_{\left[ABrs  \right]}V^{[4]}_{\left[rs gg  \right]}\right) \times
\nonumber\\&&
\times
\left(2p_{C}\cdot p_{D}-p_{C}\cdot p_{L}-p_{C}\cdot p_{T}-p_{D}\cdot p_{L}-p_{D}\cdot p_{T}+2p_{L}\cdot p_{T}\right)\delta^{ab}\varepsilon_{\alpha\beta}=
\nonumber\\&=&
-\frac{8}{192}\lambda^{2} \left(2+m-n\right)^2 \int\dd^{d}q\dd^{d}k\frac{-2k^{2}-2q^{2}}{\left( q^{2}+M^{2} \right)^{2}\left( k^{2}+M^{2} \right)}\varepsilon_{\alpha\beta}=
\nonumber\\&=&
\frac{1}{12}\lambda^{2} \left(2+m-n\right)^2 \int\dd^{d}q\dd^{d}k\frac{k^{2}+q^{2}}{\left( q^{2}+M^{2} \right)^{2}\left( k^{2}+M^{2} \right)}\varepsilon_{\alpha\beta}
\ .
\end{eqnarray}

\subsubsection{Second Diagram}

The second diagrams is
{
\begin{center}
\includegraphics[scale=.3]{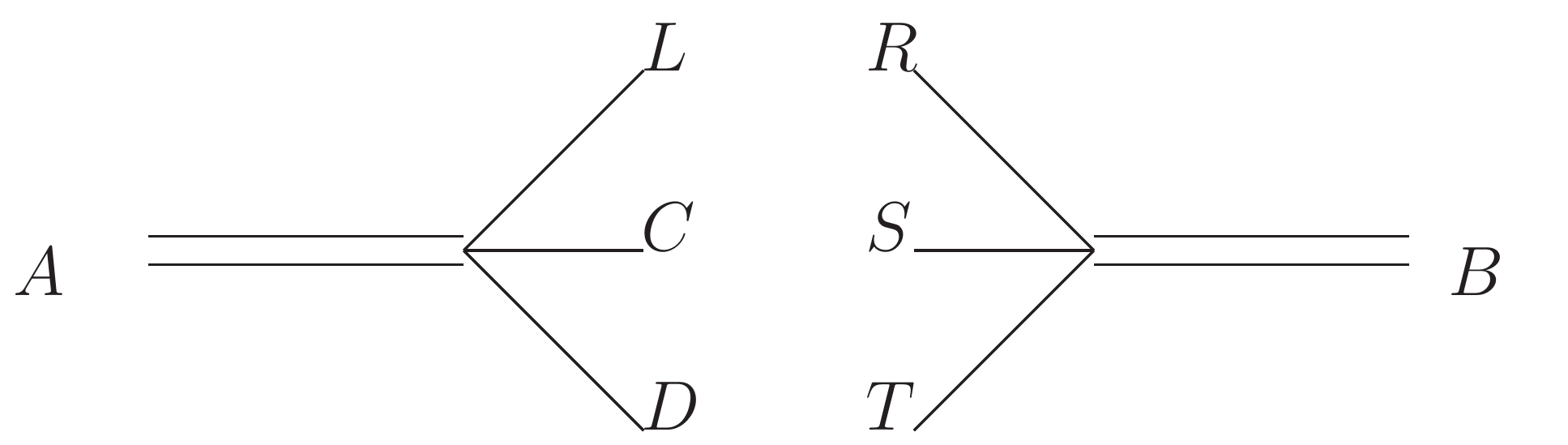}
\end{center}
}
\noindent
with the following conventions
\begin{align}
&
K_{R}=q\ ,
&
&
K_{C}=q-p-k\ ,
&
&K_{D}=k\ ,
&
\nonumber\\
&
K_{L}=-q\ ,
&
&
K_{S}=p+k-q\ ,
&
&
K_{T}=-k\ .
&
\label{BB2loop2conv}
\end{align}
We obtain
\begin{eqnarray}
&&
\left( \frac{1}{4} \right)^{3}\left( -i\lambda\frac{4}{3} \right)^{2}\left( -2V^{[3]}_{\left[Abcd  \right]}V^{[3]}_{\left[Bbcd \right]} \right)=
\nonumber\\&=&
\frac{1}{18}\lambda^{2}V^{[3]}_{\left[Abcd  \right]}V^{[3]}_{\left[Bbcd \right]}=
\nonumber\\&=&
-\frac{72}{18}\lambda^{2}\left(n+m\left(-1+2n\right)\right)
\times
\nonumber\\&&
\times
\int\dd^d q\dd^d k \frac{\left(k^2+\frac{1}{3}p^2+k\left(p-q\right)-pq+q^2\right)\delta^{ab}\varepsilon_{\alpha\beta}}{\left( q^{2}+M^{2} \right)^{2}\left( k^{2}+M^{2} \right)\left( \left( q-k-p \right)^{2}+M^{2} \right)}\ .
\end{eqnarray}
We shall use the results (\ref{UV_Pass_2loop_2}),(\ref{UV_Pass_2loop_3}),(\ref{UV_Pass_2loop_3b}) to extract explicitly the $UV$ divergent part
\begin{eqnarray}
&&
\frac{72}{18}\frac{3}{2}\lambda^{2}\left( m-n-2m n \right)\int\dd^d q\dd^d k \frac{1}{\left( q^{2}+M^{2} \right)\left( k^{2}+M^{2} \right)}+ O\left( 1 \right)=
\nonumber\\&=&
6\lambda^{2}\left( m-n-2m n \right)\int\dd^d q\dd^d k \frac{1}{\left( q^{2}+M^{2} \right)\left( k^{2}+M^{2} \right)}+ O\left( 1 \right)\ .
\end{eqnarray}

\subsubsection{Third Diagram}

The third diagrams is
{
\begin{center}
\includegraphics[scale=.3]{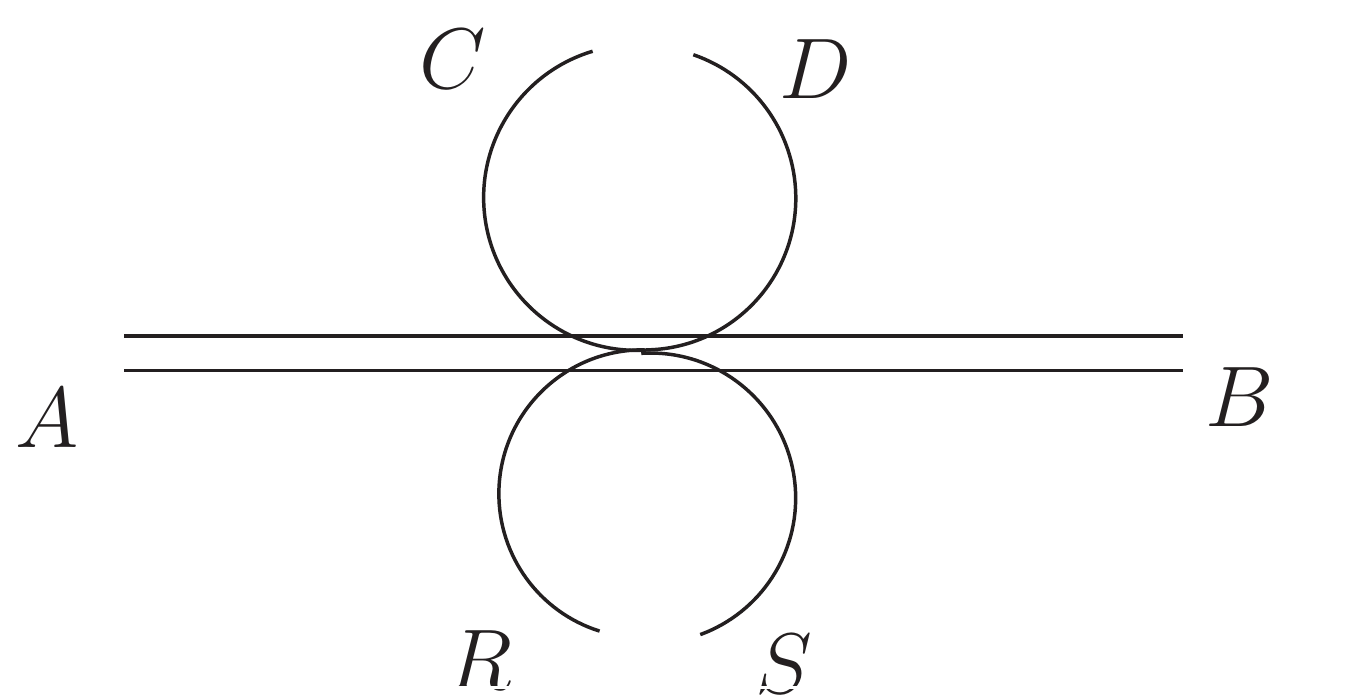}
\end{center}
}
\noindent
with the following conventions
\begin{align}
&
K_{R}=k \ , && K_{C}=-q\ ,&
\nonumber\\&
K_{S}=-k\ , && K_{D}=q\ .&
\label{BB2loop3conv}
\end{align}
We obtain
\begin{eqnarray}
&&
\left( \frac{1}{4} \right)^{2}\left( \frac{1}{12}\lambda^{2} \right)\left( -V^{[6]}_{\left[Abccdd  \right]}\right)=
\nonumber\\&=&
-\frac{1}{192}\lambda^{2}V^{[6]}_{\left[Abccdd  \right]}=
\nonumber\\&=&
-\frac{96}{192}\lambda^{2}\left( 4+m^{2}+m\left( 7-8n \right)-7n +n^{2} \right)
\int\dd^d q\dd^d k \frac{1}{\left( q^{2}+M^{2} \right)\left( k^{2}+M^{2} \right)}=
\nonumber\\&=&
-\frac{1}{2}\lambda^{2}\left( 4+m^{2}+m\left( 7-8n \right)-7n +n^{2} \right)
\int\dd^d q\dd^d k \frac{1}{\left( q^{2}+M^{2} \right)\left( k^{2}+M^{2} \right)}
\ .
\end{eqnarray}
\subsubsection{Results}

To compute the total correction to  $BB$ $2$-point function we combine the three partial results, obtaining
\begin{eqnarray}
\Gamma^{XX}_{2loop}
&=& 
\frac{\lambda^{2}}{12}\left( m+2-n \right)\int\dd^d q\dd^d k \frac{1}{\left( q^{2}+M^{2} \right)\left( k^{2}+M^{2} \right)}
\ .\label{BFM2loopFIN}
\end{eqnarray}
This confirms the conformal property of $OSp(m+2|m)/SO(m+2)\times Sp(m)$ coset models at $2$-loops.

\newpage
\chapter*{Conclusions of Part I}
\addcontentsline{toc}{chapter}{Conclusions of Part I}

We discuss some aspects of fermionic T-duality from the quantum point of view. 
For that purpose we decided to adopt  the fermionic cosets 
introduced in \cite{Berkovits:2007zk} as a new limit of the $AdS_n \times S^m$ string theory 
models as a playground. They have the advantage that the large amount of isometries permits an easy, 
even though not straightforward, computation of the quantum corrections at higher loops. 
In addition, for that model we can easily point out some of the obstructions in the T-dual 
construction. 

We start by considering three different techniques to build this coset models based on 
the underlying superalgebra, on the nilpotency of the supercharges  in terms of vielbeins and connections. In particular we discuss 
the pricipal $\sigma$-model based on orthosymplectic group $OSp(n|m)$. We discuss the constraints to be satisfied for having a T-duality in the conventional sense (namely by gauging the 
isometry group and then eliminating the original coordinates in terms of the 
Lagrange multipliers) and we show that for the fermionic T-duality there might be some obstructions due to anticommuting nature of the fundamental fields. Nonetheless, we propose a new technique based on non-abelian T-duality derived in \cite{de la Ossa:1992vc}. 
We show that it is possibile to construct the T-dual for all the models proposed and 
we give a recipe to compute the quantum corrections. Moreover, we derived the simplest terms for the dual lagrangian and we found they possess the same structure of the original model.

In the second chapter of the Thesis, we use two different methods to compute the corrections 
to the action. Using the first method, we are able to compute the first loop corrections finding that 
they vanish if the relation between the dimensions of the bosonic subgroups $SO(n)$ and $Sp(m)$ is $n=m+2$. This condition guarantees that the supergroup, viewed as a supermanifold, is a super Calabi-Yau and that implies the conformal invariance of the principal 
$\sigma$-model (as discussed also in \cite{Bershadsky:1999hk}). Using the BFM, we are able to 
push it to two-loops confirming the result at one-loop. 

There are several open issues that are not discussed in the present work and presently
are under investigation: 1) is it possibile to extend the well-known result of 
\cite{Berkovits:1999im} and \cite{Berkovits:2004xu} to all orders also for 
orthosymplectic groups? 2) is it possible to extend the fermionic T-duality  to
other models by overpassing the obstruction discussed? 3) do the WZW models 
presented in \cite{Gotz:2006qp,Mitev:2008yt,Candu:2009ep} can be T-dualized? 
4) how does the T-duality survive the quantum corrections?


\part{Supersymmetric Fluid Dynamic}


\chapter{Supersymmetric Fluid Dynamics from Black Hole Superpartners}
\label{chRew}

\section{Review of Relativistic Fluid Dynamics}\label{secFlu}


In this section we briefly review the main concepts of relativistic fluid dynamics \cite{Landau6,Son:2007vk,Rangamani:2009xk,Hubeny:2011hd,Kovtun:2012rj}.
After a general introduction, we focus on conformal fluids.

Hydrodynamics can be interpreted as an effective long--distance description for a given
classical or quantum many--body system at non--zero temperature.
Any interacting system is characterized by an intrinsic length scale: the mean free path length $l_{\textrm{mfp}}$
which is, in kinetic theory, the distance of free motion of the particles between two successive interactions.
In the hydrodynamic limit we consider a scale $L$ much larger than $l_{\textrm{mfp}}$ and this means that in a small region (compared to $L$) of our system the constituents interact by themselves several times and therefore they thermalize locally.
For this reason we can treat the long--distance system as a fluid and we can study it by the analysis of thermodynamical quantities and the laws of hydrodynamics.

We refer to relativistic hydrodynamics when the microscopic components of the fluid are constrained by Lorentz symmetry, as in relativistic quantum field theories. 
Thus, the subject of study of relativistic hydrodynamics is not limited to collective motion of particles which move with speed similar to light.

Due to the presence of dissipative terms, hydrodynamics is generally formulated not by an action principle but by the analysis of equations of motion.\footnote{We briefly discuss the action principle for non--dissipative hydrodynamics in sec.~\ref{bosonlag}.}
To determine these equations, we consider the Noether currents associated with the symmetries. 
In relativistic systems the symmetries are given by translations, boost, rotations and eventually by some internal symmetries.
Therefore, the equations of motion for relativistic hydrodynamics are simply the conservation equations of the currents associated to these symmetries, {\it i.e.} the energy--momentum tensor $T^{\mu\nu}$, the ``angular momentum''  $M^{\mu\nu\rho}=x^{\mu}T^{\nu\rho}-x^{\nu}T^{\mu\rho}$ and the internal currents $J^{\mu}_{I}$. 
Notice however that the conservation $\nabla_{\rho}M^{\mu\nu\rho}$ follows from the conservation of the energy--momentum tensor.
Hence the whole dynamical content of hydrodynamics is encoded in the request of conservation of energy--momentum tensor $T^{\mu\nu}$ and charge currents $J^{\mu}_{I}$
\begin{align}
	&\nabla_{\mu} T^{\mu\nu} = 0
	\ ,
	& \nabla_{\mu} J^{\mu}_{I} = 0 \ ,
	\label{FluidCons}
\end{align}
where $I=\{1,\cdots\}$ labels the charges which characterise the fluid.
Focusing on the energy--momentum tensor, note that in $d$ dimension the number of components of $T^{\mu\nu}$ is $\frac{1}{2}d(d+1)$ and then for $d>2$ the variables are more than the equations, which are $d$.
To close the system of equations, we need to reduce the number of independent components of $T^{\mu\nu}$. 
As we have already stressed out, if the perturbations are larger than $l_{\textrm{mfp}}$, the system can be consider in local thermal equilibrium and hence at a given time is determined by the temperature $T(\vec{x})$ and the local fluid velocity $u^{\mu}(\vec{x})$. 
If we normalize the velocity to $-1$, {\it i.e} $u_{\mu}u^{\mu}=-1$, the total number of independent field is $d$ and then we have the same number of equation and variables.\footnote{If currents are taken into account, we add to the variables also the chemical potentials $\mu_{I}$, one for each conserved current.}
The dependence of $T^{\mu\nu}$ in terms of $T(x)$ and $u^{\mu}(x)$ is given by {\it constitutive relations}.
Since we assume the deviation from equilibrium to be small, we expect that the contribution of the terms at higher order in derivatives of the variables $T$ and $u^{\mu}$ is increasingly subdominant in the hydrodynamic limit.
For that, we write the constitutive relations in a derivative expansion
\begin{align}
	&T^{\mu\nu} = \sum_{n=0}^{\infty} T^{\mu\nu}_{n}
	\ ,
	&J^{\mu}_{I} = \sum_{n=0}^{\infty} (J^{\mu}_{I})_{n}
	\ ,
	\label{FluidT2}
\end{align}
where the $l$--term is the $n^{\textrm{th}}$ order in the derivatives of the fluid fields. 
One can estimate the contribution of the $n^{\textrm{th}}$--order term as $(l_{\textrm{mfp}}/L)^{n}$ with $L$ is the scale of temperature and velocity fields, as we defined above.


At zero order in the derivatives, we retrieve the perfect fluid dynamics: the energy momentum tensor is determined by Pascal law, which in the rest frame ($u^{0}=1$, $u^{i}=0$ with $i$ spatial coordinates) provides the following energy--momentum tensor
\begin{align}
	T^{\mu\nu} = \textrm{diag}\left[ \rho,p,p,p \right] 
	\ ,
	\label{FluidPascal}
\end{align}
where the component longitudinal to the flow $\rho$ is fluid energy density and the spatial components transverse to the fluid flow $p$ are the fluid pressure.
By acting with a finite boost transformation on (\ref{FluidPascal}) with velocity $u^{\mu}$ we obtain the energy--momentum tensor in covariant formalism
\begin{align}
	T^{\mu\nu} 
	= 
	p g^{\mu\nu} 
	+ 
	( \rho + p ) u^{\mu} u^{\nu}
	\ .
	\label{FluidT0}
\end{align}
Notice that the velocity can be defined as
\begin{align}
	u^{\mu} = \left\{ \frac{1}{\sqrt{1-v^{2}}}\,,\, \frac{\vec{v}}{\sqrt{1-v^{2}}} \right\}
	\ ,
	\label{FluBoost}
\end{align}
with $v^{2}=v_{i}v^{i}$.
Conserved currents takes the simple form
\begin{align}
	J^{\mu}_{I} = n_{I} u^{\mu}
	\ ,
	\label{FluidJ0}
\end{align}
with $n_{I}$ conserved charges.
From now on we reduce the number of conserved currents to $1$ in order to suppress the index $I$.
In relativistic models it refers to baryon number conservation, since the particle number is not conserved (it is always possible to create a particle--antiparticle pair).
Please note that equations (\ref{FluidT0}\ref{FluidJ0}) are true in global equilibrium with $n$, $u^{\mu}$, $p$ and $\rho$ constants.
In hydrodynamics we promote these quantities as local functions of spacetime coordinates.

Energy--momentum conservation leads to energy conservation and Euler equation
\begin{align}
	\nabla_{\mu} T^{\mu\nu} 
	& =
	\nabla_{\mu} (\rho + p) u^{\mu} u^{\nu} 
	+ 
	(\rho + p)\nabla_{\mu} u^{\mu} u^{\nu} 
	+
	(\rho + p) u^{\mu} \nabla_{\mu} u^{\nu}
	+ 
	g^{\mu\nu} \nabla_{\mu} p
	= 0
	\ ,
	\label{FluidCon1}
\end{align}
its projection along the fluid flow read
\begin{align}
	u_{\nu} \nabla_{\mu} T^{\mu\nu} 
	& =
	- u^{\mu} \nabla_{\mu} \rho
	-
	( \rho + p ) \nabla_{\mu} u^{\mu}=0
	\ ,
	\label{FluidCon2}
\end{align}
and the transversal projection is
\begin{align}
	P_{\alpha\nu} \nabla_{\mu} T^{\mu\nu} 
	& =
	( \rho + p) u^{\nu} \nabla_{\nu} u_{\alpha} 
	+ 
	P_{\alpha}{}^{\nu} \nabla_{\nu} p = 0
	\ ,
	\label{FluidCon3}
\end{align}
where the projector orthogonal to fluid velocity is defined as
\begin{align}
	P^{\mu\nu} = u^{\mu} u^{\nu} + g^{\mu\nu}
	\ .
	\label{FluidProj}
\end{align}
To understand the meaning of (\ref{FluidCon2},\ref{FluidCon3}) we consider the non--relativistic limit $|v|<<1$.
In this limit we have
\begin{align}
	&u^{\mu}\sim\left\{ 1,\vec{v} \right\} \ ,&
	u^{\mu}\nabla_{\mu} \sim \partial_{0} + \vec{v}\cdot\vec{\partial}\ ,
	\label{Flu31}
\end{align}
moreover, equations of state are characterized by $\rho>>p$ and the energy density is dominated by matter.
With these assumptions, eq.~(\ref{FluidCon2}) reduces to
\begin{align}
	\partial_{0}\rho + {\partial_{i}}(\rho{v^{i}}) = 0
	\ ,
	\label{Flu32}
\end{align}
which is the non--relativistic continuity equation.
Eq.~(\ref{FluidCon3}) has two different set of components: for $\alpha=0$ we have
\begin{align}
	\vec{v}\cdot\vec{\partial}p=0\ ,
	\label{CFlu33}
\end{align}
and for $\alpha=i$ we have 
\begin{align}
	\rho \partial_{0}{v^{i}} + \rho (\vec{v}\cdot \vec{\partial}){v^{i}} + \partial_{i} p + {v_{i}}\partial_{0} p = 0
	\ .
	\label{CFlu34}
\end{align}
Contracting with $v_{i}$ it reduces to eq.~(\ref{CFlu33}).
In non--relativistic limit, the last term of (\ref{CFlu34}) is neglected and we obtain
\begin{align}
	\partial_{0}{v_{i}} +  (\vec{v}\cdot \vec{\partial}){v_{i}} = -\frac{1}{\rho}\partial_{i} p 
	\ ,
	\label{CFlu35}
\end{align}
which is the non--relativistic Euler equation.

As expected, at zero order (ideal fluid) we have no dissipative contribution.
A way to see it is considering the conservation of energy--momentum tensor projected along the velocity
\begin{align}
	u_{\nu} \nabla_{\mu} T^{\mu\nu} = - u^{\mu} \partial_{\mu} \rho - (\rho + p) \nabla_{\mu} u^{\mu} = 0
	\ ,
	\label{CFluS1}
\end{align}
with the thermodynamic relations
\begin{align}
	& \rho = T s - p + \mu n \ ,
	&\dd \rho = T \dd s + \mu \dd n \ ,&
	\label{CFluS2}
\end{align}
where $s$ is the entropy density and $\mu$ is the chemical potential. We get
\begin{align}
	\mu \nabla_{\mu} ( n u^{\mu} ) + T \nabla_{\mu} ( s u^{\mu} ) = 0
	\ .
	\label{CFlu3}
\end{align}
Assuming the conservation of the matter current $j^{\mu} = n u^{\mu}$, eq.~(\ref{CFlu3}) reduces to entropy conservation along the fluid flow.
\begin{align}
	\nabla_{\mu} J^{\mu}_{s} = \nabla_{\mu} ( s u^{\mu} ) = 0\ .
	\label{CFlu4}
\end{align}

To add dissipative contribution (viscosity and thermal conduction) to energy momentum tensor we have to go beyond the zero order in derivative expansion.
We decompose the velocity derivative $\nabla^{\mu} u^{\nu}$ into irreducible representations:
first, we split it in the contribute along the fluid velocity (the acceleration $a^{\mu}$) and in a transverse part, which can in turn be decomposed into symmetric traceless tensor (shear $\sigma^{\mu\nu}$), antisymmetric tensor (vorticity $\omega^{\mu\nu}$) and trace part (the divergence $\theta$)
\begin{align}
	\nabla^\mu u^\nu
	=
	- a^\mu u^\nu + \sigma^{\mu\nu} + \omega^{\mu\nu} + \frac{1}{3} \theta P^{\mu\nu} 
	\ ,
	\label{FluidUdec}
\end{align}
where the various components are defined as
\begin{align}
	\theta & = \nabla_{\mu} u^{\mu} \ ,
	\nonumber \\
	a^{\mu} & = u^{\nu} \nabla_{\nu} u^{\mu} \ ,
	\nonumber \\
	\sigma^{\mu\nu} & = P^{\mu\alpha} P^{\nu\beta} \nabla_{\left( \alpha \right.} u_{\left. \beta \right)}
	- \frac{1}{d-1}\theta P^{\mu\nu}
	\ ,\nonumber\\
	\omega^{\mu\nu} & = P^{\mu\alpha} P^{\nu\beta} \nabla_{\left[ \alpha \right.} u_{\left. \beta \right]}
	\ .
	\label{FluidDefGrad}
\end{align}
Notice that due to the normalization of the velocity $u_{\mu}u^\mu=-1$ the following identities hold
\begin{align}
	&u^{\mu} \nabla_{\rho} u_{\mu} = 0\ ,&
	&\nabla_{\rho} u^{\mu} \nabla_{\sigma} u_{\mu} = - u^{\mu} \nabla_{\rho} \nabla_{\sigma} u_{\rho}
	\ ,&	
	u_{\mu} a^{\mu} = 0
	\ ,&
	\label{FluDefId}
\end{align}
where the last one states the tracelessness of $a^\mu u^\nu$.

In relativistic fluid it is not possible to distinguish between mass and energy flux, one flux involving necessarily the other.
For this reason it is convenient to fix the velocity field in an unambiguous, even if arbitrary, way. 
There are different conventions and we choose the so--called {\it Landau frame}, in which at equilibrium the components of energy--momentum tensor which are longitudinal to the velocity are associated with the energy density
\begin{align}
	u_{\mu} T^{\mu\nu} = - \rho u^{\nu}
	\ ,
	\label{CFluLF}
\end{align}
as a consequence, the dissipative contributions to energy momentum tensor must satisfy
\begin{align}
	u_{\mu} T^{\mu\nu}_{n} = 0 
	\ ,
	\qquad\qquad n>0
	\ .
	\label{CFluLF2}
\end{align}
It is possible to change frame by performing an appropriate field redefinition. 
In particular, with this choice we have no energy flow in the local rest frame.


To determine the first order contribution to energy momentum tensor, we consider all the symmetric two--indices tensor respecting the Landau frame condition eq.~(\ref{CFluLF}).
We get
\begin{align}
	T_{1}^{\mu\nu} = - 2 \eta \sigma^{\mu\nu} - \zeta \theta P^{\mu\nu}
	\ ,
	\label{Fluid1ord}
\end{align}
where we introduced two new parameters, the {\it shear viscosity} $\eta$ and the {\it bulk viscosity} $\zeta$. 
Notice that the Landau frame condition (\ref{CFluLF2}) is satisfied.
Hence, the energy--momentum tensor at first order in gradient expansion reads
\begin{align}
	T^{\mu\nu} = p g^{\mu\nu} 
	+ 
	( \rho + p ) u^{\mu} u^{\nu} - 2 \eta \sigma^{\mu\nu} - \zeta \theta P^{\mu\nu}	
	\ .
	\label{FluTmunu2}
\end{align}
In the same way we obtain the correction to conserved current. Since this result is not needed in the followings we do not report the result. We remand the interested reader to \cite{Rangamani:2009xk}.

Imposing the conservation of the energy--momentum tensor up to first order  $T_{0}^{\mu\nu}+T_{1}^{\mu\nu}$ we obtain the Navier--Stokes equations. As before, we project the conservation equation first along the fluid and then along the transverse direction, obtaining 
\begin{align}
	u_{\nu}\nabla_{\mu} T^{\mu\nu} 
	& =
	- u^{\mu}\nabla_{\mu} \rho 
	- (\rho + p) \nabla_{\mu} u^{\mu} 
	+
	u_{\nu} \nabla_{\mu} T_{1}^{\mu\nu}
	= 0
	\ ,
	\label{Flu41}
\end{align}
and
\begin{align}
	P_{\alpha\nu} \nabla_{\mu} T^{\mu\nu}
	& =
	(\rho + p) u^{\mu} \nabla_{\mu} u_{\alpha} 
	+
	\nabla_{\alpha} p 
	+ u_{\alpha} u^{\mu} \nabla_{\mu} p
	+ P_{\alpha\nu} \nabla_{\mu} T_{1}^{\mu\nu} 
	= 0
	\ .
	\label{Flu42}
\end{align}
The spatial component of (\ref{Flu42}) represent the generalization of Euler equation in presence of dissipative effects, that is the Navier--Stokes equation for compressible relativistic fluids. 
In non--relativistic limit and assuming the incompressible condition $\partial_{i}v^{i}=0$ it is possible to show that (\ref{Flu42}) reduces to
\begin{align}
	\partial_{0}{v_{i}} +  (\vec{v}\cdot \vec{\partial}){v_{i}} = -\frac{1}{\rho}\partial_{i} p + \frac{\eta}{\rho} \partial_{j}\partial^{j} v_{i}
	\ ,
	\label{Flu43}
\end{align}
which is the incompressible non--relativistic Navier--Stokes equation.

Since we introduced dissipative corrections, we expect that the second law of thermodynamics is respected. 
Using the same arguments as before, it is possible to demonstrate that the divergence of the entropy current $J^{\mu}_{s}$ introduced in (\ref{CFlu4}) is always positive.
Writing explicitly the energy--momentum tensor conservation equation along the fluid flow eq.~(\ref{Flu41}) we have
\begin{align}
	u_{\nu} \nabla_{\mu} T^{\mu\nu} 
	= &
	- u^{\mu} \partial_{\mu} \rho - (\rho + p) \partial_{\mu} u^{\mu} 
	- \eta \left( 
		u_{\nu} \Box u^{\nu} 
		- a_{\nu} a^{\nu} 
		- \nabla_{\mu} u^{\rho} \nabla_{\rho} u^{\mu} 
	\right)
	-
	\left( \frac{2\eta}{d-1} - \zeta \right) \theta^{2} = 0
	\ .
	\label{CFluS1a}
\end{align}
Comparing this result with the shear tensor contracted with itself
\begin{align}
	2 \sigma_{\mu\nu} \sigma^{\mu\nu} 
	& =
	- u_{\nu} \Box u^{\nu} + \nabla_{\mu} u_{\nu} \nabla^{\nu} u^{\mu} + a_{\nu} a^{\nu} - \frac{2}{d-1} \theta^{2}
	\ ,
	\label{CFluS1b}
\end{align} 
we have
\begin{align}
	u_{\nu} \nabla_{\mu} T^{\mu\nu} 
	= &
	- u^{\mu} \partial_{\mu} \rho - (\rho + p) \partial_{\mu} u^{\mu}
	+ \frac{\eta}{2} \sigma_{\mu\nu} \sigma^{\mu\nu}
	+
	\zeta \theta^{2} = 0
	\ ,
	\label{CFluS1c}
\end{align}
and using the thermodynamic relations (\ref{CFluS2}) as in (\ref{CFlu3}) we get
\begin{align}
	\nabla_{\mu} J^{\mu}_{s} 
	& =
	2 \frac{\eta}{T} \sigma_{\mu\nu} \sigma^{\mu\nu} + \frac{\zeta}{T} \theta^{2}
	\ ,
	\label{FluidS2}
\end{align}
therefore, with $\zeta$ and $\eta$ positive numbers, the divergence of entropy current is always non--negative.


In the present work we will consider the long wave limit of conformal theories and therefore we are interested in conformal fluid dynamics.
Conformal fluids enjoy different simplifications: first of all, due to scale invariance, 
the energy momentum tensor for conformal theories is traceless.
Imposing this condition on energy--momentum tensor with first--order dissipative effects (\ref{FluTmunu2}) we get
\begin{align}
	&p = \frac{1}{d - 1}\rho \ ,
	& \zeta = 0 \ ,&
	\label{CFlu0}
\end{align}
hence, conformal fluids have no bulk viscosity.
Hence, for conformal perfect fluid the energy--momentum tensor reads
\begin{align}
	T^{\mu\nu} = \rho \left( g^{\mu\nu} + d u^{\mu} u^{\nu} \right) - 2 \eta \sigma^{\mu\nu} 
	\ .
	\label{CFlu0a}
\end{align}
Second simplification, $T^{\mu\nu}$ must transform covariantly respect to Weyl transformation.
This will restrict the possible structures to correct $T^{\mu\nu}$ at higher order in the derivative expansion.
In particular this will play a key role in deriving the second order dynamic.
Last, since there are no dimensionful scales involved, the dependence of the various hydrodynamics quantities on the temperature are determined by dimensional analysis.

To understand the conformal properties of the energy--momentum tensor we consider Weyl transformation of the  metric and its inverse
\begin{align}
	&g_{\mu\nu} \rightarrow \tilde g_{\mu\nu} = \Omega^{2} g_{\mu\nu}\ ,
	&g^{\mu\nu} \rightarrow \tilde g^{\mu\nu} = \Omega^{-2} g^{\mu\nu}\ ,
	\label{CFlu1}
\end{align}
where the second equation is determined by the definition of inverse metric $g_{\mu\nu} g^{\nu\rho} = \delta_{\mu}^{\rho}$.
We require that in a conformal theory the dynamical equations are invariant after conformal transformations.
In general, an equation involving a generic field $\Psi$ is conformal invariant
if there exists a number $w\in \mathbb R$ such that $\Psi$ is a solution of the equation with metric $g_{\mu\nu}$ if and only if $\tilde \Psi = \Omega^{w}\Psi$ is solution of the equation with metric $\tilde g_{\mu\nu}$. The number $w$ is called the {\it conformal weight} of the field. 
For fluid dynamics the conformal invariant conditions read
\begin{align}
	&\tilde\nabla_{\mu} \tilde T^{\mu\nu} = \Omega^{k_{1}} \nabla_{\mu} T^{\mu\nu}
	\ ,
	&
	\tilde\nabla_{\mu} \tilde j^{\mu} = \Omega^{k_{2}} \nabla_{\mu} j^{\mu}
	\ ,
	\label{CFlu2}
\end{align}
for some $k_1, k_2 \in \mathbb R$ and for $\tilde T^{\mu\nu} = \Omega^{w} T^{\mu\nu}$, $\tilde j^{\mu} = \Omega^{p} j^{\mu}$.
Notice that due to metric rescaling, the covariant derivative is modified. We have that the affine connection transforms as
\begin{align}
	\Gamma_{\mu\nu}^{\rho}
	\rightarrow
	\tilde 	\Gamma_{\mu\nu}^{\rho}
	& =
	\Gamma_{\mu\nu}^{\rho}
	+
	\Pi_{\mu\nu}^{\rho}
\ ,\nonumber\\
	\Pi_{\mu\nu}^{\rho}
	&
	=
	\Omega^{-1}
	\left( 
	-  g_{\mu\nu} \partial^{\rho}\Omega + \delta^{\rho}_{\mu} \partial_{\nu} \Omega  + \delta^{\rho}_{\nu} \partial_{\mu} \Omega
	\right)
	\ ,
	\label{CFlu}
\end{align}
To derive the conformal weight for energy--momentum tensor we have
\begin{align}
	\tilde\nabla_{\mu} ( \Omega^{w} T^{\mu\nu} )
	& =
	\nabla_{\mu} ( \Omega^{w} T^{\mu\nu} )
	+
	\Omega^{w} \Pi_{\mu\lambda}^{\mu} T^{\lambda\nu} 
	+
	\Omega^{w} \Pi^{\nu}_{\mu\lambda} T^{\mu\lambda}
	=
	\nonumber\\
	& =
	\Omega^{w} \nabla_{\mu} T^{\mu\nu} 
	+
	( w + d + 2 ) \Omega^{w-1} T^{\mu\nu} \partial_{\mu} \Omega
	-
	\Omega^{w-1} T \partial^{\nu} \Omega
	\ ,
	\label{CFlu3}
\end{align}
where $T=g_{\mu\nu} T^{\mu\nu}$. Notice that since the theory is conformal, the trace of energy--momentum tensor vanishes. 
Hence, the equation of motion of fluid dynamic is conformal invariant if $w=-d-2$, that is the energy--momentum tensor transforms as
\begin{align}
	T^{\mu\nu} \rightarrow \tilde T^{\mu\nu} = \Omega^{-d-2} T^{\mu\nu} \ ,  
	\label{CFluTmunu}
\end{align}
In the same way, we compute the conformal weight for a conserved current $j^{\mu}$
\begin{align}
	\tilde\nabla_{\mu} ( \Omega^{p} j^{\mu} )
	& =
	\nabla_{\mu} ( \Omega^{p} j^{\mu})
	+
	\Omega^{p} \Pi_{\mu\lambda}^{\mu} j^{\lambda}
	=
\nonumber \\
	& =
	\Omega^{p} \nabla_{\mu} j^{\mu} 
	+ 
	( p + d )\Omega^{p-1} j^{\mu} \partial_{\mu} \Omega
	\ ,
	\label{CFlu4}
\end{align}
that is, the current scales with $p=-d$
\begin{align}
	j^{\mu} \rightarrow \tilde j^{\mu} = \Omega^{-d} j^{\mu} \ ,  
	\label{CFluJmu}
\end{align}

In the followings, we determine the scaling behaviour of the characteristic fields of conformal fluids. 
As we said above, this information permits to express hydrodynamics quantities in terms of the temperature.

The scaling invariance of the normalization of the velocity $u_{\mu}u^{\mu}=-1$ fixes the conformal weight for $u^{\mu}$
\begin{align}
	u^{\mu} \rightarrow\tilde u^{\mu} = \Omega^{-1} u^{\mu}
	\ ,
	\label{CFluu}
\end{align}
form this and eq.~(\ref{CFluJmu}) we get the scaling transformation of the entropy density
\begin{align}
	s \rightarrow \tilde s = \Omega^{1 - d} 
	\ ,
	\label{CFus}
\end{align}
and using (\ref{CFluTmunu},\ref{CFluu}) we get the conformal weight of the energy density $\rho$
\begin{align}
	\rho \rightarrow\tilde \rho = \Omega^{- d} \rho
	\ .
	\label{CFurho}
\end{align}
Finally, from the first thermodynamical relation of (\ref{CFluS2}) we derive the scaling behaviour of the temperature
\begin{align}
	T \rightarrow \tilde T = \Omega^{-1} T
	\ ,
	\label{CFlT}
\end{align}
In the same way it is possible to derive the conformal weight for chemical potentials $\mu_{I}$ and other conserved charges $n^{I}$.
The scaling of the shear tensor is obtained by considering the Weyl transformation of the covariant derivative of the velocity
\begin{align}
	\tilde\nabla_{\mu}\tilde u^{\nu}
	& =
	\nabla_{\mu} ( \Omega^{-1} u^{\nu} ) + \Omega^{-1} \Pi^{\nu}_{\mu\lambda} u^{\lambda}
	=
\nonumber\\
	& =
	\Omega^{-1} \nabla_{\mu} u^{\nu}
	+
	\Omega^{-2}
	\left( 
	\delta^{\nu}_{\mu} u^{\lambda} \partial_{\lambda} \Omega
	-
	u_{\mu} \partial^{\nu} \Omega
	\right)
	\ .
	\label{CFlu1}
\end{align}
Remarkably, using definitions (\ref{FluidDefGrad}) we have
\begin{align}
	\tilde\sigma^{\mu\nu} & =  \tilde P^{\mu\alpha} \tilde P^{\nu\beta}\tilde \nabla_{\left( \alpha \right.}\tilde u_{\left. \beta \right)}
	- \frac{1}{d-1}\tilde\theta\tilde P^{\mu\nu}
	=
\nonumber \\
	& =
	\Omega^{-3} \sigma^{\mu\nu}
	\ ,
	\label{CFlusigma}
\end{align}
that is, the shear tensor transforms homogeneously. From this relation, the scaling of $T^{\mu\nu}$ eq.~(\ref{CFluTmunu}) and eq.~(\ref{CFlu0a}) we have
\begin{align}
	\eta \rightarrow \tilde\eta = \Omega^{1-d}\eta\ .
	\label{CFlueta}
\end{align}

Since the temperature $T$ scales with conformal weight $-1$, we can use the obtained Weyl transformation to express the other thermodynamic variables as powers of the temperature itself. As we said above this is possible in conformal field theories, since no dimensionful scales are present.
For instance we have
\begin{align}
	\rho = \rho' T^{d} \ ,&
	&\eta = \eta' T^{d-1} \ ,&
	&s = s' T^{d-1} \ ,
	\label{CFlurhoT}
\end{align}
where the constants of proportionality $\rho',\eta',s'$ are dimensionless numbers.
Therefore, energy--momentum tensor (\ref{CFlu0a}) becomes
\begin{align}
	T^{\mu\nu} =  
	\rho' T^{d} \left( g^{\mu\nu} + d u^{\mu} u^{\nu} \right) - 2 \eta' T^{d-1} \sigma^{\mu\nu}
	\ .
	\label{CFluTmunu2}
\end{align}

\section{Remarks on $AdS_{d+1}$}\label{secAdS}

Anti--de Sitter space in $d+1$ dimensions is the maximally symmetric solution of Einstein equations with negative cosmological constant $\Lambda$
\begin{align}
	S_{grav} = \int \dd^{d+1} x \sqrt{-g} \left( R - 2 \Lambda \right)
	\ ,&
	&
	R_{AB} -  \frac{1}{2} R\, g_{AB} + \Lambda g_{AB}= 0
	\ .
	\label{AdS01}
\end{align}
The curvature scalar is then
\begin{align}
	R = 2 \frac{d+1}{d-1} \Lambda \ ,
	\label{AdS02}
\end{align}
and this permits to write the Ricci tensor as proportional to the metric
\begin{align}
	R_{AB} = \frac{\Lambda}{d-1} g_{AB} \ ,
	\label{AdS03}
\end{align}
that is, $AdS_{d+1}$ is an Einstein space. Moreover, It is possible to write the Riemann tensor as
\begin{align}
	R_{ABCD} = \frac{\Lambda}{d(d-1)}\left( g_{AC} g_{BD} - g_{AD}g_{BC} \right) \ ,
	\label{AdS04}
\end{align}
which is the condition of being a maximally symmetric space.
It can be defined as the coset space
\begin{align}
	AdS_{d+1}=\frac{SO(2,d)}{SO(1,d)} \ ,
	\label{AdS05}
\end{align}
From the definition, we can write $AdS_{d+1}$ as an embedding by the quadratic algebraic equation
\begin{align}
	-
	X_{0}^{2} 
	- 
	X_{d+1}^{2}  
	+
	\sum_{i=1}^{d}X_{i}^{2}
	=
	-R_{AdS}^2
\ ,
\label{AdSdef}
\end{align}
in flat space ${\mathbb{R}}^{2,d}$ with metric
\begin{align}
	\dd s^{2} = 
		-
	\dd X_{0}^{2} 
	- 
	\dd X_{d+1}^{2}  
	+
	\sum_{i=1}^{d}\dd X_{i}^{2}
	\ ,
	\label{AdS06}
\end{align}
For completeness we express scalar curvature and cosmological constant in (\ref{AdS02},\ref{AdS03}) in terms of $AdS_{d+1}$ radius
\begin{align}
	R=-\frac{d(d+1)}{R_{AdS}^{2}}
	\ , &
	&
	\Lambda=-\frac{d(d-1)}{2 R_{AdS}^{2}}
	\ .
	\label{AdS07}
\end{align}
In order to obtain a metric for the $AdS$ surface, we can solve equation (\ref{AdSdef}) for $X_{d+1}$
\begin{align}
X_{d+1}
&=
\sqrt{
	R_{AdS}^{2}
	-
	X_{0}^{2}
	+
	\sum_{i=1}^{d}X_{i}^{2}
}
~,
\label{AdSx5}
\end{align}
and substituting it in the $\mathbb{R}^{2,d}$ metric eq.~(\ref{AdS06}) we obtain
\begin{align}
\dd s^{2}
&=
\left[
\eta_{KL}
+
\frac{
X^{I}X^{J}
}{R_{AdS}^{2} - X^{R}\eta_{RS}X^{S}}
\eta_{IK}\eta_{JL}
 \right]
\dd X^{K}\dd X^{L}
~,
\label{AdSmetric2}
\end{align}
where $\eta_{IJ}=\textrm{diag}\{-,+,\cdots,+\}$ and capital Latin indices run over $\{0,\cdots,d\}$.
For our purposes tt is convenient to express $AdS_{d+1}$ in other coordinates systems. 
The typical description of global $AdS_{d+1}$ spaces is given by setting
\begin{align}
	X_{0} = R \cosh \rho\, \cos\tau \ ,
	&
	&
	X_{d+1} = R \cosh \rho \, \sin \tau \ ,
	&
	&
	X_{i} = R \sinh \rho\, \Omega_{i}\ ,
	\label{AdS08}
\end{align}
where
\begin{align}
	\sum_{i=1}^{d} \Omega_{i}^{2}=1
	 \ .
	\label{AdS09}
\end{align}
In this coordinates chart $AdS_{d+1}$ metric reads
\begin{align}
	\dd s^2
	& =
	R_{AdS}^{2} 
	\left( 
	- \cosh^{2}\rho\,\dd \tau^{2}
	+
	\dd\rho^{2}
	+
	\sinh^{2}\rho\,\dd \Omega^{2}
	\right)
	\ ,
	\label{AdS10}
\end{align}
where $\dd \Omega^{2}$ is the metric of the $d-1$--sphere. 
Note that the time $\tau$ is periodic, hence closed time curves are present. 
In order to remove them, we unwrap the $\tau$ coordinates, by setting $\tau\in \mathbb{R}$ without identifying $\tau$ with $\tau+2\pi$. 
The obtained space is the universal covering of $AdS$. 
It is a common habit to refers to the universal covering as simply $AdS$ space.
This coordinates system covers entirely $AdS$ space, hence are global coordinates. 
Notice that the boundary is $\mathbb{R}\times S^{d-1}$.
It is possible to recast metric (\ref{AdS10}) in another form by performing the following change of coordinates
\begin{align}
	\rho = \textrm{arcsinh}\, r
	\ ,
	&
	&
	\tau = t \ ,
	&
	&
	x^{i} = x^{i}
	\ .
	\label{AdS10a}
\end{align}
The resulting global metric is 
\begin{align}
	\dd s^{2} 
	=
	-
	\left( 1+r^{2} \right) \dd t^{2}
	+
	\frac{1}{1+r^{2}} \dd r^{2}
	+
	r^{2} \dd x_{i} \dd x^{i}
	\ .
	\label{AdS10b}
\end{align}

In this work, we are most interested in the $AdS$--spaces described in Poincar\'e patch, which are defined by
\begin{align}
	& X_{0} = \frac{1}{2r}\left( 1 + r^{2} (R_{AdS}^{2} + \vec{x}^{2} - t^{2}  ) \right) \ ,
	&
	X_{d+1} = R_{AdS}\, r\, t \ ,
	\nonumber\\
	&
	X_{d} = \frac{1}{2r}\left( 1 - r^{2} (R_{AdS}^{2} + \vec{x}^{2} - t^{2}  ) \right) \ ,
	&
	X_{i}= R_{AdS}\, r\, x^{i}\ , 
	\label{AdS11}
\end{align}
where $\vec{x}=\{x^{i}\}$ with $i=\{1,2,3\}$.
The metric reads
\begin{align}
	\dd s^{2} 
	=
	R_{AdS}^{2} 
	\left( 
	\frac{\dd r^{2}}{r^{2}} 
	+ r^{2} \dd x_{\mu} \dd x^{\mu}
	\right)\ ,
	\label{AdS12}
\end{align}
where $x^{\mu}=\{t,x^{i}\}$ and the contraction is made with the mostly plus flat metric $\eta_{\mu\nu}=\{-,+,\cdots,+\}$.
Metric~(\ref{AdS12}) covers only half of $AdS$ but contains Minkowski slices for fixed $r$.

From the definitions (\ref{AdS05},\ref{AdSdef}) it is easy to see that  $AdS_{d+1}$ is invariant under $SO(2,d)$ group and that for metric the killing vectors are
\begin{align}
	J_{\tilde{A} \tilde B} = X_{\tilde A} \frac{\partial}{\partial X^{\tilde B}}	+ X_{\tilde B} \frac{\partial}{\partial X^{\tilde A}}
	\ .
	\label{AdS13}
\end{align}
where the $X_{\tilde A}$ are the $d+2$ coordinates $\{X_{0},\cdots,X_{d+1}\}$. The number of isometries is then
\begin{align}
	\textrm{dim}\left( SO(2,d) \right) = \frac{1}{2}(d+1)(d+2)\ .	
	\label{AdS14}
\end{align}
To obtain the Killing vectors for the desired coordinates patch, we have to perform the correct change of coordinates.
For instance the Killing vectors for $AdS_{5}$ in Poincar\'e patch metric~(\ref{AdS12}) read
\begin{align}
K^{t} = &
-
\left( \frac{t^{2}}{2} + \frac{1}{2r^{2}} \right) c
-
t\left( x_{j} e^{j} + e \right)
-
\frac{1}{2} x_{j}x^{j} c
+
b_{j} x^{j}
+
b
\ ,\nonumber\\
K^{r} = &
r t c+ r\left( x_{j}e^{j} + e \right)
\ ,\nonumber\\
K^{i} = &
\left(\frac{1}{2r^{2}} - \frac{t^{2}}{2} \right) e^{i}
-
t x^{i} c
+
t b^{i}
+
\frac{1}{2}x_{j}x^{j} e^{i}
-
x^{i} x_{j} e^{j}
-
x^{i} e
+
w^{ij} x_{j}
+
h^{i}
\ ,
\label{KV0}
\end{align}
where the $15$ infinitesimal parameters are interpreted as follows: $\left\{ b_{i} \right\}$ are the boundary boost parameters,
$\left\{ b, h_{i} \right\}$ represent translations in $\left\{t, x^{i}\right\}$ directions,
$e$ is the dilatation,
$\left\{ c, e_{i} \right\}$ are associated to conformal transformations
and $\left\{w_{ij}\right\}$ is the antisymmetric tensor responsible of the $3$ rotations in $\left\{x_{i}\right\}$.

\section{Boundary Energy--Momentum Tensor}\label{secBY}

In this section we briefly present the procedure to define the quasilocal energy--momentum tensor in the boundary of a given spacetime region. 
This quantity will be a key ingredient to derive the boundary fluid dynamics.
The first definition of quasilocal boundary energy--momentum tensor was proposed by Brown and York \cite{Brown:1992br}, but in this work we use the one described by Kraus and Balasubramanian \cite{Balasubramanian:1999re} which is well suited to deal with asymptotically anti--de Sitter spaces.

At first we present the gravitational action with cosmological constant $\Lambda$ (\ref{AdS01}) with boundary Gibbons--Hawking--York term \cite{York:1972sj,Gibbons:1976ue,Wald:1984rg}
\begin{align}
	S_{grav} & = 
	\frac{1}{16\pi G}
	\int_{\mathcal{M}} \dd^{d+1} x
	\sqrt{-g} \left( R - 2\Lambda \right)
	-
	\frac{1}{8\pi G}
	\int_{\partial\mathcal{M}}\dd^{d}x \sqrt{-\gamma} \Theta
	\ ,
	\label{BK1}
\end{align}
where $\mathcal{M}$ is a $d+1$ spacetime, $\partial\mathcal{M}$ is its boundary and $G$ is the Newton constant. Notice that in the followings we set  $8\pi G=1$ and $R_{AdS} = 1$.
We refer to $\gamma_{\mu\nu}$ as the induced metric on the boundary of the considered spacetime region $\partial\mathcal{M}$ and to $\Theta$ as the trace of the extrinsic curvature $\Theta_{\mu\nu}$.
Note that Greek indices runs over the $d$ boundary directions and the uppercase Latin ones label the $d+1$ spacetime directions.

To compute $\gamma_{\mu\nu}$ we define the constraint $\Phi$ which characterizes the surface on which $\gamma_{\mu\nu}$ is defined.
Since we want to compute the quasilocal energy--momentum tensor for the boundary of anti--de Sitter space, we consider slices at constant $r$
\begin{equation}
\Phi = r - c = 0	
\ ,
	\label{BYconstraint}
\end{equation}
with $c \in \mathbb{R}$. 
The outward-pointing normal vector to the boundary $\partial\mathcal{M}_{r = c}$ is defined as\footnote{This definition is valid as long as the surface is not null-like. In that case, the outward pointing normal will be $k_{M} = - \partial_{M} \Phi$.}
\begin{equation}
n_{M} = \frac{\partial_{M}\Phi}{\sqrt{g^{R S}\partial_{R}\Phi\partial_{S}\Phi}}	
\ .
	\label{BYnormal}
\end{equation}
Using $n_{M}$ we define the induced boundary metric $\gamma$:
\begin{equation}
	{\gamma}_{M N} = g_{M N} - n_{M} n_{N}
\ .
	\label{BYboundarymetric}
\end{equation}
In order to obtain a $d$--dimensional metric we delete from ${\gamma_{MN}}$ the $r$--column and the $r$--row, obtaining the $d$--dimensional induced boundary metric $\gamma_{\mu\nu}$.
In a similar way we calculate the extrinsic curvature $\Theta_{MN}$ and then $\Theta_{\mu\nu}$:
\begin{equation}
\Theta_{M N} = -\frac{1}{2} \left(\nabla_M n_N + \nabla_N n_M \right) \ .
	\label{BYextrinsicCurv}
\end{equation}
Moved by action principle and Hamilton--Jacobi formalism, Brown and York \cite{Brown:1992br} proposed a quasi--local energy--momentum tensor defined locally on the boundary of a given region of the spacetime.
Introducing $\gamma_{\mu\mu}$ as boundary metric, the Brown--York energy--momentum tensor is defined as
\begin{align}
	T^{\mu\nu} \propto \frac{1}{\sqrt{-\gamma}} \frac{\delta S_{grav}}{\delta\gamma_{\mu\nu}} \ ,
	\label{BY0}
\end{align}
with $S$ defined in (\ref{BK1}). Explicit computation give
\begin{align}
	T^{\mu\nu} = \Theta^{\mu\nu} - \Theta \gamma^{\mu\nu}
	\ .
	\label{BY21}
\end{align}
In general this energy--momentum tensor diverges if the boundary is taken at the infinity. 
In order to renormalize it, Brown and York modified the action by adding a new boundary term $S^{ct}$.
Being a boundary term, it does not modify the equations of motion. 
To fix $S^{ct}$, they proposed to embed the boundary metric in a reference spacetime and then to set $S^{ct}$ as the gravitational action for the resulting spacetime region.

However, in general it is not possible to obtain the embedding needed and therefore the renormalization procedure is not well defined.
Balasubramanian and Kraus in \cite{Balasubramanian:1999re} proposed an alternative way to compute $S^{ct}$ for asymptotically $AdS$ spaces.
They write $S^{ct}$ as a series of counterterms made by scalars constructed from the boundary metric $\gamma_{\mu\nu}$ and fix the coefficients by requiring the divergence cancellation.
In various dimensions, the counterterm actions $S^{ct} = \int_{\partial \mathcal{M}_{r}} L^{ct}$  read
\begin{align}
	L^{ct}_{AdS_{3}}  = - \sqrt{-\gamma} \ ,&
	& L^{ct}_{AdS_{4}}  = - 2 \sqrt{-\gamma} \left( 1 - \frac{1}{4} R \right) \ ,&
	& L^{ct}_{AdS_{5}}  = - 3 \sqrt{-\gamma} \left( 1 - \frac{1}{12} R \right) \ ,
	\label{BY31}
\end{align}
therefore the divergence--free boundary energy momentum tensors are
\begin{align}
	T^{\mu\nu}_{AdS_{3}} & = \Theta^{\mu\nu} - \Theta \gamma^{\mu\nu} - \gamma^{\mu\nu}
	\ ,
\nonumber \\
	T^{\mu\nu}_{AdS_{4}} & = \Theta^{\mu\nu} - \Theta \gamma^{\mu\nu} - 2 \gamma^{\mu\nu} + G^{\mu\nu}
	\ ,
\nonumber\\
	T^{\mu\nu}_{AdS_{5}} & = \Theta^{\mu\nu} - \Theta \gamma^{\mu\nu} - 3 \gamma^{\mu\nu} + \frac{1}{2}G^{\mu\nu}
	\ ,
	\label{BY32}
\end{align}
where $G^{\mu\nu}$ is the Einstein tensor\footnote{A careful reader may have noticed a change of sign in front of the Einstein tensor with respect to \cite{Balasubramanian:1999re}. This is just a matter of convention in the definition of the Riemann tensor.} build from $\gamma^{\mu\nu}$.

As last remark, we comment the case of general relativity coupled to matter.
It is possible to demonstrate that if the matter action does not contain metric derivatives -- hence the matter is minimally coupled with gravity -- the energy--momentum tensor is unchanged \cite{Brown:1992br}.

To analyse the boundary terms it is useful to cast the metric in the form prescribed by ADM decomposition\cite{Arnowitt:1962hi,Wald:1984rg}:
\begin{align}
	\dd s^2 = N^{2} \dd r^{2} + \g_{\mu\nu} 
	\left( \dd x^{\mu} + V^{\mu} \dd r \right)
	\left( \dd x^{\nu} + V^{\nu} \dd r \right)
	\ ,
	\label{BKADM}
\end{align}
where $N$ is called lapse function and $V^{\mu}$ is the shift function.

Having computed the boundary energy--momentum tensor, it is easy to compute the conserved charges associated with the isometries of the considered spacetime.
To perform the computation the boundary metric $\gamma_{\mu\nu}$ is cast in ADM--like form \cite{Arnowitt:1962hi,Wald:1984rg}
\begin{align}
	\gamma_{\mu\nu} \dd x^{\mu} \dd x^{\nu}
	& =
	- N_{\Sigma}^{2} \dd t^{2}
	+
	\sigma_{ab}
	( 	\dd x^{a}  + V^{a}_{\Sigma} \dd t )^{2}
	\ ,
	\label{ADM1}
\end{align}
where $\Sigma$ is the $d-1$--dimensional surface at constant time.
Hence, the conserved charges associated to the Killing vectors $\xi$ are defined as
\begin{align}
	Q_{\xi} & =
	\lim_{r\rightarrow\infty}\int_{\Sigma} \dd x^{d-1} \sqrt{\sigma} u^{\mu} T_{\mu\nu} \xi^{\nu}
	\label{charges1}
\end{align}
where  $u^{\mu}=N_{\Sigma}^{-1}\delta^{\mu t}$ is the timelike unit vector normal to $\Sigma$.

\section{Fluid/Gravity Correspondence}\label{sezFG}

This section briefly reviews the basic concepts of fluid/gravity correspondence.
For a complete treatment, the interested reader is referred to \cite{Bhattacharyya:2008jc,Rangamani:2009xk,Hubeny:2011hd}.

\subsection{The Correspondence}

As it is well known, the first example of gauge/gravity correspondence states an equivalence between type IIB superstring theory on $AdS_{5}\times S^{5}$ space and a conformal field theory (CFT) defined on the boundary of $AdS_{5}$: supersymmetric $\mathcal{N}=4$ Yang--Mills theory (SYM).
In $\mathcal{N}=4$ SYM  two dimensionless parameters are present: the rank of the gauge group $SU(N)$ and the 't Hooft coupling  $\lambda = g_{YM}^{2} N$.
In the string side, $g^{2}_{YM}$ corresponds to the string coupling constant while $\lambda$ represents the ratio between the radius of $AdS$ and the string length.

In the limit in which both $\lambda$ and  $N$ go to infinity in such a way that their ratio remains constant, the massive modes in the string sector decouple and Type IIB superstring reduces to classical supergravity (Type IIB).
On the dual side, in this limit the boundary gauge theory becomes strongly coupled.
Therefore using gauge/gravity correspondence it is possible to derive quantites for non--perturbative $CFT$ from classical supergravity computations.

An interesting feature of type IIB supergravity is that it admits consistent truncations, the simplest of them being the pure Einstein gravity with negative cosmological constant (\ref{AdS01})
\begin{align}
	S_{grav} = \int \dd^{d+1} x \sqrt{-g} \left( R - 2 \Lambda \right)
	\ ,
	\label{FGgrav}
\end{align}
which contains the metric $g_{AB}$ as the only variable. 
Without entering into the details of the dictionary between gauge theory and gravity, we remind that $g_{AB}$ is associated to the energy--momentum tensor of the boundary theory.
As a consequence, 
it must exist a sector of the dual CFT which in a particular limit corresponds to Einstein gravity with negative cosmological constant.

One can observe that an interacting system near thermal equilibrium at sufficiently high temperature can be described by hydrodynamics effective theory.
Therefore, we could expect that in some long wavelength limit the pure gravity theory corresponds to the hydrodynamic limit of strongly coupled gauge theory,
The resulting map is named fluid/gravity correspondence. 

Moved by these facts, we consider stationary solutions of Einstein equations in asymptotically $AdS_{5}$, which are best suited to deal with fluids in thermal equilibrium. 
For instance, the vacuum of the gauge theory is associated to empty $AdS_{5}$, while $AdS$--Schwarzschild black hole (here written in Poincar\'e patch)
\begin{align}
	&
	\dd s^{2} 
	=
	- r^{2} f (b r) \dd t^{2} 
	+
	\frac{\dd r^{2}}{r^{2} f(b r)}
	+
	r^{2} \delta_{ij} \dd x^{i} \dd x^{j}
	\ ,
	&
	f(r) = 1 - \frac{1}{r^{4}}
	\ ,
	\label{FG08}
\end{align}
corresponds to gauge theory states in thermal equilibrium whose temperature corresponds to the black hole one
\begin{align}
	T = \frac{1}{\pi b} = \frac{r_{+}}{\pi}
	\ .
	\label{FG08a}
\end{align}

The regime of validity of the hydrodynamic description is that the scale of fluctuation of the fields must be larger than the mean free path length $l_{\textrm{mfp}}$ (long wavelength regime).
In conformal field theory we have
\begin{align}
	l_{mfp} \sim \frac{1}{T}
	\ ,
	\label{FG09}
\end{align}
which in the gravity dual side translates in
\begin{align}
	r_{+} >> R_{AdS}
	\ ,
	\label{FG10}
\end{align}
that is, the hydrodynamic limit of the strongly coupled boundary field theory is obtained by large (respect to the $AdS$ radius $R_{AdS}$) black holes in the bulk.

To determine the fluid dual of the solution of Einstein equations, we compute the boundary energy--momentum tensor via the Brown--York procedure described in sec.~[\ref{secBY}].
Hence for the Schwarzschild black hole eq.~(\ref{FG08}) we have
\begin{align}
	T^{\mu\nu} \propto \frac{1}{b^{4}}\,{\textrm{diag}} \left[ 3,\,1,\,1,\,1 \right]
	\ ,
	\label{FG11}
\end{align}
which represents a conformal fluid in the rest frame with temperature (\ref{FG08a}).

Performing a boost along the boundary directions $x^{\mu}=\{t,x^{i}\}$ on (\ref{FG08}) we obtain the so called boosted black hole
\begin{align}
	\dd s^{2}
	=
	\frac{1}{r^{2}f( b r)}\dd r^{2}
	+
	r^{2}\left[ 
		P_{\mu\nu}
		-f( b r ) u_{\mu}u_{\nu} 
		\right]\dd x^{\mu}\dd x^{\nu}
	\label{FGcorr1}
\end{align}
with
\begin{align}
	u^{\mu}
	= 
	\left\{ \frac{1}{\sqrt{1-\beta^{2}}}\,,\,\frac{\beta^{i}}{\sqrt{1-\beta^{2}}} \right\}
	\ ,
	&
	&
	\beta^{2} = \delta_{ij}\beta^{i}\beta^{j}
	\ ,
	&
	&
	P_{\mu\nu}=u_{\mu}u_{\nu}+\eta_{\mu\nu}
	\ ,
	\label{FGcorr2}
\end{align}
where $P_{\mu\nu}$ is the projector along the orthogonal to fluid velocity.
In this case, the associate boundary fluid is described by 
\begin{align}
	T^{\mu\nu}
		  \propto 
	    \frac{1}{b^{4}}
		   \left( 4u^{\mu} u^{\nu} +\eta^{\mu\nu} \right)
		   \ ,
	\label{FG13}
\end{align}
that is, the usual covariant form of energy--momentum tensor for ideal conformal fluids.

Having set up the fluid/gravity dictionary for static quantities, we can argue what happens in near equilibrium systems.
Promoting parameters $b,\beta^{i}$ as slowly varying functions of the boundary coordinates $x^{\mu}$, the boundary fluid dynamic is governed by the conservation of energy--momentum tensor
\begin{align}
	\nabla_{\mu} T^{\mu\nu} = 0 \ ,
	\label{FG12}
\end{align}
where in general $T^{\mu\nu}$ is modified by adding derivatives of the parameters, that is by expressing it in gradient expansion as was presented in [\ref{secFlu}].
This means that the metric (\ref{FGcorr1}) with local parameters is no longer solution to the Einstein equations. 
Since we assume slowly variations of the parameters, we are able to construct a metric solution of the Einstein equations in a derivative expansion
This perturbative procedure will be described in the next section.

Hence, fluid/gravity correspondence gives a natural framework to derive the fluid dynamic energy--momentum tensor from a computation in pure gravity.
This procedure determines precisely the various order transport coefficients in an unambiguous way.

\subsection{Perturbative Procedure}\label{sezPP}

In this section we briefly describe the perturbative procedure 
developed in order to solve the Einstein equations in a boundary derivative expansion.

As we said previously, if the parameters $\{b,\beta^{i}\}$ are local functions of the boundary coordinates 
$\{t,x^{i}\}$, the metric does not satisfy the Einstein equations anymore.
However, it is possible to construct a new solution by modifying the starting metric 
and by constraining the local parameters.
The boundary dual of resulting metric has the interpretation of 
a fluid (with dissipative contribution for $d>2$) and the parameters $\{b,\beta\}$ are connected with fluid 
temperature and velocity.

To implement the procedure, it is useful to consider the parameters as function of the rescaled boundary coordinates $\varepsilon x^{\mu}$
\begin{align}
	&
	b(x^{\mu}) \rightarrow b(\varepsilon x^{\mu})
	\ ,
	&
	\beta^{i}(x^{\mu}) \rightarrow \beta^{i}(\varepsilon x^{\mu})
	\ ,
	\label{PerProc1}
\end{align}
then the symbol $\varepsilon$ counts the number of derivatives.
To reconstruct the solution of Einstein equations we need to correct the metric order by order in $\varepsilon$
\begin{align}
	g & =
	g^{(0)}(b,\beta^{i})
	+
	\varepsilon g^{(1)}(b,\beta^{i})
	+
	\varepsilon^{2} g^{(2)}(b,\beta^{i})
	+ 
	O\left( \varepsilon^{3} \right)
	\ ,
	\label{PerProc2}
\end{align}
where $g^{(0)}(b,\beta)$ represent the boosted black hole metric (\ref{FGcorr1}) with local parameters and $g^{(n)}$ for $n>0$ are the corrections to the metric.
\footnote{We dropped the dependence on boundary coordinates for convenience of notation.}
Since the equations of motion satisfied by the parameters change order by order, we need to add corrections also to the parameters
\begin{align}
	&
	b = b^{(0)} + \varepsilon b^{(1)} + O\left( \varepsilon^{2} \right)
	\ ,
	&
	\beta_{i} = \beta_{i}^{(0)} + \varepsilon \beta_{i}^{(1)} + O\left( \varepsilon^{2} \right)
	\ .
	\label{PerProc3}
\end{align}
The procedure we want to describe is an iterative one: 
Having the solution at order $(n-1)$, {\it i.e.}  $g^{(m)}$ for $m\le n-1$ and ${b,\beta^{i}}^{(m)}$ for $m\le n-2$,
we can solve the Einstein equations for the $n$--order quantities.
We plug the corrected quantities (\ref{PerProc2},\ref{PerProc3}) into the Einstein equations and we consider the $\varepsilon^{n}$ coefficient of the perturbative expansion.
We obtain a set of differential equations which schematically read
\begin{align}
	\mathbb{H} 
	\left[ 
		g^{(0)}\left(\b^{(0)}(x) , \beta_{i}^{(0)}(x) \right)
	\right]
	g^{(n)}(x) = S^{n}
	\ ,
	\label{PerProc4}
\end{align}
where $\mathbb{H}$ is a linear operator of second order which depends only on the variable $r$.
Indeed $g^{(n)}$ is already at order $\varepsilon^{n}$ and hence a boundary derivative would produce
an higher order term.
Notice also that $\mathbb{H}$ depends only on the value of the parameters at $x^{\mu}$ and not on their boundary derivatives.
Thus, we can consider $\mathbb{H}$ as a ultralocal operator in the boundary direction and this allows us to
solve the equations point by point on the boundary.
In particular, we choose to work in a neighborhood of the boundary coordinate origin $x^{\mu}=0$. As initial value for 
the parameters, we set
\begin{align}
	&
	b(0) = 1 
	\ ,
	&
	\beta_{i}(0) = 0 
	\ ,
	\label{PerProc5}
\end{align}
accordingly to the interpretation of dilatation and boost parameters. 
With this choice, it is useful to write (\ref{PerProc3}) as
\begin{align}
	&
	b(x) = 1+  \left( \varepsilon x^{\mu} \partial_{\mu} b^{(0)} (x) \right)|_{x^{\mu}=0} + O (\varepsilon^{2})
	\ ,
	&
	\beta_{i}(x) = \left( \varepsilon x^{\mu} \partial_{\mu} \beta_{i}^{(0)} (x) \right)|_{x^{\mu}=0} + O (\varepsilon^{2})
	\ ,
	\label{PerProc6}
\end{align}
furthermore the metric (\ref{PerProc4}) read
\begin{align}
	g & =
	g^{0}\left(b=1 ,\beta_{i}=0\right)
	+
	\varepsilon\left[ g^{0}\left(x^{\mu}\partial_{\mu}b^{\left( 0 \right)} , x^{\mu}\partial_{\mu}\beta_{i}^{\left( 0 \right)}\right)+g^{1}\left(\beta_{i}=0,b=1 \right) \right]
	+ O (\varepsilon^{2})
	\ .
	\label{PerProc32}
\end{align}
The r.h.s of (\ref{PerProc32}) is an expression of order $n$ made of derivatives of the parameters $\{b,\beta^{i}\}^{(k)}$ with $k\le n-1$.

In general, being the Einstein tensor $E_{MN}$  a $d$--dimensional symmetric tensor,
(\ref{PerProc4}) describes $\frac{d (d+1)}{2}$ independent differential equations.
They naturally splits into two different sets. 
We consider the ones with only one $r$--derivative as {\it constraint equations}.
These are obtained as
\begin{align}
	E^{\left( c \right)}_{M} = E_{MN} \xi^{N}
	\ ,
	\label{PerProc43}
\end{align}
where $\xi_{N}$ is the one--form normal to the boundary. In this work it is simply $\xi_{M}=\dd r$ and thus $\xi^{N}=g^{N\,r}$.
Considering the boundary directions of $E^{\left( c \right)}_{M}$ we obtain a set of $d-1$ equations which involve only the parameters and not the unknown metric correction. 
These reproduce the equations of conservation of boundary energy--momentum tensor at order $n-1$ and then
we can interpret the parameters as fluid quantities.
The other $\frac{1}{2}(d^{2}-d+2)$ differential equations ($E^{\left( c \right)}_{r}$ and the so--called {\it dynamical equations}) generically could depend on both parameters and $g^{(n)}$ components.
These are used to obtain the metric correction $g^{(n)}$ in terms of the derivatives of the parameters.
Notice that being $g^{(n)}$ a  $n^{\textrm{th}}$--order contribution, it will depend on  $n$ derivatives of the parameters.
In general, it is convenient to classify dynamical equations according to the representations of the little group $SO(1,d-1)$.

To simplify the computation of $g^{(n)}$ it is possile to choose a gauge for the metric. 
There are different convenient choices, for instance in paper \cite{Rangamani:2009xk} the author set
\begin{align}
	&
	g_{rr}=0 
	\ ,
	&
	g_{r\mu} = u_{\mu}
	\ .
	\label{PerProc44}
\end{align}
With this choice, the curves of constant $x^{\mu}$ are null geodesic parametrized by the affine parameter $r$.

\subsection{First Order Results and Finite Metric Reconstruction}

The technique reviewed in the last section permits to construct a metric solution of Einstein equations for a given order in boundary derivative expansion near the boundary coordinates origin $x^{\mu}=0$.
However, since it is defined in the neighborhood of $x^{\mu}=0$, the metric is not global.

However, to compute the full form of the boundary stress--energy tensor we need a finite metric, {\it i.e.} a metric defined in any point.
Due to the ultralocality, to obtain the finite metric we can postulate the most general covariant metric which is function of $u^{\mu}$ and $b$ and then, by expanding it about $x^{\mu}=0$ we fix the various coefficients.

Once we have the global metric up to first order in derivative expansion, we compute the Brown--York boundary energy--momentum tensor
\begin{align}
	T^{\mu\nu} \propto
	\frac{1}{b^{4}} \left( \eta^{\mu\nu} + 4 u^{\mu} u^{\nu} \right)
	-
	\frac{2}{b^{3}} \sigma^{\mu\nu}
	\ ,
	\label{MT01}
\end{align}
which corresponds to conformal fluid in presence of dissipative effects (\ref{CFluTmunu2}). 
Using the thermodynamical relation (\ref{CFluS2})  and the definition of fluid temperature (\ref{FG08a}) it follows that the entropy to shear viscosity ratio is
\begin{align}
	\frac{\eta}{s} = \frac{1}{4\pi}\ ,
	\label{MT02}
\end{align}
as expected.

It worth mentioning that the constraint equations (\ref{PerProc43}) up to first order are \begin{align}
	\partial_{t} b^{(0)} = \frac{1}{3} \partial_{i} \beta_{i}^{(0)}
	\ ,
	& 
	&
	\partial_{i} b^{(0)} = \partial_{t} \beta_{i}^{(0)}
	\ .
	\label{MT03}
\end{align}
As we have already said, these are the conservation equations of (\ref{MT01}) expanded at first order about the origin.

\section{Supergravity Generalization}\label{generalization}

As pointed out in \cite{Bhattacharyya:2008jc} the $d$--parameter family of black holes parametrized by $r$--dilatation $b$ and boost $\beta^{i}$ can be obtained by acting on a Schwarzschild black hole with a set of isometries of empty $AdS$.

A first question could arise: what happens if we act with the whole $AdS_{d+1}$ isometry group $SO(2,d)$ on the black hole (\ref{FG08})?
Obviously, there exists a subset of $AdS_{d+1}$ isometries which still preserves the black hole: 
the translations in the boundary directions $x^{\mu}$ and the rotations in  the spatial boundary volume. 
In $AdS_{5}$ this means an amount of $7$ preserved isometries over the total $15$.
The other $8$ transformations ($r$--dilatation, the $3$ boosts and $4$ ``conformal'' transformations)
generate a deformation of the black hole metric $g_{BH}$.
If $\xi=\xi^{\mu}\partial_{\mu}$ are the Killing vectors associated to the desired $AdS_{d+1}$ isometries, the black hole deformation is the Lie derivative of $g_{BH}$ along $\xi$
\begin{align}
	\delta g_{BH} = \mathcal{L}_{\xi} (g_{BH})
	\ .
	\label{GE00}
\end{align}
Therefore, one could argue what kind of boundary fluid is associated with this new metric.

Another interesting point is the extension of the original fluid/gravity correspondence, which considers pure Einstein gravity, to include couplings with other fields.
In particular, it has been shown how to get the
Navier--Stokes equations from different solutions of general relativity \cite{Banerjee:2008th,Bhattacharyya:2008ji} 
and more recently how to get back to Einstein equation starting from a boundary fluid \cite{Bredberg:2011jq,Lysov:2011xx}.

Nevertheless, the complete AdS/CFT correspondence can be only fully
established between the supergravity extension of the general relativity and its holographic
dual.
In the present work we propose to include fermionic fields in the computation and this can be done using supergravity theory as a playground.
In particular, we consider supergravity theories in $AdS$ spaces for $5$, $4$ and $3$ dimensions.

One of the key concepts of fluid/gravity correspondence is the metric deformation:
acting with a transformation which preserves the reference solution (empty $AdS$) on a static solution of the equations of motion (the background solution) we generate a metric deformation which depends upon the parameters of the transformation.
The resulting finite transformation corresponds via the Brown--York technique to the boundary fluid energy--momentum tensor.

As we will see, we can apply the same procedure in a supergravity context.
In particular, we have to recall that $AdS$ spaces are endowed with
superisometries which introduce new constant parameters in the solution.
We present the main concepts without entering into details by introducing the minimal supergravity multiplet, composed by vielbein $e_{M}^{A}$ and gravitino $\psi_{M}$.\footnote{The indices $\{M,\cdots\}$ are curved ones while $\{A,\cdots\}$ are flat ones.}
In this toy model, we define the following supersymmetry transformations with parameter $\epsilon$
\begin{align}
	\delta_{\epsilon}e_{M}^{A} = \bar\epsilon \Gamma^{A} \psi_{M}
	\ ,
	&
	&
	\delta_{\epsilon}\psi_{M} = \mathcal{D}_{M}\epsilon
	\ ,
	\label{GE01}
\end{align}
where $\Gamma^{A}$ are the gamma matrices and $\mathcal{D}_{M}$ is the covariant derivative.
If the gravitino is different from zero the metric is in general modified by (\ref{GE01}).
However, we need a transformation which preserves the empty $AdS$ solution
\begin{align}
	e_{M}^{A} = (e_{M}^{A})_{AdS}
	\ ,
	&
	&
	\psi_{M} = 0
	\ ,
	\label{GE02}
\end{align}
that is we have to solve the so--called Killing spinor equation
\begin{align}
	\delta_{\epsilon}\psi_{M} = (\mathcal{D}_{AdS})_{M}\epsilon = 0
	\ ,
	\label{GE03}
\end{align}
which is the supersymmetric counterpart of the Killing vector equation.
The solution $\epsilon$ is called Killing spinor and it can be written as
\begin{align}
	\epsilon = \Xi \zeta
	\ ,
	\label{GE04}
\end{align}
where $\Xi$ is a coordinates--depending matrix which can be expanded on the basis of gamma matrices and $\zeta$ is a spinor whose components are constant Grassmann numbers.

Having found the analogous to $AdS$ Killing vectors, we perform a susy variation on a static non--trivial solution of the supergravity equations of motion
\begin{align}
	e_{M}^{A} = (e_{M}^{A})_{BH}
	\ ,
	&
	&
	\psi_{M} = 0
	\ ,
	\label{GE05}
\end{align}
where $(e_{M}^{A})_{BH}$ is the vielbein associated to $AdS$--Shwarzschild black hole. 
Since the gravitino is zero, at first order we have no metric deformation
\begin{align}
	\delta^{1}_{\epsilon}e_{M}^{A} = 0
	\ ,
	&
	&
	\delta_{\epsilon}\psi_{M} = \mathcal{D}_{M}\epsilon
	\ ,
	\label{GE06}
\end{align}
anyway, if the chosen black hole breaks supersymmetries a gravitino is generated.
In this work we focus only on non--BPS black holes to have the largest number of degrees of freedom for the gravitino.
Thus, performing a second order susy transformation 
\begin{align}
	\delta^{2}_{\epsilon}e_{M}^{A} = \delta_{\epsilon}^{1} \left(  \bar\epsilon\, \Gamma^{A}  \psi_{M}  \right)
	=
	\bar\epsilon\, \Gamma^{A}  \mathcal{D}_{M}\epsilon
	\neq 0
	\ ,
	\label{GE07}
\end{align}
we generate a metric deformation called black hole superpartner \cite{Behrndt:1998ns,Behrndt:1998jd,Burrington:2004hf}, which depends on bilinears in the $\zeta$ spinors
\begin{align}
	\left\{ 
	\bar\zeta\,\zeta 
	\ ,\ 
	\bar\zeta\,\Gamma^{A}\zeta 
	\right\}
	\ .
	\label{GE08}
\end{align}
Notice that as in the bosonic Killing vector case, the deformed metric is still a solution of the equations of motion.

At this point there are two alternatives. The first one, according
to \cite{Bhattacharyya:2008jc}, is promoting the fermionic zero modes $\zeta$ to local fermions; then one is forced to study
the equations of motion for the gravitino. 
Namely, one has to see if certain conditions on $\zeta$, promoted to local function, lead to dynamical equations for boundary degrees of freedom. 
The second alternative is to maintain $\zeta$ constant and to promote only the bilinears to local functions.
In that case only the equations of motion for the vielbein are affected since we assume that the equations for the gravitino are preserved at that order since the background has no fermions.
In order to obtain the fermionic correction to hydrodynamic, we combine the transformations generated by both Killing vectors and Killing spinors.
Therefore we could proceed with the fluid/gravity perturbative procedure, modifying the metric order by order to solve the equations of motion and then analysing the boundary fluid dynamic.

By considering the constraint equations it is possibile to derive the fermionic corrections to Navier-Stokes equations in terms of fermion bilinears. The latter may acquire a non-vanishing expectation value yielding physical modifications of the fluid dynamics.

However, the equations computed from the perturbative procedure are written as expansion about $x^{\mu}=0$.
To derive the global Navier--Stokes equations we have to consider the full boundary energy--momentum tensor,
which is obtained via the Brown--York procedure from a finite boosted black hole metric.
Next section refers to the computation of the finite superpartner (named ``{\it wig}" to recall that 
all fermionic hairs are resummed into a complete non--linear solution) for a given metric.

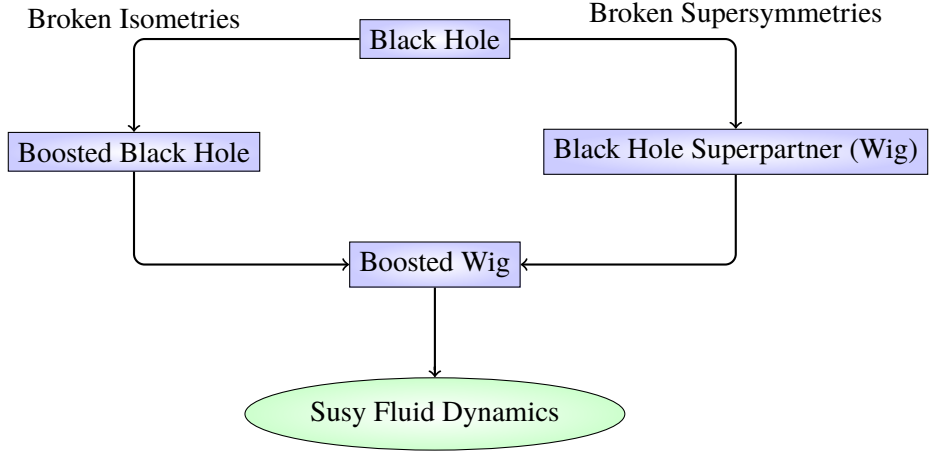
\begin{figure}[tp]
	\centering
\begin{tikzpicture}[auto, thick]


    \path[->] (4.5,0) node[format] (BH) {Black Hole};

    \path[->] node[format, left of=BH, left=3cm, below=1.5cm, anchor=center] (BBH) {Boosted Black Hole};

    \path[->] node[format, right of=BH, right=3cm, below=1.5cm, anchor=center] (BHS) {Black Hole Superpartner (Wig)};

    \path[->] node[format, below of=BH, below=2cm, anchor=center] (BW) {Boosted Wig};

    \path[->] node[medium, below of=BW, below=.5cm] (SFD) {Susy Fluid Dynamics};

    \path[->, draw, rounded corners, above] (BH) -| node {Broken Isometries} (BBH);
    
    \path[->, draw, rounded corners, above] (BH) -| node {Broken Supersymmetries}  (BHS);

    \path[->, draw, rounded corners, above] (BBH) |- (BW);

    \path[->, draw, rounded corners, above] (BHS) |- (BW);

    \path[->, draw, rounded corners] (BW) |- (SFD);

\end{tikzpicture}
\caption{Fluid/supergravity correspondence in brief.}
\label{figura}
\end{figure}

We summarize the whole procedure described in this section in figure \ref{figura}.

\subsection{The Wig}

To compute the full boundary energy--momentum tensor it is necessary to have a finite metric transformation,
namely we need the analogous of the boosted black hole (\ref{FGcorr1}) for susy transformation.
This is the key ingredient to obtain the supersymmetric counterpart of the covariant boundary energy--momentum tensor for ideal conformal fluid (\ref{FG13}).

Starting from an infinitesimal transformation, it is in general a cumbersome problem finding the associated finite transformation by resuming the infinite series.
However, in this case due to the anticommuting nature of supersymmetry parameters, the series
truncates after few steps. 
The unconventional nature of fermionic hair prompted us to
adopt the word {\it wigs} to denote the Schwarzschild solution decorated with fermionic zero
modes. The number of needed steps depends upon the number of independent fermionic
parameters entering the supersymmetry transformations, therefore in our case it depends
upon the number of the independent parameters of the $AdS$ Killing spinors.

Our analysis is based on the papers by Aichelburg and Embacher \cite{Aichelburg}
where the Schwarzschild solution for $\mathcal{N} = 2$, $D = 4$ supergravity in flat space has been lifted to a
full non--linear solution of the equations of motion including the fermionic zero modes. The fields
are constructed iteratively starting from the pure bosonic expressions and acting on them with the
supersymmetry.

The wig for a generic field $\Phi$ is constructed by acting with a finite supersymmetry transformation on the original field \cite{Burrington:2004hf}:
\begin{align}
	\boldsymbol{\Phi} = e^{\delta_{\epsilon}} \Phi = \Phi + \delta_{\epsilon} \Phi + \frac{1}{2} \delta_{\epsilon}^{2} \Phi + \dots
	\ .
	\label{defsuperpartner}
\end{align}
In this work we consider as background fields the AdS--Schwarzschild black hole with all the other independent fields set to zero
\begin{align}
	& e_{M}^{[0]\,A} =  \left. e_{M}^{A}\right|_{BH} \ ,
	& \textrm{other fields} = 0 \ .
	\label{back0}
\end{align}
Notice that we we do not consider the spin connection as an independent field since it is obtained in terms of the vielbein by the zero--torsion condition (the vielbein postulate).
With our choice of background (\ref{back0}), the vielbein postulate is the same as in pure gravity (\ref{AdS01})
\begin{align}
	\dd e^{A} + \omega^{A}{}_{B} \wedge e^{B} = 0 \ .
	\label{GEvielPos}
\end{align}
Of course, the wig procedure in general modifies the vielbein postulate as well as the other equations of motion.

To implement algorithms to compute the wigs for the various fields, it is more useful to deal with an expansion in powers of bilinears of $\epsilon$. This is denoted by the superscript $\left[ n \right]$, which counts the number of bilinears.
We have
\begin{align}
	B^{\left[ n \right]}&= \frac{1}{2n!}\delta_{\epsilon}^{2 n}B
	\ ,&
	F^{\left[ n \right]}&= \frac{1}{(2n-1)!}\delta_{\epsilon}^{2 n - 1}F
	\ ,&
	n>0 \ ,
	\label{defTransf}
\end{align}
where $B$ and $F$ are respectively bosonic and fermionic fields. Then, for fermionic fields $\left[ n \right]$ counts $n-1$ bilinears plus a spinor $\epsilon$ while for bosonic fields it indicates $n$ bilinears. The $n=0$ case represents the background fields (\ref{back0}).

\subsection{Killing Spinors Factorization}

To compute the solutions to Killing spinor equation (\ref{GE03}) we use gamma matrices parametrized in terms of Pauli matrices \cite{Burrington:2004hf}
\begin{align}
	\sigma_{1}
	=
	\left( \begin{array}{cc}
	0 & 1 \\
	1 & 0
	\end{array} \right)
	\ ,
	&
	&
	\sigma_{2}
	=
	\left( \begin{array}{cc}
	0 & -i \\
	i & 0
	\end{array} \right)
	\ ,
	&
	&
	\sigma_{3}
	=
	\left( \begin{array}{cc}
	1 & 0 \\
	0 & -1
	\end{array} \right)
	\ ,
		\label{Pauli}
\end{align}
which satisfies
\begin{align}
	\sigma_{i} \sigma_{j} = \delta_{ij} \sigma_{0} + i \varepsilon_{ijk} \sigma^{k}
	\ ,
	\label{Pauli2}
\end{align}
where $\sigma_{0}=\one_{2}$ and $\varepsilon_{123}=1$.
Following \cite{VanProeyen:1999ni}, we define for odd dimensional space with signature mostly plus $\{-,+,\cdots,+\}$ the gamma matrices
\begin{align}
	\Gamma_{0} & = i\sigma_{2} \otimes \sigma_{0} \otimes \sigma_{0} \otimes \cdots
	\ ,
	\nonumber\\
	\Gamma_{1} & = \sigma_{1} \otimes \sigma_{0} \otimes \sigma_{0} \otimes \cdots
	\ ,
	\nonumber\\
	\Gamma_{2} & = \sigma_{3} \otimes \sigma_{1} \otimes \sigma_{0} \otimes \cdots
	\ ,
	\nonumber\\
	\cdots & = \cdots
	\ ,
	\label{Pauli3}
\end{align}
notice that apart for the choice of $\Gamma_{0}$, the other definitions can be modified. 
When dealing with $3$--dimensional case we will consider different $\Gamma_{1}$ and $\Gamma_{2}$.
As a consequence of definition (\ref{Pauli3}), the spinors are factorized in $2$--dimensional complex spinors.

As we have said, since we are interested into the complete solution -- namely all powers of fermions -- we have to deal with the fermionic nature of the spinor fields.
Therefore, by factorizing the spinors into a product of spinors in lower dimensions, we have to declare the statistic of each part.
As a matter of fact, in the $5$--dimensional case we saw that the map between the original fermion $\epsilon$ and its decomposition $\varepsilon\otimes\eta$  spoils the correct number of degrees of freedom only if all possible choices are taken into account.
Namely, we have to choose first $\varepsilon$ to be anticommuting and $\eta$ commuting and subsequently $\varepsilon$ commuting and $\eta$ anticommuting:
\begin{align}
	\epsilon = \varepsilon|_{A} \otimes \eta|_{C} + \varepsilon|_{C} \otimes \eta|_{A}\ .
	\label{KScomment}
\end{align}
The generalization to an arbitrary number of dimensions is straightforward.
As we will see, in the present case $\varepsilon$ has only one degree of freedom. This allows us to consider
just $\epsilon=\varepsilon|_{C} \otimes \eta|_{A}$.
In the forthcoming we will drop indices $A,C$.

\subsection{Technicalities}

To compute the wigs for a generic solution in $AdS_{d+1}$ and to perform the perturbative procedure to reconstruct solutions of the supergravity equations of motions we developed a series of \verb Mathematica \textsuperscript{\textregistered} programs.

The algorithms for wig computation in generic dimensions $d$ are collected in \verb Wig.nb \  program. 
Given the vielbein for a starting metric, the Killing spinors for empty $AdS_{d+1}$ and the supersymmetry transformations, \verb Wig.nb \  computes order by order the various fields generated by susy transformations.
To do this, he uses subroutines to perform gamma matrices computations through symbolic manipulations and generates, whenever possible, the fermionic bilinears.
We 
The outputs are saved as both binary files and text files to be stored and used by other machines or other \verb Mathematica \textsuperscript{\textregistered} instances.

The fluid/supergravity procedure is implemented in the \verb SugraFlu.nb \  program.
It reads from files the finite fields the that has to be analyzed. 
For our supergravity models, they are vielbein, bilinears in the gravitino and gauge fields.
Then, it computes the perturbative expansion in power of bilinears of equations of motion, following the generalization of the procedure in sec.~\ref{sezPP}.
Then, if the transformation parameters are constant, \verb SugraFlu.nb \ checks the equations of motion, otherwise in case of parameters promoted to local functions it gives both constraints and dynamical equations.

Both of them will be made available in the future.


\chapter{Supersymmetric Fluid Dynamic from BTZ Black Hole}
\label{chBTZ}

In this chapter we analyze the supergravity extension of fluid/gravity correspondence
in a simpler situation where it is possible to compute all contributions analytically and we present them in a compact and manageable form.
For that we consider $\mathcal{N}=2$, $D=3$ supergravity theory with cosmological term \cite{Achucarro:1987vz,Banados:1992wn,Banados:1992gq,Coussaert:1993jp,Izquierdo:1994jz,Howe:1995zm,Banados:1998pi,Henneaux:1999ib}.

It has been pointed out that this theory is a topological one and therefore it does not possess any local degrees of freedom,
that is all fluctuations can be reabsorbed by gauge redefinitions.
Nevertheless the theory has non--trivial localized solution with singularitiy, named BTZ black hole, which is a trivial solution except for the fixed points of an orbifold action (the orbifold is defined in terms of a discrete subgroup of the isometry group \cite{Banados:1992wn, Banados:1992gq}).

On the boundary side, $1+1$--dimensional conformal fluids is quite trivial, as described in \cite{Bhattacharyya:2008mz}.
The main reason is that the number of independent component of energy--momentum tensor is equal to the number of equations of motion and hence it does not contain physical information beyond the equations of energy--momentum conservation.

However, thanks to the simplicity of this model, we are able to solve in detail the following problems: 1) the computation of the complete supergravity solution and 2) the computation of the equations of motion order by order after promoting the parameters of the superisometries to local fields on the boundary.

Moreover the addition of new fermionic degrees of freedom could in principle modify the structure of the models in a non trivial way and this could bring to new interesting results.
For our purposes, we consider BTZ black hole in both global and Poincar\'e coordinates.

In order to generate the complete wig we start
from the supersymmetry transformations associated to the Killing spinors of $AdS$ space. Since the BTZ black hole
is non-extremal any supersymmetry transformation will produce a change in the solution. 
Multiple applications of the supersymmetry transformation
generated by Killing spinors will result in the application of the corresponding Killing vector generating the complete
supergroup of isometries of $AdS$ space which is ${OSp}(2|2)\times{OSp(2|2)}/SO(2)\times OSp(2|2)$.\footnote{$AdS$ space considered here is actually a superspace
with 3 bosonic coordinates and 4 fermionic coordinates, it can be viewed as $OSp(2|2)/SO(2)$ (since $Sp(2)  \sim SL(2,\mathbb R)
\sim SO(1,2)$).}


Given the new solution, one can observe that some isometries of the black hole such as the translation invariance
in the time direction and in the angular coordinate (or in the space coordinate in the Poincar\'e patch) are preserved. That implies that
the mass $M$ and the angular momentum $J$ are still conserved charges. Indeed, we can compute them using the ADM formalism and
that gives a mass and an angular momentum which is shifted by fermionic bilinears. In the case of extremal black hole, $M = |J|$, the fermionic
corrections will not spoil the extremality condition. In the same way, we can compute also
the entropy of the black hole which is modified by fermionic bilinears.


Having set up the stage for the computation, we promote the fermionic parameters of the superisometries to
local parameters on the boundary. Then, by inserting the fields in the supergravity equations we find
two sets of new equations which should be satisfied: Navier-Stokes equations and new differential equations for the fermionic degrees of freedom.
In order to interpret the result obtained we also perform the bosonic isometries associated to the dilatation
and to the translation on the boundary reproducing the usual linearized
version of relativistic Navier-Stokes equations. On the other side, by inserting the
solution in the gravitino equation, we finally derive a new set of partial differential equations for
the fermionic degrees of freedom which we interpret as Dirac-type equation for the fluid.


With the complete metric, we can finally compute the extrinsic curvature and, using
Brown-York  procedure \cite{Brown:1992br,Balasubramanian:1999re} we derive the boundary
energy-momentum tensor. The form of the latter resemble the tensor for a perfect fluid, except
for a term (violating the chirality). Nonetheless a redefinition of the velocity of the fluid takes the
energy momentum tensor to the standard formula for a perfect fluid and the temperature is shifted by terms dependent on bilinears.
The computation has been performed at the first level in the isometry parameters and it shows the absence of dissipative effects, as expected from a conformal fluid in $1+1$ dimensions.
To see the emergence of new structures in the fluid energy--momentum tensor one needs a complete second order computation.


\section{Setup}

\subsection{Action and Equations of Motion}

As mentioned in the introduction, we consider the supersymmetric $\mathcal{N}=2$, $D=3$ of \cite{Achucarro:1987vz,Izquierdo:1994jz} whose field content is described by the vielbein $e^{A}$, the gravitino (complex) $\psi$, an abelian gauge field $A$ and the spin connection $\omega^{AB}$.
Those are the gauge fields of the diffeomorphism, the local supersymmetry, the local $U\left( 1 \right)$ transformations and of the Lorentz symmetry.
The gauge symmetry can be used to gauge out all local degrees of freedom and the remaining d.o.f. are localized singular solutions \cite{Banados:1992wn,Banados:1992gq,Henneaux:1999ib,Henneaux:1984ei}.

The invariant action has the following form
\begin{align}
	S = & \int_{\mathcal{M}} \left(
		R^{AB} \wedge e^{C} \varepsilon_{ABC}	
		-
		\frac{\Lambda}{3} e^{A} \wedge e^{B} \wedge e^{C} \varepsilon_{ABC}
		-
		\bar \psi \wedge {\cal D} \psi
		-
		2 A \wedge \dd A
		\right)
		\ ,
	\label{action0}
\end{align}
where the curvature $2$--form is defined as $R^{AB}= \dd \omega^{AB} + \omega^{A}{}_{C}\wedge \omega^{CB}$ and $\mathcal{M}$ is a $3d$ manifold.
In components, the action reads
\begin{align}
	S= \int\! d^3x\, \left[
		e \left( R - 2 \Lambda \right)
		- \bar\psi_{M} {\cal D}_{N} \psi_{R} \varepsilon^{MNR}
		- 2 A_{M} \partial_{N} A_{R} \varepsilon^{MNR}
	\right]\ ,
\label{action1}
\end{align}
where $e$ is the vielbein determinant and $R$ is the Ricci scalar\footnote{$\left\{ A,B,\dots \right\}$ label flat indices and $\left\{ M,N,\dots \right\}$ refer to curved ones}.
The action, is invariant under all gauge transformations and it can be cast in a Chern--Simons form\cite{Achucarro:1987vz}\footnote{Note that $AdS_{3}$ radius is set to one and $\left(8 \pi G_3\right)^{-1} = 1$.}.
The covariant derivative ${\cal D}_M$ is defined as
\begin{align}
	{\cal D}_M = D_M + i A_M - \frac{\Lambda}{2}e_{\phantom{A}M}^A \Gamma_A \ ,
\label{action2}
\end{align}
where $D= \dd + \frac{1}{4}\omega^{AB} \Gamma_{AB}$ is the usual Lorentz-covariant differential.
It can be easily shown that (\ref{action1}) is invariant under the local supersymmetry transformations
\begin{align}
\delta_\epsilon\psi &= {\cal D}\epsilon \ ,&
\delta_\epsilon e^A &= {\frac14}\left( \bar\epsilon\Gamma^A\psi - \bar\psi\Gamma^A\epsilon \right)
\ ,&
\delta_\epsilon A &= {\frac{i}{4}} \left( \bar\epsilon \psi - \bar\psi\epsilon \right) \ .
\label{action4}
\end{align}
The spin connection transforms accordingly when the vielbein postulate is used to compute $\omega^{AB}$.
The signature for the flat metric $\eta_{AB}$ is
$(-,+,+)$ and the gamma
matrices $\Gamma^A$ are real
\begin{align}
	\Gamma_{0} = &\  i \sigma_{2}
	\ , &
	\Gamma_{1} = &\  \sigma_{3}
	\ , &
	\Gamma_{2} = &\  \sigma_{1}
	\ , &
	\left\{ \Gamma_{A}\,,\,\Gamma_{B} \right\} = & 2 \eta_{AB}
	\ .
	\label{gammaBTZ}
\end{align}
From (\ref{action1}) we deduce the following equations of motion
\begin{align}
	& \cD \psi = 0
	\ , \nonumber \\
	& \dd A = \frac{i}{4} \bar \psi \wedge \psi
	\ , \nonumber \\
	& \dd e^{A} + \omega^{A}_{\phantom{A}B} \wedge e^B = {\frac14}\bar\psi\wedge\Gamma^A \psi
	\ ,\nonumber \\	
	& d \omega^{AB} + \omega^{A}{}_{C}\wedge \omega^{CB} - \Lambda e^{A} \wedge e^{B} = - \frac{\Lambda}{4} \varepsilon^{AB}{}_{C} \bar\psi\Gamma^{C}\psi
	\ .
	\label{EoMs1}
\end{align}
The third equation is the vielbein postulate, from which the spin connection $\omega^{AB}$ is computed.
It is possible to check the above equations against the Bianchi identities.
Note that the theory, being topological, can be written in the form language.\footnote{
Using the forms, the gauge symmetries are obtained by shifting all fields
 $e^{A}\rightarrow e^A + \xi^{A}$, $\psi\rightarrow\psi+\eta$, $\omega\rightarrow\omega^{AB} + k^{AB}$ and $A\rightarrow A+C$
and consequently the differential operator $\dd\rightarrow \dd+s$. $\xi^{A}$, $\eta$, $k^{AB}$ and $C$ are the ghosts associated to diffeomorphism, supersymmetry, Lorentz symmetry and $U\left( 1 \right)$ transformation, respectively and $s$ is the BRST differential associated to those gauge symmetries.}
The gravitino equation simply implies the vanishing of its field strength.
The second equation fixes the field strength of the gauge field and the fourth one fixes the Riemann tensor.
Note that for $AdS_{3}$, the cosmological constant is $\Lambda=-1$.

\subsection{$AdS_3$ and BTZ Black Hole}

The supergravity equations of motion admit as solution the $AdS_{3}$ space
\begin{align}
	& g_{MN} = (g_{AdS})_{MN} \ , &
	& A_{M} = 0 \ ,&
	\psi_{M} = 0 \ ,	
	\label{emptyAdS}
\end{align}
where the $AdS_{3}$ metric in global coordinates reads
\begin{align}
	\dd s^{2} =
	- f^{2} \dd t^{2} 
	+ f^{-2} \dd r^{2}
	+ r^{2} \dd \phi^{2}
	\ ,
	\label{AdSCglobal1}
\end{align}
where $f^{2}= r^{2} + 1$. 
The associated vielbeins are
\begin{align}
	  e^{0} = &\  f \dd t \ ,
	& e^{1} = &\  f^{-1} \dd r \ ,
	& e^{2} = &\  r \dd \phi 
	\ ,
\label{vielb1}
\end{align}
and the spin connection components read
\begin{align}
	\omega^0{}_{1} = &\ r \dd t \ ,
	& \omega^{0}{}_{2} = &\  0 \ ,
	& \omega^2{}_{1} = &\  f \dd \phi \ .
	\label{spinC1}
\end{align}
As usual, the flat metric $\eta_{AB}$ is mostly plus  $\left( -,+,+ \right)$.

Another solution is the so called BTZ black hole\footnote{We refer the reader to the vast literature on the subject for the geometry of this solution.} whose global metric reads:
\begin{align}
	\dd s^{2} =
	-N^{2} \dd t^{2}
	+
	N^{-2} \dd r^{2}
	+
	r^{2} \left( N^{\phi}\dd t + \dd \phi \right)^{2}
	\ ,
	\label{BTZ1}
\end{align}
where $N$ and $N^{\phi}$ are defined as
\begin{align}
	N = & \sqrt{- M_{0} + r^{2} + \frac{J_{0}^{2}}{4 r^{2}}}
	\ ,
	& N^{\phi} = & - \frac{J_{0}}{ 2 r^{2}}
	\ .
	\label{BTZ2}
\end{align}
The non--zero vielbein components  are
\begin{align}
	e^{0} = & N \dd t \ ,
	&
	e^{1} = & N^{-1} \dd r
	\ , &
	e^{2} = & r N^{\phi} \dd t + r \dd \phi
	\ ,
	\label{BTZviel}
\end{align}
and the non--zero spin connection components read
\begin{align}
	\omega^{0}{}_{1}= &
	r \dd t - \frac{J_{0}}{2 r} \dd \phi
	\ , &
	\omega^{0}{}_{2} = & - \frac{J_{0}}{2 r^{2} N} \dd r
	\ , &
	\omega^{1}{}_{2} = & - N \dd \phi
	\ .
	\label{BTZspinC}
\end{align}
The parameter $M_{0}$ is to be identified with the mass of the black hole while $J_{0}$ represents its angular momentum.
Notice that neither $N$ nor $N^{\phi}$ depends on coordinate $t$ which means that the solution is stationary. 

The existence of a horizon is constrained by \cite{Banados:1992wn,Banados:1992gq}
\begin{align}
	M_{0}&>0 \ , & \left| J_{0}\right| & \leq M \ ,
	\label{horiz1}
\end{align}
the case $J_{0}=0$, $M_{0}=-1$ reduces to empty $AdS_{3}$ (\ref{AdSCglobal1}) while the case $-1 < M_{0} < 0$ can be excluded from the physical spectrum since it corresponds to a naked singularity. 
Note also that for $r\rightarrow\infty$ the metric approaches the empty $AdS_{3}$ solution (\ref{AdSCglobal1}). 
It exists also an extremal solution for $M_{0}=\left| J_{0} \right|$, which preserves two of the four supersymmetries of $AdS_{3}$. 
In the present case we want to focus on the non--extremal case where all supersymmetries are broken.

To analyze the boundary fluid dynamic using fluid/gravity technique it is convenient to consider $AdS_{3}$ metric written in Poincar\'e patch
\begin{align}
	\dd s^{2} & =
	- r^2 \dd t^{2}
	+
	\frac{1}{r^2} \dd r^{2}
	+
	r^{2} \dd x^{2}
	\ .
	\label{AdSPP}
\end{align}
As in \cite{Bhattacharyya:2008jc,Rangamani:2009xk} we perform an ultralocal analysis and then we also use Poincar\'e patch for BTZ black hole metric
\begin{align}
	\dd s^{2} & =
	-\left( r^{2} - M_{0} \right) \dd t^{2}
	+
	\frac{1}{r^2 - M_{0}} \dd r^{2}
	+
	r^{2} \dd x^{2}
	\ .
	\label{BTZPP}
\end{align}
Notice that in this case  the $AdS_{3}$ metric (\ref{AdSPP}) is obtained by setting $M_{0}=0$. The form of the metric is similar to (\ref{BTZ1}) but it will cover just a sector of the entire $AdS$ space.
As we will show in the next section, after a finite boost transformation the metric (\ref{BTZPP}) can be cast as in (\ref{BTZ1}), with mass and angular momentum depending on the boost parameters and $M_{0}$.

\subsection{Killing Vectors}\label{KilVec3}

In this section we compute the Killing vectors for $AdS_{3}$ for both global metric (\ref{AdSCglobal1}) and Poincar\'e patch (\ref{AdSPP}). 
As we will discuss later, we consider the isometries of $AdS_3$ space to generate orbits of the black hole solution. This is obtained by acting with the generators of $AdS_3$ isometries on the black hole metric.

The Killing vectors are solutions to the equations
\begin{equation}
	\mathcal{L}_{\xi} (g_{AdS}) =  0
	\ ,
\end{equation}
where
\begin{equation}
	\xi = \xi^t \partial_t + \xi^r \partial_r + \xi^{\phi} \partial_{\phi}
	\ ,
\end{equation}
and, for global $AdS_{3}$ (\ref{AdSCglobal1}), they are
\begin{align}
\xi^t & = \frac{r}{\sqrt{1+r^2}} \partial_t A\left(t,\phi\right) + e_0 \ ,
\nonumber\\
\xi^r & = \sqrt{1+r^2} A\left(t,\phi\right) \ , \nonumber
 \\
\xi^\phi & = \frac{\sqrt{1+r^2}}{r} \partial_{\phi} A\left(t,\phi\right) + f_0 \ ,
\label{KVglobal}
\end{align}
where the function $A\left(t,\phi\right)$ is defined as
\begin{equation}
A(t,\phi) = a_0 \cos\left(t+\phi\right)+ b_0 \cos\left(t-\phi\right) + c_0 \sin\left(t+\phi\right)+ d_0 \sin\left(t-\phi\right) \ .
\end{equation}
The solution depends upon the $6$ free parameters $\{a_{0},b_{0},c_{0},d_{0},e_{0},f_{0}\}$, associated to the $AdS_3$-isometry group, namely, $SO\left(2,2\right)$.

The Killing vectors for $AdS_{3}$ in Poincar\'e patch, defined as $K=K^t \partial_t + K^r \partial_r + K^x \partial_x$ are
\begin{align}
	K^{t}
	& =
	- \frac{c_{1}}{2}
	\left( \frac{1}{r^{2}} +t^{2} + x^{2} \right)
	-
	c_{2} t x
	-
	b t + w x + t_{0}
	\ , \nonumber \\
	K^{r}
	& =
	r \left(
	c_{1} t + c_{2} x + b
	\right)
	\ , \nonumber \\
	K^{x}
	& =
	- \frac{c_{2}}{2}
	\left( - \frac{1}{r^{2}} +t^{2} + x^{2} \right)
	-
	c_{1} t x
	+
	w t - b x + x_{0}
	\ .
	\label{KVPP}
\end{align}
The $6$ real infinitesimal constant parameters describe the $6$--parameters isometry group of $AdS_{3}$: $b$ is associated with dilatation, $w$ is the boost parameter, $c_{1}$ and $c_{2}$ are related to conformal transformations and $t_{0}$ and $x_{0}$ parameterize the $t-$ and $x-$translations.

In order to complete the procedure outlined in \cite{Bhattacharyya:2008jc,Rangamani:2009xk,Gentile:2011jt,Gentile:2012tu} we perform a finite boost on the
BTZ solution in the $t$-$x$ plane, namely
\begin{align}
	& t \rightarrow \frac{t - w x}{\sqrt{1 - w^{2}}}
	\ ,
	&
	x \rightarrow \frac{x - w t}{\sqrt{1 - w^{2}}}
	\ ,
	&
	\label{boost1}
\end{align}
where $w$ is the boost parameter.

We now perform a finite dilatation of the BTZ black hole. This transformation will allow us to define a parameter for the temperature of the fluid in the same fashion as \cite{Bhattacharyya:2008jc}.
The correct dilatation weights can be obtained by redefining the coordinates as follows
\begin{equation}
\label{dilatate}
r \rightarrow \hat b r
\ , \qquad \qquad
t \rightarrow \hat b^{-1} t
\ , \qquad \qquad
\phi \rightarrow \hat b^{-1} \phi
\ .
\end{equation}
The infinitesimal dilatation are given by
$\hat b = 1 + b + O(b^{2})$,
where $b$ is the infinitesimal parameter introduced in eq.~(\ref{KVPP}).

The boosted and dilatated metric can be recast in the form (\ref{BTZ1}) by replacing
\begin{align}
	&
	M_{0} \rightarrow M = \frac{1 + w^{2}}{1 - w^{2}} \frac{M_{0}}{\hat b^{2}}
	\ ,
	&
	J_{0} \rightarrow J =
	- \frac{2 w}{1 - w^{2}}\frac{M_{0}}{\hat b^{2}}
	\ ,
	&
	\label{boost2}
\end{align}
and
\begin{align}
	r^{2} \rightarrow R^{2}
	=
	r^{2}
	+
	\frac{w^{2}}{1-w^{2}} \frac{M_{0}}{\hat{b}^2}
	\ .
	\label{boost3}
\end{align}
Note that the boost transformations can be applied to the global BTZ metric (\ref{BTZ1}) to generate a new set of solutions, as described in \cite{Martinez:1999qi} (see eq.~(\ref{boost1}) with $x$ substituted by the angular coordinate $\phi$). In this case the replacing rules for mass, angular momentum and radius coordinate will be
\begin{subequations}
\begin{align}
&
	M_{0} \rightarrow M = \frac{1 + w^{2}}{1 - w^{2}} M_{0} - \frac{2 w}{1 - w^{2}} J_{0} \ , \\
	&
	J_{0} \rightarrow J = \frac{1 + w^{2}}{1 - w^{2}} J_{0} - \frac{2 w}{1 - w^{2}} M_{0} \ , \\
	&
	r^{2} \rightarrow R^{2}
	=
	r^{2}
	-
	\frac{w}{1-w^{2}} \left( J_{0} - w M_{0}  \right)
	\ .
\end{align}
\end{subequations}
Defining
\begin{align}
	&
	\gamma = \frac{w^{2} + 1}{w^{2} - 1}
	\ ,
	&
	\beta = - \frac{2 w}{ w^{2} + 1}
	\ ,
	&
	\label{boost4}
\end{align}
the metric for the new global BTZ solutions can be written modifying mass and angular momentum in the following Lorentz-like form, \textit{i.e.}
\begin{align}
M& = \gamma M_0 - \beta \gamma J_0 \ , \nonumber\\
J& = \gamma J_0 - \beta \gamma M_0 \ , \nonumber\\
R^2& = r^2 -\frac{1}{2} \left[\beta J_0 - \left(\gamma +1\right)M_0\right]
\ .
\end{align}


\subsection{Killing Spinors}

Now, we need to construct the Killing spinors of $AdS_3$ and the isometries generated by them\footnote{We remind the reader that Killing vectors can be obtained constructing Killing spinors bilinears such as $\xi^{\mu}= \bar{\epsilon}\Gamma^{\mu}\epsilon$. By construction they will indeed satisfy the Killing vectors equation.}.
To construct the BTZ wig we compute the Killing spinors $\epsilon $ for $AdS_{3}$ first in global coordinates and then in Poincar\'e patch. As we have said in  they are defined from  Killing spinors equation
\begin{align}
	\mathcal{D}_{AdS} \epsilon = 0 \ .
	\label{KSeq}
\end{align}
In global coordinates, it reads
\begin{align}
	& \partial_{r} \, \epsilon + \frac{1}{2 f} \Gamma_{1} \, \epsilon	= 0 \ ,
	\nonumber\\
	& 
	\partial_{t} \, \epsilon
	+
	\frac{1}{2} \left( - r \Gamma_{2}  + f \Gamma_{0} \right) \, \epsilon = 0 \ ,
	\nonumber\\
	& 
	\partial_{\phi} \, \epsilon
	+
	\frac{1}{2}  \left( r \Gamma_{2}  - f \Gamma_{0} \right) \, \epsilon = 0 \ .
	\label{KilEq2BTZ}
\end{align}
The index of gamma matrices is flat since the vielbein is written explicitly.
Note that $\left( f \Gamma_{1} + r \Gamma_{0} \right)^{2} = \one$. 
To solve eqs.~(\ref{KilEq2BTZ}), we define the projected spinors
\begin{align}
	\epsilon_{\pm} = \pm \Gamma_{1} \epsilon_{\pm}
	\ ,
	\label{KilEq3}
\end{align}
hence, equations (\ref{KilEq2BTZ}) read
\begin{align}
	&
	\partial_{r} \epsilon_{+} + \frac{1}{2 f} \epsilon_{+} =  0
	\ ,
	&
	\partial_{r} \epsilon_{-} - \frac{1}{2 f} \epsilon_{-} =  0
	\ , & \nonumber \\
	&
	\partial_{t} \epsilon_{+} + \frac{1}{2} \left( f - r \right) \epsilon_{-} =  0
	\ ,
	&
	\partial_{t} \epsilon_{-} - \frac{1}{2} \left( f + r \right) \epsilon_{+} =  0
	\ , & \nonumber \\
	&
	\partial_{\phi} \epsilon_{+} + \frac{1}{2} \left( r - f \right) \epsilon_{-} =  0
	\ ,
	&
	\partial_{\phi} \epsilon_{-} + \frac{1}{2} \left( r + f \right) \epsilon_{+} =  0
	\ .
	\label{eps2BTZ}
\end{align}
Solving the $r$--equations we have
\begin{align}
	\epsilon_{+} = & \left( r + f \right)^{-1/2} \eta_{+}\left( t,\phi \right) 
	\ , 
	& \epsilon_{-} = & \left( r + f \right)^{1/2}  \eta_{-}\left( t,\phi \right)
	\ ,
	\label{eps21BTZ}
\end{align}
thus the $t$--and $\phi$--equations reduce to
\begin{align}
	\partial_t \eta_+ + \frac{1}{2} \eta_- = & 0
	\ ,
	&
	\partial_t \eta_- - \frac{1}{2} \eta_+ = & 0
	\ ,\nonumber\\
	\partial_\phi \eta_+ - \frac{1}{2} \eta_- = & 0
	\ ,
	&
	\partial_\phi \eta_- + \frac{1}{2} \eta_+ = & 0
	\ .
	\label{eps22BTZ}
\end{align}
The solutions read
\begin{align}
	\epsilon_{+} = & \left( r + f \right)^{-1/2} \left( \lambda_{1} \cos \left[ \frac{t - \phi}{2} \right] - \lambda_{2} \sin \left[ \frac{t - \phi}{2} \right]  \right)
	\ , \nonumber \\
	\epsilon_{-} = & \left( r + f \right)^{1/2} \left( \lambda_{2} \cos \left[ \frac{t - \phi}{2} \right] + \lambda_{1} \sin \left[ \frac{t - \phi}{2} \right]  \right)
	\ ,
	\label{eps3BTZ}
\end{align}
that is
\begin{align}
	\epsilon
	= &
	\frac{1}{2} 
	\left[
	\left( \sqrt{r + f} + \frac{1}{\sqrt{r + f}}  \right) \one
	-
	\left( \sqrt{r + f} - \frac{1}{\sqrt{r + f}}  \right) \Gamma_{1}	
	\right]\times
	\nonumber\\
	&\ \times
	\left( 
	\cos \left[ \frac{t - \phi}{2} \right] \one 
	-
	\sin \left[ \frac{t - \phi}{2} \right] \Gamma_{0}
	\right) \lambda
	\ ,
	\label{eps4BTZ}
\end{align}
where $\lambda$ is a spinor with two complex constant Grassmann components $\lambda_{1}$ and $\lambda_{2}$.

In the same way,we compute the Killing spinors for the BTZ black hole in Poincar\'e patch.
In components the Killing spinor equation reads
\begin{align}
	&
	\partial_{t}\epsilon + \frac{r}{2}\Gamma_{0}\left( \Gamma_{1}+\one \right)\epsilon=0\ ,
	\nonumber\\
	&
	\partial_{r}\epsilon+\frac{1}{2r}\Gamma_{1}\epsilon=0\ ,
	\nonumber\\
	&
	\partial_{x}\epsilon + \frac{r}{2} \Gamma_{2}\left( \Gamma_{1}+\one \right)\epsilon=0\ .
	\label{ksEQ1btz}
\end{align}
Solving this we have
\begin{equation}
	\epsilon
	 =
	\left[
	\frac{1}{2\sqrt{r}} \left( \one - r x^{\mu} \Gamma_{\mu}\right)
	\left( \one +\Gamma_{1} \right)
	+\frac{\sqrt{r}}{2} \left( \one - \Gamma_{1} \right)
	\right]
	\zeta\ ,
	\label{ciao}
\end{equation}
where as in the previous case $\zeta$ is a Dirac spinor with $2$ complex constant Grassmann components $\zeta_{1}$ and $\zeta_{2}$
\begin{align}
	\partial_{R} \zeta = 0
	\ .
	\label{constantZeta}
\end{align}


\section{Fermionic Wig}

We now proceed to the construction of the fermionic wig (\textit{i.e.} the complete solution in the fermionic zero modes) associated with a boosted and dilatate BTZ black hole in Poincar\'e patch.\footnote{Note that our Killing spinors (or anti-Killing spinors as defined in \cite{Kallosh:1996qi}) are not time independent but the fermionic black hole superpartner does not depend on $t$.}
As explained in  sec.~\ref{KilVec3}, the boost and the dilatation shift the mass and angular momentum of the black hole.
Therefore, to get the complete solution, we
first compute the wig for the BH as described in sec.~\ref{generalization} and then we replace $M_0$ and $J_0$ with $M$ and $J$ as defined in (\ref{boost2}).
As we have already pointed out, our background solution is the $AdS$--Schwarzschild black hole with all the other fields set to zero
\begin{align}
	e^{\left[ 0 \right]}{}_{M}^{A} & =  \left. e_{M}^{A}\right|_{BTZ} \ ,
	& \psi^{[0]}{}_{M} = 0 \ ,&
	& A^{[0]}{}_{M} = 0 \ .&
	\label{back1}
\end{align}
We define the real bilinears
\begin{align}
	\bB_{0} = &
	- i \zeta^{\dagger}\zeta
	\ ,
	& \bB_{1} = &
	- i \zeta^{\dagger}\sigma_{1}\zeta
	\ ,	
	& \bB_{2} = &
	i\zeta^{\dagger}\sigma_{2}\zeta
	\ ,
	& \bB_{3} =  &
	- i \zeta^{\dagger}\sigma_{3}\zeta
	\ ,
	\nonumber\\
		\bC_{0} = &
	- i \lambda^{\dagger}\lambda
	\ ,
	& \bC_{1} = &
	- i \lambda^{\dagger}\sigma_{1}\lambda
	\ ,	
	& \bC_{2} = &
	i\lambda^{\dagger}\sigma_{2}\lambda
	\ ,
	& \bC_{3} =  &
	- i \lambda^{\dagger}\sigma_{3}\lambda
	\ ,
	\label{realBil0}
\end{align}
due to the anticommutative nature of $\zeta_{1}$ and $\zeta_{2}$, these identities hold
\begin{align}
	&\bB_{1}^{2}=\bB_{2}^{2}=\bB_{3}^{2}=-\bB_{0}^{2}\ ,&
	&\bB_{i}\bB_{j}=0\,,\, i\neq j\ ,&
	&\bB_{0}^{n}=0\,,\, n>2
	\ ,&
	\nonumber\\
	&\bC_{1}^{2}=\bC_{2}^{2}=\bC_{3}^{2}=-\bC_{0}^{2}\ ,&
	&\bC_{i}\bC_{j}=0\,,\, i\neq j\ ,&
	&\bC_{0}^{n}=0\,,\, n>2
	\ ,&
	\label{fierz1}
\end{align}
In the formalism described in (\ref{defTransf}), from susy transformations (\ref{action4}) we derive algorithms to compute iteratively the various fields
\begin{align}
	\psi^{\left[ n \right]}{}_{M}
	& = \frac{1}{(2n-1)} {\cal D}^{\left[ n \right]}{}_{M} \epsilon
\ , \nonumber \\
	e^{\left[ n \right]}{}^{A}_{M}
	& = 
	\frac{1}{4 (2n)}  \bar\epsilon\, \Gamma^{A} \psi^{\left[ n \right]}{}_{M} + h.c.
\ , \nonumber \\
	A^{\left[ n \right]}{}_{M}
	& =
	\frac{i}{4(2n)}\bar\epsilon\, \psi^{\left[ n \right]}{}_{M} + h.c.
	\ .
	\label{algVielbein}
\end{align}
Then, the wig order by order is written as\footnote{Note that $e_{\left( M \right.}e_{\left. N \right)}=\frac{1}{2}\left( e_{M}e_{N}+e_{N}e_{M} \right)$.}
\begin{align}
	g^{\left[ n \right]}{}_{M N}
	=  &
	\sum_{p=0}^{n} e^{\left[ p \right]}{}^{A}_{\left( M \right.} \, e^{\left[ n-p \right]}{}^{B}_{\left.  N \right)} \eta_{AB}
	\ .
	\label{algMetric}
\end{align}
In order to implement the computation in \verb Mathematica \textsuperscript{\textregistered}, we developed algorithms to derive at each order the inverse vielbein and the spin connection.
To compute the inverse vielbein $e_{A}^{M}$, we use the definition
\begin{equation}
	e_{M}^{A} e_{B}^{M} = \delta^{A}_{B}
	\ ,
	\label{AlgEI0}
\end{equation}
expanding the vielbeins we get
\begin{equation}
	\left(
	e^{\left[ 0 \right]}{}^{A}_{M}
	+
	e^{\left[ 1 \right]}{}^{A}_{M}
	+
	e^{\left[ 2 \right]}{}^{A}_{M}
	\right)
	\left(
	e^{\left[ 0 \right]}{}^{M}_{B}
	+
	e^{\left[ 1 \right]}{}^{M}_{B}	
	+
	e^{\left[ 2 \right]}{}^{M}_{B}	
	\right)
	= \delta^{A}_{B}
	\ .
	\label{AlgEI1}
\end{equation}
We then obtain one equation for each perturbative order
\begin{align}
	\delta^{A}_{B} = & e^{\left[ 0 \right]}{}^{A}_{M} e^{\left[ 0 \right]}{}^{M}_{B}
	\ ,
	\nonumber\\
	0 = &
	e^{\left[ 0 \right]}{}^{A}_{M} e^{\left[ 1 \right]}{}^{M}_{B}
	+
	e^{\left[ 1 \right]}{}^{A}_{M} e^{\left[ 0 \right]}{}^{M}_{B}
	\ ,
	\nonumber\\
	0 = &
	e^{\left[ 0 \right]}{}^{A}_{M} e^{\left[ 2 \right]}{}^{M}_{B}
	+
	e^{\left[ 1 \right]}{}^{A}_{M} e^{\left[ 1 \right]}{}^{M}_{B}
	+
	e^{\left[ 2 \right]}{}^{A}_{M} e^{\left[ 0 \right]}{}^{M}_{B}
	\ .
	 \label{AlgEI2}
\end{align}
The first one is solved as usual by inverting the vielbein $e_{M}^{A}$.
The other equations are solved by
\begin{align}
	e^{\left[ 1 \right]}{}^{M}_{B}
	= &
	- e^{\left[ 0 \right]}{}^{M}_{A}
		\left[
			e^{\left[ 1 \right]}{}^{A}_{ R}
			e^{\left[ 0 \right]}{}^{R}_{B}
		\right]
	\ ,\nonumber\\
	e^{\left[ 2 \right]}{}^{M}_{B}
	= &
	- e^{\left[ 0 \right]}{}^{M}_{A}
		\left[
			e^{\left[ 1 \right]}{}^{A}_{ R}
			e^{\left[ 1 \right]}{}^{R}_{B}
			+
			e^{\left[ 2 \right]}{}^{A}_{ R}
			e^{\left[ 0 \right]}{}^{R}_{B}
		\right]
	\ .
	\label{AlgEI3}
\end{align}
In general we have, for $n > 1$
\begin{align}
	e^{\left[ n \right]}{}^{M}_{B}
	= &
	- e^{\left[ 0 \right]}{}^{M}_{A}
	V^{\left[ n \right]}{}^{A}_{B}
	\ ,\nonumber\\
	V^{\left[ n \right]}{}^{A}_{B}
	= &
	\sum_{p=1}^{n}
	e^{\left[ p \right]}{}^{A}_{ R}
	e^{\left[ n-p \right]}{}^{R}_{B} \ .
	\label{AlgEIfin}
\end{align}
The spin connection $\omega_{M}^{A B}$ is defined by the vielbein postulate (\ref{EoMs1})
\begin{equation}
	\dd e^{A} + \omega_{\phantom{a}B}^{A}\wedge e^{B} = \frac{i}{4} \bar\psi \Gamma^{A} \psi\ .
	\label{AlgSpinC0}
\end{equation}
Extracting the $1$--form basis $\left\{ \dd x^{M} \right\}$, it becomes
\begin{eqnarray}
	\partial_{\left[ M \right.} e^{A}_{\left.  N \right]}
	+
	\omega^{A B}_{\left[ M \right.}\,\eta_{B C}\, e^{C}_{\left.  N \right]}
	=
	\frac{i}{4}
	\bar\psi_{\left[ M \right.} \Gamma^{A} \psi_{\left.  N \right]} \ .
	\label{AlgSpinC1}
\end{eqnarray}
As in the case of inverse vielbein, we expand in perturbative order.  We obtain the following result
\begin{align}
	\omega^{\left[ n \right]}{}^{DC}_{M}
	& =
	e^{\left[ 0 \right]}{}_{M\,A}\left[
		\Omega^{\left[ n \right]}{}^{D C\,,A}
		-
		\Omega^{\left[ n \right]}{}^{C A\,,D}
		-
		\Omega^{\left[ n \right]}{}^{A D\,,C}
		\right]
		\ ,\nonumber\\
	\Omega^{\left[ n \right]}{}^{D C\,,A}
	& = 
	e^{\left[ 0 \right]}{}^{N \left[ D \right.} e^{\left[ 0 \right]}{}^{M \left. C \right]}
		\left[
			\partial_{\left[ M \right.} e^{\left[ n \right]}{}^{A}_{\left. N \right]}
			+
			\sum_{p=1}^{n}\omega^{\left[ n-p \right]}{}^{A B}_{\left[ M \right.}\,\eta_{B C}\,e^{\left[ p \right]}{}^{C}_{\left. N \right]}
			-
			\frac{i}{4} \sum_{p=1}^{n}\eta^{A B}\, \bar\psi^{\left[ p \right]}{}_{\left[ M \right.}\Gamma_{B}\psi^{\left[ n-p+1 \right]}{}_{\left. N \right]}
		\right] \ .
	\label{AlgSpinCfin}
\end{align}

\subsection{Wig in Global Coordinates}

From the Killing spinors  and the definitions (\ref{realBil0})  it is useful to compute the following bilinears
\begin{align}
	\bar\epsilon \epsilon 
	& = 
	\bC_{2}
	\ ,\nonumber\\
	\bar\epsilon \Gamma_{0} \epsilon 
	& = 
	i \left[ - f \bC_{0} + r \left( c \bC_{3} -s \bC_{1} \right) \right]
	\ ,\nonumber\\
	\bar\epsilon \Gamma_{1} \epsilon 
	& = 
	- i \left[ c \bC_{1} + s \bC_{3}\right]
	\ ,\nonumber\\	
	\bar\epsilon \Gamma_{2} \epsilon 
	& = 
	i \left[ - r \bC_{0} + f \left( c \bC_{3} -s \bC_{1} \right) \right]
	\ ,
	\label{BTZbilinears1}
\end{align}
where $c=\cos\left( t - \phi \right)$ and $s=\sin\left( t - \phi \right)$.
At first order in bilinears the gravitino $1$--form reads
\begin{align}
	\psi^{\left[ 1 \right]}
	= &
	\frac{1}{2} \left[ 
	\left( N - f \right) \Gamma_{0}
	- \frac{J}{2 r} \Gamma_{2}
	\right] \epsilon
	\left( \dd t - \dd \phi \right)
	+
	\frac{1}{2} \left( \frac{1}{N} - \frac{1}{f} - \frac{J}{2 r^{2} N} \right)
	\Gamma_{1} \epsilon \, \dd r
	\ .
	\label{grav1}
\end{align}
The first order wig is
\begin{align}
	g^{\left[ 1 \right]} 
	=\  &
	\frac{1}{4}  
	\left[ 
		M - r^{2} + f N
	\right] \bC_{2} \,
	\dd t^{2}
	-
	\frac{1}{8 r^{2} N^{2} f}
	\left[ 
		2 r^{2} \left( N - f \right) + f J
	\right] \bC_{2}	\,
	\dd r^{2} 
	+
	\nonumber\\ &
	-
	\frac{1}{8} 
	\left[ 
	J + 2 M - 2 r^{2} + 2 f N
	\right] \bC_{2}	\,
	\dd t \dd \phi
	+
	\frac{J}{8} \bC_{2} \,
	\dd \phi^2
	\ .
	\label{BTZwig1}
\end{align}
The gauge field is\footnote{$c$ and $s$ are defined as in (\ref{BTZbilinears1})}
\begin{align}
	A^{\left[ 1 \right]}
	=\ &
	\frac{1}{8} \left[ 
	\bC_{0} 
	\left( 
	f \left( N - f \right)
	- \frac{J}{2}
	\right)
	+
	\left( c \bC_{3} -s \bC_{1} \right)
	\left( 
	- r \left( N - f \right)
	+
	\frac{f J}{2 r}
	\right)
	\right]
	\left( \dd t - \dd \phi \right)	
	\nonumber\\
	&
	+
	\frac{1}{8}\left( \frac{1}{N} - \frac{1}{f} - \frac{J}{2 r^{2} N} \right)\left( c \bC_{1} + s \bC_{3}\right)\dd r
	\ .
	\label{BTZgauge1}
\end{align}
The first order of the spin connection is obtained from the vielbein postulate. 
Using standard coordinates transformation, the singularity in $N=0$, being apparent as in the zero order metric (\ref{BTZ1}), can be removed. Notice that all dangerous $1/N$ terms appear along the $\dd r$ component. The pattern repeats itself at the second order.

Iterating the procedure, namely by inserting the first order corrections in (\ref{algVielbein}), we derive the second order results. The gravitino is  
\begin{align}
	\psi^{\left[ 2 \right]}
	& =
	\frac{1}{96 r (r+f)^{3/2}}
	\Big[ 
		\Big(
			(J - 2 r f) \left( \bC_{0} - s \bC_{1} + c \bC_{3} \right)
+  \nonumber \\ & \quad\quad
			-
			\left( 
				r^{2} J + 2 r f (J + r^{2}) + f^{2} (J + 4 r^{2}) + 2 r f^{3} - 2 r N (r + f)^{2}
			\right)
\times	  \nonumber \\ & \quad\quad	\times
			\left( 	 \bC_{0} + s \bC_{1} - c \bC_{3}	\right)
		\Big)
		\big( 
		( \one + \Gamma_{1} ) + ( r + f )( \one - \Gamma_{1} )		
		\big)
+  \nonumber \\ & \quad\quad
		+
		\Big( 
			( - J - 2 r N + 2 r f ) ( \one - \Gamma_{1} )
			+
			( r J - 2 r N (r + f) 
+  \nonumber \\ & \quad\quad			
			+
			f ( J + 2 r^{2} + 2 r f)) ( \one + \Gamma_{1} )
		\Big)
		(r + f) \Gamma_{0} \bC_{2}
	\Big] 
\times \nonumber \\ & \quad \times
	\left( 
		\sin \left[ \frac{t - \phi}{2} \right]  \one - \cos \left[ \frac{t - \phi}{2} \right]  \Gamma_{0} 
	\right)
	\epsilon \, \left( \dd t - \dd \phi \right)
+
\nonumber \\ &
\quad +
	\frac{1}{96 r^{2} N f ( r + f)^{1/2}}
	\left( 2 r^{2} N + f (J - 2 r^{2}) \right)
	\times
\nonumber \\ & \quad\quad
	\times
	\Big( - (\one + \Gamma_{1}) + (r + f) (\one - \Gamma_{1}) \Big) 
	\left( \cos \left[ \frac{t - \phi}{2} \right]  \one - \sin \left[ \frac{t - \phi}{2} \right]  \Gamma_{0} \right) \bC_{2} \epsilon\, \dd r
	\ .
	\label{BTZgrav2}
\end{align}
The second order wig reads
\begin{align}
	g^{\left[ 2 \right]} 
	& =
	\frac{1}{128 r^{2}} 
	\left[ 
		J^{2} 
		-
		2 r^{2} \left( N - f \right)
		\left( 2 N - f \right)
	\right]
	\bC_{2}^{2}\,\dd t^{2}
	+
	\nonumber\\&
	-
	\frac{1}{256 r^{2}} 
	\left[ 
		 J\left( 3 r^{2} + 2 J \right)
		-
		4 r^{2} 
		\left( N - f \right)
		\left( 3 N - 2 f \right)
	\right]
	\bC_{2}^{2}\,\dd t \dd \phi
	+
	\nonumber\\&
	+
	\frac{1}{256 r^{2} N^{2} f^{2}}
	\left( 
	J f + 2 r^{2}\left( N - f \right)
	\right)
	\left( J f + 2 r^{2}\left( N - 2 f \right) \right)
	\bC_{2}^{2}\,\dd r^{2}
	+
	\nonumber\\&
	+
	\frac{1}{256 r^{2}} 
	\left[ 
		J\left( J + 2 r^{2} \right)
		- 4 r^{2} \left( N - f \right)^{2}
	\right]
	\bC_{2}^{2}\,\dd \phi^{2}
	\ .
	\label{BTZwig2}
\end{align}
The second order gauge field is zero.

We note that first order wig (\ref{BTZwig1}) is proportional to the bilinear $\bC_{2}$ only, while the second order one (\ref{BTZwig2}) depends on $\bC_{2}^2$. In addition, for $J\rightarrow0$ and $M\rightarrow-1$ we recover the $AdS_{3}$ solution since in that case the supersymmetry preserves the solution and then there is no wig at all. 
The complete wig does not depend on $t$ and $\phi$, therefore the isometries of the BH are preserved. In the next section we will compute the associated conserved charges, namely the mass and the angular momentum.
The gauge field, which is zero at the bosonic level, is generated at the first order, but it receives no contribution at higher orders.

Notice that it is possible to recast the complete metric $\mathbf{g} = g^{\left[ 0 \right]} + g^{\left[ 1 \right]} + g^{\left[ 2 \right]}$ in the following form
\begin{align}
	\dd s^{2} 
	& =
	- \widetilde N^{2} \dd t^{2}
	+
	\rho^{2} \left( \widetilde N^{\phi} \dd t + \dd \phi \right)^{2}
	+
	\frac{R^{2}}{\widetilde N^{2} \rho^{2}} \dd R^{2}
	\ ,
	\label{recastedWig}
\end{align}
where we defined
\begin{align}
	\rho^{2} & = \mathbf{g}_{\phi\phi}\ ,
	& \widetilde N^{\phi} = \frac{\mathbf{g}_{t\phi}}{\mathbf{g}_{\phi\phi}}&\ ,
	& \widetilde N^{2} = - \frac{\textrm{det} G_{red}}{\mathbf{g}_{\phi\phi}}&\ ,
	& R^{2} = \int_{0}^{r}\sqrt{-\mathbf{g}_{rr}\textrm{det} G_{red}}&\ ,
	\label{recastedWig2}
\end{align}
with $G_{red}$ reduced metric obtained cutting out the $r$--components of $\mathbf{g}$.

\subsection{Wig in Poincar\'e Patch}

In the following, $M,J$ are defined as in (\ref{boost1}) and we replace $\zeta \rightarrow \boldsymbol{\zeta}$ to highlight the fermionic contributions. The gravitino reads
\begin{align}
	\psi^{\left[ 1 \right]}
	= &
	\frac{1}{8 r \sqrt{r}}
	\left[
		\sigma_{1} \left[ - J \left( 1 + r \right) - 2 r \left( r - 1 \right)\left( r - N \right) \right]
\acI	\quad	
		+
		i\sigma_{2} \left[ - J \left( r - 1 \right) - 2 r \left( r + 1 \right)\left( r - N \right) \right]
\acI	\quad
		+
		\left( \sigma_{0} + \sigma_{3} \right) r \left( t - x \right)\left( - J - 2 r^{2} + 2 r N \right)
		\right] \boldsymbol{\zeta} \ \left( \dd t - \dd x \right)
\ac
		+
	\frac{1}{ 8 r^{2} \sqrt{r} N }
	\left( J - 2 r^{2} + 2 r N \right)
	\left[
		\sigma_{0} \left( r - 1 \right)
		-
		\sigma_{3} \left( r + 1 \right)
		+
		\left( \sigma_{1} - i \sigma_{2} \right) r \left( t - x \right)
		\right] \boldsymbol{\zeta} \ \dd r
	\ ,
	\label{grav}
\end{align}
and
\begin{align}
	\psi^{\left[ 2 \right]}
	= &
	\frac{1}{192 r^{2} \sqrt{r}}
	\left[
	i \bB_{3}
		\left(
			\left[ 1 - r^{2} \left( -1 + (t-x)^{2} \right) \right]
			-
			2 r( r - N)
			\left[ 1 + r^{2} \left( -1 + (t-x)^{2} \right) \right]
		 \right)
		 \times
\accI		
		 \qquad\times
		 \left(
		 	\sigma_{3} (1-r)
		 	+
		 	\sigma_{0} (1+r)
		 	+
		 	(\sigma_{1}-i \sigma_{2}) r (t-x)
		 \right)
\acI
	 +
	 i \bB_{0}
		 \left(
			\left[ 1 - r^{2} \left( 1 + (t-x)^{2} \right) \right]
			-
			2 r( r - N)
			\left[ 1 + r^{2} \left( 1 + (t-x)^{2} \right) \right]
		 \right)
		 \times
\accI
		 \qquad\times
		 \left(
		 	\sigma_{3} (1-r)
		 	+
		 	\sigma_{0} (1+r)
		 	+
		 	(\sigma_{1}-i \sigma_{2}) r (t-x)
		 \right)
\acI
	-
	2 \bB_{2} r
		\left(
			\sigma_{1}
				\left[
				J (1+r)
				+
				2 r (r-1) (r-N)
				\right]
			+
			i \sigma_{2}
				\left[
				J (r-1)
				+
				2 r (r+1) (r-N)
				\right]
\acII
			\qquad+
			(\sigma_{0}+\sigma_{3})
			r
			\left( J + 2 r^{2} - 2 r N \right)(t - x)
		\right)
\acI
	+
	2 i \bB_{1} r^{2}
	\left( - J - 2 r^{2} + 2 r N \right)
		\left(
		 	\sigma_{3} (1-r)
		 	+
		 	\sigma_{0} (1+r)
		 	+
		 	(\sigma_{1}-i \sigma_{2}) r (t-x)
		 \right)
	\right] \,\boldsymbol{\zeta}\, \left( \dd t - \dd x \right)
\ac	
	+
	\frac{J - 2 r^{2} + 2 r N}{96 r^{2} N \sqrt{r}}
	\left[
		i \left( - \bB_{1} + (\bB_{0} - \bB_{3}) (t-x) \right)
		\times
\accI
		\qquad\times
		\left(
		\sigma_{3} (1-r)
		+
		\sigma_{0} (1+r)
		+
		\left( \sigma_{1} -i \sigma_{2} \right) r (t-x)
		\right)
\acI
		+
		\bB_{2}
		\left(
		\sigma_{0}(r-1)
		-
		\sigma_{3}(1+r)
		+
		\left( \sigma_{1}-i\sigma_{2} \right)
		r
		(t-x
		\right)		
		\right]
		\,\boldsymbol{\zeta}\,\dd r
		\ .
        \label{grav2}
\end{align}
The metric corrections are
\begin{align}
	g^{\left[ 1 \right]}
	& =
	\frac{1}{4}
	\left(
	M - r^{2} + r N
	\right) \bB_{2} \dd t^{2}
	-
	\frac{1}{8}
	\left(
	J + 2 M -2 r^{2} + 2 r N
	\right) \bB_{2} \dd t \dd x
+ \nonumber \\ &
	+
	\frac{1}{8} J \bB_{2} \dd x^{2}
	-
	\frac{1}{8 r^{2} N^{2}}
	\left(
	J - 2 r^{2} + 2 r N
	\right) \bB_{2} \dd r^{2}
	\ ,
	\label{wig1}
\end{align}
and
\begin{align}
	g^{\left[ 2 \right]}
	& =
	\frac{1}{192}
	\left( 7M -10 r^{2} + 10 r N \right) \bB_{2}^{2} \dd t^{2}
	+
	\frac{1}{192}
	\left(
	2 J + 3 M - 6 r^{2} + 6 r N
	\right) \bB_{2}^{2} \dd x^{2}
+ \nonumber \\ &
	-
	\frac{1}{96}
	\left(
	J + 5 M - 8 r^{2} + 8 r N
	\right) \bB_{2}^{2} \dd t \dd x
+ \nonumber \\ &
	+
	\frac{1}{384 r^{4} N^{2}}
	\left[
		3 J^{2} - 6 r^{2} M
		+ 20 r^{3} \left( r - N \right)
		- 2 J r \left( 5 r -3 N \right)
		\right] \bB_{2}^{2} \dd r^{2}
	\ .
	\label{wig2}
\end{align}
The gauge field one--form is
\begin{align}
	A^{\left[ 1 \right]} = &
	\frac{1}{32 r^{2}}
	\left[
		\left( J - 2 r^{2} + 2 r N \right) \left( \bB_{3} + \bB_{0} \right)
		+
		r^{2} \left( J + 2 r^{2} - 2 r N \right)
		\left(
		\left( 1 - r^{2}\left( t-x \right)^{2} \right) \bB_{3}
\acII
		-
		\left( 1 + r^{2} \left( t - x \right)^{2} \right) \bB_{0}
		-
		2 \left( t - x \right) \bB_{1}
		\right)
	\right] \left( \dd t - \dd x \right)
\ac
	-
	\frac{1}{16 r N}
	\left( J - 2 r^{2} + 2 r N \right)
	\left( \bB_{1} + \left( \bB_{0} + \bB_{3} \right) \left( t-x \right) \right)
	\dd r
	\ ,
	\label{deltaGauge}
\end{align}
at second order, the gauge field is zero. Notice that in the large $r$ expansion $A^{\left[ 1 \right]}_{r} = O\left( \frac{1}{r^{3}} \right)$. As expected, the fermionic corrections collapse in the $AdS_{3}$ limit $M \rightarrow 0$, $J \rightarrow 0$. Note that the metric correction (wig) does not depend upon the boundary coordinates $x$,$t$.
Moreover, there is no off--diagonal corrections in the $rt$ and $rx$ components.
Last remark: Notice that the metric does not depend on boundary coordinates $t$ and $x$, that is the two translational isometries of BTZ black hole are preserved by the wig. This allows to define the wig's mass and the angular momentum.


\subsection{Large \boldmath{$r$} Results in Poincar\'e Patch}

Here we present the obtained results in large $r$ expansion.
To simplify the notation, we define the following expressions
\begin{align}
	\bF & =
	\left[ 1 + ( t - x )^{2} \right] \bB_{0}
	+
	\left[ - 1 + ( t - x )^{2} \right] \bB_{3}
	+
	2 ( t - x ) \bB_{1}
	\ , \quad\quad \bF^2 = 0\,.
	\label{LargeR1}
\end{align}
and
\begin{align}
	\bH = \frac{1}{8} \bB_{2} + \frac{1}{96} \bB^{2}_{2}
	\ .
	\label{LargeR1a}
\end{align}
The gravitino reads
\begin{align}
	\psi & \sim
	\frac{J+M}{192}
	\left[
		\sqrt{r}\, \bF
		\left(
		i \sigma_{0} + i\sigma_{3}
		-
		( i\sigma_{1} + \sigma_{2} ) ( t - x )
		\right)
\acI	
		-
		\frac{1}{\sqrt{r}}
		\left(
		2 ( 12 + \bB_{2})\left( \sigma_{1} + i \sigma_{2} \right)
		+
		\left(
		2 ( 12 + \bB_{2} ) ( t - x ) + i \bF
		\right)
		\left( \sigma_{0} + \sigma_{3} \right)
		\right)
	 \right] \boldsymbol{\zeta} \left( \dd t - \dd x \right)
\ac	
	 +
	 \frac{J-M}{96 r^{2} \sqrt{r}}
	 \left[
	 12 + \bB_{2} - i \bB_{1}
	 -
	 i \left( \bB_{0} + \bB_{3} \right)( t - x )		
	\right]
	\left[ \sigma_{0} - \sigma_{3}
	+
	(\sigma_{1} - i \sigma_{2})( t - x )
	\right]\boldsymbol{\zeta} \, \dd r
	\ .
	\label{LargeRgrav}
\end{align}
The full metric at large $r$ is
\begin{align}
	g \sim &
	-
	\left[ r^{2} - M \left( 1 + \bH  \right) \right] \dd t^{2}
	-
	\left[ J + (M+J)\bH  \right] \dd t \dd x
	\ac
	+
	\left[ r^2 + J \bH \right]  \dd x^{2}
	+
	\frac{1}{r^{2}}
	\left[ 1+ \frac{1}{r^{2}} \left( M -(M-J)\bH  \right) \right]\dd r^{2}
	\ ,
	\label{glarger}
\end{align}
that is
\begin{align}
	g \sim &
	-(r^{2} -M) \dd t^{2}
	- J \dd t \dd x
	+ r^{2} \dd x^{2}
	+\frac{1}{r^{2}}  \left( 1+\frac{M}{r^{2}} \right) \dd r^{2}
\ac
	+
	\bH
	\left[
		M \dd t^{2} - (M+J) \dd t \dd x + J \dd x^{2} - \frac{1}{r^{4}} (M-J) \dd r^{2}
		\right]
		\ .
	\label{glarger1}
\end{align}
Last, the large $r$ gauge field is
\begin{align}
	A & \sim
	-\frac{J+M}{32}\, \bF (\dd t - \dd x)
	-
	\frac{J- M}{16 r^{3}}
	\left[ \bB_{1} + ( \bB_{0} + \bB_{3} )( t - x ) \right] \dd r
	\ .
	\label{LagerRgauge}
\end{align}
In this limit we can rewrite the vielbein and the spin connection for the metric (\ref{glarger}). They read
\begin{align}
	e^0 & =
	\left( r - \frac{M}{2 r} \right) \dd t +
	\frac{M}{2 r} 	\bH (\dd x - \dd t)
\ , \nonumber \\
	e^1 & =
	\left(
		\frac{1}{r} +
		\frac{M}{2 r^{3}}
		+ \frac{M-J}{2 r^{3}} \bH
	\right)\dd r	
\ ,\nonumber  \\
	e^2 & =
	\left( r - \frac{J}{2 r} \right) \dd x
	+
	\frac{J}{2 r} \bH (\dd x - \dd t)
\ ,
\end{align}
and
\begin{align}
	\omega^{01}
	& = \left(r \dd t -\frac{J}{2 r} \dd x \right) + \bH \frac{J}{2 r} \left(\dd t - \dd x\right)
	\ ,\nonumber\\
	\omega^{02}
	& =
	- \frac{1}{2 r^{3}}
	\left[
	J + ( J - M ) \bH
	\right] \dd r
\ ,\nonumber \\
	\omega^{12}
	& =
	- \left(r - \frac{M}{2 r}\right) \dd x - \bH \frac{M}{2 r}\left( \dd t - \dd x\right)
\ .
\end{align}
The large--$r$ curvature $2$--form is computed from the definition in (\ref{action0}).
The non--zero components are
\begin{align}
	R^{01}
	& =
	\frac{M}{2 r^{2}} \bH \dd r \wedge \dd x
	+ \left( 1 - \frac{J}{2 r^{2}}  \bH \right) \dd r \wedge \dd t
\ , \nonumber \\
	R^{02}
	& =
	\left[
		r^{2} + \frac{J}{2} \bH - \frac{M}{2}\left( 1 + \bH \right)
	\right] \dd x \wedge \dd t
\ , \nonumber \\
	R^{12}
	& =
	\frac{J}{2 r^{2}} \left( 1 + \bH \right) \dd r \wedge \dd t
	-
	\left[
		1 + \frac{M}{2 r^{2}} \left( 1 + \bH \right)
	\right] \dd r \wedge \dd x
	\ .
	\label{LargeRcurvature}
\end{align}
It is easy to show that the equations of motion (\ref{EoMs1}) are satisfied.
In particular, in the large $r$ limit, the term $- \frac{\Lambda}{4} \varepsilon^{AB}{}_{C} \bar\psi\Gamma^{C}\psi$ is subleading order, hence it does not contribute to the equations of motion.


\section{Linearized Boundary Equations}

We refer to sec.~\ref{sezFG} to compute the Navier-Stokes equations dual to Einstein equations, for a boosted and dilatate BTZ.
However our method is slightly different: our fermionic degrees of freedom induce a non--zero torsion that must be taken into account to
verify Einstein's equations. Moreover, we derive a new set of equations of motion which emerges from the gravitino field equation.

Technically for computing the Riemann tensor we use the spin connection formalism:
\begin{equation}
R^{ab} = \dd \omega^{ab} + \omega^{a}_{\phantom{a}c} \wedge \omega^{cb}
\ .
\end{equation}
In the form language it is easy to check that -- working at first order and expanding $\hat b$ around $1$ (no dilatation) and $w$ around $0$ (no boost) -- the boosted metric together with the boosted wig, satisfies (\ref{EoMs1}).

As explained in sec.~\ref{generalization} when we promote the parameters to local functions of the boundary coordinates the obtained metric is not a solution of the equations of motion anymore.
In order to reconstruct a solution, we must constrain the parameters to fulfill some equations which represent the equations of motion for the boundary fluid and we also need to add corrections to the metric.
Consequently, also the parameters must be modified accordingly.\footnote{The interested reader shall refer to \cite{Bhattacharyya:2008jc,Rangamani:2009xk} for further details.}
Since we work in a perturbative procedure, the metric is corrected order by order in the derivative expansion:
\begin{align}
	g \rightarrow g^{(0)} + g^{(1)} + \dots
	\ ,
	\label{gcorrection}
\end{align}
where $g^{(0)}$ represents the deformed metric and $g^{(i)}$ for $i>0$ are the metric corrections at the order $i$ in boundary derivatives.
In the following we limit our discussion at first order, namely we consider only $g^{(1)}$ correction.

As a warming--up exercise, we compute the NS equations from the constraint equations derived from the metric variation due to the $AdS_3$ isometries acting on the global BTZ black hole metric
\begin{equation}
\delta g = \mathcal{L}_{\xi}\left(g_{BH}\right)
\ ,
\end{equation}
where $g_{BH}$ is the BTZ metric (\ref{BTZ1}) and $\xi$ are defined in (\ref{KVglobal}). We observe that all isometries are broken, except the ones generated by $e_0$ and $f_0$.

We now proceed as follows. First of all, we promote all Killing vectors parameters to local functions of the boundary coordinates ($t$ and $\phi$); then we check Einstein's equations for the metric
\begin{equation}
	g = g_{BH} + \delta g + g^{(1)}
\end{equation}
which, as expected, are not satisfied. Yet, imposing them yields the following equations for the functions $b_0,d_0\ldots$ expanding near $t=\phi =0$ we get:
\begin{align}
& J_0 \left[\partial_{\phi} \left(b_0 + d_0\right) + \partial_t \left(b_0 - d_0\right)\right] - 2 \left(1 + M_0 \right) \partial_t \left(b_0 + d_0\right) = 0 \ , \nonumber \\
& J_0 \left[\partial_{t} \left(b_0 + d_0\right) + \partial_{\phi} \left(b_0 - d_0\right)\right] - 2 \left(1 + M_0 \right) \partial_{\phi} \left(b_0 + d_0\right) = 0 \ .
\label{NS1}
\end{align}
Note that these equations are computed in the global $AdS_3$; for other choices of neighborhoods, for example $t= \phi = \pi/2$, similar equations for the other parameters are obtained.
These are the Navier-Stokes equations derived by the global metric. As expected in the empty $AdS_3$ limit $J_0 \rightarrow 0, M_0 \rightarrow -1$ they are satisfied identically.

For what concerns the dynamical equations, the $3$--dimensional case is slightly different from higher dimensional cases. In fact, once the constraint equations are satisfied, no further corrections are needed and Einstein's equations are satisfied up to the first order in the derivative expansion.
Therefore $g^{(1)}$ can be set to zero. This is an important result since it implies that we are dealing with a perfect fluid with no dissipative corrections (contrary to \cite{Bhattacharyya:2008jc}, where the non-vanishing first order corrections corresponded to the shear tensor) and with second order, non--dissipative transport coefficients.

%

\subsection{Corrected NS Equations}

Having added fermionic fields to our scheme, the Navier-Stokes equations are now dual to the equations of motion derived from the $\mathcal{N}=2$, $D=3$ $AdS-$supergravity action (\ref{action1}).\footnote{Note that $\mathcal{N}=2$ supergravity Killing spinors do not suffer the problem pointed out by Gibbons in \cite{Gibbons:1981ux}. In fact, their behavior is stable even in the large $r$ limit, in contrast with in $\mathcal{N}=1$ theories.}

Once the fermionic bilinears are taken into account, imposing equations of motion (\ref{EoMs1}) and taking the large $r$ limit, we find:
\begin{align}\label{NSwig1}
&M_0\left[\partial_x b + \partial_t w - \frac{1}{16}\left(\partial_x +\partial_t\right)\bB_2\right]=0\ ,
\nonumber\\
&M_0\left[\partial_t  b + \partial_x w - \frac{1}{16}\left(\partial_x  +\partial_t\right)\bB_2\right]=0\ .
\end{align}
These are the Navier-Stokes equations for the Poincar\'e patch (cfr. (\ref{NS1})). Note that in this case they are identically satisfied if $M_0$ is set to zero.

Remarkably, as in the case of BTZ in global coordinates without fermionic wig, all the equations of motion lead to (\ref{NSwig1}).
Therefore the first order metric correction $g^{(1)}$ can be set to zero.
As in the previous section, this means that the conformal fluid on the boundary have non -- dissipative first order corrections, as expected for a two dimensional  conformal fluid.

\subsection{Dirac-type equation}

This is a truly original study, since nobody takes the deformation of Rarita-Schwinger equation in to account in the present framework. Therefore we explain carefully the technique adopted.

We proceed as follows: first we consider the solution of $\mathcal{D} \psi = 0$  where the spinor $\zetab$ is a constant field (zero mode) and we promote it to be local upon boundary coordinates. This implies that we can rewrite the gravitino field proportional to the fermionic field itself:
\begin{align}
	\psi_{M} = \Upsilon_{M} \zetab
	\ ,
	\label{RS1}
\end{align}
where $\Upsilon_{M}$ is a generic $2\times2$ matrix which depends on the coordinates $t,r,x$ (and in principles also on the bilinears) that can be decomposed on the basis of the Pauli matrices (\ref{gammaBTZ}) and the identity.
Notice that since $\psi_t = - \psi_x$ we have
\begin{align}
	\Upsilon_{x} = - \Upsilon_{t}
	\ .
	\label{RS2}
\end{align}
Consequently, the equations of motion read
\begin{align}
	\varepsilon^{MNR} {\cal D}_{N} \left(\Upsilon_{R} \zetab \right) = 0
	\ ,
	\label{EoMgrav}
\end{align}
By promoting $\zetab$ to be local on the boundary coordinates $t,x$ and using the equations of motion for the constant $\zetab$,
eqs.~(\ref{EoMgrav}) become
\begin{align}
	\varepsilon^{MNR} \Upsilon_{R} \partial_{N} \zetab (t,x) = 0
	\ .
	\label{EoMgrav2}
\end{align}
Being $\partial_{N} \zetab (t,x)$ a spinor, it can be written as a linear transformation of the spinor $\zetab(t,x)$ itself
\begin{align}
	\partial_{N} \zetab (t,x) = \Theta_{N} \zetab (t,x)
	\ ,
	\label{RS3}
\end{align}
where $\Theta_{N}$ is a $2\times2$ matrix. Notice that since $\zetab$ is not a function of the radial coordinate $r$, we have
\begin{align}
	&\Theta_{R} = \left\{\Theta_t (t,x),0,\Theta_x (t,x) \right\}
	\ .
	\label{RS5}
\end{align}
Eqs.~(\ref{EoMgrav2}) then reduce to
\begin{align}
	\varepsilon^{MNR} \Upsilon_{R} \Theta_{N} \zetab (t,x) = 0
	\ ,
	\label{EoMgrav3}
\end{align}
which in components read
\begin{align}
	\left( \Upsilon_{r} \Theta_{x} - \Upsilon_{x} \Theta_{r}  \right) \zetab = 0
	\ , \qquad
	\left( \Upsilon_{t} \Theta_{x} - \Upsilon_{x} \Theta_{t} \right) \zetab = 0
	\ , \qquad
	\left( \Upsilon_{r} \Theta_{t} - \Upsilon_{t} \Theta_{r} \right) \zetab = 0
	\ .
	\label{RS4}
\end{align}
Using (\ref{RS2}) and (\ref{RS5}) we have
\begin{align}
	&\Theta_{x} = - \Theta_{t}\ ,&
	\Upsilon_{r} \Theta_{t} \zetab = 0 \ .
	\label{RS6}
\end{align}
Thus, there is only one independent  matrix $\Theta$:
\begin{align}
	\Theta_{t} \equiv
	\Theta =
	\left(
	\begin{array}{cc}
		\theta_{11} & \theta_{12} \\
		\theta_{21} & \theta_{22} \\
	\end{array}
	\right)
	\ .
	\label{RStheta}
\end{align}
Considering only the first order gravitino (\ref{grav}), after a straightforward computation  at leading order in $r\rightarrow\infty$ expansion we get
\begin{align}
	\Upsilon_{r} \sim
	\frac{1}{4} (J - M)
	r^{-5/2}
	\left(
	\begin{array}{cc}
	- 1/r & 0 \\
	(t - x) & 1 \\
	\end{array}
	\right)
	\ ,
	\label{RS7}
\end{align}
In $r\rightarrow\infty$ asymptotic limit the matrix $\Upsilon_{r}$ is no longer invertible, therefore the second equation of (\ref{RS6}) in that limit becomes:
\begin{align}
	\left[
	\theta_{21} + \theta_{11} (t - x)
	\right] \zeta_{1}	
	+
	\left[
	\theta_{22} + \theta_{12} (t - x)
	\right] \zeta_{2}
	=
	0
	\ ,
	\label{RS8}
\end{align}
where $\zeta_{1}$ and $\zeta_{2}$ are the Grassmann components of $\zetab$.
Solving for generic $\zeta_{1},\zeta_{2}$, we obtain
\begin{align}
	&\theta_{21}  = - (t-x) \theta_{11} \ ,&
	&\theta_{22}  = - (t-x) \theta_{12} \ .&
	\label{RS9}
\end{align}
Summing up the results, eqs.~(\ref{RS3}) read
\begin{align}
	& \partial_{t} \zeta_{1} = \theta_{11} \zeta_{1} + \theta_{12} \zeta_2 \ ,
	& \partial_{t} \zeta_{2} = -( t - x ) \left( \theta_{11} \zeta_{1} + \theta_{12} \zeta_2  \right)\ ,&
	\nonumber \\
	& \partial_{x} \zeta_{1} = - \theta_{11} \zeta_{1} - \theta_{12} \zeta_2 \ ,
	& \partial_{x} \zeta_{2} = + ( t - x ) \left( \theta_{11} \zeta_{1} + \theta_{12} \zeta_2  \right)\ ,&
	\label{RS10}
\end{align}
Notice that this implies
\begin{align}
	\left( \partial_{t} + \partial_{x}  \right) \zetab = 0\ .
	\label{derBil0}
\end{align}
From the definitions (\ref{realBil0}), we compute the bilinears derivatives
\begin{align}
	\partial_{t} \bB_{0}
	& =
	\bB_{0} \left[ \textrm{Re}\theta_{11} - ( t- x ) \textrm{Re}\theta_{12} \right]
	+
	\bB_{1} \left[ \textrm{Re}\theta_{12} - ( t- x ) \textrm{Re}\theta_{11} \right]
	+
	\nonumber\\
	&
	+
	\bB_{2} \left[ \textrm{Im}\theta_{12} + ( t- x ) \textrm{Im}\theta_{11} \right]
	+
	\bB_{3} \left[ \textrm{Re}\theta_{11} + ( t- x ) \textrm{Re}\theta_{12} \right]
	\ ,
	\label{derBil1}
\end{align}
\begin{align}
	\partial_{t} \bB_{1}
	& =
	\bB_{0} \left[ \textrm{Re}\theta_{12} - ( t- x ) \textrm{Re}\theta_{11} \right]
	+
	\bB_{1} \left[ \textrm{Re}\theta_{11} - ( t- x ) \textrm{Re}\theta_{12} \right]
	+
	\nonumber\\
	&
	-
	\bB_{2} \left[ \textrm{Im}\theta_{11} + ( t- x ) \textrm{Im}\theta_{12} \right]
	-
	\bB_{3} \left[ \textrm{Re}\theta_{12} + ( t- x ) \textrm{Re}\theta_{11} \right]
	\ ,
	\label{derBil2}
\end{align}
\begin{align}
	\partial_{t} \bB_{2}
	& =
	\bB_{0} \left[ \textrm{Im}\theta_{12} + ( t- x ) \textrm{Im}\theta_{11} \right]
	+
	\bB_{1} \left[ \textrm{Im}\theta_{11} + ( t- x ) \textrm{Im}\theta_{12} \right]
	+
	\nonumber\\
	&
	+
	\bB_{2} \left[ \textrm{Re}\theta_{11} - ( t- x ) \textrm{Re}\theta_{12} \right]
	-
	\bB_{3} \left[ \textrm{Im}\theta_{12} - ( t- x ) \textrm{Im}\theta_{11} \right]
	\ ,
	\label{derBil3}
\end{align}
\begin{align}
	\partial_{t} \bB_{3}
	& =
	\bB_{0} \left[ \textrm{Re}\theta_{11} + ( t- x ) \textrm{Re}\theta_{12} \right]
	+
	\bB_{1} \left[ \textrm{Re}\theta_{12} + ( t- x ) \textrm{Re}\theta_{11} \right]
	+
	\nonumber\\
	&
	+
	\bB_{2} \left[ \textrm{Im}\theta_{12} - ( t- x ) \textrm{Im}\theta_{11} \right]
	+
	\bB_{3} \left[ \textrm{Re}\theta_{11} - ( t- x ) \textrm{Re}\theta_{12} \right]
	\ .
	\label{derBil4}
\end{align}
where
\begin{align}
	& \textrm{Re}\theta = \frac{1}{2} ( \theta + \theta^{*} ) \ ,
	&
	 \textrm{Im}\theta = \frac{1}{2 i} ( \theta - \theta^{*} )
	\ .&
	\label{derBil31}
\end{align}
The $x$--derivative of bilinears satisfies
\begin{align}
	\partial_{x} \bB_{i} = - \partial_{t} \bB_{i} \ .
	\label{derBilx}
\end{align}
The last equation has a strong implication on the linearized Navier-Stokes equations (\ref{NSwig1}), indeed this implies
that the last term there vanishes. Therefore, the two sets of equations are decoupled at the linearized level. This yields the possibility of a
clear separation of the bosonic and fermionic degrees of freedom. It would be very interesting to study the complete non-linearized version
of these equations.


\section{Physics at the Horizon and at the Boundary}

\subsection{Energy--Momentum Tensor dual to BTZ black hole}
Using the Brown--York and Kraus--Balasubramanian technique \cite{Brown:1992br,Balasubramanian:1999re} described in \ref{secBY}
we compute the boundary energy--momentum tensor $T_0^{\mu\nu}$ for the boosted metric in Poincar\'e patch. 
Notice that Greek indices labels the boundary coordinates $t,x$.

Defining the normal vector $n^{M}$ to constant $r-$slice we can compute the extrinsic curvature and then
the boundary energy--momentum tensor.
This turns out to be
\begin{align}
	T_0^{\mu\nu}
	=
	\frac{1}{2}\left(
	\begin{array}{cc}
	M & - J \\
	- J & M
	\end{array}
	\right)\ .
	\label{Tmunu1}
\end{align}
In order to get the usual form of perfect fluid energy--momentum tensor
\begin{align}
	T_0^{\mu\nu} = \eta^{\mu\nu} + 2 u^{\mu} u^{\nu}
	\ ,
	\label{Tmunu2}
\end{align}
it is sufficient to consider the case $J_{0}=0$.
Indeed the metric will acquire angular momentum due to the Lorentz transformation
as shown in (\ref{boost2}).
The fluid boundary energy--momentum tensor dual to the metric (\ref{BTZ1}) with $J_{0}=0, M_{0}\neq 0$
is the standard one for the perfect fluid in the rest frame.

Then we perform the boost transformation which switches on an angular momentum and modifies the mass parameter
\begin{align}
	&
	M =  \frac{1 + w^{2}}{1 - w^{2}} M_{0}
	\ ,
	&
	J =   \frac{2 w}{1 - w^{2}} M_{0}
	\ .
	&
	\label{boostNoJ}
\end{align}
Notice that our results are in perfect agreement with \cite{Martinez:1999qi} since we obtain the extremality condition
once we set $\left|w \right|=1$.
Starting from the boosted metric, \textit{i.e.} the metric (\ref{BTZ1}) in which $M_0$ and $J_0$ has been replaced with eqs. (\ref{boost2}) and $r$ with (\ref{boost3}), the computation of $T_0^{\mu\nu}$ yields
\begin{align}
	T_0^{\mu\nu}
	=
	M_0 \gamma \left(
	\begin{array}{cc}
	1  & \beta \\
	\beta & 1
	\end{array}
	\right)\ .
	\label{Tmunu3}
\end{align}
where $\gamma$ and $\beta$ are defined in (\ref{boost4}).
Setting
\begin{align}
	&
	u^{0} = \frac{1}{\sqrt{1 - w^{2}}}
	\ ,
	&
	u^{1} = -\frac{w}{ \sqrt{1 - w^{2}}}
	\ ,
	&
	\label{Tmunu4}
\end{align}
we find precisely (\ref{Tmunu2}) where $u^{\mu}$ is the normalized fluid velocity ({\it i.e.} $u^{\mu}u_{\mu}=-1$).

It is now straightforward to recover the variation of $T_0^{\mu\nu}$ due to a dilatation. In fact, being proportional to $M_0$, it scales as
\begin{equation}
	T_0^{\mu\nu} \rightarrow T^{\mu\nu} = \frac{1}{\hat{b}^2}T_0^{\mu\nu}
\ .
\end{equation}

Using the results obtained in \cite{Gentile:2012tu} we compute the Brown-York energy--momentum tensor dual to the BTZ black hole with fermionic wig. Note that this is an exact result since the series in the fermionic bilinears naturally truncates at second order:
\begin{align}
\label{TmunuNew0}
T^{\mu\nu} &  = \frac{M_0}{2 \hat{b}^2} \left(1+\bH\right) \left(\eta^{\mu\nu} + 2 u^{\mu} u^{\nu}\right) - \frac{M_0}{2 \hat{b}^2} ~ \bH ~ \varepsilon^{\mu\sigma}\left(\delta^{\nu}_{\phantom{\nu}\sigma} + 2 u^{\nu} u_{\sigma}\right) \ ,
\end{align}

Eq.~(\ref{TmunuNew0}) can be recast in the following form
\begin{align}
T^{\mu\nu} &  = \frac{M_0}{2 \hat{b}^2} \left(1+\bH\right) \left(\eta^{\mu\nu} + 2 u^{\mu} u^{\nu}\right) - \frac{M_0}{\hat{b}^2} ~\bH~ \varepsilon^{\left(\mu|\sigma\right.} u^{\left.\nu\right)} u_{\sigma} \ .
\label{TmunuNew1}
\end{align}
By assuming that the bilinears contained in $\bH$ are local quantities, the equations for the conservation of the energy-momentum tensor $T^{\mu\nu}$ lead to differential
equations involving also the bilinears.
At linearized level these equations reduce to eqs.~(\ref{NSwig1}).

\subsection{Redefining the Velocity}

At first glance, equation (\ref{TmunuNew1}) reveals a parity-violating term. This term has been
studied in \cite{Dubovsky:2011sk}, where anomalous fluid are considered, and they concluded that the most general form for it is
\begin{equation}
\label{deltaT}
\Delta T^{\mu\nu} = - \left[\mu^2 C + \alpha \left(T^2 + \frac{2 n T \mu}{s}\right)\right] u^{\left( \mu \right.} \tilde{u}^{\left. \nu\right)}
\ ,
\end{equation}
where $\tilde{u}^{\mu} = \varepsilon^{\mu\nu} u_{\nu}$, $C$ is the coefficient of the anomaly, $T$ is the temperature, $n$ is the fluid charge density, $s$ the entropy density, $\mu$ is the chemical potential and $\alpha$ an arbitrary integration constant.

Nevertheless, as pointed out by \cite{Jain:2012rh}, the anomaly require the following background metric and gauge field
\begin{align}
\dd s^2 &= -e^{2 \sigma} \left(\dd t + a_1 \dd x\right)^2 + g_{11} \dd x^2 \ , \\ \nonumber
A &= A_0 \dd t + A_1 \dd x \ .
\label{metric1plus1}
\end{align}
where $\sigma, a_1$ and $g_{11}$ are functions of $x,t$.
In the present case we have
\begin{align}
\dd s^2 =& -\dd t^2 +  \dd x^2 \ ,\\
A =& -\frac{1}{32}\left(M+ J\right) \left[2 \bB_1 \left(t - x\right) +
    \bB_3 \left(-1 + t^2 - 2 t x + x^2\right) \right. + \nonumber \\
    & \left. +\bB_0 \left(1 + t^2 - 2 t x + x^2\right)\right] \left(\dd t - \dd x\right) \ ,
\end{align}
and, comparing with (\ref{metric1plus1}) we get
\begin{align}
& \sigma = 0\ ,&
& a_1 = 0\ ,&
& g_{11} = 0\ , &
& F = \dd A = 0 \ .
\end{align}
Using the Poincar\'e lemma, we conclude that $A = \dd \lambda$ globally, therefore $A$ is a pure gauge and our theory is not anomalous.

Thus $C = 0$ leads to
\begin{equation}
\label{deltaTfinal}
\Delta T^{\mu\nu} = 2 \alpha T^2  u^{\left( \mu \right.} \tilde{u}^{\left. \nu\right)} \ .
\end{equation}

As explained in \cite{Dubovsky:2011sk}, in absence of an anomaly there is the freedom to add this term and it corresponds to a choice of the entropy current. In fact, it is possible to recast the energy--momentum tensor (\ref{TmunuNew1}) in the perfect fluid form
\begin{align}
	T^{\mu\nu}
	& =
	(1 + \frac{1}{8} \bB_{2} + \frac{1}{384} \bB_{2}^2) \frac{M_{0}}{2 \hat{b}^{2}}
	\left(
	2 U^{\mu} U^{\nu} + \eta^{\mu\nu}
	\right)
	\ ,
	\label{TmunuRecasted1}
\end{align}
through a redefinition of the velocity field
\begin{align}
	u^{\mu} \rightarrow U^{\mu}
	=
	\left( 1 + \frac{1}{512} \bB_{2}^{2} \right) u^{\mu}
	-
	\frac{1}{16} \left( \bB_{2} - \frac{1}{24} \bB_{2}^{2} \right) \tilde u^\mu \ .
	\label{NewU1}
\end{align}
Note that $U^{\mu}$ is correctly normalized to $-1$. Recalling the conformal thermodynamics identities \cite{Bhattacharyya:2008mz}
\begin{align}
& b = \frac{1}{2 \pi T} \ ,
&p = \rho = \frac{M_0}{2 b^2} = 2 \pi^2 T^2
\ ,
\end{align}
we immediately see that the temperature gets a shift due to the presence of bilinears
\begin{equation}
T' = T \left(1 + \frac{\langle\bB_2\rangle}{16} - \frac{\langle\bB_2^2\rangle}{1536}\right)\ ,
\end{equation}
where the brackets denotes the v.e.v. of the bilinears.

We have to make one important remark: the expression of the temperature in terms of the bilinear acquires a numerical value whenever the bilinear 
have a v.e.v. computed by path integral means (we have to recall that Grassmann numbers pertain only to the quantum realm). 
The procedure is similar to what is usually done in the case of solitons in gauge theories and supergravity 
\cite{Amati:1988ft,Konishi:1988mb} and the gravitinos condensate leads to non-vanishing v.e.v. of the bilinears interested in the previous formula. 
In the case of BTZ black hole, the gravitational action evaluated on the solution with the wig has never been computed and it will be presented elsewhere. 

\subsection{Horizon and Entropy}

In the following we present the entropy computed from the wig of the BTZ in global coordinates.
By direct computation we notice that the event horizon radius
\begin{align}
	r_{\pm}^{2}
	& =
	\frac{1}{2} \left( M \pm \sqrt{M^{2} - J^{2}}  \right) \ .	
	\label{horizon}
\end{align}
is not modified by the presence of the fermionic wig.
We can compute the entropy from Bekenstein--Hawking formula
\begin{align}
	S & = 	\frac{1}{4}A_{H} \ ,
	\label{BekensteinHawking}
\end{align}
where the area of the horizon reads
\begin{align}
	A_{H}
	=
	\int_{0}^{2\pi} \sqrt{\mathbf{g}_{\phi\phi}(r_{+})}
	\dd \phi \ ,
	\label{Area}
\end{align}
and is computed using the complete metric with the wig.
We obtain the following result
\begin{align}
	S & =
	\frac{\pi}{2}
	\left[
		r_{+}
		+
		\langle\bC_{2}\rangle \frac{J}{16 r_{+}}
		+
		\langle\bC_{2}^{2}\rangle \frac{1}{512 r^{3}_{+}}
		\left(
		J^{2} + 2 r^{2}_{+} \left( J - 2 - 2M \right)
		\right)
		\right]
	\ ,
	\label{Entropy}
\end{align}
where we take the v.e.v. for the bilinears. As can be seen the entropy of the black hole is modified by the presence of the wig confirming that
we are studying a new solution of the theory where the fermions play a fundamental r\^ole. Setting $J =0$ the
first order correction vanishes. This could also have been checked by a simple infinitesimal calculation. Nonetheless, the
second order corrections do not vanish. In particular for vanishing angular momentum the third term in the above equation
becomes proportional to $M+1$ which vanishes for $M=-1$, namely global anti-de Sitter.

By setting $J=M$ in the case of extremal solution, we find the simplified formula
\begin{align}
	S & =
	\frac{\pi}{2}  \sqrt{2 M}
	\left(
		\frac{1}{2}
		+
		\frac{1}{16} \langle\bC_{2}\rangle
		+
		\frac{M-2}{128 M} \langle\bC^{2}_{2}\rangle
		\right)
		\ ,
	\label{Entropy2}
\end{align}
showing that also in the case of extremal black hole the entropy is modified.


\subsection{Conserved Charges for Global Wig}

Here we compute the conserved charges associated with the isometries of the BTZ black hole.
We use holographic technique based on the boundary energy momentum tensor $T_{\mu\nu}$ \cite{Gentile:2011jt,Brown:1992br,Balasubramanian:1999re,Henneaux:1999ib,Henneaux:1984ei}.
To perform the computation we cast the boundary metric $\gamma_{\mu\nu}$ in ADM--like form
\begin{align}
	\gamma_{\mu\nu} \dd x^{\mu} \dd x^{\nu}
	& =
	- N_{\Sigma}^{2} \dd t^{2}
	+
	\sigma
	( 	\dd \phi + N^{\phi}_{\Sigma} \dd t )^{2}
	\ ,
	\label{ADM1}
\end{align}
where $\Sigma$ is the $2$--dimensional surface at constant time and the integration is over a circle at spacelike infinity.
The conserved charges associated to the Killing vectors $\xi$ are defined as
\begin{align}
	Q_{\xi} & =
	\lim_{r\rightarrow\infty}\int_{V} \dd x \sqrt{\sigma} u^{\mu} T_{\mu\nu} \xi^{\nu}
	\label{charges1}
\end{align}
where  $u^{\mu}=N_{\Sigma}^{-1}\delta^{\mu t}$ is the timelike unit vector normal to $\Sigma$.

In the present case, the global wig does not depend on $t$ and $\phi$. Thus, the two resulting Killing vectors are 
\begin{align}
	\xi_{1}^{\mu} &= \delta^{\mu t} 
	\ ,
	&\xi_{2}^{\mu} = - \delta^{\mu\phi} 
	\ .&
	\label{charges2}
\end{align}
The associated charges are respectively the mass $M_{tot}$ and the angular momentum $J_{tot}$. After a short computation we find
\begin{align}
	M_{tot} & =
	M 
	+
	\frac{1}{8}\left( 1 + M + J \right)
	\left( \langle\bC_{2}\rangle + \frac{1}{16} \langle\bC_{2}^{2}\rangle \right)
	\ , \nonumber \\
	J_{tot} & =
	J 
	+
	\frac{1}{8}\left( 1 + M + J \right)
	\left( \langle\bC_{2}\rangle + \frac{1}{16} \langle\bC_{2}^{2}\rangle \right)
	\ .
	\label{MandJ}
\end{align}
As in the previous section, the charges ought to be numbers with a given value, then in formula (\ref{MandJ})  the bilinear $\bC_2$  are substituted with its v.e.v. $\langle\bC_{2}\rangle$.
In that way the mass and the angular momentum $M_{tot}$ and $J_{tot}$ make sense. A vacuum with non--vanishing v.e.v. for bilinears might explicitly break supersymmetry, leading to a modified mass and angular momentum which depend on them. 

Note that $M_{tot}-J_{tot}=M-J$ and the fermionic corrections do not affect the difference between mass and angular momentum. Thus, if the extremality condition is imposed we expect that it is not lifted.

From action (\ref{action0}) we derive the conserved electric charge $q$
\begin{align}
	q & = 
	\lim_{r\rightarrow\infty}
	\frac{1}{2} \oint \dd \phi \sqrt{\sigma} N_{\sigma}\, \varepsilon^{t M N} i  \bar\psi_{M} \psi_{N}
	\ .
	\label{chargeQ}
\end{align}
Using the equations of motion (\ref{EoMs1}) we can rewrite it in terms of the field strength of gauge field $A$. The computation shows that the leading term of the integral in the large $r$ expansion is $O\left( \frac{1}{r} \right)$, thus in the $r\rightarrow\infty$ limit $q$ vanishes.

The supercharge $\mathcal{Q}$ is connected to the presence of Killing spinors \cite{Belyaev:2007bg,Belyaev:2008ex}. As we have already pointed out, the present work deals with non--extremal BTZ black hole and therefore supersymmetry is totally broken. As a consequence, no Killing spinor exists and thus there is no conserved supercharge.\\

%
%
%


\chapter{Fermionic Wigs for Higher Dimensions}
\label{chWig}

In this section we compute the complete wig for $AdS_5$-Schwarzschild black hole solution of  $\mathcal{N}=2$, $D=5$ supergravity. From this result we derive the boundary energy--momentum tensor.
Hence, we repeat the procedure for the $4$--dimensional case.

We start from a Schwarzschild--type solution, breaking
all supersymmetries. The metric depends upon the coordinate $r$
measuring the distance between the center of $AdS_5$ space and the boundary. 
To choose a flat $D=4$ boundary we consider the metric in Poincar\'e patch.
Notice that Lorentz symmetry is manifestly broken by our solutions since the time is treated differently
from $3d$ space coordinates. 
With the factorization of the metric into a $2d$ space-time $(r,t)$ and
$3d$ space $(x^i)$, we can factorize the spinors into corresponding irreducible representations.
We compute the $AdS_5$ Killing spinors and we see that there are two independent choices
which are relevant for our study. Then, we compute the variation of the gravitino fields under the supersymmetry
where the parameters are replaced by the Killing spinors. That produces the first term of the
fermionic expansion of the gravitino solutions to the Rarita-Schwinger equation of motion.
The next step is
to compute the second variation of the metric in terms of fermionic bilinears $(\lambda, \bN, \bK_i)$.
That is achieved by computing the second supersymmetry variation of the metric.
At this stage one can check whether the Einstein equations are indeed satisfied.
Already at this step, the usage of Fierz identities to rearrange the bilinears is essential to reduce all possible
terms.
The iteration proceeds until the number of independent fermions truncates the series.  In the process, the
gauge field (the graviphoton), which has been set to zero from the beginning, is generated and its field is proportional to the fermion bilinears.
We check also the Maxwell equations order-by-order.

The computation of the Killing spinors reveals that there are essentially two structures to be taken into account (in the
text we denote those contributions as $\eta_0$ and $\eta_1$). In the first case the complete solution obtained by resumming
all fermionic contributions is rather simple since the dependence upon the boundary coordinates is very mild. On the contrary
the computations of the complete metric in the case of $\eta_1$ is rather length since all possible structures are eventually generated.
In addition, the two structures, at a certain point, start to mix and therefore a long computation has to be
done. 
This is due to the fact that by breaking Lorentz invariance from the beginning all terms of the spin connection,
of the vielbeins and of the gauge fields are generated.
Therefore we cannot use covariance under Lorentz transformation to cast our computation in an elegant and compact form and, generically,
all components are different from zero.
Technically, in order to re-sum all contributions we compute the full solution using \verb Mathematica \textsuperscript{\textregistered}.
The result is provided in a form which is still difficult to read (the electronic notebook with solutions is provided
as ancillary files of the preprint publication of \cite{Gentile:2012jm}).
Nevertheless, we make some remarks regarding the results and we give the explicit formulas for the simplest cases.

We perform the computation of the wig and the boundary energy--momentum tensor also for 
$D=4$ model, since it is very similar to the $D=5$ one.

Our construction has different purposes. First of all, we will use the present results
to derive the complete non-linear Navier-Stokes equations with fermionic contributions \cite{Gentile:2011jt}.
That would be the natural final aim of the present work, but since the results  are independent from that, we decided
to present the derivation of NS equations in a separate paper. Second, the natural question is whether
the same analysis can be done also in the case of BPS solutions. For that we refer to the first
step given in \cite{Behrndt:1998jd} and we will complete their constructions by our
 algorithm. 

Another question is the case of $D=4$. In that
case a complete explicit solution is attainable and we will publish this result elsewhere.

\section{Truncated $N=2$, $D=5$ Gauged Supergravity}

We provide some useful ingredients  for our computation based on papers \cite{D'Auria:1981kq,Gunaydin:1983bi,Gunaydin:1984ak,Ceresole:2000jd,Bergshoeff:2004kh,Behrndt:1998ns,Behrndt:1998jd}.
We consider the model $N=2$, $D=5$ gauged supergravity, but we truncate the spectrum in order to deal with the simplest solution in $AdS_{5}$ for the present paper.

\subsection{Action}

The $N=2$, $D=5$ gauged supergravity action was constructed in \cite{D'Auria:1981kq,Gunaydin:1983bi,Gunaydin:1984ak,Ceresole:2000jd,Bergshoeff:2004kh}, coupling the pure supergravity multiplet with vector and tensor multiplets. In this paper we consider a consistent truncation of that action, in order to deal with Schwarzshild solution in $AdS_{5}$. We consider the pure supergravity multiplet, formed by the vielbein $e_{M}^{A}$, two gravitini $\psi^i_{M}$ and the graviphoton $A^{0}_{M}$, and $N-1$ vector multiplets composed by vector fields $A_{M}^{\widetilde{I}}$, gauginos $\lambda^{i\,\widetilde{I}}$ and scalar fields $q^{\widetilde{I}}$.\footnote{Index $i$ labels the two spinor fields in symplectic--Majorana representation.}

To gauge the $U\left( 1 \right)$ subgroup of $SU\left( 2 \right)$ $R$--symmetry group, we consider a linear combination of vector fields $A_{M}^{\widetilde{I}}$ and graviphoton $A_{M}$: $A_{M}=V_{I} A^{I}_{M}$, where $\{V_{I}\}$ are a set of constants and index $I$ labels the graviphoton and the $N-1$ vector fields.  The gauging procedure introduces a potential in the action which depends on the scalar $q^{\widetilde{I}}$. In order to simplify this $AdS_{5}$ model we set the potential and the scalars to constant, and the gauginos to zero. The resulting action is then
{\allowdisplaybreaks
\begin{align}\label{final}
e^{-1}\mathcal{L}&=
\ft1{2} R(\omega) -\ft14a_{{I}{J}}
\widehat{F}^{{I}}_{M
N}\widehat{F}^{{J}M N}
-\,\ft1{2}\bar{\psi}_R \Gamma^{R M M}\mathcal{D}_M
\psi_N
+4 g^2 \vec{P}\cdot \vec{P}
\nonumber\\&
+\,\ft{1}{6\sqrt{6}} e^{-1} \varepsilon ^{MNLRS } {\cal C}_{I J K} A_M^I \left[ F_{N L }^J F_{R S }^K + f_{FG}{}^J A_N^F A_L ^G \left(- \ft12 g F_{R S }^K + \ft1{10} g^2 f_{H G}{}^K A_R ^H A_S^G \right)\right] \nonumber\\
&  -\, \ft{1}{8} e^{-1} \varepsilon^{M N L R S}
\Omega_{I'J'} t_{IK} {}^{I'} t_{FG}{}^{J'} A_{M}^I A_N^F A_L^G
\left(-\ft 12 g  F_{R S}^K + \ft{1}{10} g^2 f_{H G}{}^K A_R^H
A_S^G \right)\nonumber\\&
-\ft{\sqrt{6}}{16 }\,{ i}
h_{I}
F^{CD I}\bar{\psi}^A\Gamma_{ABCD}\psi^B
 +
 g\sqrt{\ft38}\,{ i}
P_{ij}\bar{\psi}_A^i\Gamma^{AB}\psi_B^j
 +\ft{1}{8}\bar{\psi}_A\Gamma_B\psi^B\bar{\psi}^A\Gamma_C\psi^C\nonumber\\&
-\,\ft{1}{16}\bar{\psi}_A\Gamma_B\psi_C\bar{\psi}^A\Gamma^C\psi^B
-\ft{1}{32}\bar{\psi}_A\Gamma_B\psi_C\bar{\psi}^A\Gamma^B\psi^C
+\ft{1}{32}\bar{\psi}_A\psi_B\bar{\psi}_C\Gamma^{ABCD}\psi_D
.
\end{align}
}
where $g$ is the $U(1)$ coupling constant. Indices $\{F,\dots,K\}$ are the special geometry ones, $\{L,M,N,\dots\}$ are the curved bulk indices and $\{A,\dots, D\}$ labels flat bulk directions. The quantities $\Omega_{IJ}$, $ {\cal C}_{I J K}$, $t_{IJ}{}^{K}$, $\vec{P}$, $h_{I}$ are related to special geometry (see \cite{Gunaydin:1984ak,Ceresole:2000jd,Bergshoeff:2004kh}). Notice that when the $i$ spinorial indices are omitted, northwest-southeast contraction is understood, e.g. $\bar{\psi}_C\psi_D = \bar{\psi}_C^{i}\psi_{i\, D} $.
We define the supercovariant field strengths $\widehat F^I_{AB}$ such that
\begin{eqnarray}
&&\widehat F^I_{AB} =F_{AB}^I-\bar{\psi}_{[A}\Gamma_{B]}\psi^I
+
\frac{\sqrt{6}}{4}\,i \bar{\psi}_A\psi_B
h^I
,\nonumber\\
&&
F_{MN}^I \equiv  2
\partial_{[M} A_{N]}^I + g f_{JK}{}^I A_M^J A_N^K
\ .
\end{eqnarray}
We define also $\vec P \equiv  h^I \vec P_I$.
The covariant derivative reads
\begin{eqnarray}
\mathcal{D}_M \psi_{N} ^i&=&\left(\partial_M +\ft14
\omega_M{}^{AB}\Gamma_{AB}\right)\psi_{N }^i-g A_M^IP_{I}{}^{ij}\psi_{N
j} .
\end{eqnarray}
This action admits the following $N=2$ supersymmetry:
\begin{eqnarray}
\delta  e_M{}^A &=& \ft12 \bar\e \Gamma^A \psi_M ,\nonumber\\
\delta \psi_M^i&=&D_\mu(\hat\omega)\e^i + \ft{i} {4\sqrt6}
h_{ I} {\widehat{F}}^{ I N R}
 ({\Gamma}_{MN R} - 4 g_{M N} {\Gamma}_R) \e^i  - \ft{1}{\sqrt6} i g P^{ij}\Gamma_M  \e_j
,\nonumber\\
\delta  A_M^I&=&- \ft{\sqrt6}{4 } \,i h^{ I} \bar\e \psi_M
  .
\end{eqnarray}
We also denoted
\begin{eqnarray}
D_M(\hat\omega) \e^i &=& {\mathcal D}_M(\hat\omega) \e^i  - g A_M^I P_I^{ij}\e_j,
\end{eqnarray}
where $\hat\omega$ indicates the spin connection defined through vielbein postulate, as we will see in the forthcoming sessions.

\subsection{Spinors Relations}

For our purpose, we find convenient to work with Dirac spinors instead of symplectic--Majorana.\footnote{Dirac spinors are also used in \cite{Behrndt:1998ns,Behrndt:1998jd} while symplectic--Majorana ones are present in \cite{D'Auria:1981kq,Gunaydin:1983bi,Gunaydin:1984ak,Ceresole:2000jd,Bergshoeff:2004kh}. }
Therefore we dedicate the present subsection to illustrate and remind the reader the translation table.

For $5$ dimensions SM spinors $\lambda^{i}$ with $i=\left\{ 1,2 \right\}$, the complex conjugate is defined through
\begin{align}
	(\lambda^{i})^{*} = &\  C \Gamma_{0} \lambda^{i}\ ,
	\label{SMspinor1}
\end{align}
the bar is the Majorana bar
\begin{align}
	\bar \lambda^{i} =&\ (\lambda^{i})^{T}\mathcal{C} \ ,
	\label{SMspinor2}
\end{align}
where $\mathcal{C}$ is the charge conjugation matrix satisfying
\begin{align}
	\mathcal{C}^{T}= -\mathcal{C}\ ,&
	&
	\mathcal{C}^{*}= -\mathcal{C}\ ,&
	&
	\mathcal{C}^{2}=\mathcal{C}^{\dagger}C=I\ ,&
	\nonumber\\
	&
	\left( \mathcal{C}\Gamma_{M} \right)^{T}=-C\Gamma_{M}\ ,&
	&
	\Gamma_{M}^{T}=\mathcal{C}\Gamma_{M}\mathcal{C}^{-1}\ .&
	\label{SMspinorC}
\end{align}
Thus, the following expressions are real
\begin{align}
	i \bar\lambda^{i}\psi_{i} \ ,
	\qquad\qquad
	\bar\lambda^{i}\Gamma_{M}\psi_{i} \ .
	\label{SMspinor3}
\end{align}
Notice that the index $i$ is raised and lowered by the antisymmetric tensor $\varepsilon_{ij}$.

For our purpose, we need Dirac spinors $\epsilon$ and the bar represents the Dirac adjoint
\begin{align}
	\bar\epsilon = \epsilon^{\dagger}\Gamma_{0}\ .
	\label{Dspinor1}
\end{align}
It is possible to construct one Dirac spinor from two SM: one has $\epsilon=\lambda_{1}+i \lambda_{2}$. For consistency then we have $\bar\epsilon=\bar\lambda_{1}-i\bar\lambda_{2}$.

Using the above relations we express the quantities (\ref{SMspinor3}) in terms of Dirac spinors
\begin{align}
	i \bar\lambda^{i}\psi_{i}
	=
	\textrm{Re}\left( \bar\epsilon \psi \right)\ ,
	\qquad\qquad
	\bar\lambda^{i}\Gamma_{M}\psi_{i}
	=
	\textrm{Re}\left( -i \bar\epsilon\Gamma_{M} \psi \right)\ ,
	\label{Dspinor2}
\end{align}
where $\textrm{Re}(x)$ denotes the real part of $x$.

\subsection{Susy Transformations}

The supersymmetry transformations~(\ref{AlgSusyTransfv1}) for $N=2$, $D=5$ gauged supergravity written with Dirac spinors are
\begin{align}
	\delta_{\epsilon} e_{M}^{A}
		= &
		-\frac{1}{2} {\textrm{Re}}\left( i \bar\epsilon \Gamma^{A} \delta\psi_{M} \right)
	\ ,
	\nonumber\\
	\delta_{\epsilon} g_{MN}
		= &
	-\frac{1}{2} {\textrm{Re}}\left( i  \bar\epsilon \Gamma_{\left( M \right.} \delta\psi_{\left. N \right)} \right)
	\ ,
	\nonumber\\
	\delta_{\epsilon} \psi_{M}
		= &
		\mathcal{D}_{M}\left( \hat\omega \right)\epsilon
	+
	\frac{i}{4\sqrt{6}} e^{a}_{M} h_{I} \hat F^{I\,BC}
	\left(
		\Gamma_{ABC}-4\eta_{AB}\Gamma_{C}
	\right) \epsilon
	\ ,
	\nonumber\\
	\delta_{\epsilon}A_{M}^{I}
		= &
	-\frac{\sqrt{6}}{4}{\textrm{Re}}
	\left(  \bar\epsilon \psi_{M}h^{I}  \right)
	\ ,
	\label{AlgSusyTransf}
\end{align}
where
\begin{align}
	\hat F_{AB}^{I}
	= &
	F_{AB}^{I}
	+
	\frac{\sqrt{6}}{4}\bar\psi_{\left[ A \right.} \psi_{\left. B \right]}
	h^{I}
	\ ,
	\nonumber\\
	\mathcal{D}_{M}\left( \hat\omega \right)
	= &
	D_{M}\left( \hat\omega \right)	
	-g A_{M}^{I}P_{I}
	\ ,
	\nonumber\\
	D_{M}\left( \hat\omega \right)
	= &
	\partial_{M} +\frac{1}{4} \hat\omega_{M}^{AB} \Gamma_{AB} -
	\frac{i}{\sqrt{6}}g P \Gamma_{M} 
	\ .
	\label{AlgDef1}
\end{align}
In order to compare this with the $AdS$ covariant derivative
\begin{equation}
	D_{M}\left( \hat\omega \right)
	=
	\partial_{M} +\frac{1}{4} \hat\omega_{M}^{AB} \Gamma_{AB}
	+\frac{1}{2} e_{M}^{A}\Gamma_{A}
	\ ,
	\label{AlgCovDerTipical}
\end{equation}
we set
\begin{align}
	g P
	= &
	\frac{i}{2}\sqrt{6}
	\ .
	\label{AlgDefKillPrep}
\end{align}
From the special geometry construction, $h^{I}$ satisfies
\begin{equation}
	h_{I}h^{I} = 1
	\ ,
	\label{AlgSpecGeom}
\end{equation}
then, in our particular case, where the gauge fields are generated only from susy transformation (\ref{AlgSusyTransf}) while the zero--order is zero, we define the gauge field as
\begin{equation}
	A_{M}^{I} = A_{M} h^{I}
	\ .
	\label{AlgRedefA}
\end{equation}
Doing so, all the indices $I$ and the quantity $h^{I}$ disappear from the equations. Moreover, using  eq.~(\ref{AlgDefKillPrep}), the $A$--part in the covariant derivative becomes
\begin{align}
	-g A_{M}^{I}P_{I}
	=
	-\frac{i}{2}\sqrt{6} \, A_{M}
	\ .
	\label{AlgNewCovDerPiece}
\end{align}
Finally, the simplified susy transformations now read
\begin{align}
	\delta_{\epsilon} e_{M}^{A}
		= &
	-\frac{1}{2} {\textrm{Re}}\left(   i  \bar\epsilon \Gamma^{A} \delta\psi_{M} \right)
	\ ,
	\nonumber\\
	\delta_{\epsilon} g_{MN}
		= &
	-\frac{1}{2} {\textrm{Re}}\left( i \bar\epsilon \Gamma_{\left( M \right.} \delta\psi_{\left.N \right)} \right)
	\ ,
	\nonumber\\
	\delta_{\epsilon} \psi_{M}
		= &
		\mathcal{D}_{M}\left( \hat\omega \right)\epsilon
	+
	\frac{i}{4\sqrt{6}} e^{A}_{M}  \hat F^{BC}
	\left(
		\Gamma_{ABC}-4\eta_{AB}\Gamma_{C}
	\right) \epsilon
	\ ,
	\nonumber\\
	\delta_{\epsilon}A_{M}
		= &
	-\frac{\sqrt{6}}{4}{\textrm{Re}}
	\left(  \bar\epsilon \psi_{M} \right)
	\ ,
	\label{AlgSusyTransfv1}
\end{align}
where
\begin{align}
	\hat F_{AB}
	= &
	F_{AB}
	+
	\frac{\sqrt{6}}{4}\bar\psi_{\left[ A \right.} \psi_{\left. B \right]}
	\ ,
	\nonumber\\
	\mathcal{D}_{\mu}\left( \hat\omega \right)
	= &
	D_{\mu}\left( \hat\omega \right)	
	-\frac{i}{2}\sqrt{6} \, A_{M}\,
	\ ,
	\nonumber\\
	D_{M}\left( \hat\omega \right)
	= &
	\partial_{M} +\frac{1}{4} \hat\omega_{M}^{AB} \Gamma_{AB}
	+\frac{1}{2} e_{M}^{A}\Gamma_{A}
	\ .
	\label{AlgDef1v1}
\end{align}

As last remark, notice that torsion is not zero:
\begin{equation}
	\dd e^{A} +\omega^{A}_{\phantom{A}B}\wedge e^{B}
	=
	\frac{i}{4} \bar\psi \Gamma^{A} \psi
	\ ,
	\label{AlgTorsion}
\end{equation}
then, the spin connection $\hat\omega$ is written in terms of both vielbein and gravitino bilinears. Moreover, the abelian field strength reads
\begin{align}
	F_{MN}
	= &
	D_{M} A_{N} - D_{N} A_{M}
	=
	\partial_{M} A_{N} - \partial_{N} A_{M}
	+i \frac{1}{4} \bar\psi_{\left[ M \right.}\Gamma^{A}\psi_{\left. N \right]} A_{A}
	\ .
	\label{AlgAbelianFmunu}
\end{align}
We are left with the vielbeins, the gauge field and the Rarita-Schwinger (RS) field, which form the $N=2$, $D=5$ pure supergravity. Now, we can truncate to the bosonic sector and we consider a Schwarzschild--type solution which is asymptotically $AdS$.
Of course there are also more intricated solutions with non--constant scalar fields or gauge fields, but we do not take these cases into account in the present work.

\subsection{Background Setup}

We choose a $AdS_{5}$ solution of pure Einstein gravity as background
\begin{equation}
\dd s^{2} =
- r^{2} \dd t^{2}
+ \frac{1}{r^{2}} \dd r^{2}
+ r^{2} \sum_{i = 1}^{3} \dd x_{i}^{2}
\ ,
\qquad
A_{M}=0
\ ,
\qquad
\psi_{M}=0
\ ,
\label{AdSmetric0}
\end{equation}
where the metric is given in the Poincar\'e patch.
Notice that in this initial set up the gauge field and the Rarita--Schwinger fields are set to zero \cite{Burrington:2004hf} and  $AdS_{5}$ radius is set to $1$. The associated non-zero vielbein components are
\begin{align}\label{AdSvielbein0}
	e^{0}_{t} 	
&=
	r\ ,
&
	e^{1}_{r} 	
&=
	\frac{1}{r}\ ,
&
	e^{a}_{i} 	
&=
	r\delta_{i}^{a}
\ ;
\end{align}
while the non-zero spin connection components are
\begin{align}\label{AdSspincon0}
	\omega^{01}_{t} 	
&=
	r\ ,
&
	\omega^{a1}_{i} 	
&=
	r\delta_{i}^{a}
\ .
\end{align}
Notice that we will use capital latin letters to indicate bulk directions (i.e. $M,N$ run from $0$ to $4$) leaving greek alphabet to boundary ones (i.e. $\mu,\nu$ run from $0$ to $3$) furthermore $\left\{ t,r,i \right\}$ are {\it curved} indices and $\left\{ 0,1,a \right\}$ represent {\it flat} ones.

In presence of a uncharged, irrotational black hole eq.~(\ref{AdSmetric0}) becomes
\begin{equation}
\dd s^{2} =
- \left( r^{2}+\frac{\mu}{r^{2}} \right) \dd t^{2}
+ \frac{1}{ r^{2}+\frac{\mu}{r^{2}} } \dd r^{2}
+ r^{2} \sum_{i = 1}^{3} \dd x_{i}^{2}
\ ,
\label{AdSmetricBH}
\end{equation}
in this case the non-zero vielbein components are
\begin{align}\label{AdSvielbeinBH}
	e^{0}_{t} 	
&=
	\sqrt{ r^{2}+\frac{\mu}{r^{2}}}\ ,
&
	e^{1}_{r} 	
&=
	\frac{1}{\sqrt{ r^{2}+\frac{\mu}{r^{2}}}}\ ,
&
	e^{a}_{i} 	
&=
	r\delta_{i}^{a}
\ ;
\end{align}
and the non-zero spin connection components are
\begin{align}\label{AdSspinconBH}
	\omega^{01}_{t}
&=
	r-\frac{\mu}{r^{3}}\ ,
&
	\omega^{a1}_{i} 	
&=
	\sqrt{ r^{2}+\frac{\mu}{r^{2}}}\,\delta_{i}^{a}
\ .
\end{align}

\section{Killing Spinors}

Here we compute $AdS$ Killing spinors. We found that there are two independent solutions.
These are obtained by first factorizing the Dirac spinors into a $2d$ spinor and a $3d$ spinor in their irreducible representations.

The Killing spinor equation for $AdS_{5}$ reads
\begin{equation}
\left(
\partial_{M}
+\frac{1}{4}\omega_{M}^{ab}\Gamma_{ab}
+
\frac{1}{2}e_{M}^{a}\Gamma_{a}
 \right)\epsilon = 0
\ .
\label{KSeq0}
\end{equation}
with $\Gamma_{ab}=\frac{1}{2}\left( \Gamma_{a}\Gamma_{b}-\Gamma_{b}\Gamma_{a} \right)$. In components we have
\begin{align}
	&
	\partial_{t}\epsilon + \frac{r}{2}\Gamma_{0}\left( \Gamma_{1}+\one \right)\epsilon=0\ ,
	\nonumber\\
	&
	\partial_{r}\epsilon+\frac{1}{2r}\Gamma_{1}\epsilon=0\ ,
	\nonumber\\
	&
	\partial_{i}\epsilon + \frac{r}{2} \Gamma_{i}\left( \Gamma_{1}+\one \right)\epsilon=0\ .
	\label{ksEQ1}
\end{align}
We can divide the $5$ dimensional space in two parts: $\left\{ t,r \right\}$ and $\left\{ x^{i} \right\}$ using the following gamma matrices parametrization
\begin{align}
	\Gamma_{0} & = i\sigma_{2}\otimes\hat\sigma_{0}
\ ,
&\Gamma_{1} & = \sigma_{1}\otimes\hat\sigma_{0}
\ ,
&\Gamma_{a} & = \sigma_{3}\otimes\hat\sigma_{a}
\ ,
\label{Gamma}
\end{align}
where $\sigma_{0}$ is the identity matrix in $2d$. Hatted matrices refer to ${x^{i}}$ space. In this way, the solution of eq.~(\ref{KSeq0}) is
\begin{eqnarray}
\epsilon
=
\left( \frac{1}{\sqrt{r}}- t \sqrt{r} \sigma_{3} \right) \varepsilon_{0}\otimes \eta_{1}
-
\sqrt{r}\sigma_{3}\varepsilon_{0}\otimes\eta_{2}
\ ,
\label{KSsolution1}
\end{eqnarray}
where
\begin{align}
\eta_{2} = &
x^{k} \hat\sigma_{k} \eta_{1}+\eta_{0}\ ,
\label{KSsolution11}
\end{align}
and $\eta_{1}\,,\,\eta_{0}$ are $2$--dimensional complex spinors (and so contain $8$ real dof's) while $\varepsilon_{0}$ is a real $2$--dimensional spinor with only one dof. The total number of degrees of freedom is then $1\times8$. The solution~(\ref{KSsolution1}) can also be written as
\begin{align}
	\epsilon
	=
	\frac{1}{\sqrt{r}}\sigma_{0}\otimes\hat\sigma_{0}\, \varepsilon_{0}\otimes\eta_{1}
	-\sqrt{r}\sigma_{3}\otimes \left( t\hat\sigma_{0}+x^{i}\hat\sigma_{i} \right)\, \varepsilon_{0}\otimes\eta_{1}
	-\sqrt{r}\sigma_{3}\otimes\hat\sigma_{0}\, \varepsilon_{0}\otimes\eta_{0}
	\ .
	\label{KSsolution2}
\end{align}
Notice that $\bar\epsilon\,\Gamma^{M}\epsilon$ reproduces the Killing vectors (\ref{KV0}) as expected.

\section{Algorithms}\label{ALG}

In this section we present the algorithm to compute the wig for the Schwarzschild--like black holes in $5$ dimensions.

\subsection{Gravitino}

Using definitions (\ref{AlgSusyTransfv1}), (\ref{AlgDef1v1}),  and (\ref{AlgAbelianFmunu}) we get\begin{align}
	\delta_{\epsilon}\psi_{M}
	= &
	\left(
	\partial_{M}
	+
	\frac{1}{4} \hat\omega_{M}^{A B} \Gamma_{A B}
	+
	\frac{1}{2} e_{M}^{A}\Gamma_{A}
	-
	\frac{i}{2}\sqrt{6} \, A_{M}
	\right)\epsilon
	+\nonumber\\
	& +
	\frac{i}{4\sqrt{6}}\, e^{A}_{M} \,
	\left( \Gamma_{A B C} - 4 \eta_{A B} \Gamma_{C} \right) \epsilon
	\, \eta^{B B'} \, \eta^{C C'}
	\left[
		e_{B'}^{ R}\, e_{C'}^{ S}
	\right] \times
	\nonumber\\
	& \times
	\left[
		\partial_{ R} A_{ S} - \partial_{ S} A_{ R}
		+
		\frac{i}{4}
			\bar\psi_{\left[  R \right.}
			\,\Gamma_{A}\,
			\psi_{\left.  S \right]}
			\,\eta^{A A'}\,A_{A'}
		+
		\frac{\sqrt{6}}{4}
		\bar\psi_{\left[  R \right.} \psi_{\left.  S \right]}
	\right]
	\ .
	\label{AlgGrav1}
\end{align}
In order to compute the gravitino variation $\psi^{\left[ n \right]}_{M}$ order by order we separate the expression above in different pieces
\begin{itemize}
	\item $
	\mathcal{D}^{\left[ n \right]}_{M}\left( \hat\omega \right)\epsilon = \left(
		\partial^{\left[ n \right]}_{M}
	+
	\frac{1}{4} \hat\omega_{M}^{\left[ n-1 \right] A B} \Gamma_{A B}
	+
	\frac{1}{2} e_{M}^{\left[ n-1 \right]A}\Gamma_{A}
	-
	\frac{i}{2}\sqrt{6} \, A_{M}^{\left[ n-1 \right]}
	\right)\epsilon$:
	this part contains ``$2n-1$ spinors'' (short way to say $n-1$ bilinears and one spinor).
	Notice that $\partial^{\left[ n \right]}_{M}\epsilon$ is simply zero for $n>1$;

	\item $
		e^{\left[ n \right] A}_{M} $: contains $n$ bilinears ($2n$ spinors);
	
\item $\left( B \right)_{A B C} = 	\left( \Gamma_{A B C} - 4 \eta_{A B} \Gamma_{C} \right) \epsilon $: this term contains always only one spinor $\epsilon$;

	\item $\left( C^{\left[ n \right]}  \right)^{ R S}_{B'C'} =\left[ e_{B'}^{ R}\, e_{C'}^{ S} \right]^{\left[ n\right]}$;

	\item $\left( D_{0}^{\left[ n \right]}  \right)_{ R S} = \partial_{ R} A_{ S}^{\left[ n \right]}  - \partial_{ S} A_{ R}^{\left[ n \right]}   $;

	\item $ \left( D_{1}^{\left[ n \right]}  \right)_{ R S\, A} =-i\left[ \bar\psi_{\left[  R \right.}
			\,\Gamma_{A}\,
			\psi_{\left.  S \right]} \right]^{\left[ n \right]} $;

		\item $ \left( D_{2}^{\left[ n \right]}  \right)_{ R S} = \left[ \left( D_{1} \right)_{ R S\, A}
			\,\eta^{A A'}\,A_{A'}  \right]^{\left[ n \right]}$;

		\item $\left( D_{3}^{\left[ n \right]} \right)_{ R S} =-i\left[  \bar\psi_{\left[  R \right.} \psi_{\left.  S \right]} \right]^{\left[ n \right]}$.
\end{itemize}
With these definitions, (\ref{AlgGrav1}) becomes
\begin{align}
	\delta_{\epsilon}^{\left[ n \right]}\psi_{M}
	= &
	\frac{1}{2n-1}\mathcal{D}^{\left[ n \right]}_{M}\left( \hat\omega \right)\epsilon
	+
	\frac{1}{2n-1}\frac{i}{4\sqrt{6}}
		\left( e^{\left[ N_{e} \right]} \right)^{A}_{M}
		\left( B \right)_{A B C}
		\left( C^{\left[ N_{C} \right]}  \right)^{ R S}_{D E}
		\, \eta^{B D} \, \eta^{C E} \times
	\nonumber\\ &
	\times
	\left[ D_{0}
	-
	\frac{1}{4} D_{2}
	+
	\frac{i\sqrt{6}}{4} D_{3}
	\right]^{\left[ N_{D} \right]} _{ R S}
	\ .
	\label{AlgFrav2}
\end{align}
To obtain the correct perturbative order $\left[ n \right]$ for $\delta_{\epsilon}^{\left[ n \right]}\psi_{M}$ the quantities  $N_{e}, N_{B}, N_{C}$ and $N_{D}$ must take the values shown in the following table.
\begin{center}
\begin{tabular}{|c|c|c|c|c|}
\hline
$n$ & $N_{e}$ & $N_{B}$ & $N_{C}$ & $N_{D}$ \\
\hline
$\left[ 1 \right]\,\left( 1/2 \right) $ & $\left[ 0 \right]\, \left( 0 \right)$ & $\left( 1/2 \right)$ & $\left[ 0 \right]\, \left( 0 \right)$ & $0$ \\
\hline
$\left[ 2 \right]\,\left( 3/2 \right)$ & $\left[ 0 \right]\, \left( 0 \right)$ & $\left( 1/2 \right)$ & $\left[ 0 \right]\,\left( 0 \right)$ & $\left[ 1 \right] \left( 1 \right)$ \\
\hline
$\left[ 3 \right]\,\left( 5/2 \right)$ & $\left[ 0 \right]\, \left( 0 \right)$ & $\left( 1/2 \right)$ & $\left[ 0 \right]\,\left( 0 \right)$ & $\left[ 2 \right] \left( 2 \right)$ \\
 & $\left[ 1 \right]\, \left( 1 \right)$ & $\left( 1/2 \right)$ & $\left[ 0 \right]\,\left( 0 \right)$ & $\left[ 1 \right] \left( 1 \right)$ \\
 & $\left[ 0 \right]\, \left( 0 \right)$ & $\left( 1/2 \right)$ & $\left[ 1 \right]\,\left( 1 \right)$ & $\left[ 1 \right] \left( 1 \right)$ \\
 \hline
$\left[ 4 \right]\,\left( 7/2 \right)$ & $\left[ 0 \right]\, \left( 0 \right)$ & $\left( 1/2 \right)$ & $\left[ 0 \right]\,\left( 0 \right)$ & $\left[ 3 \right] \left( 3 \right)$ \\
 & $\left[ 1 \right]\, \left( 1 \right)$ & $\left( 1/2 \right)$ & $\left[ 0 \right]\,\left( 0 \right)$ & $\left[ 1 \right] \left( 2 \right)$ \\
 & $\left[ 0 \right]\, \left( 0 \right)$ & $\left( 1/2 \right)$ & $\left[ 1 \right]\,\left( 1 \right)$ & $\left[ 2 \right] \left( 2 \right)$ \\

 & $\left[ 1 \right]\, \left( 1 \right)$ & $\left( 1/2 \right)$ & $\left[ 1 \right]\,\left( 1 \right)$ & $\left[ 1 \right] \left( 1 \right)$ \\
 & $\left[ 2 \right]\, \left( 2 \right)$ & $\left( 1/2 \right)$ & $\left[ 0 \right]\,\left( 0 \right)$ & $\left[ 1 \right] \left( 1 \right)$ \\
 & $\left[ 0 \right]\, \left( 0 \right)$ & $\left( 1/2 \right)$ & $\left[ 2 \right]\,\left( 2 \right)$ & $\left[ 1 \right] \left( 1 \right)$ \\
\hline
\end{tabular}
\end{center}
The numbers in square brackets are the perturbative order of the various pieces  while the ones in round brackets are the numbers of bilinears in the term, with the convention that $1/2$ bilinear $=$ $1$ spinor.

Now, we have to give explicit algorithms to compute $C^{\left[ n \right]}$, $D_{1}^{\left[ n \right]}$, $D_{2}^{\left[ n \right]}$ and $D_{3}^{\left[ n \right]}$.

\subsubsection{$C^{\left[ n \right]}$}

Using the given conventions we obtain the following result
\begin{align}
	\left( C^{\left[ n \right]}  \right)^{ R S}_{B'C'}
	= &
	\left[ e_{B'}^{ R}\, e_{C'}^{ S} \right]^{\left[ n \right]}
	=
	\nonumber\\
	= &
	\sum_{p=0}^{n}
	e_{B'}^{\left[ p \right] R}\, e_{C'}^{\left[ n-p \right] S}
	\ .
	\label{algC}
\end{align}

\subsubsection{$D_{2}^{\left[ n \right]}$}

To obtain the $D_{2}^{\left[ n \right]}$ term we need $D_{1}^{\left[ n \right]}$ and the gauge field with flat index $A_{A}^{\left[ n \right]} = \left[ e_{A}^{ R} A_{ R} \right]^{\left[ n \right]}$.
For the former we have
\begin{align}
	\left( D_{1}^{\left[ n \right]}  \right)_{ R S\, A}
	= &
	- i \left[
		\bar\psi_{\left[  R \right.}
			\,\Gamma_{A}\,
		\psi_{\left.  S \right]}
	\right]^{\left[ n \right]}
	=
	\nonumber\\
	= &
	-i \sum_{p=1}^{n} 	
		\bar\psi^{\left[ p \right]}_{\left[  R \right.}
			\,\Gamma_{A}\,
		\psi_{\left.  S \right]}^{\left[ n-p+1 \right]}
		\ ,
	\label{algD1}
\end{align}
while the latter reads
\begin{align}
	A_{A}^{\left[ n \right]}
	= &
	\left[ e_{A}^{ R} A_{ R} \right]^{\left[ n \right]}
	= \nonumber \\
	= &
	\sum_{p=1}^{n} e_{A}^{\left[ p-1 \right] R} A_{ R}^{\left[ n-p+1 \right]}
	\ .
	\label{AlgAflat}
\end{align}
Then, $D_{2}^{\left[ n \right]}$ becomes
\begin{align}
	\left( D_{2}^{\left[ n \right]}  \right)_{ R S}
	= &
	\left[ \left( D_{1} \right)_{ R S\, A}
			\,\eta^{AA'}\,A_{A'}  \right]^{\left[ n \right]}
	=
	\nonumber\\
	= &
	\sum_{p=1}^{n-1}\left( D_{1}^{\left[ p \right]} \right)_{ R S\, A}
	\,\eta^{AA'}\,A^{\left[ n-p \right]}_{A'}\ .
	\label{algD2}
\end{align}

\subsubsection{$ D_{3}^{\left[ n \right]} $}

Last, in analogy with (\ref{algD1}) we have
\begin{align}
	\left( D_{3}^{\left[ n \right]}  \right)_{ R S}
	= &
	 - i \left[
		\bar\psi_{\left[  R \right.}
		\psi_{\left.  S \right]}
	\right]^{\left[ n \right]}
	=
	\nonumber\\
	= &
	-i\sum_{p=1}^{n} 	
		\bar\psi^{\left[ p \right]}_{\left[  R \right.}
		\psi_{\left.  S \right]}^{\left[ n-p+1 \right]}
		\ .
	\label{algD3}
\end{align}

\subsection{Vielbein and Metric}

The vielbein is obtained as in eq.~(\ref{AlgSusyTransfv1})

\begin{align}
	\delta_{\epsilon} e_{M}^{\left[ n \right]A}
		= &
		-\frac{1}{2} \frac{1}{2n} {\textrm{Re}}\left(  i \bar\epsilon \Gamma^{A} \psi_{M}^{\left[ n \right]} \right)
		\ ,
	\label{algVielbein}
\end{align}
then, the metric (the wig) becomes
\begin{align}
	g_{M N}^{\left[ n \right]}
	=  &
	\sum_{p=0}^{n} e^{\left[ p \right]\,A}_{\left( M \right.} \, e^{\left[ n-p \right]\,B}_{\left.  N \right)} \eta_{AB}
	\ .
	\label{algMetric}
\end{align}

\subsection{Gauge Field}

Gauge field follows directly from eq.~(\ref{AlgSusyTransfv1})
\begin{equation}
	\delta^{\left[ n \right]}_{\epsilon}A_{M}
	=
	-\frac{\sqrt{6}}{4}\frac{1}{2n} {\textrm{Re}}
	\left( \bar\epsilon \psi^{\left[ n \right]}_{M} \right)
	\ .
	\label{algGauge}
\end{equation}

\section{Results}

In this section we collect the results obtained from the algorithms described in the previous section.
First, we present the $AdS_{5}$ wigs constructed from one of the two Killing spinors $\eta_{0}$ and $\eta_{1}$.
Since each of them contains $4$ real degrees of freedom, the series truncates after the second order in bilinears.

The wig which depends only on $\eta_{0}$ turned out to be too simple: 
we show that it gives no contribution both to the  ADM mass and to the boundary stress--energy tensor.

$\eta_{1}\neq0$, $\eta_{0}=0$ case is more interesting: the explicit dependence on the boundary coordinates leads to a modification to black hole Killing vectors. Furthermore, the boundary stress--energy tensor is not trivial and it will be discussed.

In order to present the result in different ways, we give the full wig and two particular limits of it, 
expanding in one case around small $\mu$ and in the other one around large $r$.
The former limit allows us to study a simplified, but a complete, metric while the latter shows the near boundary geometry.

The most general wig, obtained taking into account both $\eta_{0}$ and $\eta_{1}$, is derived. The degrees of freedom are now $8$, 
then the algorithm has to be iterated to the fourth order in bilinears.
The full expression is really cumbersome, even in the small $\mu$ and large $r$ limits. Then, we do not write it in this work, but the interested reader can find an electronic version in the ancillary files.

We repeat the procedure described above for the $AdS_{4}$ wigs. Apart from numerical coefficients, we find no substantial differences from the $AdS_{5}$ case. For this reason we present only the simplest results, leaving the complete wigs in the ancillary files.
Last remark, all wigs computed are asymptotically $AdS$.

\subsection{Results for $D=5$: $\eta_{1}=0$ and $\eta_{0}\neq 0$}

In this section we compute the finite BH wig choosing  $\eta_{1}=0$  and $\eta_{0}\neq 0$.
We introduce the following bilinears
\begin{align}
	\bM = -i \eta_{0}^{\dagger}\eta_{0}
	\ ,\qquad\qquad
	\bV_{i} = - i \eta_{0}^{\dagger}\hat\sigma_{i}\eta_{0}
	\ ,
	\qquad\qquad
	\lambda = \varepsilon_{0}^{t} \varepsilon_{0}
	\ ,
	\label{AlgBilEta0}
\end{align}
with these definitions, $\bM$ and $\bV_{i}$ are real numbers.

\subsubsection{Complete Wig}

The metric at first order is
\begin{align}
\delta^{\left[ 1 \right]}g
	= &\
-\frac{\mu}{2r^2 h\left( r \right)} \lambda \bM \, \dd r\dd t
\ ,
	\label{AlgEta0ord1}
\end{align}
where we defined $h\left( r \right) = \sqrt{r^2+\frac{\mu}{r^{2}}}$.
The metric at second order is
\begin{align}
\delta^{\left[ 2 \right]}g
	= &\
+
\frac{1}{192 r^{4}}\left[ \mu^{2}+\mu r^{3}\left( -5 r+14 h( r) \right)+18 r^{7}\left( r-h( r ) \right) \right]	\lambda^{2}\bM^2\, \dd t^2
+
\nonumber\\&
+
\frac{1}{192 h(r)}\left[ \mu \left( 17 r-8 h(r) \right)+18 r^{4}\left( r-h(r) \right) \right]	\lambda^{2}\bM^2\, \dd \vec{x}^2
+
\nonumber\\&
+
\frac{1}{96 \left( r h(r) \right)^{3/2}} \left[ \mu r\left( 21 r-11 h(r) \right)+20 r^{5}\left( r-h(r) \right) \right]	\lambda^{2}\bM^2\,  \dd r^2
+
\nonumber\\&
-
\frac{1}{64 r } \left[\mu  \left( -r + 3 h(r) \right)+ 4 r^{4}\left( r-h(r) \right) \right]	\lambda^{2}\bM \bV_{i}\, \dd t\dd x^{i}
\ .
	\label{AlgEta0ord2}
\end{align}
The gauge field is zero at every order.

\subsubsection{Expansion}

The complete metric result is now presented here in large-$r$ expansion and this coincides with the small-$\mu$ expansion.
\begin{align}
	\dd s^{2} = &\
	-\left( r^{2}+\frac{\mu}{r^{2}} \right)\, \dd t^{2}
	+\left( \frac{1}{r^{2}} - \frac{\mu}{r^{6}} \right) \dd r^{2}
	+r^{2} \dd \vec{x}^{2}
	-\frac{\mu}{2 r^3 } \lambda \bM \, \dd r\dd t	
	+
	\nonumber\\&\
+
	\frac{41\mu^2}{768 r^4}\lambda^{2} \bM^{2}\, \dd t^{2}
-
	\frac{7\mu^2}{768 r^4}\lambda^{2} \bM^{2}\, \dd\vec{x}^{2}
-
	\frac{\mu^2}{32 r^{8}}\lambda^{2}\bM^{2}\, \dd r^{2}
-
	\frac{\mu^{2}}{32 r^4 } \lambda^{2} \bM \bV_{i} \dd t \dd x^{i}
\ .
	\label{AlgEta0Expansion}
\end{align}

\subsubsection{ADM mass}

Following the procedure outlined in \cite{Behrndt:1998jd,Horowitz:1998ha}  we compute the ADM mass for the $\eta_{0}\neq0$, $\eta_{1}=0$ case.
The ADM mass is defined as
\begin{align}
	E_{ADM} = -\frac{1}{8\pi G} \int_{\Sigma} N \left( \Theta-\Theta_{0} \right)
	\ ,
	\label{Eta0ADM5dim2}
\end{align}
where $N=\sqrt{g_{tt}}$ is the norm of the timelike Killing vector $\partial_{t}$, $\Theta$ is the trace of the extrinsic curvature of a spacelike, near--infinity surface $\Sigma$ and $\Theta_{0}$ is $\Theta$ computed in the background $AdS_5$ geometry. Using the definition of extrinsic curvature we can rewrite eq.~(\ref{Eta0ADM5dim2}) as
\begin{align}
	E_{ADM} = -\frac{1}{8\pi G}  N \left( n^{\mu} - n^{\mu}_{0} \right)\partial_{\mu} A_{\Sigma}
	\ ,
	\label{Eta0ADM5dim21}
\end{align}
where $n^{\mu}$ is the vector normal to $\Sigma$ and $A_{\Sigma}$ is the area of $\Sigma$.
In order to consider a near infinity space--like surface, we use the large--$r$ metric eq.~(\ref{AlgEta0Expansion}).  We define a new radial coordinate
\begin{align}
	\rho^2 = r^{2} + \frac{3\mu}{32}\lambda^{2} \bM^{2}
	\ ,
	\label{Eta0ADM5dim1}
\end{align}
thus, the area of $\Sigma$ is simply $\rho^{3} V_{p}$, with $V_{p}$ the coordinate volume of the surface parametrized by $x^{i}$. The ADM mass is then
\begin{align}
	E_{ADM} = - \frac{3 \mu V_{p}}{16\pi G} + O\left( \frac{1}{\rho} \right)
	\ ,
	\label{Eta0ADM5dimFIN}
\end{align}
which is the result for Schwarzschild black hole. The wig constructed by bilinears only in $\eta_{0}$ gives no contribution to the ADM mass.

\subsubsection{Boundary Stress--Energy Tensor}

Using the prescription given in sec.~\ref{secBY} we compute the stress--energy tensor for the black hole wig. The result is
\begin{align}
	T_{\mu\nu} = -\frac{\mu}{2}\left( 4 u_{\mu}u_{\nu} +\eta_{\mu\nu} \right)
	\ ,
	\label{eta0LargeRTmunu}
\end{align}
where $u^{\mu}=\left( 1,0,0,0 \right)$ is the fluid velocity in the rest frame of the fluid. In this case, we have no contribution from the BH wig.

\subsection{Results for $D=5$: $\eta_{1}\neq0$ and $\eta_{0}= 0$}

In this section we compute the finite BH wig choosing  $\eta_{1}=0$  and $\eta_{0}\neq 0$. As in the previous case, we introduce
\begin{align}
	\bN = - i \eta_{0}^{\dagger}\eta_{0}
	\ ,\qquad\qquad
	\bK_{i} = - i \eta_{0}^{\dagger}\hat\sigma_{i}\eta_{0}
	\ ,
	\qquad\qquad
	\lambda = \varepsilon_{0}^{t} \varepsilon_{0}
	\ ,
	\label{AlgBilEta1}
\end{align}
where again $\bN$ and $\bK_{i}$ are real. Notice that in order to present the results we write the first terms in the large-$r$ expansion.

\subsubsection{First order in $\mu$}\label{d5mu}

As a first check, we want to determine only the effects due to gauge field and not to bilinears in the gravitino field. For this reason we consider the first order in the expansion around $\mu=0$ neglecting the contributions coming from bilinears in the gravitinos, since they contribute to  order $O\left( \mu^{2} \right)$.

The metric at first order is
\begin{align}
\delta^{\left[ 1 \right]}g
	= &\
	\, -\frac{\mu\lambda}{2r^{2}}\left( \bN t+\bK_{i}x^{i} \right)\ \dd t^{2}
	-
	\frac{\mu\lambda}{2r^{3}}\left[ \bN\left( t^{2}+\vec{x}^{2} \right)+2tx_{i}\bK^{i} \right]\ \dd t \dd r
	+
	\nonumber\\ &
	-
	\frac{\mu\lambda}{4r^{2}}\left( t \bK_{i}+x_{i} N \right)\ \dd t \dd x^{i}
	+
	\frac{\mu\lambda}{4r^{5}}\bK_{i}\ \dd r \dd x^{i}
	-
	\frac{\mu\lambda}{4r^{2}}\left( \bN t+ \bK_{k}x^{k} \right)\delta_{ij}\ \dd x^{i} \dd x^{j}
\ .
	\label{AlgEta1ord1}
\end{align}
The metric at second order is
\begin{align}
\delta^{\left[ 2 \right]}g
	= &\
	-\frac{\mu}{48 r^{4}} \lambda^{2} \bN \left[
	2 r^{2} t x^{i} \bK_{i}\left( 4+r^{2} \left( t^{2} + \vec{x}^{2} \right) \right)
	+ \bN \left( 1+r^{2} \left( t^{2} -3  \vec{x}^{2} \right) \right)
	\right]	\, \dd t^2
+
\nonumber\\&
-\frac{\mu}{12 r^{6}}\lambda^{2} \bN \left[
\bN t^{2} + r^{2} t x^{i} \bK_{i} \left( t^{2}+\vec{x}^{2} \right)
\right]\, \dd r^2
+
\nonumber\\&
-\frac{\mu}{24 r^{5}} \lambda^{2} \bN \left[
t \bN \left( 2+ r^{2} \left( t^{2} -  \vec{x}^{2} \right) \right)
+ x^{i}\bK_{i} \left( 1 + 2 r^{2} \left(  3 t^{2} +\vec{x}^{2} \right) \right)
\right]
\, \dd t\dd r
+
\nonumber\\&
+\frac{\mu}{96 r^{4}}\lambda^{2} \bN
\left[
-2 r^{2} x_{i} \left( 3 t \bN + 2 x^{j} \bK_{j} \right)
+
\bK_{i} \left( 1+ r^{2}\left( -3t^{2}+\vec{x}^{2} \right) \right)
\right]
\, \dd t\dd x^{i}
+
\nonumber\\&
-\frac{\mu}{48 r^{5}} \lambda^{2} \bN
\left[
x_{i}\left( \bN+8 r^{2} t x^{j} \bK_{j} \right)
+
t \bK_{i} \left( -1+2r^{2}\left( t^{2} -\vec{x}^{2} \right) \right)
\right]
\, \dd r\dd x^{i}
+
\nonumber\\&
+\frac{\mu}{96 r^{4}}\lambda^{2} \bN
\left[
\bN  \left( -1 +r^{2} \left( t^{2} +3 \vec{x}^{2} \right) \right) \delta_{i j}
+
4 r^{2} t x^{k} \bK_{k} \left( -1 + r^{2} \left( t^{2}+\vec{x}^{2} \right) \right)\delta_{i j} +
\right.
\nonumber\\&
\left.\qquad\qquad\quad
-2 \bN r^{2} x_{i} x_{j} + r^{2} t x_{\left( i \right.} \bK_{\left. j \right)}
\right]
\, \dd x^{i} \dd x^{j}
	\ .
	\label{AlgEta1ord2}
\end{align}
In this limit, the gauge field is zero at each order.

\subsubsection{Large $r$ expansion}

Here we compute the large-$r$ expansion of the metric corrections.
At first order, the wig coincides to eq.~(\ref{AlgEta1ord1}). The metric at second order is
\begin{align}
\delta^{\left[ 2 \right]}g
	= &\
	-\frac{\mu}{24} \lambda^{2} \bN t x^{i} \bK_{i} \left( t^{2} +\vec{x}^{2} \right)
	\, \dd t^2
+
\nonumber\\&
-\frac{\mu}{12 r^{4}} \lambda^{2} t x^{i} \bN \bK_{i} \left( t^{2} +\vec{x}^{2} \right)
	\, \dd r^{2}
+
\nonumber\\&
-\frac{\mu}{24 r^{3}} \lambda^{2} \bN
\left[
t \bN \left( t^{2} - \vec{x}^{2} \right) +2 x^{i} \bK_{i} \left( 3 t^{2} +\vec{x}^{2} \right)
\right]
	\, \dd t\dd r
+
\nonumber\\&
+\frac{\mu}{96 r^{2}} \lambda^{2} \bN
\left[
-2 x_{i} \left( 3 \bN t + 2 x^{k} \bK_{k} \right) +
\bK_{i} \left( - 3  t^{2} +\vec{x}^{2} \right)
\right]
	\, \dd t\dd x^{i}
+
\nonumber\\&
-\frac{\mu}{24 r^{3}} \lambda^{2} \bN
\left[
4 t^{2} x_{i} x^{k} k_{k}
+
\bK_{i} \left( t^{2} - \vec{x}^{2} \right)
\right]
	\, \dd r\dd x^{i}
+
\nonumber\\&
+
\frac{\mu}{24} \lambda^{2} \bN
\left[
t x^{k }\bK_{k} \left( t^{2} +\vec{x}^{2} \right)\delta_{ij}
+
\frac{1}{2r^{2}} \left(
t x_{\left( i \right.} \bK_{\left. j \right)} -2 \bN x_{i} x_{j}
\right)
\right]
	\, \dd x^{i} \dd x^{j}
	\ .
	\label{AlgEta1ord2LargeR}
\end{align}
The only non--zero components of the gauge field are the $A^{\left[ 2 \right]}_{i}$
\begin{align}
	A^{\left[ 2 \right]}_{i}
	=&\
	\frac{\sqrt{3}\mu^{2}}{512\sqrt{2}  r^{6}}\lambda^{2}
	\varepsilon_{ijk} x^{j} \bN \bK^{k}\left( t^{2}+\vec{x}^{2} \right) \ .
	\label{gaugeFieldlargeR4}
\end{align}

\subsubsection{Complete}

Here we present the complete wig depending on $\eta_{1}$ bilinears. The first order is
\begin{align}
\delta^{\left[ 1 \right]}g
	= &\
	-
	\frac{\mu}{2r^{3}} \lambda h\left( r \right)
	\left[
	t \bN + x^i \bK_i	
	\right]
	\ \dd t^{2}
	-
	\frac{\mu}{2r^{2}h\left( r \right)} \lambda
	\left[
		\left( t^{2}-\vec{x}^{2} \right) \bN
		+2 t x^{i} \bK_{i}	
		\right]
	\ \dd t \dd r
	+
	\nonumber\\ &
	+
	\frac{1}{2}\lambda r \left( r- h\left( r \right) \right)
	\left[ t \bK_{i}+x_{i} \bN \right]
	\ \dd t \dd x^{i}
	-
	\frac{1}{2 r h\left( r \right)}\left( r-h\left( r \right) \right) \bK_{i}
	\ \dd r \dd x^{i}
	+
	\nonumber\\ &
	+
	\frac{1}{2}\lambda r \left( r-h\left( r \right) \right)
	\left[
	t \bN + x^i \bK_i	
	\right] \delta_{ij}
	\ \dd x^{i} \dd x^{j}
\ .
	\label{AlgEta1ord1full}
\end{align}
The second order is
{\allowdisplaybreaks
\begin{align}
\delta^{\left[ 2 \right]}g
	= &\
	\frac{\bN\lambda^2}{192 r^{8}}
	\left[
	+2 r^9\left(-2 x^{i} \bK_{i} t\left(13+6 r^2\left( t^2+\vec{x}^{2}\right)\right)
	+\right.\right.\nonumber\\& \left.\left.
	+\bN\left(-9 r^2 t^4-\vec{x}^{2}\left(17+9 r^2\vec{x}^{2}\right)+ t^2\left(83+18 r^2\vec{x}^{2}\right)\right)\right)\left(- r+ h\left(r\right)\right)
	+\right.\nonumber\\& \left.+		
	\mu  r^4\left(-\bN\left(5 r^4\left( t^2-\vec{x}^{2}\right)^2- r^2\left(-85 t^2+31\vec{x}^{2}+14 r\left( t^2-\vec{x}^{2}\right)^2 h\left(r\right)\right)
	+\right.\right.\right.\nonumber\\& \left.\left.\left.
	+2\left(2+ r\left( t^2+\vec{x}^{2}\right) h\left(r\right)\right)\right)
	+\right.\right.\nonumber\\& \left.\left. +
	2 x^{i} \bK_{i} r t\left(-18 r^3\left( t^2+\vec{x}^{2}\right)-2 h\left(r\right)+ r\left(-1+20 r\left( t^2+\vec{x}^{2}\right) h\left(r\right)\right)\right)\right)
	+\right.\nonumber\\& \left.+
	\mu ^2\left(\bN\left(-4-15 r^2\left( t^2-3\vec{x}^{2}\right)+ r^4\left( t^2-\vec{x}^{2}\right)^2\right)
	+\right.\right.\nonumber\\& \left.\left.+
	18 x^{i} \bK_{i} r^2 t\left(-3+2 r^2\left( t^2+\vec{x}^{2}\right)\right)\right)
	\right] \ \dd t^{2}
	+
	\nonumber\\ &
	-\frac{\bN\lambda^2}{384 r^{7} (h\left( r \right))^{2}}
\left[
+2 r^7\left(\bN t\left(-83-53 r^2\left( t^2-\vec{x}^{2}\right)\right)
+\right.\right.\nonumber\\& \left.\left.
+5 x^{i} \bK_{i}\left(5+3 r^2\left(3 t^2+\vec{x}^{2}\right)\right)\right)\left(- r+ h\left(r\right)\right)
+\right.\nonumber\\& \left.
+\mu ^2\left(6\bN t\left(-1+4 r^2\left( t^2-\vec{x}^{2}\right)\right)
+2 x^{i} \bK_{i}\left(-7+20 r^2\left(3 t^2+\vec{x}^{2}\right)\right)\right)
+\right.\nonumber\\& \left.
+\mu  r^3\left(\bN t\left(130 r^3\left( t^2-\vec{x}^{2}\right)-61 h\left(r\right)+ r\left(160-69 r\left( t^2-\vec{x}^{2}\right) h\left(r\right)\right)\right)
+\right.\right.\nonumber\\& \left.\left.
+ x^{i} \bK_{i}\left(10 r^3\left(3 t^2+\vec{x}^{2}\right)+47 h\left(r\right)- r\left(64+9 r\left(3 t^2+\vec{x}^{2}\right) h\left(r\right)\right)\right)\right)
	\right]
\ \dd t \dd r
	+
	\nonumber\\ &
	-
	\frac{\bN\lambda^2}{96 r^8\left( h\left(r\right)\right)^{4}}
\left[
+2r^7\left(16 x^{i} \bK_{i} r^2 t-\bN\left(13+8 r^2\left( t^2-\vec{x}^{2}\right)
\acIII+
+10 r^4\left( t^2-\vec{x}^{2}\right)^2\right)\right)\left(- r+ h\left(r\right)\right)
+\right.\nonumber\\& \left.
+\mu ^2\left(\bN\left(13+12 r^2\left( t^2-\vec{x}^{2}\right)+11 r^4\left( t^2-\vec{x}^{2}\right)^2\right)
\acII
+4 x^{i} \bK_{i} r^2 t\left(-4+3 r^2\left( t^2+\vec{x}^{2}\right)\right)\right)
+\right.\nonumber\\& \left.
+\mu \left(\bN r^3\left(31 r^5\left( t^2-\vec{x}^{2}\right)^2-26 h\left(r\right)+ r\left(39-4 r\left(3 t^2-5\vec{x}^{2}\right) h\left(r\right)\right)
\acIII
-7 r^3\left( t^2-\vec{x}^{2}\right)\left(-4+3 r\left( t^2-\vec{x}^{2}\right) h\left(r\right)\right)\right)
+\right.\right.\nonumber\\& \left.\left.
-4 x^{i} \bK_{i} r^5 t\left(-3 r^3\left( t^2+\vec{x}^{2}\right)-8 h\left(r\right)+ r\left(12+ r\left( t^2+\vec{x}^{2}\right) h\left(r\right)\right)\right)\right)
	\right]
	\ \dd r^{2}
	+
	\nonumber\\ &
	+
	\frac{\bN\lambda^2}{384 r^{5}}
\left[
2 r^5\left(\bK_{i}\left(11+ r^2\left(27 t^2-7\vec{x}^{2}\right)+6 r^4\left( t^4-\vec{x}^{4}\right)\right)
\acII
+2 r^2x_{i}\left(-\bN t+ x^{k} \bK_{k}\left(17+6 r^2\left(3 t^2+\vec{x}^{2}\right)\right)\right)\right)\left( r- h\left(r\right)\right)
\acI
-\mu  r\left(\bK_{i} r\left(-22- r^2\left(27 t^2+17\vec{x}^{2}\right)+3 r^4\left( t^4-\vec{x}^{4}\right)\right)
\acII
+2 r^3x_{i}\left(29\bN t+ x^{k} \bK_{k}\left(-5+3 r^2\left(3 t^2+\vec{x}^{2}\right)\right)\right)
\acII
+\left(\bK_{i}\left(9+ r^2\left(6 t^2+22\vec{x}^{2}\right)-9 r^4\left( t^4-\vec{x}^{4}\right)\right)
\acIII
-2 r^2x_{i}\left(22\bN t+ x^{k} \bK_{k}\left(8+9 r^2\left(3 t^2+\vec{x}^{2}\right)\right)\right)\right) h\left(r\right)\right)
	\right]
	\ \dd t \dd x^{i}
	+
	\nonumber\\ &
	-
	\frac{\bN\lambda^2}{384 r^{6}\left( h\left(r\right)\right)^3}
\left[
2 r^7\left(x_{2}\left(66 x^{k} \bK_{k} r^2 t+\bN\left(-27+37 r^2\left( t^2-\vec{x}^{2}\right)\right)\right)
\acII
+\bK_{2} t\left(29+3 r^2\left(5 t^2-7\vec{x}^{2}\right)\right)\right)\left( r- h\left(r\right)\right)
\acI
+2\mu ^2\left(3x_{2}\left(18 x^{k} \bK_{k} r^2 t+\bN\left(-4+5 r^2\left( t^2-\vec{x}^{2}\right)\right)\right)
\acII
+\bK_{2} t\left(14+ r^2\left(13 t^2-15\vec{x}^{2}\right)\right)\right)
\acI
+\mu  r^3\left(2 r\left(x_{2}\left(120 x^{k} \bK_{k} r^2 t+13\bN\left(-3+4 r^2\left( t^2-\vec{x}^{2}\right)\right)\right)
\acIII
+\bK_{2} t\left(43+4 r^2\left(7 t^2-9\vec{x}^{2}\right)\right)\right)
\acII
+\left(x_{2}\left(-142 x^{k} \bK_{k} r^2 t+\bN\left(55-67 r^2\left( t^2-\vec{x}^{2}\right)\right)\right)
\acIII
+\bK_{2} t\left(-61+ r^2\left(-33 t^2+43\vec{x}^{2}\right)\right)\right) h\left(r\right)\right)
		\right]
	\ \dd r \dd x^{i}
	+
	\nonumber\\ &
	+\frac{\bN\lambda^2}{192 r^3 h\left( r \right)}
	\left[
\left[
2 r^4\left(-58 x^{k} \bK_{k} r^2 t
\acIII
+
\bN\left(7-32 r^2\left( t^2-\vec{x}^{2}\right)+9 r^4\left( t^2-\vec{x}^{2}\right)^2\right)\right)\left( r- h\left(r\right)\right)
\acII
+\mu \left(-2 x^{k} \bK_{k} r^3 t\left(55+2 r^2\left( t^2+\vec{x}^{2}\right)\right)
\acIII
+\bN r\left(14+17 r^4\left( t^2-\vec{x}^{2}\right)^2+ r^2\left(-63 t^2+71\vec{x}^{2}\right)\right)
\acIII
-\bN\left(9+ r^2\left(-33+8 r^2\left( t^2-\vec{x}^{2}\right)\right)\left( t^2-\vec{x}^{2}\right)\right) h\left(r\right)
\acIII
+4 x^{k} \bK_{k} r^2 t\left(11+3 r^2\left( t^2+\vec{x}^{2}\right)\right) h\left(r\right)\right)
\right]	\delta_{ij}
\acI
-2 r^5\left(8\left(2\bN+3 x^{k} \bK_{k} r^2 t\right)x_{\left( i \right.}x_{\left. j \right)}+3x_{\left( i \right.}\bK_{\left. j \right)} t\left(-5+2 r^2\left( t^2-\vec{x}^{2}\right)\right)
\acII
+3\bK_{3} tx_{2}\left(-5+2 r^2\left( t^2-\vec{x}^{2}\right)\right)\right)\left( r- h\left(r\right)\right)
\acI
-\mu  r\left(6 r\left(\left(7\bN+8 x^{k} \bK_{k} r^2 t\right)x_{\left( i \right.}x_{\left. j \right)}
+2x_{\left( i \right.}\bK_{\left. j \right)} t\left(-2+ r^2\left( t^2-\vec{x}^{2}\right)\right)
\acIII
+2\bK_{\left( i \right.} tx_{\left. j \right)}\left(-2+ r^2\left( t^2-\vec{x}^{2}\right)\right)\right)
\acII
+\left(-2\left(11\bN+12 x^{k} \bK_{k} r^2 t\right)x_{\left( i \right.}x_{\left. j \right)}+x_{\left( i \right.}\bK_{\left. j \right)} t\left(7-6 r^2\left( t^2-\vec{x}^{2}\right)\right)
\acIII
+\bK_{\left( i \right.} tx_{\left. j \right)}\left(7-6 r^2\left( t^2-\vec{x}^{2}\right)\right)\right) h\left(r\right)\right)
	\right]
	\ \dd x^{i} \dd x^{j}
\ .
	\label{AlgEta1ord2full}
\end{align}
}

\subsection{Results for $D=4$: $\eta_{1}=0$ and $\eta_{0}\neq 0$}

The $AdS_{4}$ model is very similar to $AdS_{5}$ one. For our purpose, the only relevant difference is the Schwarzschild BH metric
\begin{align}
	\dd s^{2} = f(r)^2 \dd t^{2} + f(r)^{-2} \dd r^{2} + r^{2} \dd \vec{x}^{2}\ ,
	\label{AdSmetric}
\end{align}
where $f(r)=\sqrt{r^{2}+\frac{\mu}{r}}$.
Due to the fact that $2$-- and $3$--dimensions spinors have the same number of degrees of freedom, our algorithm can be applied with no modifications. Notice also that the Killing spinors are written in the same way of eq.~(\ref{KSsolution2}), where $x^{i}$ denotes only $x_{1}$ and $x_{2}$.

Last remark, in $4d$  $\Gamma_{5}$  is defined by dimensional reduction from $5d$ as
\begin{align}
	\Gamma_{5} = &\ \sigma_{3}\otimes\hat\sigma_{3}
	\ ,
	\label{gamma5in4d}
\end{align}
then, bilinears in $\eta$ with $\hat\sigma_{3}$ are still present.

\subsubsection{Complete Wig}

The first order is
\begin{align}
\delta^{\left[ 1 \right]}g
=&\
-\frac{3\mu}{4 r f\left( r \right)} \lambda \bM
\ \dd r \dd t
\ .
\end{align}
The second order is
\begin{align}
\delta^{\left[ 2 \right]}g
=&\
\frac{1}{384 r^{2}} \lambda^{2} \bM^{2}
\left[ 
	3\mu^{2} + \mu r^{2} \left(  3 r + 11 f\left( r \right) \right)
	+28 r^{5} \left( r-f\left( r \right) \right)\right]\
\dd t^2
+\nonumber \\
&
- \frac{1}{769 \left( f\left( r \right) \right)^{2}}\lambda^{2} \bM^{2}
\left[
	25 \mu^{2} + +\mu r^{2} \left( 81 r - 53 f\left( r \right) \right) + 28 r^{5} \left( r-f\left( r \right) \right)
\right]
\delta_{ij} \dd x^i \dd x^j
+ \nonumber \\
&
-
\frac{1}{768}\lambda^{2} \bM \bV_{i}
\left[
	\mu \left(  5 r + 23 f\left( r \right) \right)
	+ 56 r^{5} \left( r - f\left( r \right) \right) 
\right]
\dd t \dd x^i +
\nonumber\\
&
+
\frac{1}{128 r f\left( r \right)^{3}} \lambda^{2} \bN^{2}
\left[
	\mu \left( 21 r - 11 f\left( r \right) \right) + 20 r^{3} \left( r-f\left( r \right) \right)
\right]\
\dd r^2
\ .
\end{align}
Notice that, as in the $5$--dimensional case, the gauge field is zero at every order.

\subsection{Results for $D=4$: $\eta_{1}\neq0$ and $\eta_{0}= 0$}

In this section we compute the finite wig choosing  $\eta_{1}=0$  and $\eta_{0}\neq 0$.
We introduce the following bilinears
\begin{align}
	\bN =-i \eta_{0}^{\dagger}\eta_{0}
	\ ,\qquad\qquad
	\bK_{i} = -i \eta_{0}^{\dagger}\hat\sigma_{i}\eta_{0}
	\ ,
	\qquad\qquad
	\lambda = \varepsilon_{0}^{t} \varepsilon_{0}
	\ ,
	\label{AlgBilEta11}
\end{align}
with these definitions, $\bN$ and $\bK_{i}$ are real quantities.

\subsubsection{First order in $\mu$}

As in [\ref{d5mu}], we focus on effects due to gauge field and not to bilinears in the gravitino field, considering only the first order in the expansion around $\mu=0$.
The metric at first order is
\begin{eqnarray}
\delta^{\left[ 1 \right]}g
= &-&\frac{\l \m}{4r^2} \left(\bN t + \bK_i x^i \right) \dd t^2 
-\frac{\l \m}{8r^4}\left[6r^2 t \left( \bK_i x^i\right)+ \bN\left(-1+3r^2\left(t^2+\vec{x}^{2}\right)\right)\right] \dd t \dd r 
+
\nonumber \\
&-&\frac{\l \m}{4r}\left(\bK_i t + \bN x_i\right)\dd t \dd x^i
+ \frac{\l \m}{4r^4} \bK_i \dd r \dd x^i 
-\frac{\l \m}{4r} \left(\bN t +\bK_k x^k\right) \delta_{ij} \dd x^i \dd x^j
\ .
\nonumber\\
	\label{AlgEta1ord14dim}
\end{eqnarray}
The metric at second order is
\begin{align}
\delta^{\left[ 2 \right]}g
	= &-\frac{\m}{192 r^3}
	\lambda^{2} \bN
	\left[2r^2 t \left(\bK_i x^i\right)\left[7+3r^2\left(t^2 + \vec{x}^{2}\right)\right]+\bN\left[1+r^2\left(t^2-5 \vec{x}^{2}\right)\right]\right] \dd t^2
+
\nonumber \\
&- \frac{3 \m t}{32 r^5} \lambda^{2} \bN \, t
\left[\bN t +\left(\bK_k x^k\right)\left[-1 + r^2\left(t^2 + \vec{x}^{2}\right)\right]\right] \dd r^2
+
\nonumber \\
&-\frac{\m}{192 r^4}\lambda^{2} \bN
\left[\bN t \left[5+3r^2\left(t^2-x_i x^i\right)\right]+2\left(\bK_i x^i\right)\left[1+3r^2\left(3t^2+\vec{x}^{2}\right)\right]\right] \dd t \dd r
+
\nonumber \\
&- \frac{\m}{96 r} \lambda^{2} \bN
\left[ x_i \left(3 \bN t + 2 \bK_k x^k\right) + 2 \bK_i t^2\right] \dd t \dd x^i
+
\nonumber \\
&-\frac{\m}{96 r^4} \lambda^{2} \bN
\left[ \left( \bN x_i - t \bK_i \right) + r^2 t \left[  3 \bK_i  \left(   t^2-\vec{x}^2 \right)   +  12 x_i x^j \bK_j \right]\right] \dd r \dd x^i
+
\nonumber \\
&-
\frac{\mu}{192 r^{3}} \lambda^{2} \bN \left[
	\left( -1 + r^{2}\left( t^{2}+5\vec{x}^{2} \right) \right)\delta_{ij}
	+6 r^{2} t x^{k} \bK_{k} \left[ -1+r^{2}\left( t^{2}+\vec{x}^{2} \right) \right]\delta_{ij}
	+
	\right.
	\nonumber\\&
	\left.\qquad\quad
	-4 r^{2} \bN x_{i}x_{j}
	+2 r^{2} t x_{\left( i \right.} \bK_{\left. j \right)}
\right]
\ \dd x^{i} \dd x^{j}
	\ .
	\label{AlgEta1ord24dim}
\end{align}
For both orders, the gauge field is zero.

\subsubsection{Large-$r$ expansion}

Here we compute the large-$r$ expansion of the metric corrections.

The first order metric is
\begin{align}
\delta^{\left[ 1 \right]}g
= &-\frac{\l \m}{4r}\left(\bN t + \bK_i x^i\right) \dd t^2 
-\frac{3 \l \m}{8 r^2}\left[2t \left(\bK_i x^i \right) +\bN\left(t^2 + \vec{x}^{2}\right)\right] \dd t \dd r
	+ \nonumber \\ &-
	\frac{\l \m}{4r}\left(\bK_i t + \bN x_i\right) \dd x^i \dd t
	-
	\frac{\l \m}{4r^4} \bK_i \dd x^i \dd r
	+ \nonumber \\ &-
	\frac{\l \m}{4r}\left(\bN t +\bK_i x^i \right) \delta_{ij}\dd x^i \dd x^j
\ .
	\label{AlgEta1LargeR4dim}
\end{align}
The second order is
\begin{align}
\delta^{\left[ 2 \right]}g
	= &-\frac{\mu}{32}\l^2  \bN r t\left(\bK_i x^i\right)\left(t^2+\vec{x}^{2}\right) \dd t^2
	-
	\frac{\mu}{32r^3} \l^2 \bN t \left(\bK_i x^i\right)\left(t^2 +\vec{x}^{2}\right) \dd r^2
	+ \nonumber \\
&-\frac{  \m}{64 r^2}\l^2 \bN \left[\bN t\left(t-\vec{x}^{2}\right)+2\left(\bK_i x^i\right)\left(3t^2+\vec{x}^{2}\right)\right] \dd t \dd r
+ \nonumber \\
&-\frac{\mu}{96 r}\l^2 \bN\left[2\bK_i t^2 + x_i\left(\bK_k x^k  + 3 \bN t \right)\right] \dd t \dd x^i
+ \nonumber \\ &-
\frac{ \mu}{32 r^2} \l^2 \bN t \left[4x_i\left(\bK_k x^k\right)-\bK_i\left(t^2 +\vec{x}^{2}\right)\right] \dd r \dd x^i
+ \nonumber \\
&+ \frac{\mu}{32} \l^2 \bN \left[3r t \left(\bK_k x^k\right)\left(t^2+\vec{x}^{2}\right) \delta_{ij} + \frac{2}{r}\left(t \bK_{\left( i \right.} x_{\left. j \right)} - 2 \bN x_i x_j\right)\right]\dd x^i \dd x^j
	\ .
	\label{AlgEta1ord2LargeR4dim}
\end{align}
The  non--zero components of the gauge field are the $A^{\left[ 2 \right]}_{i}$
\begin{align}
	A^{\left[ 2 \right]}_{i}
	=&\
	\frac{\mu^{2}}{256 \sqrt{6} r^{4}}\lambda^{2}
	\varepsilon_{ij} x^{j} \bN \bK_{3}\left( -t^{2}+\vec{x}^{2} \right)
	\ ,
	\label{gaugeFieldlargeR5}
\end{align}
where $\varepsilon_{ij}$ is the $2d$ antisymmetric tensor, with $\varepsilon_{12}=1$.

\subsection{Stress--Energy Tensor for $AdS_{5}$}

Using the prescription given in the previous section we present the result obtained in $AdS_5$. The first-order corrections are the same both at 4 and 5 dimensions, while at second-order one they are different. We decompose the contribution to the stress-energy tensor in the perturbative form as
 \begin{align}
	T_{\mu\nu} = -\frac{\mu}{2}\left( 4 u_{\mu}u_{\nu} +\eta_{\mu\nu} \right) + \lambda\mu\mathcal{T}^{\left[ 1 \right]}_{\mu\nu} + \lambda^{2}\mu\mathcal{T}^{\left[ 2 \right]}_{\mu\nu}
	\ ,
	\label{eta1LargeRTmunu5dim}
\end{align}
where $u^{\mu}=\left( 1,0,0,0 \right)$ is the fluid velocity in the rest frame of the fluid as usual, $\bB^{\m}=\left(-\bN,\bK_i\right)$ is the bilinear $4-$vector. As usual,
we define the projectors
\begin{align}
	& P^{\shortparallel}_{\mu\nu}= \eta_{\mu\nu} + u_{\mu} u_{\nu}
	 \ ,
	 &P^{\bot}_{\mu\nu}= - u_{\mu} u_{\nu}
	 \ .
	\label{EMT}
\end{align}
The first order of $T_{\m\nu}$ is
\begin{eqnarray}
	\mathcal{T}^{\left[ 1 \right]}_{\mu\nu} &=& -\frac{d}{8}  \left[(\bB \cdot x)( \eta_{\mu\nu}+ d\,  u_\mu u_\nu )  +  4 P^{\shortparallel}_{\left(  \mu \right| \rho}  P^{\bot}_{\left. \nu \right) \sigma} \bB^{\left[  \rho  \right.} x^{\left. \sigma \right]}\right] \ ,
	\label{eta1BY2}
\end{eqnarray}
where $d$ refers to $AdS_{d+1}$. Notice that the second term in eq.~(\ref{eta1BY2}) resembles a vorticity term. Actually, the relativistic vorticity term is defined as
\begin{equation}
\Delta_{\mu\nu} = P^{\shortparallel}_{\mu\lambda}P^{\shortparallel}_{\nu\tau} \nabla^{\left[\lambda\right.}u^{\left.\tau\right]} \ .
\end{equation}
In our case the second spatial projector is actually an orthogonal projector, that in fact, mixes space and time components as a result of the supersymmetry.
$\bB$ may be seen as a ``super-correction'' to fluid velocity. However, a deeper analysis is due.

The second order reads
\begin{align}	
\mathcal{T}^{\left[ 2 \right]}_{\mu\nu} &= -\frac{1}{4} P^{\bot}_{\mu\nu} \left\{ \left(x \bB \right)^{\bot} \left(x \bB \right)^{\shortparallel}+ \frac{1}{12} x^2 \left(\bB \bB\right)^{\bot}+2 \left(\bB \cdot x\right) \left[ \left(x \bB \right)^{\bot}  + \frac{11}{12} \left(x \bB \right)^{\shortparallel}\right]  \right\} + \nonumber\\
&+ \frac{1}{4} P^{\shortparallel}_{\mu\nu} \left\{15 \left(x \bB \right)^{\bot} \left(x \bB \right)^{\shortparallel}+ 2 x^2 \left(\bB \bB\right)^{\bot}- \frac{1}{6} \left(\bB \cdot x\right) \left(x \bB \right)^{\shortparallel} + \frac{7}{12} x^2 \left(\bB \bB \right)^{\shortparallel}  \right\} \nonumber \\
&+ \frac{1}{2} \left( \frac{d}{4} \right)^{2} \left(\bB \cdot x\right)^2 \left(\eta_{\mu\nu} + 4 u_{\mu} u_{\nu}\right) + \frac{1}{24} \left(\bB \bB \right)^{\shortparallel} x_\mu x_\nu - \frac{1}{12} \left(x \bB \right)^{\shortparallel} \bB_{\left(\mu\right.} x_{\left.\nu\right)} + \nonumber \\
&+ P^{\shortparallel}_{\left(\mu\right|\rho} P^{\bot}_{\left.\nu\right)\sigma} \left\{x^{\left[\rho\right.}\bB^{\left.\sigma\right]}\left[\bB \cdot x -\frac{1}{3} \left(x \bB \right)^{\bot}\right] + \frac{1}{36} x^{\rho}x^{\sigma} \left(\bB \bB \right)^{\bot} - \frac{1}{24} \bB^{\rho} \bB^{\sigma} x^2\right\}
\end{align}
with $d=4$ and
\begin{equation}
(VW)^{\bot}= P^{\bot}_{\mu\nu}V^{\mu}W^{\nu} \ , \qquad \qquad (VW)^{\shortparallel}= P^{\shortparallel}_{\mu\nu}V^{\mu}W^{\nu} \ ,
\end{equation}
and we have used Fierz identities to substitute
\begin{equation}
\left(\bB\cdot x\right)^2 = \left[2\left(x\bB\right)^{\bot}\left(x\bB\right)^{\shortparallel} +\frac{1}{3} x^2 \left(\bB \bB\right)^{\bot}\right] \ .
\end{equation}
We can analyse the coefficient associated to the tensor  $\left( 4 u_{\mu}u_{\nu} +\eta_{\mu\nu} \right)$. For the perfect fluid, this coefficient is related to temperature $T$
\begin{align}
	T_{\mu\nu} \propto T^{d} \left( 4 u_{\mu}u_{\nu} +\eta_{\mu\nu} \right)
	\ .
	\label{BYmho00}
\end{align}
We have
\begin{align}
	-\frac{\mu}{2}\left[ 1 + \frac{d}{4} \lambda\left(\bB\cdot x\right) + \frac{1}{2} \left( \frac{d}{4}  \right)^{2} \lambda^{2} \left(\bB\cdot x\right)^{2} \right]
	\propto
	T^{d} \, \textrm{exp}\left[\frac{d}{4}\lambda \left( \bB \cdot x \right)\right]
	\ ,
	\label{BYmnho1}
\end{align}
where we reconstructed the series in the bilinears $\bB$. Doing this, the temperature of the fluid is modified as follows
\begin{align}
	T \longrightarrow T ~ \textrm{exp}\left[\frac{1}{4}\lambda \left( \bB \cdot x \right)\right]
	\ .
	\label{BYmho3}
\end{align}

\subsection{Stress--Energy Tensor for $AdS_4$}

The computation for the $AdS_{4}$ case is similar to the previous one. We consider the perturbative expansion
\begin{align}
	T_{\mu\nu} = -\frac{\mu}{2}\left( 3 u_{\mu}u_{\nu} +\eta_{\mu\nu} \right) + \lambda\mu\mathcal{T}^{\left[ 1 \right]}_{\mu\nu} + \lambda^{2}\mu\mathcal{T}^{\left[ 2 \right]}_{\mu\nu}
	\ ,
	\label{eta1LargeRTmunu4dim}
\end{align}
where we have defined $\mathcal{T}^{\left[ 1 \right]}$ as before and 
\begin{align}	
\mathcal{T}^{\left[ 2 \right]}_{\mu\nu} &=- \frac{1}{8}  P^{\bot}_{\mu\nu} \left\{\frac{37}{8} \left(x \bB \right)^{\bot} \left(x \bB \right)^{\shortparallel} + x^2 \left(\bB \bB\right)^{\bot} + \frac{1}{16} \left(\bB \cdot x\right) \left[9 \left(x \bB \right)^{\bot}  + 5 \left(x \bB \right)^{\shortparallel}\right] \right\} + \nonumber\\
&+\frac{1}{2} P^{\shortparallel}_{\mu\nu} \left\{\frac{7}{4} \left(x \bB \right)^{\bot} \left(x \bB \right)^{\shortparallel}+ \frac{7}{64} x^2 \left(\bB \bB\right)^{\bot}+  \left(\bB \cdot x\right)^{2}  \right\} \nonumber \\
&+ \frac{1}{2} \left( \frac{d}{4} \right)^{2}\left(\bB \cdot x\right)^2 \left(\eta_{\mu\nu} + 3 u_{\mu} u_{\nu}\right) + \frac{1}{64} \left(\bB \bB \right)^{\bot}x_\mu x_\nu - \frac{1}{16} \left(x \bB \right)^{\shortparallel} \bB_{\left(\mu\right.} x_{\left.\nu\right)} + \nonumber \\
&+ P^{\shortparallel}_{\left(\mu\right|\rho} P^{\bot}_{\left.\nu\right)\sigma} \left\{x^{\left[\rho\right.}\bB^{\left.\sigma\right]}\left[\frac{5}{8}\bB \cdot x -\frac{1}{4} \left(x \bB \right)^{\bot}\right] + \frac{3}{64} x^{\rho}x^{\sigma} \left(\bB \bB \right)^{\bot} - \frac{1}{32} \bB^{\rho} \bB^{\sigma} x^2\right\}
\ ,
\end{align}
with $d=3$.


\chapter{Supersymmetric Fluid Dynamics from Lagrangian}
\label{chFlu}

In the previous sections we dealt with two method to derive the fluid equations of motion at first postulating a suitable energy momentum tensor and then by  fluid/gravity correspondence.
In addition to them, if we are interested in non--dissipative fluids only, it is possible to introduce a Lagrangian and derive from it the equations of motion, which in particular will result in the Euler equations.
In the following we will address to Euler equations as Navier--Stokes equations, since the former are a particular case of the latter, as we explain in sec.~\ref{secFlu}.

Since the Lagrangian method is not suited to study viscous effects, it is not useful for ordinary fluid dynamics.
However, it can be used to derive the ``perfect'' supersymmetric fluid dynamics.
This could offer a different point of view to describe the interaction between fermionic bilinears and hydrodynamics quantities.

The construction of supersymmetric Lagrangian leading to Navier-Stokes equations has been discussed in the literature. 
We have to recall works  \cite{Nyawelo:2003bv,Nyawelo:2003bw} where a possible supersymmetric action has been proposed. There, the bosonic degrees of freedom are parametrized by a conserved current $j^\mu$, the dynamics is encoded into a function $f(j^2)$ and 
the corresponding equations of motion are obtained with the help of an auxiliary 
field $a_\mu$ coupled to the current.  In such supersymmetric generalization, both the current $j^\mu$ and the auxiliary field $a_\mu$ are embedded into two distinct real superfields, $V$ and $A$, whose 
lowest components are two scalar fields. The function $f(j^2)$  is replaced by a function $F(V)$ of the 
real superfield $V$. Expanding the action, we find that it cannot describe a generic fluid whose 
dynamics is described by the function  $f(j^2)$, namely it does not reduce to any generic bosonic models, but only to specific ones. On the other side, works \cite{Hoppe:1993gz,Jackiw:2000cc,Jackiw:2000mm,Bergner:2001db,Jackiw:2004nm} lead to generic supersymmetric models in 
lower dimensions and we have not been able to adapt them to our scopes. That non-covariant approach 
in lower dimensions seems to be suitable to study the AdS/Condensed Matter correspondence. 

In contrast to  \cite{Nyawelo:2003bv,Nyawelo:2003bw}, we observed that the conserved current 
can be better viewed as the middle component of a real linear superfield $J$.  The 
linearity of that superfield implies the conservation of the current and it does not contain auxiliary 
fields  \cite{West:1990tg,Wess:1992cp,Gates:1983nr,Weinberg:2000cr}. To overcome the problem of describing a generic 
model reducing to any bosonic Navier-Strokes system, we constructed a derived superfield 
${\cal J}_\mu$ which is a linear, real vector superfield and a linear function of $J$.

As mentioned above, the equations of motion of the fluid, namely the Navier-Stokes equations, 
are derived with the help of an auxiliary field $a_\mu$.  
However, in \cite{Nyawelo:2003bv,Nyawelo:2003bw}, $a_\mu$ has been replaced  with a 
 K\"ahler potential implementing the so-called Clebsch paremetrization (see also  \cite{Jackiw:2004nm,Zyoshi:2010,clebsch} for a complete discussion 
on the Clebsch parametrization) in a very convenient way for supersymmetric generalizations. 
We show in detail that the two choices, namely the conventional 
Clebsch parametrization and the use of the K\"ahler potential, are indeed equivalent locally. 
The origin of that potential has to be traced out into supergravity models as advocated in  \cite{D'Auria:1988qm}, and for a forthcoming analysis in a generic supergravity background, 
we adopt it in the present work. It is worth mentioning also the discussion in \cite{Binetruy:2000zx}. Finally, in 
terms of $J$, of the derived superfield ${\cal J}_\mu$ we are able to provide a general action 
whose bosonic truncation lead to any generic bosonic fluid. 

One important issue is the dependence of the K\"ahler potential. 
We provide an argument showing that the choice of the K\"ahler potential does not affect 
the physics, but we are convinced that the implementation of local supersymmetry 
invariance coupling it to supergravity, might clarify this issue. 
 
We provide the complete 
Lagrangian by expanding the superfields in components and integrating over the $\theta$'s. Due to 
this expansion, the number of possible terms increases and the Lagrangian is really cumbersome. In order to grasp the meaning of it, we derive the superfield equations of motion and we compute their bosonic sector. The energy-momentum tensor for the Lagrangian restricted to the physical 
field $C$ (the lowest component of the superfield $J$) is computed and some considerations are proposed.

\section{Bosonic Lagrangian}
\label{bosonlag}

\subsection{Action and Equations of Motion}\label{bosact}
We first discuss the bosonic Lagrangian and we derive the equations of motion. The model is characterized 
by a divergenceless current $j^\mu$ and an auxiliary field $a_\mu$ coupled to a worldvolume metric $g_{\mu\nu}$. The gauge invariance under $a_\mu \rightarrow a_\mu + \p_\mu \lambda$ 
is guaranteed by the conservation of $j^\mu$. The model is considered in 4d. There are two ways to get the equations of motion: 
the first one is by computing the energy-momentum tensor $T^{\mu\nu}$ and  requiring the vanishing of its divergence. The second method is 
requiring the invariance of the action under certain isometries. 

Let the action be
\begin{equation}
  \label{bola}
  {\mathcal{L}} = \sqrt{-g} \left( j^\mu a_\mu + f(j^2)\right)\,, \quad \quad   j^2 = j^\mu j^\nu g_{\mu\nu}\,.
\end{equation}
Note that the equation of motion obtained by taking the functional derivative w.r.t. an unconstrained 
$a_\mu$ yields $j^\mu=0$. 
The function $f$ is completely generic.
Therefore, the correct equations of motion are obtained as follow: 
varying w.r.t. $j^\mu$ and $g_{\mu\nu}$ leads to 
\begin{equation}
a_\mu = -2 f'(j^2) j_\mu\,, 
\quad \quad 
T^{\mu\nu} = f'(j^2) \left( j^\mu j^\nu - g^{\mu\nu} j^2 \right) + \frac{1}{2} f(j^2) g^{\mu\nu}\,, 
\label{deltag}
\end{equation}
and the vanishing of the divergence of energy-momentum tensor implies
\begin{equation}
  \label{cons}
  \partial^\mu T_{\mu\nu} = 0 \longrightarrow  j^\mu [ f''(j^2) \left( j_\mu \partial_\nu j^2 - j_\nu \partial_\mu j^2 \right) + f'(j^2) \left( \partial_\nu j_\mu - \partial_\mu j_\nu \right) ] = 0\,.
\end{equation}
These are the usual NS equations which, together with the conservation of the current $j^\mu$, yield the complete information on the fluid dynamics. 

Since we are primarily interested into $AdS/CFT$ correspondence, we recall that the fluid on the dual side must be a conformal one. That forces $f\left( j^{2} \right)$ to be equal to $C \left( j^{2} \right)^{2/3}$, where $C$ is a constant. This can be obtained by imposing the tracelessness of $T^{\mu\nu}$ or by studying the dilatation properties of the action, assuming that $j^{\mu}$ has dimension $3$ in 4d.

Notice that equation eq. (\ref{cons}) can also be obtained in the following way: consider the field-strength associated to the abelian vector $a_\mu$,
$F_{\mu\nu} =  \left( \partial_\mu a_\nu - \partial_\nu a_\mu\right)$;
using the first of (\ref{deltag}) into $F$ and upon contraction with $j^\mu$ we get
\begin{equation}\label{EQMO}
j^\mu F_{\mu\nu} = \partial^\mu T_{\mu\nu} = 0 \,.
\end{equation}
It should be noticed that, in both ways, the auxiliary field $a_\mu$ drops off the equations. 

Equation (\ref{EQMO}) calls for an explanation. First of all, we observe that, being $j^\mu$ a divergenceless current, 
action (\ref{bola}) is invariant under the gauge symmetry $\delta a_\mu = \partial_\mu \lambda$. Lert us  perform an isometry transformation leaving the current $j^\mu$ invariant. 
In the form language, given ${\cal A} = a_\mu dx^\mu$, ${\cal J} = j^\mu \p_\mu$ and ${\cal X} = X^\mu \p_\mu$, we have
\begin{equation}\label{isozero}
{\cal L}_{\cal X}( {\cal A} ) = \iota_{\cal X} d {\cal A} + d ( \iota_{\cal X} {\cal A} )\,, 
\quad
{\cal L}_{\cal X}( {\cal J} ) = \Big[{\cal X}, {\cal J}\Big]=0\,,
\quad
 \end{equation}
$$
{\cal L}_{\cal X} ( g) = (\nabla_\mu X_\nu + \nabla_\nu X_\mu) dx^m \otimes dx^\n\,,
$$
 and in components 
\begin{equation}\label{iso}
\delta a_\mu = -F_{\mu\nu} X^\nu + \partial_{\mu} (a_\nu X^\nu)\,, \quad 
\delta j_\mu =0\,, \quad
\end{equation} 
$$
\delta g_{\mu\nu} = g_{\mu\rho}\partial_\nu X^\rho + g_{\nu\rho}\partial_\nu X^\rho + X^\rho \p_\rho g_{\m\n}=0 \,,
$$
where $X^\mu$ are the components of the Killing vector generating the isometry commuting 
with the current ${\cal J}$. 
Requiring the invariance of the action under such an isometry, one gets eqs.   (\ref{EQMO}). 

The condition 
$\delta j^\mu =0$ (if  $g_{\m\n} = \eta_{\m\n}$) 
can be reformulated as follows: given the vector field $X = X^\m \partial_\m$, the infinitesimal variation of $j^\mu$ can be expressed as
\begin{equation}\label{AWb}
\delta j^\mu = X^\n \partial_\n j^\m - j^\n \partial_\n X^\m  \,, \quad
\end{equation}
where the first term is a traslation parametrized by the coefficients $X^\nu$ and second term 
is a rotation with the parameter $\Lambda_{\m\n} = \half (\p_\m X_\n - \p_\n X_\m)$ due to 
Killing equation in (\ref{iso}). Condition (\ref{AWb}) can be rewritten as follows
\begin{equation}\label{AWc}
\Delta_X  j^\mu\equiv X^\n \p_\n j^\m = \Lambda^\m_{~\rho} \, j^\rho\,,
\end{equation}
which implies that the translation of the current $j^\m$ is compensated by a rotation. 
In the same way, the variation of $a_\mu$ can be cast in the form 
\begin{equation}\label{AWd}
\delta a_\mu = \Delta_X a_\mu + R_\mu^{~\nu} a_\nu \equiv 
 X^\n \p_\n a_\mu + \Lambda_{\mu}^{~\rho} a_\rho\,. 
\end{equation}
 Then, computing the variation of the action under a translation, we  have 
 \begin{eqnarray}\label{AWe}
\Delta_X S &=& \int  \Big( \Delta_X j^\mu a_\mu + j^\mu \Delta_X a_\mu + \Delta_X f(j^2)\Big)\nonumber \\
&=& \int  \Big( \Lambda_\mu^{~\nu} j^\mu a_\nu + j^\mu X^\nu \p_\nu a_\mu \Big)
\nonumber \\
&=& \int \Big( j^\mu \delta a_\mu \Big)
\nonumber \\
&=& \int  \Big(  j^\mu \left(- F_{\mu\nu} X^\nu + \p_\mu (a_\rho X^\rho) \right)\Big)\,. 
\end{eqnarray}
In the first line we have used eq.~(\ref{AWc}) and the Lorentz invariance of $f(j^2)$. From the second line 
to the third line, we have used the definition of the variation of the gauge potential $a_\mu$ under isometry (\ref{iso}) combined with a gauge variation. 
Thus, the second term vanishes because $j^\mu$ is divergenceless and from the first term, 
comparing with the definition of the energy-momentum tensor obtained by the N\"other theorem 
$\Delta_X S = \int  X^\mu \partial^\nu \, T_{\mu\nu}$, it yields 
\begin{equation}\label{AWf}
j^\mu F_{\mu\nu} = \partial^\mu T_{\mu\nu} =0\,.
\end{equation}
As a consistency condition, we must have $j^\nu \partial^\mu T_{\mu\nu} =0$, which can be easily verified using its explicit form (\ref{cons}). 

\subsection{Clebsch Parametrization of $a_\mu$}\label{boscl}

One may wonder why we adopt the above derivation of NS equations instead of computing directly the equations of motion by functional derivatives. 
Actually, it is possible to obtain them by means of variational principles, considering the auxiliary field $a_\m$ as parametrized by a set of potentials.
Moreover, since $a_{\mu}$ is an auxiliary field we have to avoid any non-trivial solution for it, then we impose the constraint 
\begin{equation}\label{CSa}
F\wedge F =0\,, 
\end{equation}
where $F = d\, {\cal A}$ (${\cal A} \equiv a_\mu dx^\mu$) which, in components, becomes 
$\e^{\mu\n\rho\sigma} F_{\mu\n} F_{\rho\sigma} =0$. This constraint is 
equivalent to ${\cal A} \wedge F = d \Omega$ where $\Omega$ is a generic 2-form. 
It can be easily shown \cite{Zyoshi:2010} that the most general solution in 4d to (\ref{CSa}) is 
\begin{equation}\label{CSb}
{\cal A} = d\lambda + \alpha\, d \beta\,,
\end{equation}
where $\lambda, \a$ and $\b$ are zero forms. This implies that $F = d\alpha \wedge d\beta$ and 
the constraint (\ref{CSa}) follows immediately. This means that out of the four components of $a_\m$ only 3 of them survive the constraint and inserting them in the Lagrangian ({\ref{bola}}) 
we get 
\begin{equation}
  \label{bolalaa}
  {\mathcal{L}} =\Big( j^\mu (\p_\mu\lambda + \alpha \p_\mu \beta) + f(j^2)\Big) \,.
\end{equation}
The equations of motion are 
\begin{eqnarray}\label{bolalab}
\partial_\mu j^\m&=&0\,, \nonumber \\
j^\mu \partial_\mu \b &=&0\,, \nonumber \\
j^\mu \partial_\mu \a &=&0\,, \nonumber \\
\p_\mu\lambda + \alpha \p_\mu \beta + 2 j_\mu f'(j^2)&=&0\,.
\end{eqnarray}
With simple algebraic manipulations, one derives NS equations (\ref{EQMO}). 

 There is another way to parameterize the solution of (\ref{CSa}). Introducing one complex field $\phi$ and a 
 real function $K(\phi, \bar\phi)$, consequently $a_\mu$ becomes
 \begin{equation}\label{CSd}
a_\mu = \p_\mu \lambda + i (\partial K \partial_\mu \phi - \bar\partial K \partial_\mu \bar\phi)\,.
\end{equation}
If $K$ is identified with a K\"ahler potential for the complex manifold spanned by $\phi$, the second term in 
$a_\mu$ is the K\"ahler connection. Computing the field strength $F$ we get 
\begin{equation}\label{CSe}
F = - 2 i\, \partial \bar\partial K d \phi \wedge d\bar\phi\,.
\end{equation}
Namely, the manifold is a Hodge manifold where the $U(1)$ connection is related to the canonical 
$2$-form of the complex manifold. By the Bianchi identity, it follows that the canonical $2$-form 
$2 i\, \partial \bar\partial K d \phi \wedge d\bar\phi$, must be closed and therefore the space is K\"ahler. Notice that for a 
one dimensional complex manifold, no constraint on $K$ is due to its closure. 

The two parametrizations (\ref{CSb}) and (\ref{CSd}) are equivalent. This can be verified by 
assuming that $\alpha$ and $\beta$ are real functions of $\phi$ and $\bar\phi$. It yields 
\begin{equation}\label{CSf}
\alpha \partial \beta = i \partial K\,, \quad\quad
\alpha \bar\partial \beta = - i \bar\partial K\,.
\end{equation}
By dividing both equations by $\alpha$ and by computing the derivative we get 
\begin{equation}\label{CSg}
2 \partial \bar\partial K = (\partial K \bar\partial +  \bar\partial K \partial) \ln \alpha\,.
\end{equation}
This equation can be brought to quadrature. 
For example, assuming that $\alpha$ and $K$  are functions of the modulus $|\phi|^2 $, one can easily bring the above equation to an integral form. If 
$K(\phi, \bar\phi) = |\phi|^2$, then we get $\alpha = |\phi|^2$ and $\beta = i \ln (\phi/ \bar \phi)$. On the other hand, if $K(\phi, \bar\phi) = \ln (1 +|\phi|^2)$, then we get 
$\alpha =|\phi|^2/ (1+ |\phi|^2)$ and $\beta = i \ln (\phi/ \bar \phi)$. See also \cite{Jackiw:2004nm} for a 
discussion on this point.


\section{Supersymmetric Lagrangian}
\label{complete}

\subsection{Superfields, Action and Superfield Expansion}

We are now ready for the supersymmetrized version of the Lagrangian. We first construct the action  reproducing the usual bosonic action (\ref{bola}) in the limit in which the fermions and the additional bosonic field are set to zero. 
A conserved current is a component of a linear multiplet in 4d and therefore we introduce a superfield $J$ for it. 
The auxiliary field $a_\mu$ is a component of the vector multiplet and we introduce a real superfield $A$.
Again we face with the problem of deriving the equations of motion since the superfield $A$ is constrained and, 
for that, we adopt a Clebsch parameterization. In the present case, it becomes natural to identify the abelian 
real superfield $A$ with a K\"ahler potential \cite{Nyawelo:2003bv} which is a real function of a chiral superfield $\phi$. 

$J$ and $A$ are defined as follows\footnote{In the following we use Weinberg notation
\cite{Weinberg:2000cr}. Nevertheless, we recall that in the language of \cite{Gates:1983nr} a linear superfield is defined as $D^2 J = 0$ and $\bar D^2 \bar J =0$. If $J$ is a real linear superfield, $\bar J = J$ then the second condition follows from the first one.} 
\begin{equation}\label{LS}
\bar D D J = 0\,,  \hspace{2cm} \bar A = A\,,
\end{equation}
where $D = - \frac{\partial}{\partial \bar \theta} - (\gamma^\mu \theta) \partial_\mu$ and 
$\bar D = \frac{\partial}{\partial \theta} + (\gamma^\mu \bar\theta) \partial_\mu$  are the superderivatives. 
Using a linear superfield $J$, we automatically implement the conservation of the current $j^\mu$ which  is 
its $\theta^2$ component. The component expansion is given by 
\begin{equation}
J = C - i \bar \theta \gamma_5 \omega + \frac{i}{2} \bar \theta \gamma_5 \gamma_\mu \theta j^\mu + \frac{i}{2} \bar \theta \gamma_5 \theta \bar \theta \gamma^\mu \partial_\mu \omega + \frac{1}{8} (\bar \theta \gamma_5 \theta)^2 \square C,\label{eq:Jdef}
\end{equation}
and for the real superfield  in the Wess-Zumino gauge 
\begin{equation}
A = \frac{i}{2} \bar \theta \gamma_5 \gamma^\mu \theta a_\mu - i \bar \theta \gamma_5 \theta \bar \theta \lambda - \frac{1}{4} (\bar \theta \gamma_5 \theta)^2 D.\label{eq:Adef}
\end{equation}
The linear superfield contains one constrained vector $j^\m$, one scalar field $C$ and one Majorana spinor $\omega$. 
The vector can be dualized as $j^\mu = \epsilon^{\mu\nu\rho\sigma} H_{\nu\rho\sigma}$ where 
$H_{\mu\nu\rho}$ is the field strength of a 2-form potential $B_{\mu\nu}$. The latter can be further dualized into a scalar and therefore the linear multiplet has the same d.o.f. of an on-shell Wess-Zumino multiplet. 

Supersymmetry transformations are given by $\delta \Phi = \bar\alpha Q \Phi$ or, in component
\begin{align}\label{susy}
\delta j^\mu &= - \bar \alpha \gamma^{\mu\nu} \partial_\nu \omega\,, & \delta a_\mu &= \bar \alpha \gamma_\mu \lambda\,,  \nonumber\\
\delta \omega &= \left( - i \gamma_5 \gamma^\mu \partial_\mu C + \gamma_\mu j^\mu \right) \alpha\,, &
\delta \lambda &= - \left( i \, D \gamma_5 + F_{\mu\nu} \gamma^{\mu\nu} \right) \alpha\,,  \nonumber \\
\delta C &= i \, \bar \alpha \gamma_5 \omega\,, & \delta D &= i \, \bar \alpha \gamma_5 \gamma^\mu \partial_\mu \lambda\,.  \nonumber
\end{align}

Using the properties listed in the Appendix A it is possible to show that
\begin{equation}
\int d^4 x \int d^4 \theta [-JA] = \int d^4 x [j^\mu a_\mu + \bar \omega \lambda - CD]\,,
\end{equation}
which is the supersymmetric generalization of (\ref{bola}). In order to reproduce 
also the second term in (\ref{bola}), we need to introduce a new superfield defined as
\begin{equation}\label{newsuss}
{\cal J}_\mu =  \frac{1}{4 \, i}  (\bar D \gamma_5 \gamma_\mu D)J \,,
\end{equation}
which contains $j^\mu$ as the first component and its expansion is 
\begin{eqnarray}\label{newEXA} 
{\cal J}_\mu = \frac{1}{4 \, i}  (\bar D \gamma_5 \gamma_\mu D)J &=& j_\mu + \bar\theta \gamma_{\mu\nu} \partial^\n \om - \frac{i}{2} \bar\theta \g_5 \g^\n \theta \left( \p_\m \p_\n C - g_{\m\n} \square C \right) +\nonumber \\
&-& \half \bar\theta \g_5 \theta \bar\theta \g_5 \g^\nu (g_{\mu\n} \square \om - \p_\m \p_\n \omega) + \frac{1}{8} (\bar\theta \g_5 \theta)^2 \square j_\mu\,.
\end{eqnarray}
It should be noted that all the terms in the above expansion are divergenceless. This can also be 
proven directly by the $D$-algebra and because of the linearity of the superfield $J$. Moreover, 
the new superfield ${\cal J}_\mu$ is itself a linear superfield. This can be seen by observing that 
each component of the superfield ${\cal J}_\mu$ is in the same relation with higher terms of the expansion 
as the components of the superfield $J$, or it can be checked by direct use of superderivatives.
 
Therefore the complete supersymmetric action is given by 
\begin{equation}\label{susyLAG}
S = \int d^4x \int d^4 \theta \Big( - J A +  F({\cal J}_\mu {\cal J}^\mu) \, J^2  \Big) \,.
\end{equation}
The minus sign in front of the first term is choosen to reproduce the normalization of the bosonic Lagrangian. 
The coefficients are chosen in order that eq. (\ref{susyLAG}) coincides with the normalization of the bosonic Lagrangian where $f(j^2) = F(j^2) j^2$. 
 The argument of $F$, namely ${\cal J}_\mu {\cal J}^\mu$, is a dimensionful superfield and therefore it would be convenient to rescale it by a 
 dimensionful parameter. In the following, we will discard that parameter and we set it to 1. 

As discussed above, we would like to deal with superconformal fluid. 
For that, we require the theory to be conformal and supersymmetric, thus superconformal invariance follows. 
In particular, we first impose the dilatation properties of $F$ and it turns out  that $F\left( x \right)=C x^{-1/3}$.
That guarantees the conformal invariance of the action. The superconformal transformation rules for $J$ are deduced by its geometrical properties.

 To compute the component action, we need the expansion of 
 ${\cal J}_\mu {\cal J}^\mu$ and, using (\ref{newEXA}) we get 
 \begin{eqnarray}\label{EXAcalJ2}
{\cal J}_\mu {\cal J}^\mu 
	&=& j^2 
	+ 2 \, \bar\theta j_\m \g^{\m\n} \partial_\n \omega 
        + \bar \theta \theta \left( - \half \p_\m \bar \om \g^{\m\n} \p_\n \om - \frac{3}{4} \p_\m \bar \om \p^\m \om \right) + \nonumber \\
	&+& (\bar\theta \g_5 \theta) \left( - \half \p_\m \bar \om \g_5 \g^{\m\n} \p_\n \om - \frac{3}{4} \p_\m \bar \om \g_5 \p^\m \om \right) + \nonumber \\
        &+& \bar\theta \g_5 \g^\m \theta \left( i j_\m \square C - i j\cdot \p \p_\m C + \p_\m \bar \om \g_5 \not\!\p \om - \frac{1}{4} \p^\n \bar \om \g_5 \g_\m \p_\n \om \right) + \nonumber \\
	&+& (\bar\theta \g_5 \theta) \bar\theta \g_5 \left( j \cdot \p \not\!\p \om - j \cdot \g \square \om \right) + \bar\theta \g_5 \theta \bar\theta \left(2 i \not\!\p \omega \square C + i \g^\m \p^\n \om \p_\m \p_\n C \right) \nonumber \\   
&+& \frac{1}{4} (\bar\theta \g_5 \theta)^2 \left( j_\m \square j^\m + \p_\m \p_\n C \p^\m \p^\n C + 2 \square C \square C + \right. \nonumber \\
&+& \left. \p_\m \bar \omega \p^\m \omega + \p^\m \bar \omega \g_{\m\n} \p^\n \omega -2 \square \bar \omega \not\!\p \om \right) \,, 
\end{eqnarray}
similarly, for $J^2$, we have
 \begin{eqnarray}\label{EXAJ2}
J^2 &=& C^2 - 2 \, i \, C \bar\theta \g_5 \om + \nonumber \\
&+& \frac{1}{4}(\bar\theta \theta) \bar\om\om + \frac{1}{4} (\bar\theta  \g_5 \theta) \bar\om  \g_5 \om +  \bar\theta \g_5 \g_\m \theta ( i \, C j^\mu  + \frac{1}{4} \bar\om \g_5 \g^\m \om)  + \nonumber \\
&+& i \, \bar\theta  \g_5 \theta \bar\theta\!\!\not\!\partial \omega C + \,  \bar\theta  \g_5 \g_\mu \theta \bar\theta  \g_5 \om j^\m \nonumber \\
&+& \frac{1}{4} (\bar\theta \g_5 \theta)^2 ( C \square C + j^2 - \bar\om \!\!\not\! \p\om) \,. 
 \end{eqnarray}

Notice that the choice $f(j^2) = F(j^2) j^2$ does not spoil the generality of (\ref{susyLAG}) 
since it coincides with bosonic Lagrangian if $f(j^2)$ is defined up to an unessential constant. 
Action (\ref{susyLAG}) is chosen such that, by setting $C$ and $\omega$ to zero, it exactly reproduces the bosonic 
Lagrangian (\ref{bola}) and the corresponding NS equations. The presence of two different superfields, namely $J$ and 
${\cal J}_\mu$ in the Lagrangian is needed because of dimensional reasons or, equivalently, because $J$'s lowest component is not $j^\mu$. 

In components the supersymmetric Lagrangian turns out to be
\begin{align}
  \int d^4 x \left[j^\mu a_\mu + \bar \omega \lambda - CD + \int d^4 \theta J^2  \sum_{i=0}^4 \frac{1}{i!}F^{(i)}(j^2) ({\cal J}_\mu 
  {\cal J}^\mu - j^2)^i  \right]\,, \label{ssl}
\end{align}
where we expanded the function $F$ around the first bosonic component of ${\cal J}_\mu {\cal J}^\mu$. 
The first term in the expansion reproduces the bosonic Lagrangian, while the other terms 
are classified according to their dimensions. Notice that the computation of the component action 
is made unhandy by the fact that there is a product of two or more superfields $({\cal J}_\mu 
  {\cal J}^\mu - j^2)^i J^2$. After the $\theta$-expansion is taken, one needs to compute all Fierz identities to simplify 
  the expressions and, finally, the integration over the $\theta$ variables can be taken. 

The first two terms in the expansion of $FJ^2$ are
\begin{align}
&  \int d^4\theta \int d^4 x  \left[ F^{(0)}(j^2) J^2  + F^{(1)}(j^2) ({\cal J}_\mu {\cal J}^\mu - j^2) J^2  \right] = \nonumber\\
& = \int d^4 x  \Big\{\left[ F^{(0)}(j^2) (C \square C + j_\mu j^\mu - \bar\omega \gamma_\mu \p^\mu \omega) \right] + \nonumber \\
& + \Big[ F^{(1)}(j^2) 
\Big( 
 - C^2 [ j_\m \square j^\m + (\p_\m \p_\n C \p^\m\p^\n C + 2 \square C \square C )]  + 4 C j^\m j^\n \left(\p_\m \p_\n - \eta_{\m\n} \square \right) C + \nonumber\\
& - C^2  \p_\m \bar \omega \p^\m {\not\!\p} \omega 
+ 2 C^2 \square \bar \omega {\not\!\p} \omega - 2 i C j^\m \p_\m \bar \omega \g_5 {\not\!\p} \omega  - i C j^\m \p_\n \bar \omega \g_5 \g_\m \p^\n \omega + 4 C \square C \bar \omega {\not\!\p} \omega + \nonumber\\
& +2 C \p_\m\p_\n C \bar \omega \g^\m \p^\n \omega + 2 C j_\m \p_\n \bar \omega \g_\rho \p_\s \omega \varepsilon^{\m\n\rho\s}  - 2 i C j^\m \bar \omega \g_5 \p_\m {\not\!\p} \omega  + 2 i C j^\m \bar \omega \g_5 \g_\m \square \omega  
+ 2 j^2 \bar \omega {\not\!\p} \omega 
+ \nonumber\\
&- 2 j^\m j^\n \bar \omega \g_\m \p_\n \omega - i j^\m \left(\p_\m \p_\n - \eta_{\m\n} \square \right) C \bar \omega \g_5 \g^\n \omega 
- \frac{3}{4} \bar \omega \omega \p_\m \bar \omega \p^\m \omega  - \frac{1}{2} \bar \omega \omega \p_\m \bar \omega \g^{\m\n} \p_\n \omega + \nonumber\\
&+ \frac{3}{4} \bar \omega \g_5 \omega \p_\m \bar \omega \g_5 \p^\m \omega 
+ \frac{1}{2} \bar \omega \g_5 \omega \p_\m \bar \omega \g_5 \g^{\m\n} \p_\n \omega - \bar \omega \g_5 \g^\m \omega \p_\m \bar \omega \g_5 {\not\!\p} \omega 
+ \frac{1}{4} \bar \omega \g_5 \g_\m \omega \p_\n \bar \omega \g_5 \g^\m \p^\n \omega
 \Big) \Big]\Big\} \,.\label{18June1}
 \end{align}
As can be seen from this expression, they contain the interaction between the current $j^\mu$ and the 
fields $C$ and $\omega$. The part proportional to $F^{(1)}$ contains terms with four fields $\omega$ and therefore their 
self-interactions. In the forthcoming section, we will discuss the implications of those terms. Even though the action might seem 
bulky, it is a good starting point for the perturbation theory since the expansion is done in terms of higher derivative terms. 

Since the resulting action is rather cumbersome, we find convenient also to provide its bosonic truncation 
\begin{align}\label{bostrunch}
& \int d^4 x \left[ j^\mu a_\mu - CD + F^{(0)}(j^2) (C \square C + j_\mu j^\mu) \right] + \nonumber \\
& + \int d^4 x  \Big[ F^{(1)}(j^2) \Big( - C^2 j_\m \square j^\m - C^2 \left( \p_\m \p_\n - g_{\m\n} \square \right) C \left(\p^\m \p^\n - g^{\m\n} \square \right) C + \nonumber \\
& + 4 C j^\m j^\n \left(\p_\m \p_\n - g_{\m\n} \square \right) C \Big) \Big] + \nonumber \\
& + \int d^4 x  \Big[ F^{(2)}(j^2) \Big( - 4 C^2 j^\m j^\n \left( \p_\m \p_\rho - g_{\m\rho} \square \right) C \left(\p_\n \p^\rho - \d_\n^\rho \square \right) C \Big) \Big]\,.
  \end{align}
The bosonic action truncates at the second order in $F$, since all other terms are purely fermionic. This is due to the fact that 
in the expansion of the third power and of the fourth power, only those terms with a single $\theta$ contribute to the expansion since 
we have decided to expand around $j^\mu$. This simplifies the derivation of the energy-momentum tensor for the bosonic sector as we are going to 
discuss in the forthcoming section. In appendix B, all other terms are given. 


\subsection{Clebsch Parametrization for the Supersymmetric Case}
\label{susyclebsch}

We discuss here the Clebsch parameterization for the supersymmetric case. Here, 
the gauge field $a_\mu$ is replaced by the real superfield 
$A$ and therefore we have to parametrize it using a Clebsch parametrization as above. 
As suggested in \cite{Binetruy:2000zx} and in \cite{Nyawelo:2001fk}  we identify  
\begin{equation}\label{suCLEa}
A = \chi + \bar \chi + K(\phi, \bar \phi)\,,
\end{equation}
where $\chi, \phi$ and $\bar\chi, \bar\phi$ are chiral and anti-chiral fields, respectively. $K(\phi, \bar\phi)$ 
is a K\"ahler potential represented by a real function of the superfields $\phi$ and $\bar\phi$. 
The condition for the complex manifold spanned by $\phi$ and $\bar\phi$ to be K\"ahler is $d {\cal K}=0$, where ${\cal K}$ is the canonical 2-form. Since the complex manifold is one dimensional, 
no interesting condition emerges from this constraint. 

The identification in (\ref{suCLEa}) implies that the Fayet-Ilioupoulos term induced by the abelian gauge field $A$ 
is given by 
\begin{equation}
S_{F-I} = \int d^4x d^4\theta A = \int d^4x d^4\theta K(\phi, \bar\phi) \,,
\end{equation}
and it generates the dynamical equations of motion for the chiral fields (see for example \cite{Binetruy:2000zx}). 
In our case, however, this term is replaced by 
\begin{equation}\label{suCLEb}
S = \int d^4x d^4\theta (-J A + \dots)  = \int d^4x d^4\theta (-J (K(\phi,\bar \phi) + \chi + \bar \chi) + \dots)\,. 
\end{equation}
So that a naive kinetic term for $\phi$ and $\bar\phi$ is absent, being replaced by the 
superfield expansion of $JK$. The chiral field $\chi$ and the antichiral field $\bar\chi$ implement the linearity condition on $J$. 

Let us now consider the first term of action (\ref{suCLEb}) which, after Berezin integration reads 

\begin{align}
S = \int d^4 x &\frac{1}{2} K(\varphi , \bar \varphi) \square C + \nonumber \\
& - \p K \left(i j^\m \p_\m \varphi + \frac{1}{2} C \square \varphi
- i \frac{\sqrt{2}}{2} \bar \psi_L {\not\!\p} \lambda 
+ i \frac{\sqrt{2}}{2} \bar \lambda {\not\!\p} \psi_L \right) + c.c. +
\nonumber\\
& - \frac{1}{2} \p^2 K \left( C \p_\m \varphi \p^\m \varphi
- \sqrt{2} i \p_\m \varphi \bar \psi_L \g^\m \lambda  \right) + c.c. +
\nonumber\\
&- \p \bar \p K \left( 2 |P|^2 C
- C \p_\m \varphi \p^\m \bar \varphi 
+ \sqrt{2} i P \bar \psi_R \lambda  
- \sqrt{2} i \bar P \bar \psi_{L} \lambda 
+ \right. \nonumber \\
&\left. - C \bar \psi_{L} {\not\!\p} \psi_{R}
- C \bar \psi_{R} {\not\!\p} \psi_{L}
+ i j^\m \bar \psi_{L} \g_\m \psi_{R} 
+ \frac{\sqrt{2}}{2} i \p_\m \bar \varphi \bar \psi_{L} \g^\m \lambda
- \frac{\sqrt{2}}{2} i \p_\m \varphi  \bar\psi_{R} \g^\m \lambda  \right)
\nonumber\\
&- \frac{1}{3} \p^2 \bar \p K \left( 
- 2 C \bar P \bar \psi_L \psi_{L}
+ 2 C \p_\m \varphi \bar \psi_{L} \g^\m \psi_{R} 
- \sqrt{2} i \bar \psi_{L} \psi_{L} \bar \psi_{R} \lambda  \right) + c.c. +
\nonumber\\
&- \frac{1}{2} \p^2 \bar \p^2 K C \bar \psi_{R} \psi_{R} \bar \psi_{L} \psi_{L}\,,
\end{align}
where the chiral and antichiral superfields $\phi$ (respectively $\bar\phi$) are defined by conditions
\begin{align}
&\frac{1-\g_5}{2} D \phi = 0\,, &\frac{1+\g_5}{2} D \bar\phi = 0\,,
\end{align}
and their components include a left-chiral spinor field $\psi_L = \left( \frac{1 + \g_5}{2} \right) \psi$ (respectively right-chiral $\psi_R$) and two scalar complex fields $\varphi$ and $P$ (respectively $\bar \varphi$ and $\bar P$).
The expression $Q_\m \equiv i \left( \p K \p_\m \varphi - \bar \p K \p_\m \bar \varphi \right) - i \p \bar \p K \bar \psi_{L} \g_\m \psi_{R} $ is known as the K\"ahler connection. Action (\ref{susyLAG}) contains a piece which depends upon the superfield $A$. Inserting the above expressions into (\ref{ssl}), we get 
an action which depends upon the components $\varphi, \psi_L$ and $F$ of the superfield $\phi$ (and their conjugated). Differentiation w.r.t. those fields, leads to the equations of motion. Truncating the action to its bosonic part, the first term in (\ref{susyLAG}) reads
\begin{align}
S = \int d^4 x & \left[ \frac{1}{2} K \square C - i j^\m \left( \p K \p_\m \varphi - \bar \p K \p_\m \bar \varphi \right) - \frac{1}{2} C \left( \p K \square \varphi + \bar \p K \square \bar \varphi \right) + \right. \nonumber \\
& - \left. \frac{1}{2} C \left( \p^2 K \p_\m \varphi \p^\m \varphi + \bar \p^2 K \p_\m \bar \varphi \p^\m \bar \varphi \right) - C \p \bar \p K \left( 2 |P|^2 - \p_\m \varphi \p^\m \bar \varphi \right) \right]\,,
\end{align}
where the K\"ahler potential $K$ is evaluated on $\varphi$ and its conjugated. 
Notice that, if we integrate by parts $K \square C$, the above expression considerably simplifies and becomes
\begin{align}
S = \int d^4 x \left[ i j^\m \left( \bar \p K \p_\m \varphi - \p K \p_\m \bar \varphi \right) - 2 C \p \bar \p K \left( |P|^2 - \p_\m \varphi \p^\m \bar \varphi \right) \right]\,.
\end{align}
The Lagrangian is diagonal in the auxiliary fields $P$, $\bar P$ and their equations of motion (at the lowest level) imply either $C=0$ (fluid dynamics approximation) or 
$P= \bar P=0$ (which is the supersymmetric dynamics). 

To compute the equations of motion we recall the expansion of $F$, given in (\ref{bostrunch}).
Varying w.r.t. $\varphi$ we get
\begin{equation}
	\begin{split}
		2\partial\bar\partial K \partial_{\mu} \bar\varphi\left( 
		i j^{\mu}-\partial^{\mu}C
		\right)
		-
		2C\partial\bar\partial K 
		\square \bar\varphi
		-
		2C\partial\bar\partial K
		\partial_{\mu}\bar\varphi\partial^{\mu}\bar\varphi
		=
		0
		\,.
	\end{split}
	\label{eomvarphi1}
\end{equation}
Analogously, we can get the equation of motion for $\bar\varphi$. The one for $j^{\mu}$ reads
\begin{equation}
	\begin{split}
	&
	i\left( 
	\bar\partial K \partial_{\mu}\bar\varphi
	-\partial K \partial_{\mu}\varphi
	\right)
	+
	2F^{\left( 1 \right)} j^{\mu}
	\left( 
	C\square C +j^{2}
	\right)
	+2F j_{\mu}+
	\\&
	-
	2F^{\left( 2 \right)} C^{2} j_{\mu} j_{\nu}\square j^{\nu}
	-
	\square\left( 
	F^{\left( 1 \right)} C^{2} j_{\mu}
	\right)
	+\\&
	-
	F^{\left( 1 \right)}
	C^{2} \square j_{\mu}
	+8 F^{\left( 1 \right)} j^{\nu}\left( 
	\partial_{\mu}\partial_{\nu}
	-g_{\mu\nu}\square
	\right) C
	-8 F^{\left( 3 \right)}C^{2} j_{\mu}
	+\\&
	-4F^{\left( 2 \right)} C^{2} j^{\nu}\left( 
	\partial_{\mu}\partial_{\rho}
	-g_{\mu\rho}\square
	\right) C
	\left( \partial_{\nu}\partial^{\rho} 
	-\delta_{\nu}^{\rho}\square
	\right)C
	= 0
	\,,
	\end{split}
	\label{eomjmu}
\end{equation}
finally, the one for $C$ is
\begin{equation}
	\begin{split}
		&
		2\partial\bar\partial K \partial_{\mu} \varphi \partial^{\mu}\bar\varphi
		+
		F\square C
		+
		\square F C
		-
		2F^{\left( 1 \right)} C j_{\mu}\square j^{\mu}
		+
		\\&
		+
		2 F^{\left( 1 \right)} \partial_{\mu}\partial_{\nu} C \partial^{\mu} \partial^{\nu} C
		+
		2
		\partial_{\mu}\partial_{\nu} \left( 
		F^{\left( 1 \right)}C^{2} \partial^{\mu} \partial^{\nu} C
		\right)
		+
		\\&
		+
		4F^{\left( 1 \right)} C \left( \square  C \right)^{2}
		+
		4\square \left( 
		F^{\left( 1 \right)}C^{2}\square C
		\right)
		+4 F^{\left( 1 \right)} j^{\mu} j^{\nu} 
		\left( \partial_{\mu}\partial_{\nu} -g_{\mu\nu} \square \right)
		 C
		+
		\\&
		+
		4 
		\left( \partial_{\mu}\partial_{\nu} -g_{\mu\nu} \square \right)
		\left( 
		F^{\left( 1 \right)} j^{\mu}j^{\nu} C
		\right) 
		+
		\\&
		-
		4 F^{\left( 2 \right)} C j^{\mu} j^{\nu} 
		\left( \partial_{\mu}\partial_{\rho} -g_{\mu\rho} \square \right)
		C 
		\left( \partial_{\nu}\partial^{\rho} -\delta_{\nu}^{\rho} \square \right)
		C
		+
		\\
		&
		-
		4 
		\left( \partial_{\mu}\partial_{\rho} -g_{\mu\rho} \square \right)
		\left( 
		F^{\left( 2 \right)} C^{2} j^{\mu} j^{\nu} 
		\left( \partial_{\nu}\partial^{\rho} -\delta_{\nu}^{\rho} \square \right)
		C
		\right)
		=
		0
		\,.
	\end{split}
	\label{eomC}
\end{equation}


\subsection{Superfield Equations}

Action (\ref{susyLAG}) is written in terms of 
a linear superfield $J$ and a real superfield $A$. For those superfields, the usual 
functional derivative cannot be used and therefore we cannot obtain the equations of motion 
by usual means (see \cite{Gates:1983nr} for a complete discussion). 
To overcome such a problem, we add two auxiliary generic superfields $Z, S^\mu$, one chiral superfield $\chi$ 
and one antichiral superfield $\bar\chi$. 

The following action
\begin{eqnarray}\label{SFa}
S &=& \int d^4xd^4\theta \left( -J ( A + \chi + \bar \chi) \Big) +  
F\Big[ \Big(\frac{1}{4i}(\bar D \gamma_5 \gamma_\mu D) J\Big)^2\Big] J^2 \right)
\nonumber \\
&=& \int d^4xd^4\theta \left( -J ( A + \chi + \bar \chi) \Big) +
F[ {\cal J}^2] J^2 + 
S^\mu \Big[ \frac{1}{4i} (\bar D \gamma_5 \gamma_\mu D) J - {\cal J}_\mu\Big]  \right),
\end{eqnarray}
turns out to be equivalent to (\ref{susyLAG}). The chiral and antichiral superfields $\chi, \bar\chi$ impose the linearity condition on the superfield $J$. 

As already discussed above, in order to get the correct equations of motion, we replace the superfield $A$ with the K\"ahler 
potential. Then, we have 
\begin{equation}\label{KactA}
S_K= \int d^4xd^4\theta \left( - J ( K(\phi, \bar\phi) + \chi + \bar \chi) + 
F[ {\cal J}^2] J^2 + S^\mu \Big[ \frac{1}{4i} (\bar D \gamma_5 \gamma_\mu D) J - {\cal J}_\mu\Big]  \right)\,,
\end{equation}
from which we can get the equations of motion by taking the functional (unconstrained) derivatives with 
respect to superfields $J, \phi, \bar\phi, S^\mu, \chi, \bar\chi$ to get
\begin{eqnarray}\label{KactB}
\bar D D J &=&0\,, \nonumber \\
{\cal J}_\mu - \frac{1}{4i} (\bar D \gamma_5 \gamma_\mu D) J &=&0\,, \nonumber \\
S^\mu + 2 {\cal J}^\mu F'[ {\cal J}^2] J^2 &=&0\,, \nonumber \\
\bar D D( J \frac{\partial K}{\partial \phi} ) &=& 0\,, \nonumber \\
\bar D D( J \frac{\partial K}{\partial \bar\phi} ) &=& 0\,, \nonumber \\
K(\phi, \bar\phi) + \chi +\bar \chi - 2 J\, F[{\cal J}^2] - \frac{1}{4i} (\bar D \gamma_5 \gamma_\mu D) S^\mu&=&0\,.
\end{eqnarray}
 
 To study the above equations, we proceed as follows. The first eq. in (\ref{KactB}) implies the 
linearity of $J$ (and therefore its $\theta$ expansion is given by (\ref{eq:Jdef})). Then, we plug $J$ into the 
second equation f
to compute the vector superfield ${\cal J}_\mu$. Subsequently, we plug ${\cal J}_\mu$ 
into the third equation to evaluate $S^\mu$ and finally, using all those results, we can express $K$ in terms of the superfields $\phi$ and $\bar\phi$. 
Given that, eqs. (\ref{KactB}) become the new NS equations, written in terms 
of the linear superfield $J$ which contains the physical degrees of freedom of the 
super-fluid.  

\subsection{Bosonic Sector}

In the present section, we study the model by setting to zero the fermions. We first write the Lagrangian as a function 
of the fields $j^\mu$ and $C$ and then we provide a  new Lagrangian with new auxiliary fields which simplifies the 
derivation of the energy-momentum tensor.  

The bosonic part of the Lagrangian is (up to a factor $\sqrt{-g}$)
\begin{equation}\label{24JuneLbos}
\begin{split}
	\mathcal{L}_{bos}
	&=
	j^{\mu}a_{\mu} - C\,D + 
	\\&
	+ F^{(0)}\left( j^{2} \right)\left( C\square C +j_{\mu}j^{\mu}\right) +
	\\&+F^{(1)}\left( j^{2} \right)
	\left[ 
	-C^{2}j_{\mu}\square j^{\mu}
	+ C^{2}\partial_{\mu}\partial_{\nu}C\partial^{\mu}\partial^{\nu}C\right.
	\left.+2C^{2}\left( \square C \right)^{2}
	+4j^{\mu}j^{\nu} C \left( \partial_{\mu}\partial_{\nu}-g_{\mu\nu}\square \right)C
	\right]+
	\\&+
	\frac{1}{2}F^{(2)}\left( j^{2} \right)
	\left[ 
	-4C^{2}j^{\mu}j^{\nu}\left(  \partial_{\mu}\partial_{\rho}-g_{\mu\rho}\square \right)C
	\left(  \partial^{\rho}\partial_{\nu}-\delta^\rho_{\nu}\square \right)C
	\right]\,.
\end{split}
	\end{equation}
We define the quadratic differential operator
\begin{eqnarray}
	M_{\mu\nu} =  \partial_{\mu}\partial_{\nu}-g_{\mu\nu}\square\,, \quad\quad 
	\partial^\mu M_{\mu\nu} = 0\,, \quad\quad \square = - \frac{1}{3} g^{\mu\nu} M_{\mu\nu}\,. 
	\label{24JuneMmunu}
\end{eqnarray}
and we rewrite (\ref{24JuneLbos}) with the Lagrangian multiplier $S^{\mu\nu}$
\begin{equation}
	\begin{split}
		\mathcal{L}_{bos}
	&=
	j^{\mu}a_{\mu} - C\,D+
	\\&+
	F^{(0)}\left( j^{2} \right)\left( 
	-\frac{1}{3}g^{\mu\nu}B_{\mu\nu} C +j_{\mu}j^{\mu}
	\right)+ 
	\\&
	+F^{(1)}\left( j^{2} \right)
	\left[ 
	-C^{2}j_{\mu}\square j^{\mu}
	+ C^{2} B_{\mu\nu}B_{\rho\sigma}g^{\mu\rho}g^{\nu\sigma}
	+4 C j^{\mu}j^{\nu}B_{\mu\nu}
	\right]+
	\\&+
	\frac{1}{2}F^{(2)}\left( j^{2} \right)
	\left[ 
	-4C^{2}j^{\mu}j^{\nu}B_{\mu\rho}B_{\nu\sigma} g^{\rho\sigma}
	\right]+
	\\&+
	S^{\mu\nu}\left( 
	B_{\mu\nu}-M_{\mu\nu}C
	\right)
	\,.
	\end{split}
	\label{24JuneLbos2}
\end{equation}
In this way, we restrict the covariantization of the differential operator $M_{\mu\nu}$ in a single term and the derivation 
of the energy-momentum tensor is greatly simplified. 
We now compute the equations of motion for $C$, $B_{\mu\nu}$ and $j^{\mu}$ respectively
\begin{equation}
	\begin{split}
	D=&
	-2F^{(1)}C+
	2F^{(1)}B_{\mu\nu}B^{\mu\nu}
	+
	4 F^{(1)}j^{\mu}j^{\nu}B_{\mu\nu}
	+\\&-
	4F^{(2)} Cj^{\mu}j^{\nu}B_{\mu\rho}B_{\nu \sigma}g^{\rho\sigma}
	-
	\frac{1}{3}F^{(0)}g^{\mu\nu}B_{\mu\nu}
	-
	M_{\mu\nu}S^{\mu\nu}\,,
	\end{split}
	\label{24JuneEomD}
\end{equation}
\begin{equation}
	\begin{split}
		S^{\mu\nu}
		=&
		-2F^{(1)} B^{\mu\nu} C^{2}
		-
		4C j^{\mu}j^{\nu}
		+
		4 F^{(2)} C^{2}j^{\mu}j^{\rho}B_{\rho\sigma}g^{\nu\sigma}
		+
		\frac{1}{3}F^{(0)} Cg^{\mu\nu}\,,
	\end{split}
	\label{24JuneEomB}
\end{equation}
\begin{equation}
	\begin{split}
		a_{\mu} =&
		-
		F^{(2)} j_{\mu}N_{\left[ 0 \right]}
		+
		F^{(1)} C^{2}\square j^{\mu}
		+\\&+
		\square\left( F^{(1)} C^{2}j^{\mu} \right)
		-
		8\,F^{(1)}\,C\,B_{\mu\nu}j^{\nu}
		-
		F^{(3)} N_{\left[ 1 \right]}
		+\\&+
		4F^{(2)}C^{2}B_{\mu\rho}B_{\nu\sigma}g^{\rho\sigma}j^{\nu}
		-
		2F^{(1)}j_{\mu}N_{\left[ 2 \right]}
		-	
		2F^{(0)}j_{\mu}\,,
	\end{split}
	\label{24JuneEomj}
\end{equation}
where $N_{\left[ 0 \right]}$, $N_{\left[ 1 \right]}$ and $N_{\left[ 2 \right]}$ are the terms in (\ref{24JuneLbos2}) proportional to $F^{(0)}$, $F^{(1)}$, and $F^{(2)}$, respectively.

In the case $j^\mu =0$, the Lagrangian (\ref{24JuneLbos2}) coupled to worldline metric is (we set $F^{(0)} = F^{(1)} = 1$)
\begin{equation}
	\mathcal{L}_{bos}{\big |}_{j=0} 
	=
	\sqrt{-g}\left[ 
	C^{2}
	g^{\mu\rho} g^{\nu\sigma}B_{\mu\nu}B^{\rho\sigma}
	-
	CD
	-
	\frac{1}{3} C g^{\mu\nu} B_{\mu\nu}
	+
	S^{\mu\nu}\left( B_{\mu\nu}-M_{\mu\nu} C \right)
	\right]\,.
	\label{24JLj0}
\end{equation}
The equations of motion for $C$ and $B_{\mu\nu}$ are
\begin{equation}
	D 
	=
	2C	g^{\mu\rho} g^{\nu\sigma}B_{\mu\nu}B^{\rho\sigma}
	-
	\frac{1}{3}g^{\mu\nu}B_{\mu\nu}
	-
	M_{\mu\nu}S^{\mu\nu}
	\,,
	\label{24JEoMCj0}
\end{equation}
and
\begin{equation}
	S^{\mu\nu}
	=
	-
	2C^{2}B^{\mu\nu}
	+
	\frac{1}{3}Cg^{\mu\nu}
	\,.
	\label{24JEoMBj0}
\end{equation}
Finally, for this simplified Lagrangian we derive the energy momentum tensor. We obtain
\begin{equation}
	\begin{split}
		T^{\mu\nu} = &
		-
		g^{\mu\nu} C\,\square C 
		-
		\frac{1}{2}g^{\mu\nu} \partial^{\rho}C\,\partial_{\rho} C
		+
		\partial^{\mu}C\,\partial^{\nu}C
		+\\&
		-
		\frac{5}{2} g^{\mu\nu} C^{2}\, \nabla^{\rho}\partial^{\sigma}C\,\nabla_{\rho}\partial_{\sigma}C
		-
		7g^{\mu\nu}C^{2}\,\square C\,\square C
		+\\&
		-
		2g^{\mu\nu}C\,\partial_{\rho}C\,\partial_{\sigma}C \nabla^{\rho}\partial^{\sigma}C
		-
		14
		g^{\mu\nu}C^{2}\,\partial_{\rho}C\,\partial^{\rho}\square C
		+\\&
		-
		3g^{\mu\nu} C^{3}\,\square\square C
		-
		8g^{\mu\nu}C\,\partial_{\rho}C\,\partial^{\rho}C\,\square C
		+\\&
		-
		C^{2}\,\square C\, \nabla^{\mu}\partial^{\nu}C
		+
		4C\,\partial_{\rho} C\, \nabla^{\rho}\partial^{\left( \mu \right.}C\,\partial^{\left.\nu \right)}C
		+\\&
		-2C\,\partial^{\rho} C\,\partial_{\rho}C\, \nabla^{\mu}\partial^{\nu}C
		+
		6C^{2}\,\partial^{\left( \mu \right.}C\,\nabla^{\left.\nu \right)}\square C
		+\\&
		-
		C^{2}\, \nabla^{\rho}\nabla^{\mu}\partial^{\nu}C\, \partial_{\rho}C
		+
		8C\,\partial^{\mu}C\,\partial^{\nu}C\,\square C
		\,.
	\end{split}
	\label{30JEnergyMomTensor}
\end{equation}

We prefer to analyze only the equations of motion with the Clebsch parametrization and in the case $\omega=0$. This 
gives novel dynamical equations.

\subsection{Dependence on the K\"ahler Potential}
\label{kpot}

We have to discuss the dependence of the equations of motion 
upon the K\"ahler potential. For that, we discuss only the bosonic sector 
and we observe the following identity 
\begin{equation}\label{depA}
-j^\mu F_{\mu\nu} + C \partial_\nu D = -4 \partial_\mu \left[ \partial \bar \partial K \, C ( \partial^\mu \bar\varphi  \partial_\nu \varphi  +  \partial^\mu \varphi  \partial_\nu \bar\varphi)\right]\,,
\end{equation}
where the r.h.s. can be also be written as $\partial_\mu (C G^{\mu\nu})$ where $G^{\mu\nu}$ is the inverse of the K\"ahler metric. It appears as a 
total derivative. However, we cannot discard such term. The reason is that it does not follows directly from the action, namely it is not a total derivative term 
derived from the action. Nevertheless, we can show that it is harmless and, at least in the rigid case, can be discarded. 

The left hand side of (\ref{depA}) can be obtained by the same method as in sec. (2.1). Indeed, by requiring the invariance under an isometry 
and using the same equations as above we get a new equation of the form 
\begin{equation}\label{depB}
\int d^4x X^\nu \left(-j^\mu F_{\mu\nu} + C \partial_\mu D \right)= 
- 4 \int d^4x X^\nu \partial_\mu \left[ \partial \bar\partial K \, C ( \partial^\mu \bar\varphi  \partial_\nu \varphi  +  \partial^\mu \varphi  \partial_\nu \bar\varphi)\right]\,.
\end{equation}
Now, we can use the integration by parts in the r.h.s. and by using the fact that $X^\mu$ must be a Killing vector for the 
flat metric we can easily conclude that the l.h.s. of (\ref{depA}) is effectively a total derivative and it can be discarded. A complete proof 
of this statement would be very interesting since itwould show that the dynamical equations of motion are independent of the parametrization of 
the gauge field $A$.


\chapter*{Conclusions of Part II}
\addcontentsline{toc}{chapter}{Conclusions of Part II}

The second part of the present work deals with the study of supersymmetric extension of fluid dynamics.
The task is performed in two ways: by a supergravity generalization of fluid/gravity correspondence
 and by proposing a suitable action written in superfield formalism.

We construct the complete non linear solution of supergravity $\mathcal{N}=2$, $D=3$ equations of motion starting from the BTZ black hole and generating the fermionic corrections by a finite supersymmetry transformation. 
Due to the simplicity of the framework we are able to give analytic expression for the metric, the gravitino and the gauge field at the highest order in the fermionic expansion.

The aim of this work is to prepare the ground for a complete computation of gravity/fluid correspondence type and derive the non--linear supersymmetric Navier--Stokes equations for the complete solution.
By following the rules of the fluid/gravity correspondence, we derive the boundary equations of motion for a supersymmetric fluid. This means
a set of bosonic equations of motion, but also some Dirac-type equation for the supersymmetric long range d.o.f. of the fluid. The computation is performed at the
first order in the bosonic parameters.

The analysis of the boundary fluid energy--momentum tensor is performed by computing the finite metric associated to the transformation generated by Killing spinors and Killing vectors.
We denote the finite metric as the fermionic wig, to remind the
reader the anticommuting nature of these ``hairs''.
We provide the exact analytical solutions for both global coordinates and Poincar\'e patch BTZ black hole.
These are computed by
automatic computation and for that we describe the algorithms based on iterative solution of supergravity equations.
With this results we provide the energy-momentum tensor which is cast in a form from which one can read the thermodynamic quantities.

Moreover, we analyze in detail the structure of the fermionic wig for the global BTZ black hole. We derive the fermionic corrections to the mass and to the angular momentum
of the BTZ black hole. In addition, we compute the entropy of the black hole which also shows new terms depending on the vev's of the fermionic bilinears.
We present the $r$-large expressions for the several geometrical quantities in the presence of the fermionic corrections.

Using the developed programs we compute the wig for a more complicated model.
We consider a  Schwarzschild black holes in $AdS_{5}$ which is a solution of truncated $N=2$, $D=5$ gauged supergravity.
The structure of the model is more rich than the BTZ black hole, the Killing spinors have eight real components and hence the wig is computed by a four--step iterative algorithm (we remind that the algorithm counts the number of fermionic bilinears).
We present the result in different limits and in some of them we check the supergravity equations of motion.
After that, we compute the boundary energy--momentum tensor and we briefly comment the results. 
We apply the same techniques also for black hole in $AdS_{4}$.

We propose a new supersymmetric action for supersymmetric fluid dynamics, discussing some of its aspects, such as the new Navier--Stokes equations and their derivation. A discussion on the Clebsch parametrization is proposed and the derivation of the superfield equations is done in that framework. In this context, there are several open issues: what is the complete dynamics described by the present action and what is the role of the boson $C$? 
Moreover, a fluid described only in terms of fermionic field 
can be discussed by setting to zero both $j^\mu$ and $C$. 
We believe that the  study of the present system in the context of supergravity might shed some light on the coupling 
with the worldvolume metric.


\part{Appendices}

\newpage
\appendix


\chapter{$\mathfrak{osp}(n|m)$ Algebra}\label{appendixA}

The generators of the $\mathfrak{osp}(n|m)$ algebra satisfy the following (anti)commutator relations
\begin{eqnarray}
  &&
  \left[ T^{ab},T^{cd} \right]=\delta^{bc}T^{ad}+\delta^{ad}T^{bc}-\delta^{ac}T^{bd}-\delta^{bd}T^{ac}
  \ ,\nonumber\\&&
  \left[ T_{\alpha\beta},T_{\gamma\delta} \right]=
  -\left( -\varepsilon_{\beta\gamma}T_{\alpha\delta} -\varepsilon_{\alpha\delta}T_{\beta\gamma} -\varepsilon_{\alpha\gamma}T_{\beta\delta} -\varepsilon_{\beta\delta}T_{\alpha\gamma} \right)
  \ ,\nonumber\\&&
  \left[ T^{ab},T_{\alpha\beta} \right]=0
  \ ,\nonumber\\&&
  \left[ T^{ab},Q^{c}_{\gamma} \right]=\delta^{bc}Q^{a}_{\gamma}-\delta^{ac}Q^{b}_{\gamma}
  \ ,\nonumber\\&&
  \left[ T_{\alpha\beta},Q^{c}_{\gamma} \right]=-\varepsilon_{\gamma\alpha}Q^{c}_{\beta}-\varepsilon_{\gamma\beta}Q^{c}_{\gamma}
  \ ,\nonumber\\&&
  \left\{ Q^{a}_{\alpha},Q^{b}_{\beta} \right\}=\varepsilon_{\alpha\beta}T^{ab}+\delta^{ab}T_{\alpha\beta}
  \ .\label{appendixalgebra1}
\end{eqnarray}
We can now choose the following matrix form for the fundamental representation of $\mathfrak{osp}(n|m)$
\begin{eqnarray}
  &&
  \left( Q^{a}_{\alpha} \right)^{I}_{\phantom{I} B}=\delta^{aI}\varepsilon_{\alpha B}+\delta^{a}_{\phantom{a} B}\varepsilon^{\phantom{\alpha}I}_{\alpha}
  \ ,\nonumber\\&&
  \left( T^{ab} \right)^{I}_{\phantom{I} B}=\delta^{aI}\delta^{b}_{\phantom{a} B}-\delta^{bI}\delta^{a}_{\phantom{a}B}
  \ ,\nonumber\\&&
  \left( T_{\alpha\beta} \right)^{I}_{\phantom{I} B}=\varepsilon_{\alpha}^{\phantom{a}I}\varepsilon_{\beta B}+\varepsilon_{\beta}^{\phantom{a}I}\varepsilon_{\alpha B}
  \ .
  \label{appendixgenfund}
\end{eqnarray}

To compute the supertraces of the generators we use the fundamental representation instead of the adjoint one\footnote{The trace of generators in the adjoint representation corresponds to the Killing metric.}. The reason for this is that for a particular choice of $2n$ and $2m$ the dual Coxeter number is zero and so the Killing metric is totally degenerate. We obtain
\begin{eqnarray}
  Str\left( T_{\alpha\beta}T_{\rho\sigma} \right) &=& -2\varepsilon_{\alpha\sigma}\varepsilon_{\beta\rho}-2\varepsilon_{\alpha\rho}\varepsilon_{\beta\sigma}
  	\ ,\nonumber\\
	\phantom{a}
	\nonumber\\
	Str\left( T^{ab}T^{rs} \right)&=& -2\delta^{ar}\delta^{bs}+2\delta^{as}\delta^{br}
	\ ,\nonumber\\
	\phantom{a}
	\nonumber\\
	Str\left( Q_{\alpha}^{a}Q_{\beta}^{b} \right)&=& 2 \delta^{ab}\varepsilon_{\alpha\beta} 	
	\ .
  \label{bfmsupertraces}
\end{eqnarray}

\chapter{Details on $OSp\left( 2|2 \right)/SO\left( 2 \right)\times Sp\left( 2 \right)$ Construction}\label{appendixAA}

In this appendix we show the complete derivation of the left invariant $1$-form for the coset model $OSp\left( 2|2 \right)/SO\left( 2 \right)\times Sp\left( 2 \right)$ described in sec.~\ref{OSP22etc}.

From (\ref{OSPMNr}) we extract the non trivial structure constants 
\begin{equation}
 \begin{array}{c}
\phantom{\bigg |}
 C_{01}^{\phantom{01}1}=1\ ,\quad\quad C_{02}^{\phantom{01}2}=1\ ,\quad\quad C_{03}^{\phantom{03}3}=-1\ ,\quad\quad C_{04}^{\phantom{04}4}=-1
\ ,\\
\phantom{\bigg |}
 C_{1'2}^{\phantom{01}1}=-2\ ,\quad\quad C_{1'4}^{\phantom{01}3}=-2\ ,\quad\quad C_{3'1}^{\phantom{03}2}=2\ ,\quad\quad C_{3'3}^{\phantom{04}4}=2
\ ,\\
\phantom{\bigg |}
 C_{2'3}^{\phantom{01}3}=1\ ,\quad\quad C_{2'4}^{\phantom{01}4}=-1\ ,\quad\quad C_{2'1}^{\phantom{03}1}=1\ ,\quad\quad C_{2'2}^{\phantom{04}2}=-1
\ ,\\
\phantom{\bigg |}
 C_{13}^{\phantom{01}1'}=1\ ,\quad\quad\quad C_{14}^{\phantom{01}0}=-1\ ,\quad\quad\quad C_{14}^{\phantom{03}2'}=1
\ ,\\
\phantom{\bigg |}
 C_{23}^{\phantom{01}2'}=1\ ,\quad\quad\quad C_{23}^{\phantom{01}0}=1\ ,\quad\quad\quad C_{24}^{\phantom{03}3'}=1
\ .\end{array}
\end{equation}
The constants from the first three lines are antisymmetric respect the exchange of the lower indices, the other are otherwise symmetric. The reduced Killing metric is then ($A=\{i,i'\}$)
\begin{equation}\label{OSP22KMR1}
\k_{AB}= 4\left(\begin{array}{cccc}
  0 & 0 & 0 & 1 \\
0 &  0 & -1 & 0 \\
 0 &  1 & 0 & 0 \\
-1 & 0 & 0 & 0  
\end{array}\right)\ .
\end{equation}
The representative is chosen as in (\ref{MIOosp12eq1})
\begin{equation}\label{OSP22LX}
 L(\t)=e^{\t_{1}Q_{1}}\,e^{\t_{2}Q_{2}}\,e^{\t_{3}Q_{3}}\,e^{\t_{4}Q_{4}}\ ,
\end{equation}
and, expanding in series, we obtain
\begin{equation}\label{OSP22L1X}
 L(\t)=\left(1+\t_{1}Q_{1} \right)\left(1+\t_{2}Q_{2} \right) \left(1+\t_{3}Q_{3} \right) \left(1+\t_{4}Q_{4} \right) \ .
\end{equation}
To construct the left-invariant $1$-form we compute
\begin{equation}\label{OSP22L-1}
 L^{-1}=\left(1-\t_{4}Q_{4} \right)\left(1-\t_{3}Q_{3} \right) \left(1-\t_{2}Q_{2} \right) \left(1-\t_{1}Q_{1} \right)
\end{equation}
and
\begin{eqnarray}\label{OSP22dL}
 \dd L&=&
\dd\t_{1}\,Q_{1}\left(1+\t_{2}Q_{2} \right) \left(1+\t_{3}Q_{3} \right) \left(1+\t_{4}Q_{4} \right) +\phantom{\Big |}
{}\nonumber\\
	&& {}
\phantom{\Big |}
+\left(1+\t_{1}Q_{1} \right)\dd\t_{2}\,Q_{2}\left(1+\t_{3}Q_{3} \right) \left(1+\t_{4}Q_{4} \right) +
{}\nonumber\\
	&& {}
\phantom{\Big |}
+\left(1+\t_{1}Q_{1} \right)\left(1+\t_{2}Q_{2} \right) \dd\t_{3}\,Q_{3}  \left(1+\t_{4}Q_{4} \right) +
{}\nonumber\\
	&& {}
\phantom{\Big |}
+\left(1+\t_{1}Q_{1} \right)\left(1+\t_{2}Q_{2} \right) \left(1+\t_{3}Q_{3} \right) \dd\t_{4}\,Q_{4} \ .
\end{eqnarray}
Finally, the left-invariant $1$-form reads
\begin{eqnarray}\label{OSP22conti1}
 L^{-1}\dd L&=&
\phantom{\Big|}
\left(1-\t_{4}Q_{4} \right)\left(1-\t_{3}Q_{3} \right) \left(1-\t_{2}Q_{2} \right) \left(1-\t_{1}Q_{1} \right)\times
{}\nonumber\\
	&& {}\phantom{\Big|}
\times \dd\t_{1}\,Q_{1}\left(1+\t_{2}Q_{2} \right) \left(1+\t_{3}Q_{3} \right) \left(1+\t_{4}Q_{4} \right)+
{}\nonumber\\\phantom{\Big|}
	&& {}
+
\left(1-\t_{4}Q_{4} \right)\left(1-\t_{3}Q_{3} \right) \left(1-\t_{2}Q_{2} \right)\dd\t_{2}\,Q_{2}\left(1+\t_{3}Q_{3} \right) \left(1+\t_{4}Q_{4} \right)+
{}\nonumber\\\phantom{\Big|}
	&& {}
+
\left(1-\t_{4}Q_{4} \right)\left(1-\t_{3}Q_{3} \right) \dd\t_{3}\,Q_{3}  \left(1+\t_{4}Q_{4} \right)+\left(1-\t_{4}Q_{4} \right)\dd\t_{4}\,Q_{4} 
\ .
\end{eqnarray}
As we have already told in sec.~\ref{OSP22etc}, the vielbeins receive contribution only from terms with a even number of commutators between coset generators $Q$. 

A single $Q$ is obtained only from $\dd\t$
\begin{equation}
\dd\t_{1}Q_{1}+\dd\t_{2}Q_{2}+\dd\t_{3}Q_{3}+\dd\t_{4}Q_{4}\ .
\end{equation}
Three $Q$'s come from $\t_{i}\t_{j}\dd\t_{k}$
\begin{equation}
  \begin{array}{c}
\phantom{\Big|}
-\t_{4}\dd\t_{1}\t_{2}Q_{4}Q_{1}Q_{2}-\t_{4}\dd\t_{1}\t_{3}Q_{4}Q_{1}Q_{3}-\t_{3}\dd\t_{1}\t_{2}Q_{3}Q_{1}Q_{2}-\t_{3}\dd\t_{1}\t_{4}Q_{3}Q_{1}Q_{4}+
\\\phantom{\Big|}
-\t_{2}\dd\t_{1}\t_{3}Q_{2}Q_{1}Q_{3}-\t_{2}\dd\t_{1}\t_{4}Q_{2}Q_{1}Q_{4}-\t_{1}\dd\t_{1}\t_{2}Q_{1}Q_{1}Q_{2}-\t_{1}\dd\t_{1}\t_{3}Q_{1}Q_{1}Q_{3}+
\\\phantom{\Big|}
-\t_{1}\dd\t_{1}\t_{4}Q_{1}Q_{1}Q_{4}+\t_{4}\t_{3}\dd\t_{1}Q_{4}Q_{3}Q_{1}+\t_{4}\t_{2}\dd\t_{1}Q_{4}Q_{2}Q_{1}+\t_{4}\t_{1}\dd\t_{1}Q_{4}Q_{1}Q_{1+}
\\\phantom{\Big|}
+\t_{3}\t_{2}\dd\t_{1}Q_{3}Q_{2}Q_{1}+\t_{3}\t_{1}\dd\t_{1}Q_{3}Q_{1}Q_{1}+\t_{2}\t_{1}\dd\t_{1}Q_{2}Q_{1}Q_{1}+\dd\t_{1}\t_{2}\t_{3}Q_{1}Q_{2}Q_{3}+
\\\phantom{\Big|}
+\dd\t_{1}\t_{2}\t_{4}Q_{1}Q_{2}Q_{4}+\dd\t_{1}\t_{3}\t_{4}Q_{1}Q_{3}Q_{4}-\t_{4}\dd\t_{2}\t_{3}Q_{4}Q_{2}Q_{3}-\t_{3}\dd\t_{2}\t_{4}Q_{3}Q_{2}Q_{4}+
\\\phantom{\Big|}
-\t_{2}\dd\t_{2}\t_{3}Q_{2}Q_{2}Q_{3}-\t_{2}\dd\t_{2}\t_{4}Q_{2}Q_{2}Q_{4}+\t_{4}\t_{3}\dd\t_{2}Q_{4}Q_{3}Q_{2}+\t_{4}\t_{2}\dd\t_{2}Q_{4}Q_{2}Q_{2}+
\\\phantom{\Big|}
+\t_{3}\t_{2}\dd\t_{2}Q_{3}Q_{2}Q_{2}+\dd\t_{2}\t_{3}\t_{4}Q_{2}Q_{3}Q_{4}-\t_{3}\dd\t_{3}\t_{4}Q_{3}Q_{3}Q_{4}+\t_{4}\t_{3}\dd\t_{3}Q_{4}Q_{3}Q_{3}
\ .
\end{array}
\end{equation}
Finally, the five generators contribute
\begin{equation}
  \begin{array}{c}
\phantom{\Big|}
-\t_{1}\dd\t_{1}\t_{2}\t_{3}\t_{4}Q_{1}Q_{1}Q_{2}Q_{3}Q_{4}+
\t_{4}\t_{1}\dd\t_{1}\t_{2}\t_{3}Q_{4}Q_{1}Q_{1}Q_{2}Q_{3}+
\\\phantom{\Big|}
+\t_{3}\t_{1}\dd\t_{1}\t_{2}\t_{4}Q_{3}Q_{1}Q_{1}Q_{2}Q_{4}+
\t_{2}\t_{1}\dd\t_{1}\t_{3}\t_{4}Q_{2}Q_{1}Q_{1}Q_{3}Q_{4}+
\\\phantom{\Big|}
-\t_{4}\t_{3}\t_{1}\dd\t_{1}\t_{2}Q_{4}Q_{3}Q_{1}Q_{1}Q_{2}
-\t_{4}\t_{2}\t_{1}\dd\t_{1}\t_{3}Q_{4}Q_{2}Q_{1}Q_{1}Q_{3}+
\\\phantom{\Big|}
-\t_{3}\t_{2}\t_{1}\dd\t_{1}\t_{4}Q_{3}Q_{2}Q_{1}Q_{1}Q_{4}
\ .
\end{array}
\end{equation}
Due to nilpotency and the choice of the representative (\ref{OSP22L}), all the previous terms are zero. We then have 
\begin{equation}
  \begin{array}{c}
\phantom{\Big|}
\t_{2}\t_{4}\dd\t_{1}\left[ 
-Q_{4}Q_{1}Q_{2}+Q_{2}Q_{1}Q_{4}+Q_{1}Q_{2}Q_{4}-Q_{4}Q_{2}Q_{1}
\right] 
\ ,\\\phantom{\Big|}
\t_{3}\t_{4}\dd\t_{1}\left[ 
-Q_{4}Q_{1}Q_{3}+Q_{3}Q_{1}Q_{4}-Q_{4}Q_{3}Q_{1}+Q_{1}Q_{3}Q_{4}
\right] 
\ ,\\\phantom{\Big|}
\t_{2}\t_{3}\dd\t_{1}\left[ 
-Q_{3}Q_{1}Q_{2}+Q_{2}Q_{1}Q_{3}-Q_{3}Q_{2}Q_{1}+Q_{1}Q_{2}Q_{3}
\right] 
\ ,\\\phantom{\Big|}
\t_{3}\t_{4}\dd\t_{2}\left[ 
-Q_{4}Q_{2}Q_{3}+Q_{3}Q_{2}Q_{4}-Q_{4}Q_{3}Q_{2}+Q_{2}Q_{3}Q_{4}
\right] \ ,
\end{array}
\end{equation}
that is
\begin{equation}
  \begin{array}{c}
\phantom{\Big|}
\t_{2}\t_{4}\dd\t_{1}\left[ 
\left\lbrace Q_{1} \,,\, Q_{2}\right\rbrace \,Q_{4}
\right] = 0
\ ,\\\phantom{\Big|}
\t_{3}\t_{4}\dd\t_{1}\left[ 
\left\lbrace Q_{1} \,,\, Q_{3}\right\rbrace \,Q_{4}
\right] = -2\t_{3}\t_{4}\dd\t_{1}\,Q_{3}
\ ,\\\phantom{\Big|}
\t_{2}\t_{3}\dd\t_{1}\left[ 
\left\lbrace Q_{1} \,,\, Q_{2}\right\rbrace \,Q_{3}
\right] = 0
\ ,\\\phantom{\Big|}
\t_{3}\t_{4}\dd\t_{2}\left[ 
\left\lbrace Q_{2} \,,\, Q_{3}\right\rbrace \,Q_{4}
\right] = -2\t_{3}\t_{4}\dd\t_{2}\,Q_{4}\ .
\end{array}
\end{equation}
Summing up all the contributions, the left-invariant $1$-form is
\begin{equation}
 L^{-1}\dd L = Q_{1}\dd\t_{1}+Q_{2}\dd\t_{2}+Q_{3}\left(-2\t_{3}\t_{4}\dd\t_{1}+\dd\t_{3} \right) +Q_{4}\left(-2\t_{3}\t_{4}\dd\t_{2}+\dd\t_{4} \right) + \Omega^{I}H_{I}\ .
\end{equation}

\chapter{Non Linear Isometry for $2\theta$ Actions}\label{AppNonLinIsom}

We find a generic non--linear isometry transformation for a generic $2\theta$ action
\begin{equation}
S\propto \int_{\Sigma}\left( 1+B\theta_{1}\theta_{2} \right)\dd\theta_{1}\wa\dd\theta_{2}\ ,
\label{app2action}
\end{equation}
where $B$ is a generic constant. Due to the nilpotent behaviour of fermionic fields $\theta$, the generic non--linear transformation is
\begin{eqnarray}
&&
\theta_{i}\rightarrow \theta_{i}+\left( 1+A_{i} \theta_{1}\theta_{2} \right)\varepsilon_{i}\ ,
\label{appvariat}
\end{eqnarray}
where $\varepsilon$ is a fermionic constant and $A$ is a generic constant.
Imposing the invariance of the action we find a constraint for $A$ and $B$
\begin{eqnarray}
S&\ra&\int_{\Sigma}
\Big( 
1+B\left[ \t_{1}+\left(1+A_{1}\t_{1}\t_{2} \right)\e_{1} \right]
\left[ \t_{2}+\left(1+A_{1}\t_{1}\t_{2} \right)\e_{2} \right]
\Big)\times\phantom{\bigg |}
{}\nonumber\\
	&& {}\phantom{\bigg |}
\times  \dd \left[ \t_{1}+\left(1+A_{1}\t_{1}\t_{2} \right)\e_{1} \right]\wa
\dd\left[ \t_{2}+\left(1+A_{1}\t_{1}\t_{2} \right)\e_{2} \right]=
{}\nonumber\\
	&=&\int_{\Sigma} {}\phantom{\bigg |}
\Big( 
1+B\t_{1}\t_{2}+B\t_{1}\left(1+A_{1}\t_{1}\t_{2} \right)\e_{2}+
B\left(1+A_{1}\t_{1}\t_{2} \right)\e_{1} \t_{2} 
\Big)\times
{}\nonumber\\
	&& {}\phantom{\bigg |}
\times  \left[ \dd \t_{1}+A_{1}\dd\t_{1}\t_{2}\e_{1}+A_{1}\t_{1}\dd\t_{2}\e_{1} \right]\wa
\left[ \dd \t_{2}+A_{2}\dd\t_{1}\t_{2}\e_{2}+A_{2}\t_{1}\dd\t_{2}\e_{2} \right]=
{}\nonumber\\
	&=& \int_{\Sigma}{}\phantom{\bigg |}
\Big( 
1+B\t_{1}\t_{2}+B\t_{1}\e_{2}+
B\e_{1} \t_{2} 
\Big)\times
{}\nonumber\\
	&& {}\phantom{\bigg |}
\times 
\Big(
\dd \t_{1}\wa\dd \t_{2}+
A_{2}\dd\t_{1}\wa(\t_{1}\dd\t_{2}\e_{2})+A_{1}(\dd\t_{1}\t_{2}\e_{1})\wa\dd\t_{2}+
{}\nonumber\\
	&& {}\phantom{\bigg |}
+2A_{1}A_{2}(\dd\t_{1}\t_{2})\wa(\t_{1}\dd\t_{2})\e_{1}\e_{2}
\Big)=
{}\nonumber\\
	&=&\int_{\Sigma} {}\phantom{\bigg |}
\Big(
1+B\t_{1}\t_{2}
\Big)\dd \t_{1}\wa\dd \t_{2}+
{}\nonumber\\
	&& {}\phantom{\bigg |}
+
BA_{1}\t_{1}\e_{2}\dd\t_{1}\t_{2}\e_{1}\wa\d\t_{2}+BA_{2}\e_{1}\t_{2}\dd\t_{1}\wa(\t_{1}\dd\t_{2}\e_{2})+
{}\nonumber\\
	&& {}\phantom{\bigg |}
+
2A_{1}A_{2}(\dd\t_{1}\t_{2})\wa(\t_{1}\dd\t_{2})\e_{1}\e_{2}+
{}\nonumber\\
	&& {}\phantom{\bigg |}
+B\t_{1}\e_{2}\dd\t_{1}\wa\dd\t_{2}+B\e_{1}\t_{2}\dd\t_{1}\wa\dd\t_{2}+
{}\nonumber\\
	&& {}\phantom{\bigg |}
+
A_{2}\dd\t_{1}\wa(\t_{1}\dd\t_{2}\e_{2})+A_{1}\t_{2}(\dd\t_{1}\t_{2}\e_{1})\wa\dd\t_{2}=
{}\nonumber\\
	&=&\int_{\Sigma} {}\phantom{\bigg |}
\Big(
1+B\t_{1}\t_{2}
\Big)\dd \t_{1}\wa\dd \t_{2}+
{}\nonumber\\
	&& {}\phantom{\bigg |}
+A_{1}[B-A_{2}]\t_{1}\t_{2}\dd\t_{1}\wa\dd\t_{2}\e_{1}\e_{2}+A_{2}[B-A_{2}]\t_{1}\t_{2}\dd\t_{1}\wa\dd\t_{2}\e_{1}\e_{2}+
{}\nonumber\\
	&& {}\phantom{\bigg |}
+
[B-A_{2}]\t_{1}\dd\t_{1}\wa\dd\t_{2}\e_{2}+[A_{1}-B]\t_{2}\dd\t_{1}\wa\dd\t_{2}\e_{1} \ ,
\end{eqnarray}
then, (\ref{appvariat}) is an isometry if
\begin{equation}
A_{1}=A_{2}=B \ .
\label{appconcl}
\end{equation}


\chapter{
$OSp(1|2)$ T-duality Construction}\label{detailsosp12}

Here we compute the nine different pieces that form $\bar\Omega^{\left(\alpha\beta\right)}
\left[ \hat\Pi^{-1} \right]_{\left( \alpha\beta \right)\left( \rho\sigma \right)}\Omega^{\left( \rho\sigma \right)}$.
Notice that
\begin{itemize}
\item we rewrite $\Omega$ in three parts:
 \begin{eqnarray}
\Omega^{\rho\sigma}&=& 
-i\partial\hat\phi\delta^{\rho\sigma}
-\frac{1}{1-4\hat\phi^{2}}\theta^{\left( \rho \right.}\partial\theta^{\left. \sigma \right)}
-\frac{2i\hat\phi}{1-4\hat\phi^{2}}\theta^{\left( \rho \right.}\varepsilon^{\left.\sigma  \right)\lambda}\delta_{\lambda\tau}\partial\theta^{\tau}
\ ,\nonumber\\
\bar\Omega^{\rho\sigma}&=& 
+i\bar\partial\hat\phi\delta^{\rho\sigma}
-\frac{1}{1-4\hat\phi^{2}}\theta^{\left( \rho \right.}\bar\partial\theta^{\left. \sigma \right)}
+\frac{2i\hat\phi}{1-4\hat\phi^{2}}\theta^{\left( \rho \right.}\varepsilon^{\left.\sigma  \right)\lambda}\delta_{\lambda\tau}\bar\partial\theta^{\tau}
\ .
\label{Omegaredef}
\end{eqnarray}

\item the following relation holds, where $M$ is a generic symmetric matrix:
$$M^{\left( \alpha\beta \right)}[<\varepsilon\delta>]_{\left( \alpha\beta \right)\left( \rho\sigma \right)}M^{\left( \rho\sigma \right)}=0\ .$$

\end{itemize}
The different pieces are
\begin{itemize}
\item part1A
\begin{equation}
\delta^{\alpha\beta}\left( L<\varepsilon\varepsilon>+M<\varepsilon\delta>+P<\delta\delta> \right)_{\left( \alpha\beta \right)\left( \rho\sigma \right)}\delta^{\rho\sigma}=4\left( L+P \right)\ ;
\label{part1A}
\end{equation}

\item part1B
 \begin{equation}
\delta^{\alpha\beta}\left( L<\varepsilon\varepsilon>+M<\varepsilon\delta>+P<\delta\delta> \right)_{\left( \alpha\beta \right)\left( \rho\sigma \right)}\theta^{\left( \rho \right.}\partial\theta^{\left. \sigma \right)}
=
2\left( L+P \right)\theta^{\alpha}\partial\theta^{\beta}\delta_{\alpha\beta}
\ ;\label{part1B}
\end{equation}

\item part1C
\begin{eqnarray}
&&
\delta^{\alpha\beta}\left( L<\varepsilon\varepsilon>+M<\varepsilon\delta>+P<\delta\delta> \right)_{\left( \alpha\beta \right)\left( \rho\sigma \right)}\theta^{\left( \rho \right.}\varepsilon^{\left.\sigma  \right)\lambda}\delta_{\lambda\tau}\partial\theta^{\tau}
\nonumber\\
&=& 
2\left( L+P \right)\theta^{\alpha}\partial\theta^{\beta}\varepsilon_{\alpha\beta}
\ ;\label{part1C}
\end{eqnarray}

\item part2A
\begin{equation}
\theta^{\left( \alpha \right.}\bar\partial\theta^{\left. \beta \right)}\left( L<\varepsilon\varepsilon>+M<\varepsilon\delta>+P<\delta\delta> \right)_{\left( \alpha\beta \right)\left( \rho\sigma \right)}\delta^{\rho\sigma}
=
2\left( L+P \right)\theta^{\alpha}\bar\partial\theta^{\beta}\delta_{\alpha\beta}
\ ;\label{part2A}
\end{equation}

\item part2B
 \begin{eqnarray}
&&
\theta^{\left( \alpha \right.}\bar\partial\theta^{\left. \beta \right)}\left( L<\varepsilon\varepsilon>+M<\varepsilon\delta>+P<\delta\delta> \right)_{\left( \alpha\beta \right)\left( \rho\sigma \right)}\theta^{\left( \rho \right.}\partial\theta^{\left. \sigma \right)}
\nonumber\\&=& 
\theta^{1}\theta^{2}\left( 
-4 M\bar\partial\theta^{\alpha}\partial\theta^{\beta} \delta_{\alpha\beta}
-3 L\bar\partial\theta^{\alpha}\partial\theta^{\beta}\varepsilon_{\alpha\beta}
+P \bar\partial\theta^{\alpha}\partial\theta^{\beta} \varepsilon_{\alpha\beta}
\right)
\ ;\label{part2B}
\end{eqnarray}

\item part2C
\begin{eqnarray}
&&
\theta^{\left( \alpha \right.}\bar\partial\theta^{\left. \beta \right)}\left( L<\varepsilon\varepsilon>+M<\varepsilon\delta>+P<\delta\delta> \right)_{\left( \alpha\beta \right)\left( \rho\sigma \right)}\theta^{\left( \rho \right.}\varepsilon^{\left.\sigma  \right)\lambda}\delta_{\lambda\tau}\partial\theta^{\tau}
\nonumber\\&=& 
\theta^{1}\theta^{2}\left( 
-4 M\bar\partial\theta^{\alpha}\partial\theta^{\beta} \varepsilon_{\alpha\beta}
+3 L\bar\partial\theta^{\alpha}\partial\theta^{\beta}\delta_{\alpha\beta}
-P \bar\partial\theta^{\alpha}\partial\theta^{\beta} \delta_{\alpha\beta}
\right)
\ ;\label{part2C}
\end{eqnarray}

\item part3A
\begin{eqnarray}
&&
\theta^{\left( \alpha \right.}\varepsilon^{\left.\beta  \right)\lambda}\delta_{\lambda\tau}\partial\theta^{\tau}
\left( L<\varepsilon\varepsilon>+M<\varepsilon\delta>+P<\delta\delta> \right)_{\left( \alpha\beta \right)\left( \rho\sigma \right)}\delta^{\rho\sigma}
\nonumber\\
&=& 
2\left( L+P \right)\theta^{\alpha}\partial\theta^{\beta}\varepsilon_{\alpha\beta}
\ ;\label{part3A}
\end{eqnarray}

\item part3B
\begin{eqnarray}
&&
\theta^{\left( \alpha \right.}\varepsilon^{\left.\beta  \right)\lambda}\delta_{\lambda\tau}\partial\theta^{\tau}
\left( L<\varepsilon\varepsilon>+M<\varepsilon\delta>+P<\delta\delta> \right)_{\left( \alpha\beta \right)\left( \rho\sigma \right)}
\theta^{\left( \rho \right.}\partial\theta^{\left. \sigma \right)}
\nonumber\\
&=& 
\theta^{1}\theta^{2}\left( 
+4 M\bar\partial\theta^{\alpha}\partial\theta^{\beta} \varepsilon_{\alpha\beta}
+3 L\bar\partial\theta^{\alpha}\partial\theta^{\beta}\delta_{\alpha\beta}
-P \bar\partial\theta^{\alpha}\partial\theta^{\beta} \delta_{\alpha\beta}
\right)
\ ;\label{part3B}
\end{eqnarray}

\item part3C
\begin{eqnarray}
&&
\theta^{\left( \alpha \right.}\varepsilon^{\left.\beta  \right)\lambda}\delta_{\lambda\tau}\partial\theta^{\tau}
\left( L<\varepsilon\varepsilon>+M<\varepsilon\delta>+P<\delta\delta> \right)_{\left( \alpha\beta \right)\left( \rho\sigma \right)}
\theta^{\left( \rho \right.}\varepsilon^{\left.\sigma  \right)\lambda}\delta_{\lambda\tau}\partial\theta^{\tau}
\nonumber\\
&=& 
\theta^{1}\theta^{2}\left( 
-4 M\bar\partial\theta^{\alpha}\partial\theta^{\beta} \delta_{\alpha\beta}
-3 L\bar\partial\theta^{\alpha}\partial\theta^{\beta}\varepsilon_{\alpha\beta}
+P \bar\partial\theta^{\alpha}\partial\theta^{\beta} \varepsilon_{\alpha\beta}
\right)
\ .\label{part3C}
\end{eqnarray}

\end{itemize}

\chapter{UV-divergences}\label{appendixB}

Here we summarize some important results for divergent integrals. First of all, we recall the near--zero--expansion of the Euler gamma function
\begin{eqnarray}
\Gamma(\varepsilon)=\frac{1}{\varepsilon}-\gamma+O(\varepsilon) \ ,
\label{Gammaexpansion}
\end{eqnarray}
we have that (see \cite{Collins198601,BardinPassarino199912,Smirnov200609})
\begin{eqnarray}
-i B^{\alpha}_{0}=\int\dd^{d}q\frac{1}{(q^{2}+M^{2})^{\alpha}}&=&\pi^{d/2}\frac{\Gamma\left( \alpha-\frac{d}{2} \right)}{\Gamma\left(\alpha\right)}\left( M^{2} \right)^{\left( d/2-\alpha \right)}\ ,
\label{UV_Pass_A}
\end{eqnarray}
which yields
\begin{eqnarray}
-i B^{1}_{0}=\int\dd^{d}q\frac{1}{q^{2}+M^{2}}&=&\frac{2\pi}{\varepsilon}-\pi\left( \gamma+\ln \pi+\ln M^{2} \right)+O\left( \varepsilon \right)\ .
\label{UV_Pass_A2}
\end{eqnarray}
Moreover
\begin{eqnarray}
\int\dd^{d}q\frac{q_{\mu}}{q^{2}+M^{2}}&=&0\ ,
\label{UV_Pass_A3}
\end{eqnarray}
and
\begin{eqnarray}
\int\dd^{d}q\frac{q_{\mu}q_{\nu}}{(q^{2}+M^{2})^{\alpha}}&=&-iB^{\alpha}_{0}\left( -\frac{1}{2}\frac{M^{2}}{\frac{d}{2}-\alpha+1}\gamma_{\mu\nu} \right)\ ,
\label{UV_Pass_A4}
\end{eqnarray}
from which we obtain
\begin{eqnarray}
\int\dd^{d}q\frac{q_{\mu}q_{\nu}}{q^{2}+M^{2}}&=&-\frac{M^{2}}{2}\left( \frac{2\pi}{\varepsilon}-\pi\left( \gamma+\ln \pi+\ln M^{2} \right)+O\left( \varepsilon \right) \right)\gamma_{\mu\nu}\ .
\label{UV_Pass_A5}
\end{eqnarray}
%
%

%
\noindent
With these results, we compute the following integrals
\begin{eqnarray}
I_{1}&\equiv&\int\dd^d q\dd^d k \frac{1}{\left( q^{2}+M^{2} \right)\left( k^{2}+M^{2} \right)}
=
\nonumber\\&=& 
\left( \frac{2\pi}{\varepsilon} \right)^{2}-2\pi\left( \gamma+\ln \pi+\ln M^{2} \right)\frac{2\pi}{\varepsilon}+O\left( 1 \right)\ ,
\label{UV_Pass_2loop_1}
\end{eqnarray}
and
\begin{eqnarray}
I_{2}&=&
\int\dd^d q\dd^d k \frac{q^{2}}{\left( q^{2}+M^{2} \right)^{2}\left( k^{2}+M^{2} \right)\left( \left( q-k-p \right)^{2}+M^{2} \right)}
=\nonumber\\&=&
\int\dd^d q\dd^d k \frac{q^{2}+M^{2}-M^{2}}{\left( q^{2}+M^{2} \right)^{2}\left( k^{2}+M^{2} \right)\left( \left( q-k-p \right)^{2}+M^{2} \right)}
=\nonumber\\&=&
\int\dd^d q\dd^d k \frac{1}{\left( k^{2}+M^{2} \right)\left( \left( q-k-p \right)^{2}+M^{2} \right)}+O\left( 1 \right)
=\nonumber\\&=&
\int\dd^d q\dd^d k \frac{1}{\left( q^{2}+M^{2} \right)\left( k^{2}+M^{2} \right)}+O\left( 1 \right)
\nonumber\\&=&
I_{1}+O\left( 1 \right)\ ,
\label{UV_Pass_2loop_2}
\end{eqnarray}
where we perform the shift $q\rightarrow q-k-p$.
\begin{eqnarray}
&&
\int\dd^d q\dd^d k \frac{q\cdot k}{\left( q^{2}+M^{2} \right)^{2}\left( k^{2}+M^{2} \right)\left( \left( q-k-p \right)^{2}+M^{2} \right)}
\equiv I_{3}\ .
\label{UV_Pass_2loop_3}
\end{eqnarray}
We notice that
\begin{equation}
2q\cdot k= -\left( \left( q-k-p \right)^{2}+M^{2} \right)+q^{2}+k^{2}+p^{2}+M^{2}-2p\cdot q+2p\cdot k\ ,
\label{UV_Pass_note1}
\end{equation}
so we get
\begin{eqnarray}
I_{3}&=& 
\frac{1}{2}\left( -I_{1}+I_{2}+I_{2} \right)+O\left( 1 \right)=\frac{1}{2}I_{1}+O\left( 1 \right)\ .
\label{UV_Pass_2loop_3b}
\end{eqnarray}

\chapter{Feynman Rules Conventions}\label{appendixC}

We define the Green function $G\left( x'-x \right)$ as the solution of
\begin{eqnarray}
O G\left( x'-x \right)=+\delta^{2}\left( x'-x \right)\ .
\label{FRC1}
\end{eqnarray}
Where $O$ is the operator associated to the quadratic term in the fields $\phi$ obtained by rewriting the lagrangian\footnote{For simplicity consider a single real field $\phi$.} as
\begin{eqnarray}
\mathcal{L}=\frac{1}{2}\phi O \phi\ .
\label{FRC2}
\end{eqnarray}
To solve the equation we use the Fourier transformation defined as
\begin{eqnarray}
f\left( x \right)=\int \frac{\dd^{2} p}{\left( 2\pi \right)^{2}}e^{-ip\cdot x}\tilde{f}\left( p \right)\ ,
\label{FRC3}
\end{eqnarray}
from which we have that the transformation rule for the derivative operator is
\begin{eqnarray}
\partial_{\mu}\rightarrow -i p_{\mu}\ .
\label{FRC4}
\end{eqnarray}
Now, for quantum field theory purpose, we need the vacuum expectation value of the T-product of two fields. It can be shown that the following relation holds
\begin{eqnarray}
<0|T \phi\left( x \right)\phi\left( x' \right)|0>=iG\left( x-x' \right)\ .
\label{FRC5}
\end{eqnarray} 
Although, we use the convention to define the propagator as the Green function (\ref{FRC1}).

The vertices are defined via the Gell-Mann low formula, in which is present the factor $\exp\left[ -i S \right]$, with $S$ the action of the model. Again, in spite of this we define the vertex without any factor.
\\
The {\it 1PI $2$-point function} is defined as the inverse of the propagator.

\chapter{Feynman Rules}\label{FRappendix}

We summarize here the Feynman rules.

\begin{itemize}
\item Propagator $XX$:\\
\begin{eqnarray}
\Delta^{\beta\gamma}_{cb}(\theta)=+\frac{1}{4}\frac{\varepsilon^{\gamma\beta}\delta_{cb}}{p^{2}}
\ ;\label{FR_XX}
\end{eqnarray}
\item Vertex $BX$:\\
\begin{eqnarray}
\frac{\delta^{2}\Gamma}{\delta B_{\mu b}^{\phantom{\mu} \beta}\left( p \right)\delta \theta_{c}^{\gamma}\left( -p \right)}&=&
4\lambda^{-1}\varepsilon_{\beta\gamma}\delta^{bc}\left( -i \right)q_{\mu}=
\nonumber\\&=&
-4i\lambda^{-1}\varepsilon_{\beta\gamma}\delta^{bc}\left( -p_{\mu} \right)=
\nonumber\\&=&
4i\lambda^{-1}\varepsilon_{\beta\gamma}\delta^{bc}p_{\mu} \ ;
\label{FR_BX}
\end{eqnarray}
\item Vertex $BBXX$:\\

\begin{eqnarray}
  \left[BBXX\right]^{abcd}_{\alpha\beta\gamma\delta}&=&
V^{\left[ 2 \right]}=
	\nonumber\\&=&
\left[
-4\delta^{ac}\delta^{bd}\varepsilon_{\alpha\delta}\varepsilon_{\beta\gamma}+2\delta^{ab}\delta^{cd}\varepsilon_{\alpha\delta}\varepsilon_{\beta\gamma}+4\delta^{ad}\delta^{bc}\varepsilon_{\alpha\gamma}\varepsilon_{\beta\delta} 
+  \right.
	\nonumber\\&&
  \left.
  -2\delta^{ab}\delta^{cd}\varepsilon_{\alpha\gamma}\varepsilon_{\beta\delta}+2\delta^{ad}\delta^{bc}\varepsilon_{\alpha\beta}\varepsilon_{\gamma\delta}+2\delta^{ac}\delta^{bd}\varepsilon_{\alpha\beta}\varepsilon_{\gamma\delta}
  \right]\gamma_{\mu\nu}
\ ;  \label{FR_bfmBBXX}
\end{eqnarray}

\item Vertex $BXXX$:\\
\begin{eqnarray}
  \left[BXXX\right]^{abcd}_{\alpha\beta\gamma\delta\,\mu}&=&
-i\frac{4\lambda}{3}V^{\left[ 3 \right]}=
	\nonumber\\&=&
	-i\frac{4\lambda}{3}\left(
-4p_{B}\delta^{ac}\delta^{bd}\varepsilon_{\alpha\delta}\varepsilon_{\beta\gamma}
+2p_{C}\delta^{ac}\delta^{bd}\varepsilon_{\alpha\delta}\varepsilon_{\beta\gamma}
+  \right.
	\nonumber\\&&
  \left.
+2p_{D}\delta^{ac}\delta^{bd}\varepsilon_{\alpha\delta}\varepsilon_{\beta\gamma}
+2p_{B}\delta^{ab}\delta^{cd}\varepsilon_{\alpha\delta}\varepsilon_{\beta\gamma}
+  \right.
	\nonumber\\&&
  \left.
-4p_{C}\delta^{ab}\delta^{cd}\varepsilon_{\alpha\delta}\varepsilon_{\beta\gamma}
+2p_{D}\delta^{ab}\delta^{cd}\varepsilon_{\alpha\delta}\varepsilon_{\beta\gamma}
+  \right.
	\nonumber\\&&
  \left.
+4p_{B}\delta^{ad}\delta^{bc}\varepsilon_{\alpha\gamma}\varepsilon_{\beta\delta}
-2p_{C}\delta^{ad}\delta^{bc}\varepsilon_{\alpha\gamma}\varepsilon_{\beta\delta}
+  \right.
	\nonumber\\&&
  \left.
-2p_{D}\delta^{ad}\delta^{bc}\varepsilon_{\alpha\gamma}\varepsilon_{\beta\delta}
-2p_{B}\delta^{ab}\delta^{cd}\varepsilon_{\alpha\gamma}\varepsilon_{\beta\delta}
+  \right.
	\nonumber\\&&
  \left.
-2p_{C}\delta^{ab}\delta^{cd}\varepsilon_{\alpha\gamma}\varepsilon_{\beta\delta}
+4p_{D}\delta^{ab}\delta^{cd}\varepsilon_{\alpha\gamma}\varepsilon_{\beta\delta}
+  \right.
	\nonumber\\&&
  \left.
+2p_{B}\delta^{ad}\delta^{bc}\varepsilon_{\alpha\beta}\varepsilon_{\gamma\delta}
-4p_{C}\delta^{ad}\delta^{bc}\varepsilon_{\alpha\beta}\varepsilon_{\gamma\delta}
+  \right.
	\nonumber\\&&
  \left.
+2p_{D}\delta^{ad}\delta^{bc}\varepsilon_{\alpha\beta}\varepsilon_{\gamma\delta}
+2p_{B}\delta^{ac}\delta^{bd}\varepsilon_{\alpha\beta}\varepsilon_{\gamma\delta}
+  \right.
	\nonumber\\&&
  \left.
+2p_{C}\delta^{ac}\delta^{bd}\varepsilon_{\alpha\beta}\varepsilon_{\gamma\delta}
-4p_{D}\delta^{ac}\delta^{bd}\varepsilon_{\alpha\beta}\varepsilon_{\gamma\delta}
\right)_{\mu}
\ ;  \label{FR_bfmBXXX}
\end{eqnarray}

\item Vertex $XXXX$:\\

\begin{eqnarray}
\left[XXXX\right]^{abcd}_{\alpha\beta\gamma\delta}&=& 
-\frac{1}{3}\lambda^{2}V^{\left[ 4 \right]}=
\nonumber\\&=& 
	-\frac{1}{3}\lambda^{2}\left(
-4p_{A}\cdot p_{B}\delta^{ac}\delta^{bd}\varepsilon_{\alpha\delta}\varepsilon_{\beta\gamma}+2p_{A}\cdot p_{C}\delta^{ac}\delta^{bd}\varepsilon_{\alpha\delta}\varepsilon_{\beta\gamma}
+  \right.
	\nonumber\\&&
  \left.
+2p_{A}\cdot p_{D}\delta^{ac}\delta^{bd}\varepsilon_{\alpha\delta}\varepsilon_{\beta\gamma}+2p_{B}\cdot p_{C}\delta^{ac}\delta^{bd}\varepsilon_{\alpha\delta}\varepsilon_{\beta\gamma}
+  \right.
	\nonumber\\&&
  \left.
+2p_{B}\cdot p_{D}\delta^{ac}\delta^{bd}\varepsilon_{\alpha\delta}\varepsilon_{\beta\gamma}-4p_{C}\cdot p_{D}\delta^{ac}\delta^{bd}\varepsilon_{\alpha\delta}\varepsilon_{\beta\gamma}
+  \right.
	\nonumber\\&&
  \left.
+2p_{A}\cdot p_{B}\delta^{ab}\delta^{cd}\varepsilon_{\alpha\delta}\varepsilon_{\beta\gamma}-4p_{A}\cdot p_{C}\delta^{ab}\delta^{cd}\varepsilon_{\alpha\delta}\varepsilon_{\beta\gamma}
+  \right.
	\nonumber\\&&
  \left.
+2p_{A}\cdot p_{D}\delta^{ab}\delta^{cd}\varepsilon_{\alpha\delta}\varepsilon_{\beta\gamma}+2p_{B}\cdot p_{C}\delta^{ab}\delta^{cd}\varepsilon_{\alpha\delta}\varepsilon_{\beta\gamma}
+  \right.
	\nonumber\\&&
  \left.
-4p_{B}\cdot p_{D}\delta^{ab}\delta^{cd}\varepsilon_{\alpha\delta}\varepsilon_{\beta\gamma}+2p_{C}\cdot p_{D}\delta^{ab}\delta^{cd}\varepsilon_{\alpha\delta}\varepsilon_{\beta\gamma}
+  \right.
	\nonumber\\&&
  \left.
+4p_{A}\cdot p_{B}\delta^{ad}\delta^{bc}\varepsilon_{\alpha\gamma}\varepsilon_{\beta\delta}-2p_{A}\cdot p_{C}\delta^{ad}\delta^{bc}\varepsilon_{\alpha\gamma}\varepsilon_{\beta\delta}
+  \right.
	\nonumber\\&&
  \left.
-2p_{A}\cdot p_{D}\delta^{ad}\delta^{bc}\varepsilon_{\alpha\gamma}\varepsilon_{\beta\delta}-2p_{B}\cdot p_{C}\delta^{ad}\delta^{bc}\varepsilon_{\alpha\gamma}\varepsilon_{\beta\delta}
+  \right.
	\nonumber\\&&
  \left.
-2p_{B}\cdot p_{D}\delta^{ad}\delta^{bc}\varepsilon_{\alpha\gamma}\varepsilon_{\beta\delta}+4p_{C}\cdot p_{D}\delta^{ad}\delta^{bc}\varepsilon_{\alpha\gamma}\varepsilon_{\beta\delta}
+  \right.
	\nonumber\\&&
  \left.
-2p_{A}\cdot p_{B}\delta^{ab}\delta^{cd}\varepsilon_{\alpha\gamma}\varepsilon_{\beta\delta}-2p_{A}\cdot p_{C}\delta^{ab}\delta^{cd}\varepsilon_{\alpha\gamma}\varepsilon_{\beta\delta}
+  \right.
	\nonumber\\&&
  \left.
+4p_{A}\cdot p_{D}\delta^{ab}\delta^{cd}\varepsilon_{\alpha\gamma}\varepsilon_{\beta\delta}+4p_{B}\cdot p_{C}\delta^{ab}\delta^{cd}\varepsilon_{\alpha\gamma}\varepsilon_{\beta\delta}
+  \right.
	\nonumber\\&&
  \left.
-2p_{B}\cdot p_{D}\delta^{ab}\delta^{cd}\varepsilon_{\alpha\gamma}\varepsilon_{\beta\delta}-2p_{C}\cdot p_{D}\delta^{ab}\delta^{cd}\varepsilon_{\alpha\gamma}\varepsilon_{\beta\delta}
+  \right.
	\nonumber\\&&
  \left.
+2p_{A}\cdot p_{B}\delta^{ad}\delta^{bc}\varepsilon_{\alpha\beta}\varepsilon_{\gamma\delta}-4p_{A}\cdot p_{C}\delta^{ad}\delta^{bc}\varepsilon_{\alpha\beta}\varepsilon_{\gamma\delta}
+  \right.
	\nonumber\\&&
  \left.
+2p_{A}\cdot p_{D}\delta^{ad}\delta^{bc}\varepsilon_{\alpha\beta}\varepsilon_{\gamma\delta}+2p_{B}\cdot p_{C}\delta^{ad}\delta^{bc}\varepsilon_{\alpha\beta}\varepsilon_{\gamma\delta}
+  \right.
	\nonumber\\&&
  \left.
-4p_{B}\cdot p_{D}\delta^{ad}\delta^{bc}\varepsilon_{\alpha\beta}\varepsilon_{\gamma\delta}+2p_{C}\cdot p_{D}\delta^{ad}\delta^{bc}\varepsilon_{\alpha\beta}\varepsilon_{\gamma\delta}
+  \right.
	\nonumber\\&&
  \left.
+2p_{A}\cdot p_{B}\delta^{ac}\delta^{bd}\varepsilon_{\alpha\beta}\varepsilon_{\gamma\delta}+2p_{A}\cdot p_{C}\delta^{ac}\delta^{bd}\varepsilon_{\alpha\beta}\varepsilon_{\gamma\delta}
+  \right.
	\nonumber\\&&
  \left.
-4p_{A}\cdot p_{D}\delta^{ac}\delta^{bd}\varepsilon_{\alpha\beta}\varepsilon_{\gamma\delta}-4p_{B}\cdot p_{C}\delta^{ac}\delta^{bd}\varepsilon_{\alpha\beta}\varepsilon_{\gamma\delta}
+  \right.
	\nonumber\\&&
  \left.
+2p_{B}\cdot p_{D}\delta^{ac}\delta^{bd}\varepsilon_{\alpha\beta}\varepsilon_{\gamma\delta}+2p_{C}\cdot p_{D}\delta^{ac}\delta^{bd}\varepsilon_{\alpha\beta}\varepsilon_{\gamma\delta}
\right)
\ ;\label{FR_XXXX1.0}
\end{eqnarray}
\item Vertex $BBXXXX$:\\
\begin{eqnarray}
&&\left[BBXXXX\right]^{abcdrs}_{\alpha\beta\gamma\delta\rho\sigma}=
	\frac{\lambda^{2}}{12}V^{\left[ 6 \right]}=
\nonumber\\&=&
	\frac{\lambda^{2}}{12}
\left(
-48\delta^{ad}\delta^{bs}\delta^{cr}\varepsilon_{\alpha\sigma}\varepsilon_{\beta\rho}\varepsilon_{\gamma\delta}-48\delta^{ar}\delta^{bd}\delta^{cs}\varepsilon_{\alpha\sigma}\varepsilon_{\beta\rho}\varepsilon_{\gamma\delta}-48\delta^{ac}\delta^{bs}\delta^{dr}\varepsilon_{\alpha\sigma}\varepsilon_{\beta\rho}\varepsilon_{\gamma\delta}+
  \right.
	\nonumber\\&&
  \left.
+12\delta^{ab}\delta^{cs}\delta^{dr}\varepsilon_{\alpha\sigma}\varepsilon_{\beta\rho}\varepsilon_{\gamma\delta}-48\delta^{ar}\delta^{bc}\delta^{ds}\varepsilon_{\alpha\sigma}\varepsilon_{\beta\rho}\varepsilon_{\gamma\delta}+12\delta^{ab}\delta^{cr}\delta^{ds}\varepsilon_{\alpha\sigma}\varepsilon_{\beta\rho}\varepsilon_{\gamma\delta}
+  \right.
	\nonumber\\&&
  \left.
+72\delta^{ad}\delta^{bc}\delta^{rs}\varepsilon_{\alpha\sigma}\varepsilon_{\beta\rho}\varepsilon_{\gamma\delta}+72\delta^{ac}\delta^{bd}\delta^{rs}\varepsilon_{\alpha\sigma}\varepsilon_{\beta\rho}\varepsilon_{\gamma\delta}+48\delta^{as}\delta^{bd}\delta^{cr}\varepsilon_{\alpha\rho}\varepsilon_{\beta\sigma}\varepsilon_{\gamma\delta}
+  \right.
	\nonumber\\&&
  \left.
+48\delta^{ad}\delta^{br}\delta^{cs}\varepsilon_{\alpha\rho}\varepsilon_{\beta\sigma}\varepsilon_{\gamma\delta}+48\delta^{as}\delta^{bc}\delta^{dr}\varepsilon_{\alpha\rho}\varepsilon_{\beta\sigma}\varepsilon_{\gamma\delta}-12\delta^{ab}\delta^{cs}\delta^{dr}\varepsilon_{\alpha\rho}\varepsilon_{\beta\sigma}\varepsilon_{\gamma\delta}
+  \right.
	\nonumber\\&&
  \left.
+48\delta^{ac}\delta^{br}\delta^{ds}\varepsilon_{\alpha\rho}\varepsilon_{\beta\sigma}\varepsilon_{\gamma\delta}-12\delta^{ab}\delta^{cr}\delta^{ds}\varepsilon_{\alpha\rho}\varepsilon_{\beta\sigma}\varepsilon_{\gamma\delta}-72\delta^{ad}\delta^{bc}\delta^{rs}\varepsilon_{\alpha\rho}\varepsilon_{\beta\sigma}\varepsilon_{\gamma\delta}
+  \right.
	\nonumber\\&&
  \left.
-72\delta^{ac}\delta^{bd}\delta^{rs}\varepsilon_{\alpha\rho}\varepsilon_{\beta\sigma}\varepsilon_{\gamma\delta}+48\delta^{ar}\delta^{bs}\delta^{cd}\varepsilon_{\alpha\sigma}\varepsilon_{\beta\delta}\varepsilon_{\gamma\rho}+48\delta^{ad}\delta^{br}\delta^{cs}\varepsilon_{\alpha\sigma}\varepsilon_{\beta\delta}\varepsilon_{\gamma\rho}
+  \right.
	\nonumber\\&&
  \left.
+48\delta^{ac}\delta^{bs}\delta^{dr}\varepsilon_{\alpha\sigma}\varepsilon_{\beta\delta}\varepsilon_{\gamma\rho}-12\delta^{ab}\delta^{cs}\delta^{dr}\varepsilon_{\alpha\sigma}\varepsilon_{\beta\delta}\varepsilon_{\gamma\rho}-72\delta^{ar}\delta^{bc}\delta^{ds}\varepsilon_{\alpha\sigma}\varepsilon_{\beta\delta}\varepsilon_{\gamma\rho}
+  \right.
	\nonumber\\&&
  \left.
-72\delta^{ac}\delta^{br}\delta^{ds}\varepsilon_{\alpha\sigma}\varepsilon_{\beta\delta}\varepsilon_{\gamma\rho}+48\delta^{ad}\delta^{bc}\delta^{rs}\varepsilon_{\alpha\sigma}\varepsilon_{\beta\delta}\varepsilon_{\gamma\rho}-12\delta^{ab}\delta^{cd}\delta^{rs}\varepsilon_{\alpha\sigma}\varepsilon_{\beta\delta}\varepsilon_{\gamma\rho}
+  \right.
	\nonumber\\&&
  \left.
-48\delta^{as}\delta^{br}\delta^{cd}\varepsilon_{\alpha\delta}\varepsilon_{\beta\sigma}\varepsilon_{\gamma\rho}-48\delta^{ar}\delta^{bd}\delta^{cs}\varepsilon_{\alpha\delta}\varepsilon_{\beta\sigma}\varepsilon_{\gamma\rho}-48\delta^{as}\delta^{bc}\delta^{dr}\varepsilon_{\alpha\delta}\varepsilon_{\beta\sigma}\varepsilon_{\gamma\rho}
+  \right.
	\nonumber\\&&
  \left.
+12\delta^{ab}\delta^{cs}\delta^{dr}\varepsilon_{\alpha\delta}\varepsilon_{\beta\sigma}\varepsilon_{\gamma\rho}+72\delta^{ar}\delta^{bc}\delta^{ds}\varepsilon_{\alpha\delta}\varepsilon_{\beta\sigma}\varepsilon_{\gamma\rho}+72\delta^{ac}\delta^{br}\delta^{ds}\varepsilon_{\alpha\delta}\varepsilon_{\beta\sigma}\varepsilon_{\gamma\rho}
+  \right.
	\nonumber\\&&
  \left.
-48\delta^{ac}\delta^{bd}\delta^{rs}\varepsilon_{\alpha\delta}\varepsilon_{\beta\sigma}\varepsilon_{\gamma\rho}+12\delta^{ab}\delta^{cd}\delta^{rs}\varepsilon_{\alpha\delta}\varepsilon_{\beta\sigma}\varepsilon_{\gamma\rho}-48\delta^{as}\delta^{br}\delta^{cd}\varepsilon_{\alpha\rho}\varepsilon_{\beta\delta}\varepsilon_{\gamma\sigma}
+  \right.
	\nonumber\\&&
  \left.
-48\delta^{ad}\delta^{bs}\delta^{cr}\varepsilon_{\alpha\rho}\varepsilon_{\beta\delta}\varepsilon_{\gamma\sigma}+72\delta^{as}\delta^{bc}\delta^{dr}\varepsilon_{\alpha\rho}\varepsilon_{\beta\delta}\varepsilon_{\gamma\sigma}+72\delta^{ac}\delta^{bs}\delta^{dr}\varepsilon_{\alpha\rho}\varepsilon_{\beta\delta}\varepsilon_{\gamma\sigma}
+  \right.
	\nonumber\\&&
  \left.
-48\delta^{ac}\delta^{br}\delta^{ds}\varepsilon_{\alpha\rho}\varepsilon_{\beta\delta}\varepsilon_{\gamma\sigma}+12\delta^{ab}\delta^{cr}\delta^{ds}\varepsilon_{\alpha\rho}\varepsilon_{\beta\delta}\varepsilon_{\gamma\sigma}-48\delta^{ad}\delta^{bc}\delta^{rs}\varepsilon_{\alpha\rho}\varepsilon_{\beta\delta}\varepsilon_{\gamma\sigma}
+  \right.
	\nonumber\\&&
  \left.
+12\delta^{ab}\delta^{cd}\delta^{rs}\varepsilon_{\alpha\rho}\varepsilon_{\beta\delta}\varepsilon_{\gamma\sigma}+48\delta^{ar}\delta^{bs}\delta^{cd}\varepsilon_{\alpha\delta}\varepsilon_{\beta\rho}\varepsilon_{\gamma\sigma}+48\delta^{as}\delta^{bd}\delta^{cr}\varepsilon_{\alpha\delta}\varepsilon_{\beta\rho}\varepsilon_{\gamma\sigma}
+  \right.
	\nonumber\\&&
  \left.
-72\delta^{as}\delta^{bc}\delta^{dr}\varepsilon_{\alpha\delta}\varepsilon_{\beta\rho}\varepsilon_{\gamma\sigma}-72\delta^{ac}\delta^{bs}\delta^{dr}\varepsilon_{\alpha\delta}\varepsilon_{\beta\rho}\varepsilon_{\gamma\sigma}+48\delta^{ar}\delta^{bc}\delta^{ds}\varepsilon_{\alpha\delta}\varepsilon_{\beta\rho}\varepsilon_{\gamma\sigma}
+  \right.
	\nonumber\\&&
  \left.
-12\delta^{ab}\delta^{cr}\delta^{ds}\varepsilon_{\alpha\delta}\varepsilon_{\beta\rho}\varepsilon_{\gamma\sigma}+48\delta^{ac}\delta^{bd}\delta^{rs}\varepsilon_{\alpha\delta}\varepsilon_{\beta\rho}\varepsilon_{\gamma\sigma}-12\delta^{ab}\delta^{cd}\delta^{rs}\varepsilon_{\alpha\delta}\varepsilon_{\beta\rho}\varepsilon_{\gamma\sigma}
+  \right.
	\nonumber\\&&
  \left.
-48\delta^{ar}\delta^{bs}\delta^{cd}\varepsilon_{\alpha\sigma}\varepsilon_{\beta\gamma}\varepsilon_{\delta\rho}-48\delta^{ad}\delta^{bs}\delta^{cr}\varepsilon_{\alpha\sigma}\varepsilon_{\beta\gamma}\varepsilon_{\delta\rho}+72\delta^{ar}\delta^{bd}\delta^{cs}\varepsilon_{\alpha\sigma}\varepsilon_{\beta\gamma}\varepsilon_{\delta\rho}
+  \right.
	\nonumber\\&&
  \left.
+72\delta^{ad}\delta^{br}\delta^{cs}\varepsilon_{\alpha\sigma}\varepsilon_{\beta\gamma}\varepsilon_{\delta\rho}-48\delta^{ac}\delta^{br}\delta^{ds}\varepsilon_{\alpha\sigma}\varepsilon_{\beta\gamma}\varepsilon_{\delta\rho}+12\delta^{ab}\delta^{cr}\delta^{ds}\varepsilon_{\alpha\sigma}\varepsilon_{\beta\gamma}\varepsilon_{\delta\rho}
+  \right.
	\nonumber\\&&
  \left.
-48\delta^{ac}\delta^{bd}\delta^{rs}\varepsilon_{\alpha\sigma}\varepsilon_{\beta\gamma}\varepsilon_{\delta\rho}+12\delta^{ab}\delta^{cd}\delta^{rs}\varepsilon_{\alpha\sigma}\varepsilon_{\beta\gamma}\varepsilon_{\delta\rho}+48\delta^{as}\delta^{br}\delta^{cd}\varepsilon_{\alpha\gamma}\varepsilon_{\beta\sigma}\varepsilon_{\delta\rho}
+  \right.
	\nonumber\\&&
  \left.
+48\delta^{as}\delta^{bd}\delta^{cr}\varepsilon_{\alpha\gamma}\varepsilon_{\beta\sigma}\varepsilon_{\delta\rho}-72\delta^{ar}\delta^{bd}\delta^{cs}\varepsilon_{\alpha\gamma}\varepsilon_{\beta\sigma}\varepsilon_{\delta\rho}-72\delta^{ad}\delta^{br}\delta^{cs}\varepsilon_{\alpha\gamma}\varepsilon_{\beta\sigma}\varepsilon_{\delta\rho}
+  \right.
	\nonumber\\&&
  \left.
+48\delta^{ar}\delta^{bc}\delta^{ds}\varepsilon_{\alpha\gamma}\varepsilon_{\beta\sigma}\varepsilon_{\delta\rho}-12\delta^{ab}\delta^{cr}\delta^{ds}\varepsilon_{\alpha\gamma}\varepsilon_{\beta\sigma}\varepsilon_{\delta\rho}+48\delta^{ad}\delta^{bc}\delta^{rs}\varepsilon_{\alpha\gamma}\varepsilon_{\beta\sigma}\varepsilon_{\delta\rho}
+  \right.
	\nonumber\\&&
  \left.
-12\delta^{ab}\delta^{cd}\delta^{rs}\varepsilon_{\alpha\gamma}\varepsilon_{\beta\sigma}\varepsilon_{\delta\rho}+12\delta^{as}\delta^{br}\delta^{cd}\varepsilon_{\alpha\beta}\varepsilon_{\gamma\sigma}\varepsilon_{\delta\rho}+12\delta^{ar}\delta^{bs}\delta^{cd}\varepsilon_{\alpha\beta}\varepsilon_{\gamma\sigma}\varepsilon_{\delta\rho}
+  \right.
	\nonumber\\&&
  \left.
+12\delta^{as}\delta^{bd}\delta^{cr}\varepsilon_{\alpha\beta}\varepsilon_{\gamma\sigma}\varepsilon_{\delta\rho}+12\delta^{ad}\delta^{bs}\delta^{cr}\varepsilon_{\alpha\beta}\varepsilon_{\gamma\sigma}\varepsilon_{\delta\rho}+12\delta^{ar}\delta^{bc}\delta^{ds}\varepsilon_{\alpha\beta}\varepsilon_{\gamma\sigma}\varepsilon_{\delta\rho}
+  \right.
	\nonumber\\&&
  \left.
+12\delta^{ac}\delta^{br}\delta^{ds}\varepsilon_{\alpha\beta}\varepsilon_{\gamma\sigma}\varepsilon_{\delta\rho}+12\delta^{ad}\delta^{bc}\delta^{rs}\varepsilon_{\alpha\beta}\varepsilon_{\gamma\sigma}\varepsilon_{\delta\rho}+12\delta^{ac}\delta^{bd}\delta^{rs}\varepsilon_{\alpha\beta}\varepsilon_{\gamma\sigma}\varepsilon_{\delta\rho}
+  \right.
	\nonumber\\&&
  \left.
+48\delta^{as}\delta^{br}\delta^{cd}\varepsilon_{\alpha\rho}\varepsilon_{\beta\gamma}\varepsilon_{\delta\sigma}-72\delta^{as}\delta^{bd}\delta^{cr}\varepsilon_{\alpha\rho}\varepsilon_{\beta\gamma}\varepsilon_{\delta\sigma}-72\delta^{ad}\delta^{bs}\delta^{cr}\varepsilon_{\alpha\rho}\varepsilon_{\beta\gamma}\varepsilon_{\delta\sigma}
+  \right.
	\nonumber\\&&
  \left.
+48\delta^{ad}\delta^{br}\delta^{cs}\varepsilon_{\alpha\rho}\varepsilon_{\beta\gamma}\varepsilon_{\delta\sigma}+48\delta^{ac}\delta^{bs}\delta^{dr}\varepsilon_{\alpha\rho}\varepsilon_{\beta\gamma}\varepsilon_{\delta\sigma}-12\delta^{ab}\delta^{cs}\delta^{dr}\varepsilon_{\alpha\rho}\varepsilon_{\beta\gamma}\varepsilon_{\delta\sigma}
+  \right.
	\nonumber\\&&
  \left.
+48\delta^{ac}\delta^{bd}\delta^{rs}\varepsilon_{\alpha\rho}\varepsilon_{\beta\gamma}\varepsilon_{\delta\sigma}-12\delta^{ab}\delta^{cd}\delta^{rs}\varepsilon_{\alpha\rho}\varepsilon_{\beta\gamma}\varepsilon_{\delta\sigma}-48\delta^{ar}\delta^{bs}\delta^{cd}\varepsilon_{\alpha\gamma}\varepsilon_{\beta\rho}\varepsilon_{\delta\sigma}
+  \right.
	\nonumber\\&&
  \left.
+72\delta^{as}\delta^{bd}\delta^{cr}\varepsilon_{\alpha\gamma}\varepsilon_{\beta\rho}\varepsilon_{\delta\sigma}+72\delta^{ad}\delta^{bs}\delta^{cr}\varepsilon_{\alpha\gamma}\varepsilon_{\beta\rho}\varepsilon_{\delta\sigma}-48\delta^{ar}\delta^{bd}\delta^{cs}\varepsilon_{\alpha\gamma}\varepsilon_{\beta\rho}\varepsilon_{\delta\sigma}
+  \right.
	\nonumber\\&&
  \left.
-48\delta^{as}\delta^{bc}\delta^{dr}\varepsilon_{\alpha\gamma}\varepsilon_{\beta\rho}\varepsilon_{\delta\sigma}+12\delta^{ab}\delta^{cs}\delta^{dr}\varepsilon_{\alpha\gamma}\varepsilon_{\beta\rho}\varepsilon_{\delta\sigma}-48\delta^{ad}\delta^{bc}\delta^{rs}\varepsilon_{\alpha\gamma}\varepsilon_{\beta\rho}\varepsilon_{\delta\sigma}
+  \right.
	\nonumber\\&&
  \left.
+12\delta^{ab}\delta^{cd}\delta^{rs}\varepsilon_{\alpha\gamma}\varepsilon_{\beta\rho}\varepsilon_{\delta\sigma}-12\delta^{as}\delta^{br}\delta^{cd}\varepsilon_{\alpha\beta}\varepsilon_{\gamma\rho}\varepsilon_{\delta\sigma}-12\delta^{ar}\delta^{bs}\delta^{cd}\varepsilon_{\alpha\beta}\varepsilon_{\gamma\rho}\varepsilon_{\delta\sigma}
+  \right.
	\nonumber\\&&
  \left.
-12\delta^{ar}\delta^{bd}\delta^{cs}\varepsilon_{\alpha\beta}\varepsilon_{\gamma\rho}\varepsilon_{\delta\sigma}-12\delta^{ad}\delta^{br}\delta^{cs}\varepsilon_{\alpha\beta}\varepsilon_{\gamma\rho}\varepsilon_{\delta\sigma}-12\delta^{as}\delta^{bc}\delta^{dr}\varepsilon_{\alpha\beta}\varepsilon_{\gamma\rho}\varepsilon_{\delta\sigma}
+  \right.
	\nonumber\\&&
  \left.
-12\delta^{ac}\delta^{bs}\delta^{dr}\varepsilon_{\alpha\beta}\varepsilon_{\gamma\rho}\varepsilon_{\delta\sigma}-12\delta^{ad}\delta^{bc}\delta^{rs}\varepsilon_{\alpha\beta}\varepsilon_{\gamma\rho}\varepsilon_{\delta\sigma}-12\delta^{ac}\delta^{bd}\delta^{rs}\varepsilon_{\alpha\beta}\varepsilon_{\gamma\rho}\varepsilon_{\delta\sigma}
+  \right.
	\nonumber\\&&
  \left.
+72\delta^{as}\delta^{br}\delta^{cd}\varepsilon_{\alpha\delta}\varepsilon_{\beta\gamma}\varepsilon_{\rho\sigma}+72\delta^{ar}\delta^{bs}\delta^{cd}\varepsilon_{\alpha\delta}\varepsilon_{\beta\gamma}\varepsilon_{\rho\sigma}-48\delta^{as}\delta^{bd}\delta^{cr}\varepsilon_{\alpha\delta}\varepsilon_{\beta\gamma}\varepsilon_{\rho\sigma}
+  \right.
	\nonumber\\&&
  \left.
-48\delta^{ar}\delta^{bd}\delta^{cs}\varepsilon_{\alpha\delta}\varepsilon_{\beta\gamma}\varepsilon_{\rho\sigma}-48\delta^{ac}\delta^{bs}\delta^{dr}\varepsilon_{\alpha\delta}\varepsilon_{\beta\gamma}\varepsilon_{\rho\sigma}+12\delta^{ab}\delta^{cs}\delta^{dr}\varepsilon_{\alpha\delta}\varepsilon_{\beta\gamma}\varepsilon_{\rho\sigma}
+  \right.
	\nonumber\\&&
  \left.
-48\delta^{ac}\delta^{br}\delta^{ds}\varepsilon_{\alpha\delta}\varepsilon_{\beta\gamma}\varepsilon_{\rho\sigma}+12\delta^{ab}\delta^{cr}\delta^{ds}\varepsilon_{\alpha\delta}\varepsilon_{\beta\gamma}\varepsilon_{\rho\sigma}-72\delta^{as}\delta^{br}\delta^{cd}\varepsilon_{\alpha\gamma}\varepsilon_{\beta\delta}\varepsilon_{\rho\sigma}
+  \right.
	\nonumber\\&&
  \left.
-72\delta^{ar}\delta^{bs}\delta^{cd}\varepsilon_{\alpha\gamma}\varepsilon_{\beta\delta}\varepsilon_{\rho\sigma}+48\delta^{ad}\delta^{bs}\delta^{cr}\varepsilon_{\alpha\gamma}\varepsilon_{\beta\delta}\varepsilon_{\rho\sigma}+48\delta^{ad}\delta^{br}\delta^{cs}\varepsilon_{\alpha\gamma}\varepsilon_{\beta\delta}\varepsilon_{\rho\sigma}
+  \right.
	\nonumber\\&&
  \left.
+48\delta^{as}\delta^{bc}\delta^{dr}\varepsilon_{\alpha\gamma}\varepsilon_{\beta\delta}\varepsilon_{\rho\sigma}-12\delta^{ab}\delta^{cs}\delta^{dr}\varepsilon_{\alpha\gamma}\varepsilon_{\beta\delta}\varepsilon_{\rho\sigma}+48\delta^{ar}\delta^{bc}\delta^{ds}\varepsilon_{\alpha\gamma}\varepsilon_{\beta\delta}\varepsilon_{\rho\sigma}
+  \right.
	\nonumber\\&&
  \left.
-12\delta^{ab}\delta^{cr}\delta^{ds}\varepsilon_{\alpha\gamma}\varepsilon_{\beta\delta}\varepsilon_{\rho\sigma}+12\delta^{as}\delta^{bd}\delta^{cr}\varepsilon_{\alpha\beta}\varepsilon_{\gamma\delta}\varepsilon_{\rho\sigma}+12\delta^{ad}\delta^{bs}\delta^{cr}\varepsilon_{\alpha\beta}\varepsilon_{\gamma\delta}\varepsilon_{\rho\sigma}
+  \right.
	\nonumber\\&&
  \left.
+12\delta^{ar}\delta^{bd}\delta^{cs}\varepsilon_{\alpha\beta}\varepsilon_{\gamma\delta}\varepsilon_{\rho\sigma}+12\delta^{ad}\delta^{br}\delta^{cs}\varepsilon_{\alpha\beta}\varepsilon_{\gamma\delta}\varepsilon_{\rho\sigma}+12\delta^{as}\delta^{bc}\delta^{dr}\varepsilon_{\alpha\beta}\varepsilon_{\gamma\delta}\varepsilon_{\rho\sigma}
+  \right.
	\nonumber\\&&
  \left.
+12\delta^{ac}\delta^{bs}\delta^{dr}\varepsilon_{\alpha\beta}\varepsilon_{\gamma\delta}\varepsilon_{\rho\sigma}+12\delta^{ar}\delta^{bc}\delta^{ds}\varepsilon_{\alpha\beta}\varepsilon_{\gamma\delta}\varepsilon_{\rho\sigma}+12\delta^{ac}\delta^{br}\delta^{ds}\varepsilon_{\alpha\beta}\varepsilon_{\gamma\delta}\varepsilon_{\rho\sigma}
\right)
\ .\label{FR_bfmBBXXXX}
\end{eqnarray}

\end{itemize}

\chapter{Notations for part II}
\label{notazII}

In this appendix we summarize the notations for the second part of this Thesis.

\begin{longtable}{p{0.5\textwidth}p{0.5\textwidth}}
	\Large{Expression} & \Large{Meaning} \\[1ex]
\hline
& \\
Uppercase Latin indices $\{M,N,\cdots\}$ & Curved bulk indices in $d+1$ dimensions \\[1ex]
Uppercase Latin indices $\{A,B,\cdots\}$ & Flat bulk indices in $d+1$ dimensions \\[1ex]
Lowecase Greek indices $\{\mu,\nu,\cdots\}$ & Curved boundary indices in $d$ dimensions \\[1ex]
& \\[1ex]
$\{t,r,i\}$ & Curved time, radial and spatial boundary directions \\[1ex]
$\{0,1,a\}$ & Flat time, radial and spatial boundary directions \\[1ex]
& \\[1ex]
$\eta_{MN} = \textrm{diag}[-1,1,\cdots,1]$  &  Minkowski metric \\[1ex]
$g_{MN}$ & Bulk metric in $d+1$ dimensions \\[1ex]
$\gamma_{\mu\nu}$ & Boundary metric in $d$ dimensions\\[1ex]
$T^{\mu\nu}$ & Boundary energy--momentum tensor\\[1ex]
& \\[1ex]
$e_{M}^{A}$ & Vielbein \\[1ex]
$e^{M}_{A}$ & Inverse vielbein \\[1ex]
& \\[1ex]
Symmetrization & $\left( A\,,\,B \right)=\frac{1}{2}(AB+BA)$ \\[1ex]
Antisymmetrization & $\left[ A\,,\,B \right]=\frac{1}{2}(AB-BA)$ \\[1ex]

\end{longtable}

\chapter{$AdS_{5}$ in Eddington--Finkelstein Coordinates}
\label{EFcoords}

In \cite{Bhattacharyya:2008jc,Rangamani:2009xk} the authors formalize the gauge/gravity procedure using the metric in Eddington--Finkelstein coordinates. 
Here we present some results in $5$--dimensions in the same coordinates system.

Eddington-Finkelstein coordinates for the 
$AdS_{5}$ metric (\ref{AdSmetric0})
are defined through the following change of variables:
\begin{equation}
t = v + \frac{1}{r}
\ ,
\label{EddFincoord0}
\end{equation}
by which $AdS_{5}$ metric reads
\begin{equation}
	\dd s^{2} =
- r^{2} \dd v^{2}
+ 2 \dd r \dd v
+ r^{2} \sum_{i = 1}^{3} \dd x_{i}^{2}
\ .
\label{AdSmetricEF0}
\end{equation}
The non-zero vielbein components are
\begin{align}
	e^{0}_{v} 	
&=
	r\ ,
&
	e^{0}_{r} 	
&=
	-\frac{1}{r}\ ,
&
	e^{1}_{r} 	
&=
	\frac{1}{r}\ ,
&
	e^{a}_{i} 	
&=
	r\delta_{i}^{a}
\ ;
	\label{AdSvielbeinBHef}
\end{align}
while the non-zero components of spin connection are
\begin{align}\label{AdSspinconBHef}
	\omega^{01}_{v}
&=
	r\ ,
&
	\omega^{01}_{r} 	
&=
	-\frac{1}{r}
&
	\omega^{a1}_{i} 	
&=
	r\,\delta_{i}^{a}
\ .
\end{align}
Eq.~(\ref{AdSmetricBH}) in this coordinates system is
\begin{equation}
\dd s^{2} =
- \left( r^{2}+\frac{\mu}{r^{2}} \right) \dd v^{2}
+ 2 \dd r \dd v
+ r^{2} \sum_{i = 1}^{3} \dd x_{i}^{2}
\ ,
\label{AdSmetricEF_BH}
\end{equation}
where we used the following change of coordinates
\begin{equation}
t = v - \int \frac{1}{ r^{2}+\frac{\mu}{r^{2}}}\,\dd r
\ .
\label{EddFincoordBH}
\end{equation}

The variation of the black hole metric in the Eddington-Finkelstein coordinates (\ref{AdSmetricEF_BH}), generated by these Killing vectors with all the conformal parameters set to zero reads
\begin{align}
	\dd s^{2} = &
	2 \dd v\,\dd r- h^{2}\left(r\right) \dd v^{2}
	+r^{2}\dd x_{i}\dd x^{i}
	+\nonumber\\
	&
	-2b_{i}\left( 1-\frac{r^{2}}{h^{2}\left(r\right)} \right) \dd x^{i}\, \dd r
	-2b_{i}\left( r^{2}-h^{2}\left(r\right) \right)\dd x^{i}\,\dd v
	+4 \mu \frac{b}{r^{2}} \dd v^{2}
	\ ,
	\label{MinwOurBBB}
\end{align}
where $h\left(r\right)=\sqrt{r^{2}+\frac{\mu}{r^{2}}}$.
In \cite{Bhattacharyya:2008jc,Rangamani:2009xk} a different frame has been chosen, that is achieved by setting $\mu=-1$ and by a change of coordinates
generated by the following vectors
\begin{align}
	\xi^{i} = &
	\int \frac{f^{i}\left( r \right)}{r^{2}}\dd r + \hat w^{i}{}_{j}x^{j}+\hat{d}^{i}
	\ ,\nonumber\\
	\xi^{r} = & \xi^{v} = 0 \ ,
	\label{diff1}
\end{align}
where $f^{i}\left( r \right)= 2 b^{i}\frac{r^{2}}{h^{2}\left(r\right)}\ $, $\hat w^{i}{}_{j}$ is an antisymmetric matrix and $\hat{d}^{i}$ is a constant. We get
\begin{align}
	\dd s^{2} = &
	2 \dd v\,\dd r- h^{2}\left(r\right) \dd v^{2}
	+r^{2}\dd x_{i}\dd x^{i}
	+\nonumber\\
	&
	-2b_{i} \dd x^{i}\, \dd r
	-2b_{i}\left( r^{2}-h^{2}\left(r\right) \right)\dd x^{i}\,\dd v
	-4 \frac{b}{r^{2}} \dd v^{2}
	\ .
	\label{MinwBBB}
\end{align}

\chapter{Fierz Identities}
\label{fierz}

We list here some of the properties of Majorana spinors and some useful Fierz Identities:
\begin{equation}
\label{symmetry}
\begin{split}
  \bar s_1 M s_2 &= \bar s_2 M s_1 \; \; {\rm{if}} \; M = 1,\gamma_5,
  \gamma_5 \gamma^\mu \,,\\
  \bar s_1 M s_2 &= - \bar s_2 M s_1 \; \; {\rm{if}} \; M =
  \gamma^\mu, \gamma^{\mu\nu}\,.
\end{split}
\end{equation}
The Fierz Identities for 2 identical spinors read
\begin{equation}
\label{fierzdue}
\theta \bar \theta = - \frac{1}{4} \left( \bar \theta \theta + \bar \theta \gamma_5 \theta \gamma_5 - \bar \theta \gamma_5 \gamma_\mu \theta \gamma_5 \gamma^\mu \right)\,,
\end{equation}
while those for 3 spinors are
\begin{equation}
\label{fierztre}
\begin{split}
  \theta ( \bar \theta \theta) &= - \gamma_5 \theta \bar \theta \gamma_5 \theta \,,\\
  \theta ( \bar \theta \gamma_5 \gamma_\mu \theta) &= - \gamma_\mu \theta \bar \theta \gamma_5 \theta \,.
\end{split}
\end{equation}
Using (\ref{fierztre}) it is easy to show that the following identities also hold

\begin{equation}
\label{tetaquattro}
\begin{split}
  (\bar \theta \theta)^2 &= - (\bar \theta \gamma_5 \theta)^2 \,,\\
  (\bar \theta \gamma_5 \gamma_\mu \theta) (\bar \theta \gamma_5 \gamma_\nu \theta) &= - \eta_{\mu\nu} (\bar \theta \gamma_5 \theta)^2 \,,\\
  (\bar \theta \theta) (\bar \theta \gamma_5 \theta) &= (\bar \theta \theta) (\bar \theta \gamma_5 \gamma_\mu \theta) = (\bar \theta \gamma_5 \theta) (\bar \theta \gamma_5 \gamma_\mu \theta) = 0 \,.
\end{split}
\end{equation}
Finally the integration measure for Grassmann variables is

\begin{equation}
\int d^4 \theta (\bar \theta \gamma_5 \theta)^2 = -4 \label{berezin}\,.
\end{equation}


\chapter{Complete Lagrangian}
\label{lagfinale}

Here we present the complete expansion of the supersymetric Lagrangian (\ref{susyLAG}). This can be rewritten as
\begin{equation}
	\begin{split}
		\mathcal{L}=
		\int d^4 x\int d^{4}\theta\left( 
		-
		J A
		+		
		\sum_{i=0}^{4}\frac{1}{i!}
		F^{\left( i \right)}L_{i}
		\right)\,,
\end{split}
	\label{AppsusyLagExp1}
\end{equation}
where $F^{\left( i \right)}$ is the order $i$-derivative of $ F({\cal J}_\mu {\cal J}^\mu)$ computed at $ {\cal J}_\mu {\cal J}^\mu = j_{\mu}j^{\mu}$ and
\begin{equation}
	\begin{split}
		L_{i}=&\left(  {\cal J}_\mu {\cal J}^\mu -j_{\mu}j^{\mu} \right)^{i}J^{2}\,.
	\end{split}
	\label{AppdefLi}
\end{equation}
In the following we show the explicit form of the four $L_{i}$. To perform the computation we developed a program written in \verb FORM  language (see \cite{Vermaseren:2000nd} and references therein) which, given a set of superfields expanded in components, returns as result any desired combination of these fields, integrated over $d^4 \theta$. The 
subroutine structure of the program allows us to check every intermediate passage, or to use each single procedure to perform different calculations such as Fierz identities or gamma manipulations. 

Notice that only $L_{1}$ and $L_{2}$ has purely bosonic terms (\ref{L1bos}) and (\ref{L2bos}).

\begin{subequations}
\begin{align}
L_{1} = &
- C^2 \left[ j_\m \square j^\m + \left(\p_\m \p_\n C \p^\m\p^\n C + 2 \square C \square C \right) \right] + \nonumber\\
& + 4 C j^\m j^\n \left(\p_\m \p_\n - \eta_{\m\n} \square \right) C  + \label{L1bos}\\
& - C^2  \p_\m \bar \omega \p^\m {\not\!\p} \omega + \nonumber\\
& + 2 C^2 \square \bar \omega {\not\!\p} \omega + \nonumber\\
& - 2 i C j^\m \p_\m \bar \omega \g_5 {\not\!\p} \omega + \nonumber\\
& - i C j^\m \p_\n \bar \omega \g_5 \g_\m \p^\n \omega + \nonumber\\
& + 4 C \square C \bar \omega {\not\!\p} \omega + \nonumber\\
& + 2 C \p_\m\p_\n C \bar \omega \g^\m \p^\n \omega + \nonumber\\
& + 2 C j_\m \p_\n \bar \omega \g_\rho \p_\s \omega \varepsilon^{\m\n\rho\s} + \nonumber\\
& - 2 i C j^\m \bar \omega \g_5 \p_\m {\not\!\p} \omega  + \nonumber\\
& + 2 i C j^\m \bar \omega \g_5 \g_\m \square \omega + \nonumber\\
& + 2 j^2 \bar \omega {\not\!\p} \omega + \nonumber\\
& - 2 j^\m j^\n \bar \omega \g_\m \p_\n \omega + \nonumber\\
& - i j^\m \left(\p_\m \p_\n - \eta_{\m\n} \square \right) C \bar \omega \g_5 \g^\n \omega + \nonumber\\
& - \frac{3}{4} \bar \omega \omega \p_\m \bar \omega \p^\m \omega + \nonumber\\
& - \frac{1}{2} \bar \omega \omega \p_\m \bar \omega \g^{\m\n} \p_\n \omega + \nonumber\\
& + \frac{3}{4} \bar \omega \g_5 \omega \p_\m \bar \omega \g_5 \p^\m \omega + \nonumber\\
& + \frac{1}{2} \bar \omega \g_5 \omega \p_\m \bar \omega \g_5 \g^{\m\n} \p_\n \omega + \nonumber\\
& - \bar \omega \g_5 \g^\m \omega \p_\m \bar \omega \g_5 {\not\!\p} \omega + \nonumber\\
& + \frac{1}{4} \bar \omega \g_5 \g_\m \omega \p_\n \bar \omega \g_5 \g^\m \p^\n \omega
\,,
\end{align}
\end{subequations}

\begin{subequations}
\begin{align}
L_{2} =& - 4 C^2 j^\m j^\n \left( \p_\m \p_\rho - \eta_{\m\rho} \square \right) C \left( \p_\n \p^\rho - \d_\n^\rho \square \right) C  
+ \label{L2bos}\\
& - 2 C^2 j^\m \left(\p_\m \p_\n - \eta_{\m\n} \square \right) C \left[ \p_\rho \bar \omega \g_\s \p_\tau \omega \varepsilon_{\n\rho\s\tau} - 6 i \p^\n \bar \omega \g_5 {\not\!\p} \omega + i \p_\rho \bar \omega \g_5 \g^\n \p^\rho \omega \right] \nonumber\\
& + 6 C^2 \square C j_\m \left[ \p_\n \bar \omega \g_\rho \p_\s \omega \varepsilon^{\m\n\rho\s} + 2 i  \p^\m \bar \omega \g_5 {\not\!\p} \omega - 2 i \p_\n \bar \omega \g_5 \g^\m \p^\n \omega \right] \nonumber\\
& + 2 C^2 j_\m \left( \p_\a \p_\n - \eta_{\a\n} \square \right) C \p_\rho \bar \omega \g^\a \p_\s \omega \varepsilon^{\m\n\rho\s} + \nonumber\\
& - 4 i C^2 \left( \p_\m \p_\n - \eta_{\m\n} \square \right) C j^\rho \p^\m \bar \omega \g_5 \g_\rho \p^\n \omega + \nonumber\\
& + \frac{9}{4} C^2 \p_\m \bar \omega \p^\m \omega \p_\n \bar \omega \p^\n \omega + \nonumber\\
& + 3 C^2 \p_\m \bar \omega \p^\m \omega \p_\n \bar \omega \g^{\n\rho} \p_\rho \omega + \nonumber\\
& - \frac{9}{4} C^2 \p_\m \bar \omega \g_5 \p^\m \omega \p_\n \bar \omega \g_5 \p^\n \omega + \nonumber\\
& - 3 C^2 \p_\m \bar \omega \g_5 \p^\m \omega \p_\n \bar \omega \g_5 \g^{\n\rho} \p_\rho \omega + \nonumber\\
& - 2 C^2 \p_\m \bar \omega \g_5 \g^\n \p^\m \omega \p_\n \bar \omega \g_5 {\not\!\p} \omega + \nonumber\\
& + \frac{1}{4} C^2 \p_\m \bar \omega \g_5 \g^\n \p^\m \omega \p_\rho \bar \omega \g_5 \g_\n \p^\rho \omega + \nonumber\\
& + 4 C^2 \p^\m \bar \omega \g_5 {\not\!\p} \omega \p_\m \bar \omega \g_5 {\not\!\p} \omega + \nonumber\\
& + C^2 \p_\m \bar \omega \g^{\m\n} \p_\n \omega \p_\rho \bar \omega \g^{\rho\s} \p_\s \omega + \nonumber\\
& - C^2 \p_\m \bar \omega \g_5 \g^{\m\n} \p_\n \omega \p_\rho \bar \omega \g_5 \g^{\rho\s} \p_\s \omega + \nonumber\\
& + 4 C^2 j^\m j^\n \p_\m \bar \omega \g_\n \square \omega + \nonumber\\
& - 4 C^2 j^\m j^\n \p_\m \p_\n \bar \omega {\not\!\p} \omega + \nonumber\\
& - 4 C^2 j^\m j^\n\p_\rho \bar \omega \g_\m \p_\n \p_\rho \omega + \nonumber\\
& + 4 C^2 j^2 \square \bar \omega {\not\!\p} \omega + \nonumber\\
& + 4 i C^2 j^\tau j_\m \p_\n \bar \omega \g_5 \g_\rho \p_\tau \p_\s \omega \varepsilon^{\m\n\rho\s} + \nonumber\\
& - 4 C j^2 \left[ j_\m \p_\n \bar \omega \g_\s \p_\rho \omega \varepsilon^{\m\n\rho\s} - 2 i j^\m \p_\m \bar \omega \g_5 {\not\!\p} \omega + 2i j^\n \p_\m \bar \omega \g_5 \g_\n \p^\m \omega \right] + \nonumber\\
& + 4 i C j^\m j^\n j^\rho \p_\m \bar \omega \g_5 \g_\n \p_\rho \omega + \nonumber\\
& - 8 i C j^2 j^\m \p_\m \bar \omega \g_5 {\not\!\p} \omega + \nonumber\\
& + 4 i C j^2 j^\m \p_\n \bar \omega \g_5 \g_\m \p^\n \omega + \nonumber\\
& - 8 C j^\m j^\n \left( \p_\m \p_\n - \eta_{\m\n} \square \right) C \bar\omega {\not\!\p} \omega + \nonumber\\
& + 8 C j^\m j^\n \left(\p_\n \p_\rho - \eta_{\n\rho} \square \right) C \bar\omega \g_\n \p^\rho \omega + \nonumber\\
& + 8 i C j^\tau j_\m \left(\p_\tau \p_\n - g_{\tau\n} \square \right) C \bar \omega \g_5 \g_\rho \p_\s \omega \varepsilon^{\m\n\rho\s} + \nonumber\\
& - 8 i C j^\n \bar \omega \g_\n \p_\m \omega \p^\m \bar \omega \g_5 {\not\!\p} \omega + \nonumber\\
& + 2 i C j^\rho \bar \omega \g_\rho \p_\m \omega \p_\n \bar \omega \g_5 \g^\m \p^\n\omega + \nonumber\\
& + 8 i C j^\m \bar\omega {\not\!\p} \omega \p_\m \bar \omega \g_5 {\not\!\p} \omega + \nonumber\\
& - 2 i C j^\m \bar \omega {\not\!\p} \omega \p_\n \bar \omega \g_5 \g_\m \p^\n \omega + \nonumber\\
& - 2 C j_\m \bar \omega \g_5 \g_\n \p_\rho \omega \p_\tau \bar \omega \g_5 \g_\s \p^\tau \omega \varepsilon^{\m\n\rho\s} + \nonumber\\
& + 8 C j_\m \bar \omega \g_5 \g_\n \p_\rho \omega \p_\s \bar\omega\g_5 {\not\!\p} \omega \varepsilon^{\m\n\rho\s}
+ \nonumber\\
& + 6 i C j_\m \bar \omega \g^{\m\n} \p_\n \omega \p_\rho \bar \omega \g_5 \p^\rho \omega + \nonumber\\
& + 4 i C j_\m \bar\omega\g^{\m\n} \p_\n \omega \p_\rho \bar \omega \g_5 \g^{\rho\s} \p_\s \omega + \nonumber\\
& - 6 i C j_\m \bar \omega \g_5 \g^{\m\n} \p_\n \omega \p_\rho \bar \omega \p^\rho \omega + \nonumber\\
& - 4 i C j_\m \bar \omega \g_5 \g^{\m\n} \p_\n \omega \p_\rho \bar \omega \g^{\rho\s} \p_\s \omega + \nonumber\\
& + \bar \omega \omega \left[ j^\m j^\n \p_\m \bar \omega \p_\n \omega - j^2 \p_\m \bar \omega \p^\m \omega + 2 j^\m j_\n \p_\m \bar \omega \g^{\n\rho} \p_\rho \omega - j^2 \p_\m \bar \omega \g^{\m\n} \p_\n \omega \right] + \nonumber\\
& + \bar \omega \g_5 \omega \left[ - j^\m j^\n \p_\m \bar \omega \g_5 \p_\n \omega + j^2 \p_\m \bar \omega \g_5 \p^\m \omega - 2 j^\m j_\n \p_\m \bar \omega \g_5 \g^{\n\rho} \p_\rho \omega + j^2 \p_\m \bar \omega \g_5 \g^{\m\n} \p_\n \omega \right] + \nonumber\\
& + j^\m \bar \omega \g_5 \g_\m \omega \left[ - i \varepsilon^{\n\rho\s\tau} j_\n \p_\rho \bar \omega \g_\s \p_\tau \omega + 2 j^\n \p_\n \bar \omega \g_5 {\not\!\p} \omega - 2 j^\n \p_\rho \bar \omega \g_5 \g_\n \p^\rho \omega \right] + \nonumber\\
& + \bar \omega \g_5 \g_\m \omega \left[ - i \varepsilon^{\m\n\rho\s} j^\tau j_\n \p_\rho \bar \omega \g_\tau \p_\s \omega - j^\n j^\rho \p_\n \bar \omega \g_5 \g^\m \p_\rho \omega + 2 j^\n j^\rho \p_\n \bar \omega \g_5 \g_\rho \p^\m \omega + \right. \nonumber\\
& \left. - 2 j^2 \p^\m \bar \omega \g_5 {\not\!\p} \omega + j^2\p_\n \bar \omega \g_5 \g^\m \p^\n \omega \right]
\,,
\end{align}
\end{subequations}

\begin{align}
	L_{3}
	=&
 +12 C^2 \left(\partial_{\mu}\partial_{\nu}-g_{\mu\nu}\square \right) C
 \left[ 
 j^{\mu}j^{\nu}j_{\tau}\partial_{\rho}\bar\omega\gamma_{\lambda}\partial_{\sigma}\omega\varepsilon^{\tau\rho\sigma\lambda}
 \,+ \right.
 \nonumber\nonumber\\&
 \left.
 -2i\ j^{\mu}j^{\nu}j^{\rho}\partial_{\rho}\bar\omega\gamma_{5}{\not\!\p}\omega
 +2i\ j^{\mu}j^{\nu}j^{\rho}\partial^{\sigma}\bar\omega\gamma_{5}\gamma_{\rho}\partial_{\sigma}\omega
 \,+ \right.
 \nonumber\nonumber\\&
 \left.
 +j^{\mu}j^{\rho}j^{\sigma}\partial_{\alpha}\bar\omega\gamma_{\rho}\partial_{\beta}\omega\varepsilon^{\sigma\nu\alpha\beta}
 +i\ j^{\mu}j^{\rho}j^{\sigma}\partial_{\rho}\bar\omega\gamma_{5}\gamma^{\nu}\partial_{\sigma}\omega
 \,+ \right.
 \nonumber\nonumber\\&
 \left.
 -2i\ j^{\mu}j^{\rho}j^{\sigma}\partial_{\rho}\bar\omega\gamma_{5}\gamma_{\sigma}\partial^{\nu}\omega
 +2i\ \left(j\cdot j\right) j^{\mu}\partial^{\nu}\bar\omega\gamma_{5}{\not\!\p}\omega
 \,+ \right.
 \nonumber\nonumber\\&
 \left.
 -i\ \left(j\cdot j\right)  j^{\mu}\partial^{\rho}\bar\omega\gamma_{5}\gamma^{\nu}\partial_{\rho}\omega
 \right]
+ \nonumber\\&-6 C^2j^{\nu}j^{\mu} \partial_{\mu}\bar\omega\partial_{\nu}\omega  \partial_{\rho}\bar\omega\gamma^{\rho\sigma}\partial_{\sigma}\omega \,   
 + \nonumber\\&-9 C^2j^{\mu}j^{\nu}\partial^{\rho} \bar\omega\partial_{\rho}\omega \partial_{\mu}\bar\omega\partial_{\nu}\omega  \,   
+ \nonumber\\&+9 C^2\left(j\cdot j\right) \partial^{\mu}\bar\omega\partial_{\mu}\omega \partial^{\nu}\bar\omega\partial_{\nu}\omega  \,   
+ \nonumber\\&-18 C^2j^{\nu}j_{\rho} \partial^{\mu}\bar\omega\partial_{\mu}\omega \partial_{\nu}\bar\omega\gamma^{\rho\sigma}\partial_{\sigma}\omega  \,   
+ \nonumber\\&-15 C^2\left(j\cdot j\right) \partial^{\mu}\bar\omega\partial_{\mu}\omega \partial_{\nu}\bar\omega\gamma^{\rho\nu}\partial_{\rho}\omega  \,   
+ \nonumber\\&-6 C^2j^{\mu}j^{\nu} \partial_{\mu}\bar\omega\gamma_{5}\partial_{\nu}\omega  \partial_{\rho}\bar\omega\gamma_{5}\gamma^{\sigma\rho}\partial_{\sigma}\omega \,   
+ \nonumber\\&+9 C^2j^{\mu}j^{\nu} \partial^{\rho}\bar\omega\gamma_{5}\partial_{\rho}\omega \partial_{\nu}\bar\omega\gamma_{5}\partial_{\rho}\omega  \,   
+ \nonumber\\&-9 C^2\left(j\cdot j\right) \partial^{\mu}\bar\omega\gamma_{5}\partial_{\mu}\omega \partial_{\nu}\bar\omega\gamma_{5}\partial_{\nu}\omega  \,   
+ \nonumber\\&+18 C^2j^{\nu}j_{\rho} \partial^{\mu}\bar\omega\gamma_{5}\partial_{\mu}\omega \partial_{\nu}\bar\omega\gamma_{5}\gamma^{\rho\sigma}\partial_{\sigma}\omega  \,   
+ \nonumber\\&-15 C^2\left(j\cdot j\right) \partial^{\mu}\bar\omega\gamma_{5}\partial_{\mu}\omega \partial_{\nu}\bar\omega\gamma_{5}\gamma^{\nu\rho}\partial_{\rho}\omega  \,   
+ \nonumber\\&+3 i C^2  j^{\mu}j_{\nu} \partial_{\alpha}\bar\omega\gamma_{\mu}\partial_{\rho}\omega \partial^{\sigma}\bar\omega\gamma_{5}\gamma_{\lambda}\partial_{\sigma}\omega \varepsilon^{\nu\alpha\rho\lambda}    \,   
+ \nonumber\\&-12 i C^2 j^{\mu}j_{\alpha} \partial_{\nu}\bar\omega\gamma_{\mu}\partial_{\rho}\omega \partial^{\sigma}\bar\omega\gamma_{5}\gamma_{\sigma}\partial_{\lambda}\omega \varepsilon^{\alpha\nu\rho\lambda}    \,   
+ \nonumber\\&-12 i C^2 j_{\mu}j^{\lambda} \partial_{\nu}\bar\omega\gamma_{\rho}\partial_{\sigma}\omega \partial_{\lambda}\bar\omega\gamma_{5}{\not\!\p}\omega  \varepsilon^{\mu\nu\sigma\rho}    \,   
+ \nonumber\\&+3 i C j^{\mu}j_{\nu} \partial_{\rho}\bar\omega\gamma_{\sigma}\partial_{\lambda}\omega \partial^{\alpha}\bar\omega\gamma_{5}\gamma_{\mu}\partial_{\alpha}\omega \varepsilon^{\nu\rho\lambda\sigma}   \,   
+ \nonumber\\&-6 C^2  j^{\mu} j^{\nu}  \partial_{\mu}\bar\omega\gamma_{5}\gamma_{\rho}\partial_{\rho}\omega\partial_{\nu}\bar\omega\gamma_{5}{\not\!\p}\omega \,   
+ \nonumber\\&-6 C^2 j^{\mu}j^{\nu} \partial_{\rho}\bar\omega\gamma_{5}\gamma_{\mu}\partial_{\nu}\omega \partial^{\sigma}\bar\omega\gamma_{5}\gamma_{\sigma}\partial_{\rho}\omega \,   
+ \nonumber\\&+3 C^2j^{\nu}j^{\rho}  \partial_{\mu}\bar\omega\gamma_{5}\gamma_{\nu}\partial_{\rho}\omega \partial^{\sigma}\bar\omega\gamma_{5}\gamma_{\mu}\partial_{\sigma}\omega \,   
+ \nonumber\\&-18 C^2 j^{\mu}j^{\nu} \partial_{\mu}\bar\omega\gamma_{5}{\not\!\p}\omega \partial_{\nu}\bar\omega\gamma_{5}{\not\!\p}\omega  \,   
+ \nonumber\\&+30 C^2j^{\mu} j^{\nu} \partial_{\mu}\bar\omega\gamma_{5}{\not\!\p}\omega  \partial^{\sigma}\bar\omega\gamma_{5}\gamma_{\nu}\partial_{\sigma}\omega \,   
+ \nonumber\\&-6 C^2j^{\mu}j^{\nu} \partial^{\rho}\bar\omega\gamma_{5}\gamma_{\mu}\partial_{\rho}\omega  \partial^{\sigma}\bar\omega\gamma_{5}\gamma_{\nu}\partial_{\sigma}\omega  \,   
+ \nonumber\\&-3 C^2j^{\mu}j^{\nu}  \partial^{\rho}\bar\omega\gamma_{5}\gamma^{\sigma}\partial_{\rho}\omega \partial_{\nu}\bar\omega\gamma_{5}\gamma_{\sigma}\partial_{\mu}\omega \,   
+ \nonumber\\&+3 C^2j^{\mu}j^{\nu} \partial^{\rho}\bar\omega\gamma_{5}\gamma^{\sigma}\partial_{\rho}\omega \partial_{\nu}\bar\omega\gamma_{5}\gamma_{\mu}\partial_{\sigma}\omega  \,   
+ \nonumber\\&-18 C^2\left(j\cdot j\right)  \partial^{\mu}\bar\omega\gamma_{5}\gamma^{\nu}\partial_{\mu}\omega \partial_{\nu}\bar\omega\gamma_{5}{\not\!\p}\omega \,   
+ \nonumber\\&+3 C^2 \left(j\cdot j\right) \partial^{\mu}\bar\omega\gamma_{5}\gamma^{\nu}\partial_{\mu}\omega \partial^{\rho}\bar\omega\gamma_{5}\gamma_{\nu}\partial_{\rho}\omega \,   
+ \nonumber\\&+12 C^2j^{\mu}j^{\nu}  \partial^{\sigma}\bar\omega\gamma_{5}{\not\!\p}\omega \partial_{\mu}\bar\omega\gamma_{5}\gamma_{\sigma}\partial_{\nu}\omega \,   
+ \nonumber\\&-6 C^2 j^{\mu}j^{\nu} \partial^{\sigma}\bar\omega\gamma_{5}{\not\!\p}\omega \partial_{\nu}\bar\omega\gamma_{5}\gamma_{\mu}\partial_{\sigma}\omega \,   
+ \nonumber\\&+24\left(j\cdot j\right)  C^2 \partial^{\mu}\bar\omega\gamma_{5}{\not\!\p}\omega \partial_{\mu}\bar\omega\gamma_{5}{\not\!\p}\omega \,   
+ \nonumber\\&-12 C^2 j^{\mu}j^{\nu} \partial^{\rho}\bar\omega\gamma_{5}{\not\!\p}\omega \partial_{\nu}\bar\omega\gamma_{5}\gamma_{\mu}\partial_{\rho}\omega  \,   
+ \nonumber\\&-12 C^2j_{\mu}j^{\nu} \partial_{\nu}\bar\omega\gamma^{\mu\rho}\partial_{\rho}\omega  \partial_{\sigma}\bar\omega\gamma^{\sigma\lambda}\partial_{\lambda}\omega \,   
+ \nonumber\\&+6 C^2  \left(j\cdot j\right) \partial_{\mu}\bar\omega\gamma^{\mu\nu}\partial_{\nu}\omega \partial_{\rho}\bar\omega\gamma^{\rho\sigma}\partial_{\sigma}\omega\,   
+ \nonumber\\&+12 C^2  j_{\mu}j^{\nu}\partial_{\nu}\bar\omega\gamma_{5}\gamma^{\mu\rho}\partial_{\rho}\omega \partial_{\sigma}\bar\omega\gamma_{5}\gamma^{\sigma\lambda}\partial_{\lambda}\omega \,   
+ \nonumber\\&-6 C^2\left(j\cdot j\right)  \partial_{\mu}\bar\omega\gamma_{5}\gamma^{\mu\nu}\partial_{\nu}\omega \partial_{\rho}\bar\omega\gamma_{5}\gamma^{\rho\sigma}\partial_{\sigma}\omega \,   
+ \nonumber\\&-4 C j^{\mu}j^{\nu} j_{\rho}\bar\omega\gamma_{\mu}\partial_{\nu}\omega  \partial_{\sigma}\bar\omega\gamma_{\beta}\partial_{\alpha}\omega \varepsilon^{\rho\sigma\alpha\beta} \,  \,   
+ \nonumber\\&+8 i C j^{\mu}j^{\nu} j^{\rho} \bar\omega\gamma_{\mu}\partial_{\nu}\omega  \partial_{\rho}\bar\omega\gamma_{5}{\not\!\p}\omega   \,  \,   
+ \nonumber\\&-8 i C j^{\mu}j^{\nu}j^{\rho}  \bar\omega\gamma_{\mu}\partial_{\nu}\omega  \partial^{\sigma}\bar\omega\gamma_{5}\gamma_{\rho}\partial_{\sigma}\omega    \,  \,   
+ \nonumber\\&-4 C j^{\mu}j^{\nu}j^{\alpha}  \bar\omega\gamma_{\mu}\partial_{\rho}\omega  \partial_{\sigma}\bar\omega\gamma_{\alpha}\partial_{\beta}\omega  \varepsilon^{\nu\rho\sigma\beta}\,  \,   
+ \nonumber\\&-4 i C j^{\mu}j^{\rho}j^{\nu} \bar\omega\gamma_{\mu}\partial^{\sigma}\omega  \partial_{\nu}\bar\omega\gamma_{5}\gamma_{\sigma}\partial_{\rho}\omega     \,  \,   
+ \nonumber\\&+8 i C j^{\mu}  j^{\rho}j^{\nu}  \bar\omega\gamma_{\mu}\partial^{\sigma}\omega \partial_{\nu}\bar\omega\gamma_{5}\gamma_{\rho}\partial_{\sigma}\omega   \,  \,   
+ \nonumber\\&-8 i C \left(j\cdot j\right) j^{\mu} \bar\omega\gamma_{\mu}\partial^{\nu}\omega  \partial_{\nu}\bar\omega\gamma_{5}{\not\!\p}\omega    \,  \,   
+ \nonumber\\&+4 i C \left(j\cdot j\right) j^{\mu}  \bar\omega\gamma_{\mu}\partial^{\nu}\omega \partial^{\rho}\bar\omega\gamma_{5}\gamma_{\nu}\partial_{\rho}\omega    \,  \,   
+ \nonumber\\&+4 C \left(j\cdot j\right) j_{\mu} \bar\omega{\not\!\p}\omega \partial_{\nu}\bar\omega\gamma_{\rho}\partial_{\lambda}\omega \varepsilon^{\mu\nu\lambda\rho} \,  \,   
+ \nonumber\\&-4 i C j^{\mu}j^{\nu}j^{\rho}  \bar\omega{\not\!\p}\omega \partial_{\rho}\bar\omega\gamma_{5}\gamma_{\mu}\partial_{\nu}\omega    \,  \,   
+ \nonumber\\&+4 i C \left(j\cdot j\right) j^{\mu} \bar\omega{\not\!\p}\omega \partial^{\rho}\bar\omega\gamma_{5}\gamma_{\mu}\partial_{\rho}\omega     \,  \,   
+ \nonumber\\&+8 i C  j^{\mu}  j^{\nu}j^{\rho}  \bar\omega\gamma_{5}\gamma^{\sigma}\partial_{\mu}\omega \partial_{\rho}\bar\omega\gamma_{\nu}\partial_{\sigma}\omega    \,  \,   
+ \nonumber\\&-8 i C  j^{\mu} j^{\nu}j^{\rho}  \bar\omega\gamma_{5}\gamma_{\mu}\partial^{\sigma}\omega \partial_{\rho}\bar\omega\gamma_{\nu}\partial_{\sigma}\omega    \,  \,   
+ \nonumber\\&-8 i C \left(j\cdot j\right) j^{\mu} \bar\omega\gamma_{5}\gamma_{\nu}\partial^{\rho}\omega \partial_{\rho}\bar\omega\gamma_{\mu}\partial_{\nu}\omega    \,  \,   
+ \nonumber\\&-4 C j^{\mu}j^{\nu} j_{\rho} \bar\omega\gamma_{5}\gamma_{\sigma}\partial_{\lambda}\omega \partial_{\nu}\bar\omega\gamma_{5}\gamma_{\alpha}\partial_{\mu}\omega  \varepsilon^{\rho\lambda\sigma\alpha}\,  \,   
+ \nonumber\\&+8 C j^{\mu}j^{\nu}  j_{\rho} \bar\omega\gamma_{5}\gamma_{\sigma}\partial_{\lambda}\omega \partial_{\nu}\bar\omega\gamma_{5}\gamma_{\mu}\partial_{\alpha}\omega \varepsilon^{\rho\lambda\sigma\alpha}\,  \,   
+ \nonumber\\&+4 C \left(j\cdot j\right) j_{\mu}\bar\omega\gamma_{5}\gamma_{\nu}\partial_{\rho}\omega \partial^{\sigma}\bar\omega\gamma_{5}\gamma_{\alpha}\partial_{\sigma}\omega \varepsilon^{\mu\rho\nu\alpha}  \,  \,   
+ \nonumber\\&-8 C\left(j\cdot j\right)j_{\mu} \bar\omega\gamma_{5}\gamma_{\nu}\partial_{\rho}\omega \partial_{\alpha}\bar\omega\gamma_{5}{\not\!\p}\omega \varepsilon^{\mu\rho\nu\alpha}  \,  \,   
+ \nonumber\\&-4 i C j_{\mu}j^{\nu}j^{\rho} \bar\omega\gamma^{\mu\sigma}\partial_{\sigma}\omega  \partial_{\rho}\bar\omega\gamma_{5}\partial_{\nu}\omega    \,  \,   
+ \nonumber\\&+4 i C \left(j\cdot j\right) j_{\mu}  \bar\omega\gamma^{\mu\nu}\partial_{\nu}\omega \partial^{\rho}\bar\omega\gamma_{5}\partial_{\rho}\omega     \,  \,   
+ \nonumber\\&-8 i C j_{\mu}  j_{\nu}j^{\rho} \bar\omega\gamma^{\mu\sigma}\partial_{\sigma}\omega \partial_{\rho}\bar\omega\gamma_{5}\gamma^{\nu\alpha}\partial_{\alpha}\omega     \,  \,  
+ \nonumber\\&+4 i C  \left(j\cdot j\right) j_{\mu}  \bar\omega\gamma^{\mu\nu}\partial_{\nu}\omega \partial_{\rho}\bar\omega\gamma_{5}\gamma^{\rho\sigma}\partial_{\sigma}\omega    \,  \,   
+ \nonumber\\&+4 i C j_{\mu}j^{\nu}j^{\rho}   \bar\omega\gamma_{5}\gamma^{\mu\sigma}\partial_{\sigma}\omega  \partial_{\nu}\bar\omega\partial_{\rho}\omega   \,  \,   
+ \nonumber\\&-4 i C \left(j\cdot j\right) j^{\mu}  \bar\omega\gamma_{5}\gamma_{\mu\nu}\partial_{\nu}\omega \partial^{\rho}\bar\omega\partial_{\rho}\omega    \,  \,   
+ \nonumber\\&+8 i C j_{\mu} j_{\nu}j^{\rho} \bar\omega\gamma_{5}\gamma^{\mu\sigma}\partial_{\sigma}\omega  \partial_{\rho}\bar\omega\gamma^{\nu\alpha}\partial_{\alpha}\omega     \,  \,   
+ \nonumber\\&-4 i C\left(j\cdot j\right) j_{\mu} \bar\omega\gamma_{5}\gamma^{\mu\nu}\partial_{\nu}\omega \partial_{\rho}\bar\omega\gamma^{\rho\sigma}\partial_{\sigma}\omega    \,,
\label{23Junefour}
\end{align}

\begin{align}
L_{4}
=
 &+4 C^2 j^{\mu}j^{\nu}j^{\rho}j^{\sigma} \partial_{\mu}\bar\omega\partial_{\nu}\omega \, \partial_{\rho}\bar\omega\partial_{\sigma}\omega  \,   
+ \nonumber\\&+16 C^2 j^{\mu}j^{\nu} j_{\rho}j^{\sigma} \partial_{\mu}\bar\omega\partial_{\nu}\omega  \partial_{\sigma}\bar\omega\gamma^{\rho\lambda}\partial_{\lambda}\omega \,   
+ \nonumber\\&-8 C^2  \left(j\cdot j\right)j^{\mu}j^{\nu}\partial_{\mu}\bar\omega\partial_{\nu}\omega  \partial_{\rho}\bar\omega\gamma^{\rho\sigma}\partial_{\sigma}\omega \,   
+ \nonumber\\&-8 C^2\left(j\cdot j\right) j^{\mu}j^{\nu}  \partial^{\rho}\bar\omega\partial_{\rho}\omega \partial_{\mu}\bar\omega\partial_{\nu}\omega  \,   
+ \nonumber\\&+4 C^2  \left(j\cdot j\right)^2  \partial^{\mu}\bar\omega\partial_{\mu}\omega \partial^{\nu}\bar\omega\partial_{\nu}\omega\,   
+ \nonumber\\&-16 C^2  \left(j\cdot j\right) j_{\mu}j^{\nu}\partial^{\rho}\bar\omega\partial_{\rho}\omega \partial_{\nu}\bar\omega\gamma^{\mu\sigma}\partial_{\sigma}\omega \,   
+ \nonumber\\&+8 C^2  \left(j\cdot j\right)^2\partial^{\mu}\bar\omega\partial_{\mu}\omega \partial_{\rho}\bar\omega\gamma^{\rho\sigma}\partial_{\sigma}\omega \,   
+ \nonumber\\&-4 C^2 j^{\mu}j^{\nu} j^{\rho}j^{\sigma} \partial_{\mu}\bar\omega\gamma_{5}\partial_{\nu}\omega  \partial_{\rho}\bar\omega\gamma_{5}\partial_{\sigma}\omega  \,   
+ \nonumber\\&-16 C^2 j^{\mu}j^{\nu}j_{\rho}j^{\sigma} \partial_{\mu}\bar\omega\gamma_{5}\partial_{\nu}\omega  \partial_{\sigma}\bar\omega\gamma_{5}\gamma^{\rho\lambda}\partial_{\lambda}\omega  \,   
+ \nonumber\\&+8 C^2 \left(j\cdot j\right)  j^{\mu}j^{\nu}\partial_{\mu}\bar\omega\gamma_{5}\partial_{\nu}\omega \partial_{\rho}\bar\omega\gamma_{5}\gamma^{\rho\sigma}\partial_{\sigma}\omega \, 
+ \nonumber\\&+8 C^2 \left(j\cdot j\right) j^{\mu}j^{\nu}\partial^{\rho}\bar\omega\gamma_{5}\partial_{\rho}\omega \partial_{\mu}\bar\omega\gamma_{5}\partial_{\nu}\omega  \,   
+ \nonumber\\&-4 C^2 \left(j\cdot j\right)^2 \partial^{\mu}\bar\omega\gamma_{5}\partial_{\mu}\omega \partial^{\nu}\bar\omega\gamma_{5}\partial_{\nu}\omega \,   
+ \nonumber\\&+16 C^2 \left(j\cdot j\right) j_{\mu}j^{\nu} \partial^{\rho}\bar\omega\gamma_{5}\partial_{\rho}\omega \partial_{\nu}\bar\omega\gamma_{5}\gamma^{\mu\sigma}\partial_{\sigma}\omega \,  
+ \nonumber\\&-8 C^2 \left(j\cdot j\right)^2 \partial^{\mu}\bar\omega\gamma_{5}\partial_{\mu}\omega \partial_{\nu}\bar\omega\gamma_{5}\gamma^{\nu\rho}\partial_{\rho}\omega  \,   
+ \nonumber\\&-4 C^2 \left(j\cdot j\right)j^{\mu}j^{\nu}\partial_{\mu}\bar\omega\gamma^{\rho}\partial^{\sigma}\omega  \partial_{\nu}\bar\omega\gamma_{\rho}\partial_{\sigma}\omega   \, 
+ \nonumber\\&-4 C^2 j^{\mu}j^{\nu}j^{\rho}j^{\sigma}\partial_{\mu}\bar\omega\gamma_{\nu}\partial^{\lambda}\omega  \partial_{\rho}\bar\omega\gamma_{\sigma}\partial_{\lambda}\omega  \,   
+ \nonumber\\&+4 C^2 \left(j\cdot j\right) j^{\mu} j^{\nu} \partial_{\mu}\bar\omega\gamma^{\rho}\partial^{\sigma}\omega \partial_{\nu}\bar\omega\gamma_{\sigma}\partial_{\rho}\omega \,   
+ \nonumber\\&+12 C^2  j^{\mu}j^{\nu}j^{\rho}j^{\sigma} \partial^{\lambda}\bar\omega\gamma_{\mu}\partial_{\nu}\omega \partial_{\rho}\bar\omega\gamma_{\sigma}\partial_{\lambda}\omega  \,   
+ \nonumber\\&+12 C^2 \left(j\cdot j\right) j^{\mu}  j^{\nu}  \partial^{\rho}\bar\omega\gamma^{\sigma}\partial_{\mu}\omega\partial_{\nu}\bar\omega\gamma_{\sigma}\partial_{\rho}\omega \,   
+ \nonumber\\&-12 C^2 \left(j\cdot j\right) j^{\mu}j^{\nu}   \partial^{\rho}\bar\omega\gamma^{\sigma}\partial_{\mu}\omega  \partial_{\nu}\bar\omega\gamma_{\rho}\partial_{\sigma}\omega \,   
+ \nonumber\\&-24 C^2 \left(j\cdot j\right) j^{\mu}j^{\nu}\partial^{\sigma}\bar\omega\gamma^{\rho}\partial_{\mu}\omega  \partial_{\rho}\bar\omega\gamma_{\nu}\partial_{\sigma}\omega  \,   
+ \nonumber\\&+8 C^2 \left(j\cdot j\right)j^{\mu}j^{\nu}  \partial^{\rho}\bar\omega\gamma_{\mu}\partial_{\sigma}\omega \partial_{\rho}\bar\omega\gamma^{\sigma}\partial_{\nu}\omega  \,   
+ \nonumber\\&-8 i C^2 j^{\mu}  j^{\nu}j^{\rho} j_{\sigma} \partial_{\alpha}\bar\omega\gamma_{\mu}\partial_{\beta}\omega \partial_{\rho}\bar\omega\gamma_{5}\gamma_{\lambda}\partial_{\nu}\omega\varepsilon^{\sigma\alpha\beta\lambda}    \,   
+ \nonumber\\&+16 i C^2 j^{\mu} j^{\nu}j^{\rho} j_{\sigma}  \partial_{\lambda}\bar\omega\gamma_{\mu}\partial_{\alpha}\omega  \partial_{\rho}\bar\omega\gamma_{5}\gamma_{\nu}\partial_{\beta}\omega\varepsilon^{\sigma\lambda\alpha\beta}   \,   
+ \nonumber\\&+8  iC^2 \left(j\cdot j\right)j^{\mu} j^{\nu}\partial_{\rho}\bar\omega\gamma_{\mu}\partial_{\sigma}\omega  \partial^{\lambda}\bar\omega\gamma_{5}\gamma_{\alpha}\partial_{\lambda}\omega \varepsilon^{\nu\rho\sigma\alpha}     \,   
+ \nonumber\\&-16 i C^2  \left(j\cdot j\right)  j^{\mu} j^{\nu}  \partial_{\rho}\bar\omega\gamma_{\mu}\partial_{\sigma}\omega\partial_{\alpha}\bar\omega\gamma_{5}{\not\!\p}\omega \varepsilon^{\nu\rho\sigma\alpha}    \,   
+ \nonumber\\&+8 C^2  \left(j\cdot j\right)^2  \partial^{\mu}\bar\omega\gamma^{\nu}\partial^{\rho}\omega \partial_{\mu}\bar\omega\gamma_{\nu}\partial_{\rho}\omega\,   
+ \nonumber\\&-16 C^2 \left(j\cdot j\right)^2  \partial^{\mu}\bar\omega\gamma^{\nu}\partial^{\rho}\omega \partial_{\mu}\bar\omega\gamma_{\rho}\partial_{\nu}\omega \,   
+ \nonumber\\&+8i C^2   j^{\mu}j^{\nu}j^{\rho}j^{\sigma} \partial_{\lambda}\bar\omega\gamma_{\alpha}\partial_{\beta}\omega \partial_{\rho}\bar\omega\gamma_{5}\gamma_{\mu}\partial_{\nu}\omega \varepsilon^{\sigma\lambda\beta\alpha}   \,   
+ \nonumber\\&-8i C^2  \left(j\cdot j\right)  j^{\mu} j^{\nu}\partial_{\rho}\bar\omega\gamma_{\sigma}\partial_{\lambda}\omega \partial^{\alpha}\bar\omega\gamma_{5}\gamma_{\mu}\partial_{\alpha}\omega \varepsilon^{\nu\rho\lambda\sigma}   \,   
+ \nonumber\\&+4 C^2  j^{\mu}j^{\nu} j^{\rho}j^{\sigma}\partial_{\nu}\bar\omega\gamma_{5}\gamma^{\lambda}\partial_{\mu}\omega \partial_{\sigma}\bar\omega\gamma_{5}\gamma_{\lambda}\partial_{\rho}\omega \,   
+ \nonumber\\&-8 C^2  j^{\mu}j^{\nu} j^{\rho}j^{\sigma} \partial_{\nu}\bar\omega\gamma_{5}\gamma_{\mu}\partial^{\lambda}\omega \partial_{\sigma}\bar\omega\gamma_{5}\gamma_{\lambda}\partial_{\rho}\omega\,   
+ \nonumber\\&+4 C^2 j^{\mu}j^{\nu} j^{\rho}j^{\sigma}\partial_{\nu}\bar\omega\gamma_{5}\gamma_{\mu}\partial^{\lambda}\omega  \partial_{\sigma}\bar\omega\gamma_{5}\gamma_{\rho}\partial_{\lambda}\omega \,   
+ \nonumber\\&+8 C^2j^{\mu}  j^{\nu}j^{\rho}j^{\sigma}\partial_{\mu}\bar\omega\gamma_{5}{\not\!\p}\omega  \partial_{\sigma}\bar\omega\gamma_{5}\gamma_{\nu}\partial_{\rho}\omega \,   
+ \nonumber\\&-4 C^2 \left(j\cdot j\right) j^{\mu}j^{\nu} \partial_{\mu}\bar\omega\gamma_{5}{\not\!\p}\omega  \partial_{\nu}\bar\omega\gamma_{5}{\not\!\p}\omega   \,   
+ \nonumber\\&-8 C^2 j^{\mu}j^{\nu} j^{\rho}j^{\sigma} \partial^{\lambda}\bar\omega\gamma_{5}\gamma_{\mu}\partial_{\nu}\omega \partial_{\sigma}\bar\omega\gamma_{5}\gamma_{\lambda}\partial_{\rho}\omega \,   
+ \nonumber\\&+12 C^2  j^{\mu}j^{\nu}j^{\rho}j^{\sigma}\partial^{\lambda}\bar\omega\gamma_{5}\gamma_{\mu}\partial_{\nu}\omega \partial_{\sigma}\bar\omega\gamma_{5}\gamma_{\rho}\partial_{\lambda}\omega  \,   
+ \nonumber\\&-8 C^2 \left(j\cdot j\right) j^{\mu}j^{\nu} \partial^{\rho}\bar\omega\gamma_{5}\gamma_{\mu}\partial_{\nu}\omega \partial_{\rho}\bar\omega\gamma_{5}{\not\!\p}\omega \,   
+ \nonumber\\&+8 C^2\left(j\cdot j\right)  j^{\mu}j^{\nu} \partial^{\rho}\bar\omega\gamma_{5}\gamma_{\mu}\partial_{\nu}\omega\partial^{\sigma}\bar\omega\gamma_{5}\gamma_{\rho}\partial_{\sigma}\omega  \,   
+ \nonumber\\&+8 C^2 j^{\mu} j^{\nu}j^{\rho}j^{\sigma}\partial_{\mu}\bar\omega\gamma_{5}{\not\!\p}\omega \partial_{\sigma}\bar\omega\gamma_{5}\gamma_{\nu}\partial_{\rho}\omega  \,   
+ \nonumber\\&-12 C^2\left(j\cdot j\right)j^{\mu}j^{\nu} \partial_{\mu}\bar\omega\gamma_{5}{\not\!\p}\omega  \partial_{\nu}\bar\omega\gamma_{5}{\not\!\p}\omega   \,   
+ \nonumber\\&+8 C^2 \left(j\cdot j\right)j^{\mu} j^{\nu}  \partial_{\mu}\bar\omega\gamma_{5}{\not\!\p}\omega \partial_{\sigma}\bar\omega\gamma_{5}\gamma_{\nu}\partial_{\sigma}\omega \,   
+ \nonumber\\&-16 C^2j^{\mu}j^{\nu}j^{\rho}j^{\sigma} \partial^{\lambda}\bar\omega\gamma_{5}\gamma_{\mu}\partial_{\lambda}\omega  \partial_{\sigma}\bar\omega\gamma_{5}\gamma_{\nu}\partial_{\rho}\omega  \,   
+ \nonumber\\&+8 C^2\left(j\cdot j\right)j^{\mu}j^{\nu}  \partial^{\rho}\bar\omega\gamma_{5}\gamma_{\mu}\partial_{\rho}\omega  \partial_{\nu}\bar\omega\gamma_{5}{\not\!\p}\omega  \,   
+ \nonumber\\&-8 C^2\left(j\cdot j\right) j^{\mu}j^{\nu}  \partial_{\mu}\bar\omega\gamma_{5}\gamma^{\rho}\partial_{\nu}\omega \partial^{\sigma}\bar\omega\gamma_{5}\gamma^{\rho}\partial_{\sigma}\omega \,   
+ \nonumber\\&+8 C^2\left(j\cdot j\right)j^{\mu}j^{\nu}   \partial^{\rho}\bar\omega\gamma_{5}\gamma^{\sigma}\partial_{\rho}\omega \partial_{\nu}\bar\omega\gamma_{5}\gamma_{\mu}\partial_{\sigma}\omega \,   
+ \nonumber\\&-16 C^2 \left(j\cdot j\right)^2\partial^{\mu}\bar\omega\gamma_{5}\gamma^{\nu}\partial_{\mu}\omega \partial_{\nu}\bar\omega\gamma_{5}{\not\!\p}\omega  \,   
+ \nonumber\\&+4 C^2 \left(j\cdot j\right)^2 \partial^{\mu}\bar\omega\gamma_{5}\gamma^{\nu}\partial_{\mu}\omega \partial^{\rho}\bar\omega\gamma_{5}\gamma_{\nu}\partial_{\rho}\omega  \,   
+ \nonumber\\&+16 C^2 \left(j\cdot j\right)j^{\mu}j^{\nu} \partial^{\sigma}\bar\omega\gamma_{5}{\not\!\p}\omega \partial_{\nu}\bar\omega\gamma_{5}\gamma_{\sigma}\partial_{\mu}\omega  \,   
+ \nonumber\\&-24 C^2\left(j\cdot j\right)  j^{\mu}j^{\nu} \partial^{\sigma}\bar\omega\gamma_{5}{\not\!\p}\omega \partial_{\nu}\bar\omega\gamma_{5}\gamma_{\mu}\partial_{\sigma}\omega \,   
+ \nonumber\\&+16 C^2 \left(j\cdot j\right)^2\partial^{\nu}\bar\omega\gamma_{5}{\not\!\p}\omega \partial_{\nu}\bar\omega\gamma_{5}{\not\!\p}\omega  \,   
+ \nonumber\\&+16 C^2j_{\mu}j^{\nu} j_{\rho}j^{\sigma} \partial_{\nu}\bar\omega\gamma^{\mu\lambda}\partial_{\lambda}\omega \partial_{\sigma}\bar\omega\gamma^{\rho\alpha}\partial_{\alpha}\omega  \,   
+ \nonumber\\&-16 C^2 \left(j\cdot j\right) j_{\mu}j^{\nu} \partial_{\nu}\bar\omega\gamma^{\mu\rho}\partial_{\rho}\omega  \partial_{\sigma}\bar\omega\gamma^{\sigma\lambda}\partial_{\lambda}\omega \,   
+ \nonumber\\&+4 C^2 \left(j\cdot j\right)^2 \partial_{\mu}\bar\omega\gamma^{\mu\nu}\partial_{\nu}\omega \partial_{\rho}\bar\omega\gamma^{\rho\sigma}\partial_{\sigma}\omega  \,   
+ \nonumber\\&-16 C^2 j_{\mu}j^{\nu}j_{\rho}j^{\sigma} \partial_{\nu}\bar\omega\gamma_{5}\gamma^{\mu\lambda}\partial_{\lambda}\omega  \partial_{\sigma}\bar\omega\gamma_{5}\gamma^{\rho\alpha}\partial_{\alpha}\omega  \,   
+ \nonumber\\&+16 C^2\left(j\cdot j\right)j^{\mu}j^{\nu} \partial_{\nu}\bar\omega\gamma_{5}\gamma^{\mu\rho}\partial_{\rho}\omega  \partial_{\sigma}\bar\omega\gamma_{5}\gamma^{\sigma\lambda}\partial_{\lambda}\omega  \,   
+ \nonumber\\&-4 C^2\left(j\cdot j\right)^2 \partial_{\mu}\bar\omega\gamma_{5}\gamma^{\mu\nu}\partial_{\nu}\omega \partial_{\rho}\bar\omega\gamma_{5}\gamma^{\rho\sigma}\partial_{\sigma}\omega  \,.
\label{23Junefive}
\end{align}

\vfill \eject


%
\chapter*{Acknowledgements}

I would like to thank my Ph.D. Advisor Prof.~Pietro Antonio Grassi for his support, his help and his suggestions in all these years.
I am truly grateful to Dr.~Luca Sommovigo for his kindness and patience.

A special thanks goes to Dr.~Lorenzo Giulio Celso Gentile, true friend before collaborator.

In writing my Thesis, a fundamental contribution is due to Dr.~Daniele Musso for the eighty--nine points in his vengeance list.

To Giulia a robust handshake, as a true man does.
She is my water.

I will be always in debt with my parents. Thank you for everything. Pippo deserves some thanks also.

To my Openspace fellows, Ana, Andrea, Andrea, Barbuto, Biolati, Camera, Diogo, Elisa Voera, Falcion, Gabonzo, Luca, Mattia, Marcogalli, Nic, PanciGrazia, Paolo, Roberto, Rosco, Roverto, Stefano XXVIb, Vadacchino, Yari Pochissimi O.
Thank you for these years of Actione, Velocit\`a, Suspenso, Terrore and in particular Romanza (as we were in Pampa).
With you I really feel I am in cinema di summer di 1964.

A great thank you to all the other Ph.D. students I met in these years and especially to Valentina, Agnese, Antonio, Dario Pink, Fabio, Marco, Natalia. I wish you the best for your careers.

Of course I can not forget to thank my friends Ale, il Don, Mauro, Nata and Rafino. 

Finally, I would like to express my gratitude to, in random order: Franco Franchetti and Leombruno Tosca, Elisa, The Cantina Band, Alza in alto la mano, Mr.~Dejavu, Irene, Yuri and Elisa, Frank Petriello, Deboroh, Mariana, Fabio, Utini, Falciona,  Bar, Chtulhu, Vlad Tepes PhD, Victoria, Goblin, Parthenope, Silvia, Bagni Derivativi, MarcogallInes, Claudia, Elisa (in America), all the computers I converted into stoves, Cettina, Mr.~Dejavu, Arnold PhD, Ivo, Gabriel, Olga, Antoine, Axel.


\newpage
\hfill
\thispagestyle{empty}



\end{document}